# From Co-Design to Metacognitive Laziness: Evaluating Generative AI in Vocational Education

**Amir Yunus, Gay Peng Rend, Lee Oon Teng**
Nanyang Technological University
Singapore 637718

## ABSTRACT

This study examines the development and deployment of a Generative AI proof-of-concept (POC) designed to support lecturers in a vocational education setting in Singapore. Using a user-centred, mixed-methods design process, we co-developed an AI chatbot with the lecturers to address recurring instructional challenges during exam preparation periods, including managing repeated questions and improving the efficiency of feedback delivery. The POC fulfilled its intended purpose: lecturers reported streamlined workflows, reduced cognitive load, and improved student confidence in navigating course content.

Unexpectedly, the deployment also surfaced critical insights into student learning behaviours. Despite enhanced teaching processes, performance data showed no significant improvement in overall student outcomes. Further analysis revealed patterns such as self-efficacy-driven dependency on AI, metacognitive offloading, divergent strategic versus non-strategic use, and widening performance gaps between high- and low-ability students. These findings suggest that the educational influence of generative AI extends beyond instructional efficiency to shaping cognitive engagement, self-regulation, and learner equity in complex ways.

The study, therefore, raises consequential design and pedagogical questions regarding how AI tools can minimise dependency, strengthen metacognitive development, calibrate support across ability levels, and ensure equitable learning opportunities. These questions, emerging beyond the original POC intent, underscore the need for evidence-based, context-sensitive approaches to AI implementation in education.

Overall, the research demonstrates that while generative AI can substantially enhance teaching processes, achieving meaningful learning gains requires deeper attention to learner behaviour, cognitive processes, and the equitable design of AI-supported learning environments.

### Author Keywords

Generative AI, chatbots, vocational education, personalised learning, learner behaviour, cognitive engagement, self-efficacy, AI-Supported Learning, Singapore.

## 1. INTRODUCTION

Generative Artificial Intelligence (AI) technologies enable personalised, adaptive systems in educational settings, offering potential solutions to diverse pedagogical challenges. Meta-analyses indicate medium-to-high effect sizes on achievement and learning performance, suggesting that AI-powered educational tools can enhance student engagement, improve learning outcomes, and foster critical thinking skills (Labadze et al., 2023; R. Wu & Yu, 2024).

This study examines the adoption of Generative AI in educational settings at a vocational education institute in Singapore. Singapore's vocational education sector comprises five polytechnics and three ITE colleges, serving 89,068 full-time students in 2023 (61,794 in polytechnics and 27,274 in ITE) (Ministry of Education, Singapore, 2024). These institutions prioritise practical skills and hands-on learning over theoretical knowledge, making them an ideal context for exploring AI-enhanced teaching.

The primary demographic includes lecturers with strong industry backgrounds who have completed foundational pedagogical training. Their practice-oriented expertise and varying levels of formal teaching experience influence their instructional design needs and adoption of AI-supported tools. The secondary demographic includes students enrolled in vocational and technical education programmes, characterised by diverse learning profiles and a curriculum focused on applied, hands-on skills development. This context shapes which learning support tools prove most effective.

Inclusive education is a central focus of this research. Approximately 7,000 students (roughly 8%) in Singapore's vocational education system have Special Educational Needs (SEN), including, but not limited to, Autism Spectrum Disorder (ASD), Attention Deficit Hyperactivity Disorder (ADHD), and Dyslexia (Ministry of Education, Singapore, 2025). Generative AI tools have the potential to address diverse learner needs and support equitable access to educational resources.

The primary goal of this study was to develop a Generative AI proof-of-concept (POC) with lecturers, for lecturers, using a user-centred design methodology. Recognising the flexibility of Generative AI applications, we did not predefine the POC format. Pre-implementation surveys assessed lecturers' baseline knowledge and perceptions of AI in education. Through co-development workshops, lecturers and researchers collaboratively identified the most impactful solution for their context: an AI chatbot, subsequently named StudyBuddy. Lecturers deployed and tested the chatbot with students during exam preparation periods. Post-implementation surveys evaluated changes in lecturers' experiences, and analyses of chat logs and student exam performance provided insights into tool usage and learning outcomes.

The study successfully achieved its primary objective. Lecturers reported a reduction in the repetition of questions, improved student confidence, and enhanced teaching efficiency. Students who used the chatbot achieved performance that matched or exceeded that of high-performing baseline cohorts. While student performance was not the initial focus, these observations highlight important questions for future research on the relationship between improved teaching experiences and learning outcomes.

This research contributes to understanding how Generative AI tools function across diverse educational contexts and student populations. While existing literature demonstrates the positive effects of AI in education, our findings highlight the distinctive characteristics of vocational education, the potential for AI dependency among students with varying self-efficacy, and the complex relationship between educator-focused AI interventions and student learning. These insights challenge assumptions of universal AI benefits and emphasise the need for evidence-based, context-sensitive approaches to educational AI implementation.

## 2. LITERATURE REVIEW

This literature review examines the integration of Generative AI in educational settings, with a focus on AI-powered technologies that address pedagogical challenges. Higher education institutions have integrated artificial intelligence technologies for personalised learning, administrative automation, and student support. Bahroun et al. (2023) report moderate to strong positive effects of AI-powered educational interventions on academic achievement and student engagement. Pesovski et al. (2024) provide additional evidence of improved learning outcomes and motivation in AI-enhanced educational environments. Maity & Deroy (2024) describe how Generative AI enables personalised responses based on individual learning needs. Katiyar et al. (2024) analyse AI-driven personalised learning systems for enhancing educational effectiveness.

However, AI adoption in educational settings encounters multiple implementation challenges. Chang et al. (2023) identify that many AI educational tools are developed without sufficient input from end-users, resulting in solutions that fail to address real classroom challenges. Akram et al. (2022) identify infrastructure limitations, teacher competence gaps, and insufficient institutional support as primary barriers. Kizilcec (2024) reports that educators express mixed attitudes towards AI technologies, showing optimism about their potential while voicing concerns about trust, transparency, and the need for professional development. Understanding educator perspectives is crucial for advancing the use of AI in education.

The relationship between AI tool use and learning outcomes is complex and multifaceted. Research identifies patterns of AI dependency, with Fan et al. (2025) documenting the "metacognitive laziness phenomenon" where learners become dependent on AI and offload metacognitive load, hindering self-regulation and deep learning. Zhang et al. (2025) found that AI dependency has negative relationships with critical thinking, self-confidence, and problem-solving ability. Bećirović et al. (2025) observed that while students view AI as a form of support, overreliance on it may negatively impact their performance.

Singapore's inclusive education policies have increased the diversity of student populations in higher education institutions, including students with SEN. Beamish et al. (2024) document the challenges this creates for educators who must balance multiple learning needs simultaneously. Niemi & Vehkakoski (2024) describe difficulties with traditional classroom dynamics, including non-verbal communication challenges and hyperfixation behaviours. Johnson et al. (2022) identify multiple interconnected barriers in educator practice, including time and workload constraints, resource limitations, and difficulties in providing timely, personalised feedback to diverse student populations. Perdana & Chu (2023) identify additional pressures in Singapore's higher education institutions from adapting to hybrid learning environments and managing diverse international student populations with varying English proficiency levels. The limitations of traditional support mechanisms magnify these contextual challenges. Zhou & Wolstencroft (2022) describe how office hours and email systems operate within fixed availability constraints. Huda et al. (2018) document the challenges these systems face in addressing the immediate needs of students.

Realising the benefits of Generative AI in education requires collaborative design approaches that prioritise educator involvement throughout the development process. Chang et al. (2023) describe how co-design workshops with educators create AI tools that address pedagogical challenges and incorporate user-centred design principles. Durall Gazulla et al. (2023) analyse co-design processes, finding that involving educators throughout development phases, from initial requirements gathering to prototype testing and refinement, leads to more relevant and trustworthy educational tools. However, they also note that many AI projects still lack comprehensive stakeholder engagement, particularly in later design and evaluation stages. Sillaots et al. (2024) advocate for more systematic user-centred approaches that build both technical competence and design thinking capabilities for achieving educational outcomes. Hu et al. (2020) analyse technology adoption in higher education, finding that educator perspectives and institutional contexts determine success, with perceived usefulness, ease of use, and institutional support as key predictors of adoption. Topali et al. (2025) document how user-centred design in educational technology development benefits from involving educators, students, and other stakeholders throughout all design phases. Cen et al. (2023) provide additional evidence supporting the involvement of stakeholders in creating effective AI-driven educational tools. Holstein et al. (2019) describe the co-design of tools to support teacher-AI complementarity. Renz & Vladova (2021) reinvigorate the discourse on human-centred artificial intelligence in educational technologies.

The theoretical foundation for successful implementation lies in addressing what Tsai & Chai (2012) identify as "third-order barriers," namely, the need for educators to develop design thinking skills to integrate technology effectively into their pedagogical practices. They argue that educators who possess strong design thinking capabilities can overcome contextual challenges and integrate AI technologies into their teaching practice, regardless of environmental constraints. Patel et al. (2024) describe how design thinking is applied to the implementation of educational technology, where educators creatively adapt AI to meet the specific needs of their teaching contexts and students. Cordero et al. (2024) provide frameworks for developing institutional policies regarding the use of AI in educational contexts.

Vocational education emphasises practical, hands-on skills development. Palahan (2025) found that teaching programming requires the effective conveyance of theoretical concepts and their application in coding exercises, with immediate and detailed feedback being critical. Sajja et al. (2025) emphasised the importance of practical, experience-based learning, noting that applying theoretical knowledge to solve practical problems requires hands-on learning beyond the classroom. This emphasis on practical skill development may create a mismatch with AI technologies that are primarily text-based and theoretical in nature.

Literature suggests that Generative AI technologies can improve teaching experiences for lecturers through tools that reduce educator workload, provide round-the-clock student support, and enable personalised feedback, allowing educators to focus on higher-value activities and provide better pedagogical support. However, the pathway from improved teaching experiences to student learning outcomes may be more complex than literature suggests, particularly in vocational education contexts where practical skill development may require different pedagogical approaches. A detailed table of the literature reviewed is provided in Appendix A.

## 3. METHODOLOGY

This study employed a three-phase, sequential, mixed-methods design using user-centred design approaches to explore educators' adoption of Generative AI in classroom settings. The methodology prioritises educator input throughout the development process, focusing on pedagogical requirements rather than technological capabilities.

### 3.1 Phase 1: Educator Research

Phase 1 was conducted with twenty lecturers from the vocational education system. Data collection involved pre-implementation surveys followed by semi-structured interviews with written notes. No audio recordings were made to protect participant privacy.

Pre-implementation surveys established lecturers' baseline perspectives regarding Generative AI technologies in education before POC development discussions. The survey instrument contained three sections covering teaching philosophy and practice (Q1 to Q6), familiarity and expectations of AI tools (Q7 to Q10), and philosophical orientation (Q11 to Q15). Two open-ended questions addressed key qualities and perceived risks. All Likert-scale items used a 5-point scale. We did not presume what form the Generative AI POC would take at this stage.

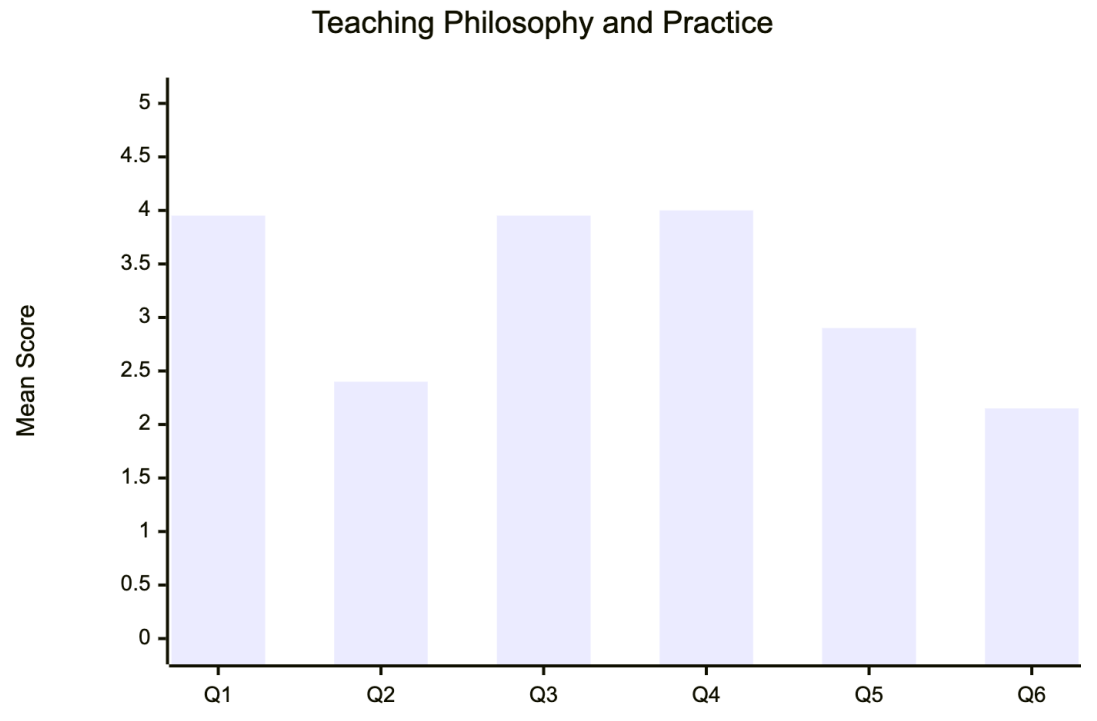

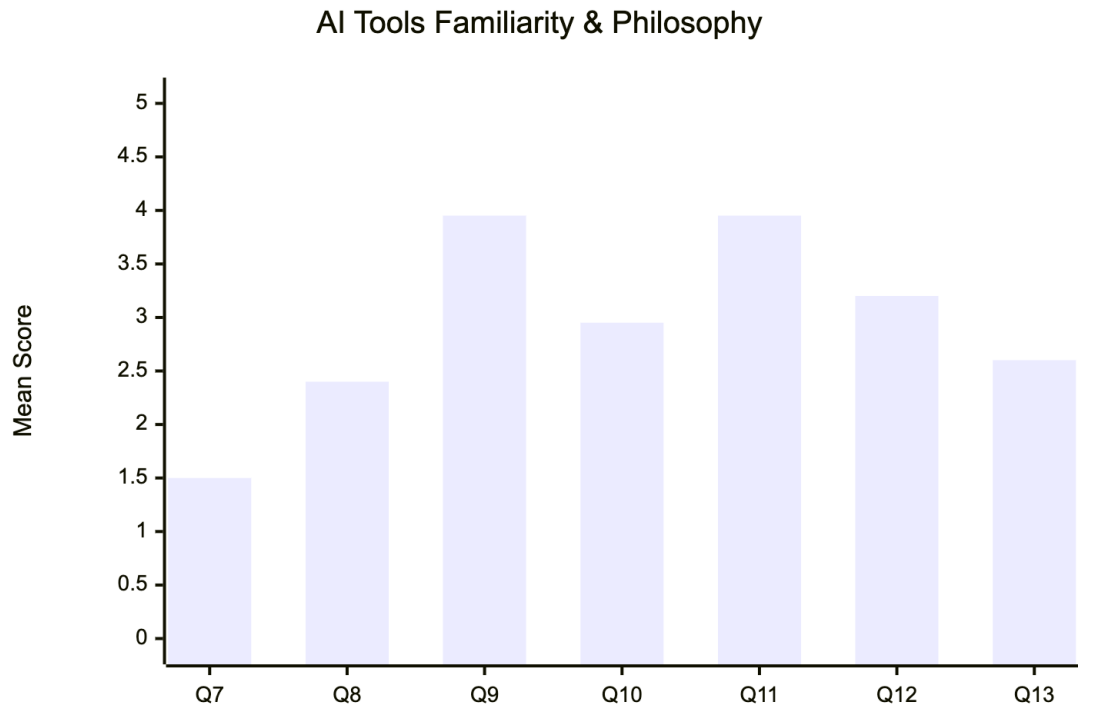

Survey responses indicated lecturers strongly agreed their role involves facilitating inquiry and independent thought (Q1 M = 3.95, range 3 to 5) and that technology augments rather than replaces human judgment (Q4 M = 4.00, range 3 to 5). They expressed lower confidence in students' ability to take ownership of learning without continuous guidance (Q2 M = 2.40, range 1-3). They acknowledged that there was insufficient time and resources for personalised feedback (Q6 M = 2.15, range 1-3), consistent with Johnson et al.'s (2022) identification of time and workload constraints as barriers in educator practice.

Lecturers reported limited prior experience with AI-based educational tools (Q7 M = 1.50, range 1-2) but expressed strong belief that real-time digital tools could support learning meaningfully (Q9 M = 3.95, range 3 to 5). They strongly agreed that students learn best through dialogue rather than content delivery (Q11, M = 3.95, range 3-5). Open-ended responses emphasised accuracy and trustworthiness as essential qualities, while concerns about accuracy and the risks of over-reliance dominated perceived risks.

Semi-structured interviews with the same twenty lecturers explored specific challenges and potential solutions through six main sections covering teaching context, reflection on current tools, support and feedback in learning, imagining new possibilities, values and concerns, and forward-thinking reflections. Each section contained main questions and follow-up control or check questions to explore responses in depth. Written notes captured responses for thematic analysis, with no audio recordings made to protect participant privacy.

Lecturers consistently identified content volume as the primary challenge during exam preparation, with students struggling to manage material across multiple documents, pages, and case studies. Students struggle to grasp underlying principles and apply knowledge to new scenarios beyond memorisation. Time management challenges arose as many students rushed to cram at the last minute. The consensus indicated that Short Answer Questions represent the most challenging component for weaker students, requiring higher-order thinking and articulation skills that many lack. Current support mechanisms face limitations, including time constraints, repeated questions, difficulty individualising at scale, and limited after-hours availability, reflecting the fixed availability constraints that Zhou & Wolstencroft (2022) describe for traditional support systems. Lecturers expected AI tools to provide continuous availability, generate practice questions, offer simplified explanations, and support personalised learning paths. They identified potential benefits for weaker students, including non-judgmental learning spaces, self-paced learning, and confidence-building, while expressing concerns about accuracy, over-reliance, exam fairness, and the impact on critical thinking.

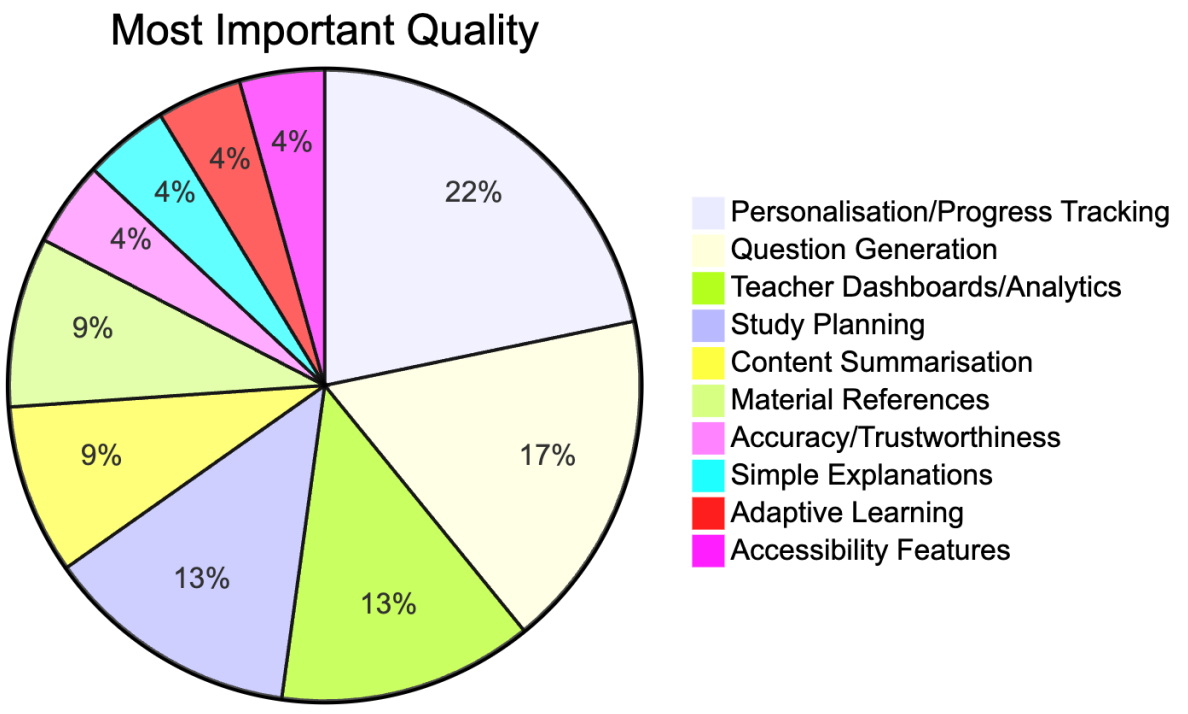

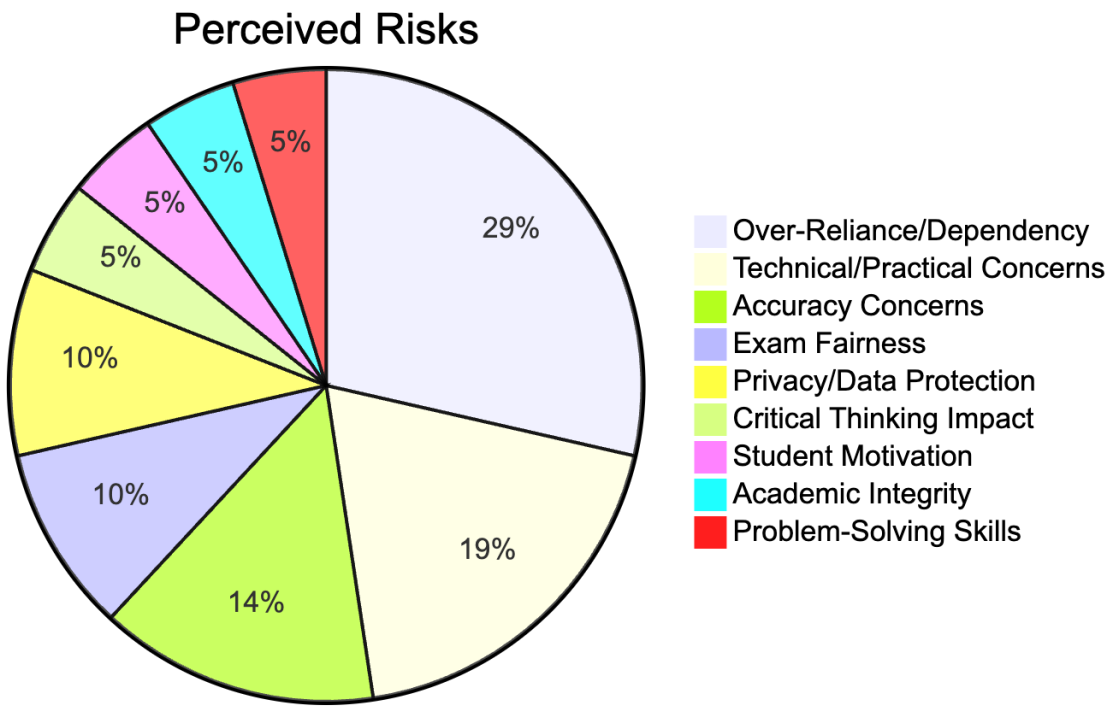

The surveys and interviews established a baseline understanding of lecturers' teaching practices, comfort levels with technology, and perspectives on Generative AI in education. We did not assume what form the Generative AI technology would take. The data captured lecturers' original state before POC development discussions. Emphasis on the practical vocational education context throughout discussions aligns with Palahan's (2025) finding that teaching programming requires the effective conveyance of theoretical concepts and their application in coding exercises, as well as Sajja et al.'s (2025) emphasis on practical, experience-based learning. Tool design must account for diverse student needs and abilities, consistent with Perdana & Chu's (2023) identification of challenges in managing diverse international student populations with varying English proficiency levels.

Detailed survey statistics, comprehensive interview findings, and thematic analysis are presented in Appendix B.

### 3.2 Phase 2: Co-Development

Educators participated in co-development workshop sessions to collaboratively explore what Generative AI technology would have the most impact on their educational context. These workshops, typically lasting 90 minutes with 4-6 educator participants, employed design thinking methodologies to systematically work through problem definition, ideation, and solution specification. Durall Gazulla et al. (2023) demonstrate that collaborative design approaches in educational technology development ensure students' needs are considered and contribute to inclusive, user-centred solutions. Treasure-Jones & Joynes (2018, as cited in Durall Gazulla et al., 2023) argue that co-design should form part of an iterative, agile development approach to support trust-building and balance tensions inherent in collaborative processes.

#### 3.2.1 Design Thinking Workshops

The co-development process followed a structured design thinking framework, incorporating human-computer interaction principles to ensure usability, accessibility, and pedagogical soundness. Through systematic problem analysis, stakeholder mapping, persona development, and iterative ideation, participants collaboratively identified that an AI chatbot would be the most impactful solution. The chatbot emerged from a comprehensive evaluation using criteria of feasibility, impact, user value, and alignment with values, scoring 18/20 points compared to alternative solutions (mobile app: 14/20, enhanced LMS: 11/20).

Labadze et al. (2023) and D.-L. Chen et al. (2025) demonstrate that AI chatbots can enhance learning experiences, improve learning outcomes, and provide on-demand support in educational contexts. Wu & Yu (2024) found that chatbots promote students' academic performance, inspire learning interest, boost learning motivation and engagement, and relieve learning anxiety. However, Labadze et al. (2023) note contradictory findings regarding critical thinking, learning engagement, and motivation, highlighting the importance of careful design and evaluation.

Key outcomes from the design thinking workshops included a refined challenge statement that addressed scalability issues and provided scaffolded support for diverse student ability levels. Participants prioritised requirements including 24/7 availability, multiple interaction modes, and scaffolded explanations. Critical boundaries were established to maintain academic integrity and prevent dependency. Both low-fidelity paper and digital PowerPoint prototypes were tested with design principles, including Don Norman's principles, Shneiderman's golden rules, and Universal Design for Learning.

The emphasis on scaffolded support aligns with Vygotsky's Zone of Proximal Development framework, where AI tools can act as a more knowledgeable other by providing personalised scaffolding, feedback, and resources tailored to learners' needs (Cai et al., 2025). When aligned with ZPD principles, AI personalised learning support provides tailored educational experiences that help learners achieve tasks they cannot complete independently but can accomplish with guidance. However, Zhai et al. (2024) and Ahmed et al. (2024) warn that excessive use of AI tools can diminish students' critical thinking, problem-solving abilities, and capacity for independent thought. These concerns informed the establishment of critical boundaries to maintain academic integrity and prevent dependency.

The refined challenge statement that emerged was to design a Generative AI-powered learning support system that provides personalised, 24/7 assistance to students during exam preparation, addressing content volume challenges and repeated questions through scaffolded support appropriate for diverse student ability levels, while maintaining academic integrity and promoting deep learning, so that lecturers can focus on higher-value teaching activities and all students receive equitable support.

Detailed documentation of the design thinking workshop process, including problem tree analysis, abstraction laddering, stakeholder mapping, user journey mapping, persona development, solution evaluation, requirements gathering, and prototyping activities, is provided in Appendix C.

### 3.2.2 Technical Architecture and Implementation

Technical architecture planning and design document creation followed the workshop sessions. StudyBuddy was built on the Open WebUI framework, selected for its modularity, built-in RAG integration, open-source flexibility, and multi-model support. The system employs a decoupled architecture, separating the web interface (Open WebUI on port 8080) from inference (vLLM server on port 8000), which enables independent scaling, resource isolation, and fault tolerance.

The vLLM framework was selected for high-throughput, low-latency LLM inference, providing 2-4x throughput improvement and approximately 50% memory reduction through PagedAttention optimisation. Domain-specific prompt engineering was implemented, utilising structured personas that support a study companion role, step-by-step teaching approaches, and encouraging, non-judgmental interaction styles aligned with constructivist learning principles. Salih et al. (2025) identify prompt engineering as crucial for efficiently utilising generative AI solutions in educational contexts, with effective prompt building encompassing objective-driven prompts, progressive prompting, and role-specific prompts. Katiyar et al. (2024) note that constructivist approaches view learning as an active, constructive process in which learners build new knowledge based on their prior experiences and existing knowledge. Personalised learning systems can support constructivist learning by providing opportunities to explore, discover, and construct knowledge.

Retrieval-Augmented Generation was implemented with hierarchical knowledge organisation from Subject to Chapter to Document Type and re-ranking to improve retrieval precision. The system includes domain-specific tools, such as an MCQ Bank Tool for generating practice questions, and supports prompt chaining for multi-stage reasoning with JSON schema enforcement. Sajja et al. (2025) demonstrate that RAG frameworks have been successfully deployed in educational contexts, enabling precise document retrieval and effective response generation tailored to user queries. Ismail et al. (2025) note that RAG combines retrieval with generation, leveraging external knowledge bases to overcome the static limitations of standalone generative models, resulting in outputs that are both context-aware and factually grounded. In contrast to fine-tuning, RAG enables more flexible and scalable solutions by allowing organisations to maintain an LLM's core language capabilities while continuously feeding it real-time information without the overhead of retraining the model.

Key technical decisions included tensor parallelism for multi-GPU distribution, floating-point 16 (float16) precision for memory efficiency, and collection-based RAG organisation for granular access. Acknowledged limitations include hardware dependency, requiring re-tuning if the infrastructure changes, prompt brittleness, and scalability constraints for very large datasets using CSV.

Detailed technical specifications, configuration parameters, implementation details, system limitations, and trade-off analyses are provided in Appendix D.

### 3.3 Phase 3: Deployment and Evaluation

During the deployment phase, lecturers deployed and tested StudyBuddy in their classrooms with students during exam preparation periods, with technical support provided as needed. Usage analytics, collected through anonymised chat logs and documented interaction patterns, revealed how lecturers and students utilised the system. Exam performance data comparison between the test group with chatbot access and the control group without chatbot access examined whether any impact on lecturers affected students' exam marks.

Following exam completion, post-implementation surveys were administered to the same 20 lecturers who participated in the pre-implementation surveys. The survey instrument was designed to enable direct comparison with pre-implementation responses. Questions 1-13 mirrored the pre-implementation survey structure, allowing quantitative comparison of changes in lecturers' perspectives. Questions 14-15 were open-ended, exploring shifts in understanding of teaching and ongoing concerns about AI in education.

Semi-structured interviews were also conducted after deployment, following a six-section framework that covered the observed impact on student learning, tool effectiveness and usage, recommendations and improvements, values and educational philosophy, reflection on teaching practice, and forward-thinking reflections. Written notes captured responses for thematic analysis, with no audio recordings made to protect participant privacy.

A comparison of pre-implementation and post-implementation survey scores reveals significant improvements across multiple dimensions. Notable improvements were observed in lecturers' confidence in supporting critical thinking with digital content (Q5: M = 2.90 to 4.20, +1.30) and their capacity to provide personalised feedback (Q6: M = 2.15 to 4.10, +1.95). Lecturers' confidence in students' ability to take ownership of learning showed improvement (Q2: M = 2.40 to 3.20, +0.80), while maintaining strong commitment to facilitating inquiry (Q1: M = 3.95 to 4.30), accommodating learning styles (Q3: M = 3.95 to 4.10), and viewing technology as augmentation (Q4: M = 4.00 to 4.20). Lecturers reported moderate to extensive active use of StudyBuddy (Q7: M = 4.10) and strong agreement that it complemented their teaching style (Q9: M = 4.30). Confidence in evaluating AI responses improved (Q8: M = 2.40 to 3.80, +1.40), while instructional design thinking showed moderate changes (Q10: M = 3.70). Lecturers maintained strong agreement that students learn best through dialogue (Q11: M = 3.95 to 4.20), with increased emphasis on curiosity over efficiency (Q12: M = 3.20 to 4.10, +0.90) and improved comfort with unpredictable outcomes (Q13: M = 2.60 to 3.90, +1.30).

Open-ended responses revealed that lecturers recognised technology's role in handling routine tasks, enabling focus on higher-value activities. Multiple lecturers noted shifts toward data-driven teaching approaches and recognition that personalised learning can be scaled through AI tools. Lecturers consistently emphasised that technology serves as a complementary tool requiring careful oversight. Despite positive experiences, concerns about dependency, the impact on critical thinking, and accuracy persisted. The most frequently mentioned concern was dependency, particularly for weaker students. Academic integrity, equity, and practical maintenance requirements were also raised.

Lecturers reported that students demonstrated increased confidence and appeared more prepared for exams, with reduced anxiety and improved consistency in study habits. A key benefit was the reduction in repeated questions directed to lecturers, allowing them to focus on higher-value interactions. Lecturers observed distinct patterns between stronger and weaker students. Stronger students used StudyBuddy strategically and verified responses, while weaker students showed intensive engagement but with less critical verification. Lecturers identified 24/7 availability, clear explanations, and the generation of practice questions as the most valuable aspects of the course. Limitations included prompt engineering challenges, concerns about over-reliance, and accuracy issues. Lecturers consistently viewed StudyBuddy as a supplement to teaching rather than a replacement, with the majority indicating it supported their teaching role and enhanced student learning.

The post-implementation surveys and interviews indicate that StudyBuddy successfully achieved its primary objective of improving teaching experiences for lecturers. Quantitative analysis shows improvements across multiple dimensions, including increased confidence in supporting critical thinking with digital content, improved capacity for personalised feedback, and greater comfort with AI systems. These improvements suggest that hands-on experience with StudyBuddy helped lecturers develop the skills and confidence necessary to integrate AI tools into their teaching practices. Qualitative analysis indicates that lecturers experienced significant benefits, particularly in reduced repetition of questions and improved teaching efficiency, enabling them to focus on higher-value activities. However, ongoing concerns about dependency, the impact on critical thinking, and accuracy persisted, indicating that positive experiences did not eliminate fundamental concerns about AI in education. The relationship between improved teaching experiences and student learning outcomes remains a critical question for future investigation, as explored in subsequent sections examining exam performance and chat log analysis.

Detailed quantitative and qualitative analyses comparing pre- and post-implementation responses are provided in Appendix E, which presents comprehensive item-by-item comparisons against the baseline data established in Appendix B.

## 4. EXAM PERFORMANCE ANALYSIS

This section presents findings from the lecturers' reports for two modules: Software Development Practices (SDP) and AI Ethics and Bias (AIEB). The analysis focuses exclusively on the examination components, which assess theoretical understanding through Multiple Choice Questions (MCQ) and Short Answer Questions (SAQ). The findings reveal nuanced patterns that align with some existing research, whilst contradicting others, highlighting the complexity of AI tool impacts on assessment performance.

### 4.1 Software Development Practices (SDP) Module

The SDP module analysis compared three groups: an AI-assisted group (103 students) and two non-AI cohorts (Groups A and B). The examination paper comprised Section A (knowledge-based, 20 marks) and Section B (application/problem-solving, 25 marks), totalling 45 marks.

| Group | Mean (marks and %) | Median (marks) | Standard Deviation (SD) | Highest / Lowest Marks | Cronbach's α | Standard Error of Measurement (SEM) | Median Paper Time (min) |
|---|---|---|---|---|---|---|---|
| AI group | 30.98 (68.83%) | 33 | 7.27 | 44 / 12 | 0.78 | 0.72 | 24.65 |
| Non-AI A | 23.11 (51.36%) | 21 | 8.77 | 44 / 4 | 0.81 | 3.80 | 23.22 |
| Non-AI B | 30.60 (68.00%) | 31.5 | 6.71 | 43 / 12 | 0.72 | 3.53 | 25.11 |

The table above shows group performance statistics. Mean scores indicate the average performance across groups. The median shows typical performance, less affected by outliers when distributions are skewed. Standard deviation quantifies score variability, with lower values indicating more consistent performance. Cronbach's alpha measures internal consistency reliability. Values above 0.70 are generally considered acceptable (Abbasi et al., 2025). High values indicate that the items measure a coherent construct and that the assessment reliably measures the intended learning outcomes.

The standard error of measurement (SEM), calculated as $SD \times \sqrt{(1 - Cronbach's\,\alpha)}$ , quantifies the expected variation in scores if the test were repeated. The AI group's SEM of 0.72 marks is substantially lower than that of Non-AI Group A (3.80 marks) and Non-AI Group B (3.53 marks), indicating better measurement precision. Lower SEM values indicate that individual scores are more stable and less affected by measurement error. The substantial reduction in SEM for the AI group (81.1% lower than Non-AI Group A and 79.6% lower than Non-AI Group B) suggests that the use of AI tools may contribute to more reliable assessment outcomes.

Median completion time indicates engagement levels. The median is preferred over the mean for time data because it is less affected by outliers.

The AI group achieved a mean score of 30.98 marks (68.83%), with a median of 33 marks (73.3%). This performance was nearly identical to Non-AI Group B, which achieved 30.60 marks (68.00%), representing a difference of only 0.38 marks (0.83 percentage points). However, both significantly outperformed Non-AI Group A, which achieved only 23.11 marks (51.36%), representing a substantial gap of 7.87 marks (17.47 percentage points) between the AI group and the lower-performing baseline.

The lecturers' report concluded that "AI group and Non-AI B perform at near-identical levels, both outperforming Non-AI A by a substantial margin. AI exposure did not inflate scores but may contribute to greater consistency in performance." The report noted that "AI support may contribute more to process-oriented benefits (consistency, scaffolding, reflection) rather than raw score gains."

This finding of performance parity rather than score inflation contradicts several studies that report significant improvements in test scores following the use of AI chatbots. For instance, Wu & Yu (2024) conducted a meta-analysis of 24 randomised studies and found that AI chatbots had a large effect on students' learning outcomes. Similarly, Fan et al. (2025) reviewed studies showing that learners using chatbots or ChatGPT achieved higher scores on final tests than those using conventional technology. However, the current findings align with research by Yin et al. (2021, as cited in Fan et al., 2025), who found no significant difference in overall learning performance between learners using AI-powered chatbots and those interacting with human tutors, despite AI chatbots providing more flexible learning environments. The emphasis on process-oriented benefits rather than score gains resonates with constructivist perspectives on AI-assisted learning, where tools enhance learning processes rather than immediate assessment outcomes (Cai et al., 2025).

The tables below present detailed section-level statistics for Section A (knowledge-based, 20 marks) and Section B (application/problem-solving, 25 marks).

**Section A (Knowledge-Based, 20 Marks)**

| Group | Mean (marks and %) | Median (marks) | Standard Deviation (SD) | Highest / Lowest Marks | Cronbach's α | Standard Error of Measurement (SEM) | Median Paper Time (min) |
|---|---|---|---|---|---|---|---|
| AI group | 14.46 (72.31%) | 15 | 2.95 | 20 / 6 | 0.63 | 1.78 | 9.87 |
| Non-AI A | 11.33 (56.65%) | 11 | 3.66 | 19 / 4 | 0.71 | 1.97 | 10.35 |
| Non-AI B | 14.01 (70.05%) | 14 | 2.92 | 19 / 5 | 0.62 | 1.81 | 10.68 |

**Section B (Application/Problem-Solving, 25 Marks)**

| Group | Mean (marks and %) | Median (marks) | Standard Deviation (SD) | Highest / Lowest Marks | Cronbach’s α | Standard Error of Measurement (SEM) | Median Paper Time (min) |
|---|---|---|---|---|---|---|---|
| AI group | 15.81 (63.26%) | 17 | 5.11 | 25 / 4 | 0.73 | 2.68 | 14.02 |
| Non-AI A | 11.78 (47.13%) | 11 | 5.82 | 25 / 0 | 0.78 | 2.70 | 11.70 |
| Non-AI B | 16.59 (66.37%) | 17 | 4.51 | 25 / 4 | 0.62 | 2.78 | 14.33 |

Section-level analysis reveals the diverse impacts of AI tools across cognitive domains. Mean scores for each section show whether AI support benefits knowledge-based recall tasks differently from application and problem-solving tasks. Research suggests that AI tools may have different effects on lower-order versus higher-order thinking skills, with some studies indicating that AI assistance may be more effective for complex problem-solving tasks than for simple recall tasks (Essien et al., 2024; Qassrawi & Al Karasneh, 2025).

Standard deviation differences across sections indicate whether AI tools reduce variability more in knowledge-based tasks compared to application tasks. Lower SD values in a particular section suggest more consistent performance on that type of question. Cronbach's alpha values for each section measure the internal consistency of items within that cognitive domain. The standard error of measurement (SEM) for each section quantifies measurement precision within that cognitive domain. Lower SEM values indicate that scores for that section are more reliable.

For Section A (knowledge-based questions), the AI group achieved 14.46 marks (72.31%), nearly identical to Non-AI Group B's 14.01 marks (70.05%), both substantially higher than Non-AI Group A's 11.33 marks (56.65%). For Section B (application and problem-solving questions), the AI group achieved 15.81 marks (63.26%), compared to Non-AI Group B's 16.59 marks (66.37%) and Non-AI Group A's 11.78 marks (47.13%). While both AI and Non-AI Group B performed well, Non-AI Group B achieved slightly higher mean scores on application questions.

The AI group's raw score distribution showed a mean of 30.98/45 (68.8%), median of 33/45 (73.3%), and standard deviation of 7.27. The distribution showed skewness of -0.54 (slightly left-skewed), kurtosis of -0.43 (slightly flatter than normal), and a Shapiro-Wilk normality test result of $p < 0.001$, indicating that the distribution deviates from perfect normality. The slight left skewness indicates clustering toward higher performance, with fewer extremely low outliers compared to Non-AI Group A (SD: 8.77). This left-skewed distribution suggests that the use of AI tools may help prevent extremely low performance.

The AI group's SD of 7.27 represents a 17.1% reduction in score variability compared to Non-AI Group A (SD: 8.77), demonstrating that AI support contributed to greater consistency in performance. However, the AI group's SD (7.27) was slightly higher than Non-AI Group B's SD (6.71), representing an 8.3% increase in variability. This suggests that whilst AI tools may improve consistency compared to lower-performing groups, they may not necessarily outperform high-performing non-AI cohorts in terms of score uniformity.

Kurtosis of -0.43 indicates a slightly flatter distribution than normal. The Shapiro-Wilk normality test ($p < 0.001$) indicates that the distribution deviates from perfect normality.

Cronbach's alpha values ranged from 0.72 to 0.81 across groups, with the AI group achieving a value of 0.78, indicating acceptable to good internal consistency reliability. Values above 0.70 are generally considered acceptable (Abbasi et al., 2025). The AI group's alpha of 0.78, whilst slightly lower than Non-AI Group A's 0.81, remains within acceptable ranges and suggests that AI exposure did not compromise test reliability. The maintained reliability suggests that AI-assisted learning does not introduce systematic biases that would undermine assessment validity.

The AI group's SEM of 0.72 marks is substantially lower than both Non-AI Group A's SEM of 3.80 marks (81.1% reduction) and Non-AI Group B's SEM of 3.53 marks (79.6% reduction), indicating better measurement precision. This lower SEM indicates that individual scores for AI group students are more reliable and less susceptible to measurement error. This improvement suggests that the use of AI tools may contribute to more reliable assessment outcomes.

Median paper completion times were nearly identical across groups (AI group: 24.65 minutes, Non-AI Group A: 23.22 minutes, Non-AI Group B: 25.11 minutes), suggesting similar effort levels. This consistency, combined with the observed performance parity, suggests that the use of AI tools did not significantly alter the time investment required for examination completion. This finding contrasts with research suggesting that AI tools can reduce the time spent on learning tasks, with some studies reporting a 25% reduction in time in AI-assisted learning contexts (Abrar et al., 2025). The current findings suggest that whilst AI tools may provide support during preparation, they do not necessarily reduce examination completion time.

### 4.2 AI Ethics and Bias (AIEB) Module

The AIEB module analysis compared a control group (without access to the AI tool) and an experimental group (with access to the AI tool). The examination paper consisted of two sections: Section A (knowledge-based and conceptual questions, 20 marks) and Section B (application-based or problem-solving questions, 25 marks), totalling 45 marks. The AIEB examination component is known for its difficulty and high failure rates, which provides context for interpreting the performance results.

The table below presents comprehensive paper-level statistics comparing the control group and AI tool group in the AIEB module examination.

| Group | Mean (marks and %) | Median (marks) | Standard Deviation (SD) | Cronbach's α | Standard Error of Measurement (SEM) | Median Paper Time (min) |
|---|---|---|---|---|---|---|
| AI Tool Group | 17.04 (37.87%) | 16 | 4.70 | 0.66 | 2.74 | 37.18 |
| Control Group | 17.43 (38.72%) | 18 | 5.15 | 0.69 | 2.87 | 33.28 |

The table above allows comparison between the control and AI tool groups. Mean scores provide the primary measure of central tendency. The nearly identical mean scores (17.43 vs. 17.04 marks) indicate that the use of AI tools did not significantly alter average performance levels, which contradicts research suggesting that AI tools consistently improve test scores.

The standard deviation reduction from 5.15 to 4.70 (8.7% decrease) indicates that the use of the AI tool contributed to more consistent performance across students. This finding aligns with research indicating that AI tools can provide consistent support, thereby reducing performance variability (Kamalov et al., 2023).

The relationship between the mean and median shows a distribution shift between groups. The AI group displays right skewness (mean 17.04, median 16), with more students scoring below the mean. The control group displays left skewness (mean 17.43, median 18), with more students scoring above the mean. This shift from a left-skewed distribution in the control group to a right-skewed distribution in the AI group indicates that the use of the AI tool altered the distribution of performance. The AI group's median of 16 marks is two marks lower than the control group's median of 18 marks, indicating lower performance in the middle of the distribution.

Cronbach's alpha values (0.69 vs. 0.66) measure internal consistency reliability, with both values above the acceptable threshold of 0.70 for most assessments. However, the slight decrease in the AI group (from 0.69 to 0.66) suggests marginally lower internal consistency. The difference in median completion time (+3.90 minutes) suggests potential changes in engagement patterns when using AI tools.

The control group achieved a mean of 17.43 marks (38.72%), compared to the AI tool group's 17.04 marks (37.87%), showing no statistically significant difference. The AI group demonstrated slightly lower mean scores (-0.39 marks, -0.85 percentage points), indicating that exposure to the GenAI tool did not directly enhance quantitative test performance for this cohort. The standard deviation was reduced from 5.15 to 4.70 in the AI group, representing an 8.7% reduction in score variability. Given the AIEB examination's reputation for difficulty and high failure rates, this finding suggests that even in challenging assessment contexts, AI tools may not automatically translate to improved scores.

The lecturer's report concluded that "no significant improvement in overall marks" was observed with the use of GenAI. Still, it highlighted positive secondary effects, including more consistent and reliable performance, as well as increased engagement and time on task. The report noted that GenAI tools enhance learning processes rather than immediate assessment outcomes, consistent with constructivist theories of AI-assisted learning.

This finding directly contradicts several studies reporting positive impacts of AI tools on academic performance. For example, Almasri (2024) found that experimental groups exposed to AI integration achieved significantly higher scores on academic tests compared to control groups in traditional learning environments. However, the current findings align with research by Asare et al. (2023, as cited in Fan et al., 2025), who found a negative influence on learners' mathematics performance after the implementation of ChatGPT, with participants noting that ChatGPT only provided solutions without analysis or explanations, thereby failing to improve understanding.

The tables below present detailed section-level statistics for Section A (knowledge-based and conceptual, 20 marks) and Section B (application-based, 25 marks).

**Section A (Knowledge-Based, 20 Marks)**

| Group | Mean (marks and %) | Median (marks) | Standard Deviation (SD) | Cronbach's α | Standard Error of Measurement (SEM) | Median Paper Time (min) |
|---|---|---|---|---|---|---|
| AI Tool Group | 12.31 (61.53%) | 12 | 2.82 | 0.54 | 1.91 | 14.24 |
| Control Group | 12.59 (62.95%) | 13 | 2.92 | 0.57 | 1.91 | 12.78 |

**Section B (Application/Problem-Solving, 25 Marks)**

| Group | Mean (marks and %) | Median (marks) | Standard Deviation (SD) | Cronbach's α | Standard Error of Measurement (SEM) | Median Paper Time (min) |
|---|---|---|---|---|---|---|
| AI Tool Group | 4.74 (18.94%) | 4 | 2.66 | 0.51 | 1.86 | 22.05 |
| Control Group | 4.84 (19.34%) | 4 | 2.73 | 0.42 | 2.08 | 19.98 |

Section-level statistics reveal distinct impacts of AI tools across various cognitive domains. Standard deviation differences indicate whether AI tools reduce variability more in knowledge-based tasks compared to application tasks. The reduction in SD for Section A (from 2.92 to 2.82, a 3.4% reduction) and Section B (from 2.73 to 2.66, a 2.6% reduction) suggests that AI tools contribute to slightly more consistent performance across both cognitive domains, though the effect is modest.

The Cronbach's alpha values for Section B measure the reliability of open-ended application questions, which typically show lower reliability than multiple-choice items. The improvement from 0.42 (control) to 0.51 (AI tool group) represents a 21.4% increase in internal consistency, suggesting that the use of AI tools may help students respond more consistently to complex application questions. The control group's alpha of 0.42 falls below the generally accepted thresholds, while the AI group's alpha of 0.51 shows improvement, albeit still below the typically accepted threshold of 0.70 for most assessments. This improvement suggests that the use of AI tools may help students approach application questions more systematically.

The standard error of measurement (SEM) for Section B shows improvement from 2.08 (control) to 1.86 (AI tool group), representing a 10.6% reduction in measurement error. This lower SEM indicates that scores for AI tool users are more precise and reliable. The reduction in SEM for Section B is significant, given the low reliability of application questions, suggesting that AI tool use may enhance measurement precision even for inherently less reliable assessment formats. The larger time increase in Section B (+2.07 minutes, 9.8% increase) compared to Section A (+1.46 minutes, 6.1% increase) suggests that AI users may engage more deeply with complex application questions.

For Section A (knowledge-based questions), the control group achieved 12.59 marks (62.95%), compared to the AI group's 12.31 marks (61.53%), a negligible difference (-0.28 marks, -1.42 percentage points). Standard deviation was slightly reduced from 2.92 to 2.82 in the AI group. For Section B (application questions), the control group achieved 4.84 marks (19.34%), compared to the AI group's 4.74 marks (18.94%), indicating a minimal difference of 0.10 marks (0.40%). Despite similar mean scores, Cronbach's alpha improved from 0.42 (control) to 0.51 (AI group), representing a 21.4% improvement in internal consistency. Standard deviation was reduced from 2.73 to 2.66 in the AI group. The standard error of measurement (SEM) decreased from 2.08 (control) to 1.86 (AI group), representing a 10.6% reduction in measurement error.

A detailed item-level analysis of Section B's open-ended questions (assessing competencies C1-C3 across taxonomic levels II-III) revealed nuanced patterns. The table below presents the summary of item-level performance.

**Summary of Item-Level Performance**

| Qn No. | Competency | Taxonomic Level | Max Mark | Mean (Control) | Mean (AI) | FV (Control) | FV (AI) | DI (Control) | DI (AI) | Median Time (Control) | Median Time (AI) |
|---|---|---|---|---|---|---|---|---|---|---|---|
| 21 | C1 | II | 5 | 0.33 | 0.29 | 0.07 | 0.06 | 0.50 | 0.46 | 3.35 | 3.52 |
| 22 | C2 | II | 5 | 1.15 | 1.11 | 0.23 | 0.22 | 0.57 | 0.46 | 3.85 | 3.68 |
| 23 | C2 | III | 5 | 1.15 | 0.83 | 0.23 | 0.17 | 0.52 | 0.47 | 4.85 | 4.82 |
| 24 | C3 | II | 5 | 0.49 | 0.43 | 0.10 | 0.09 | 0.37 | 0.62 | 3.28 | 3.88 |
| 25 | C3 | III | 5 | 1.72 | 2.07 | 0.34 | 0.41 | 0.57 | 0.60 | 3.90 | 4.00 |
| Total (B) | — | — | 25 | 4.84 | 4.74 | — | — | — | — | 19.98 | 22.05 |

The table above shows assessment quality and performance consistency metrics. Standard deviation reductions across all sections indicate whether AI tools create more uniform performance. The reductions observed (paper SD: 8.7% reduction; Section A SD: 3.4% reduction; Section B SD: 2.6% reduction) suggest that AI tools contribute to more uniform performance, potentially through consistent scaffolding that helps standardise student responses.

Cronbach's alpha measures the internal consistency of test items, with higher values indicating that items consistently measure the same construct. Values above 0.70 are generally considered acceptable for most assessments, though lower values (above 0.50) may be acceptable for open-ended questions (Abbasi et al., 2025). The improvement from 0.42 to 0.51 for Section B is noteworthy because application questions typically show lower reliability due to their open-ended nature, and this improvement suggests AI tools may help students approach these questions more consistently. The control group's alpha of 0.42 falls below acceptable thresholds, whilst the AI group's alpha of 0.51 reaches the minimum acceptable level for open-ended assessments.

The standard error of measurement (SEM) reduction from 2.08 to 1.86 for Section B represents a 10.6% improvement in measurement precision, indicating that scores for AI tool users are more reliable and less affected by measurement error. This improvement suggests that the use of AI tools enhances measurement precision, even for inherently less reliable assessment formats, thereby addressing concerns about assessment reliability in AI-assisted learning contexts (Kooli, 2023).

The discrimination index measures how well items distinguish between high and low performers, with values above 0.30 generally considered acceptable. The mean discrimination index across all items improved slightly in the AI group (from approximately 0.48 to 0.51), representing a 6.3% improvement. This reflects a more equitable spread between stronger and weaker students, suggesting that GenAI exposure helped more able learners leverage the tool effectively. In comparison, weaker students may not have benefited equally. This finding raises equity concerns about the implementation of AI tools, as the improved discrimination index suggests that they may amplify existing performance differences rather than reducing them, potentially widening achievement gaps. This pattern contradicts research suggesting that AI tools can provide personalised support that benefits all students (Cai et al., 2025). The differential benefit for stronger students aligns with research by Bećirović et al. (2025), who found that the impact of AI practical application on students' academic performance was insignificant, suggesting that students' current knowledge, skills, and abilities of AI use significantly affect outcomes. This finding suggests that the effectiveness of AI tools may depend on students' prior abilities and their level of AI literacy, potentially creating equity challenges in educational settings.

Despite similar mean scores, the AI group showed improved reliability in Section B, with Cronbach's alpha increasing from 0.42 (control) to 0.51 (AI group), representing a 21.4% improvement in internal consistency. This improvement is significant given that Section B focuses on application-based questions, which typically show lower reliability due to their open-ended nature. Standard deviations were slightly smaller across all sections for the AI group (paper SD: 4.70 vs. 5.15, an 8.7% reduction; Section A SD: 2.82 vs. 2.92, a 3.4% reduction; Section B SD: 2.66 vs. 2.73, a 2.6% reduction), suggesting more consistent performance among AI users. The examiner interpreted this as indicating more consistent performance among AI users and stabilising effects of AI tools on learning consistency.

This pattern of improved consistency without score gains distinguishes between different types of AI tool benefits. While some research emphasises score improvements (R. Wu & Yu, 2024), the current findings highlight the benefits of consistency, which may be equally valuable for educational outcomes. The stabilising effects align with research suggesting that AI tools can provide consistent scaffolding and support, reducing variability in student performance (Kamalov et al., 2023). However, concerns about the reliability and accuracy of AI tools remain valid, as scholars note that AI chatbots may provide biased responses or inaccurate information, which could mislead students (Labadze et al., 2023). The improved consistency in the current study suggests that when properly implemented, AI tools can enhance performance stability without compromising assessment validity.

The AI tool group took significantly longer overall, with a median paper completion time increasing from 33.28 minutes (control) to 37.18 minutes (AI group), representing a 3.9-minute increase (11.7% longer). Section-level time increases (Section A: +1.46 minutes, representing a 6.1% increase; Section B: +2.07 minutes, representing a 9.8% increase) support the idea of deeper engagement during cognitive tasks. The larger time increase in Section B, which focuses on application questions, suggests that AI users may engage more deeply with complex problem-solving tasks. The examiner suggested this may indicate higher engagement or reflective effort, or time spent recalling or verifying AI-learned material.

This finding of increased time-on-task aligns with research suggesting that AI tools can enhance engagement and motivation (D.-L. Chen et al., 2025). Research by Abrar et al. (2025) found that AI-powered personalised learning systems resulted in higher engagement levels, with experimental groups showing 15% higher engagement than control groups. However, the lack of corresponding score improvements creates a paradox:

increased engagement did not translate to better performance. This pattern contradicts research by Ahn et al. (2024), who found that effective use of AI tools improved understanding of academic materials, leading to better grades and higher academic performance (Ahn, 2024). The current findings suggest that engagement metrics alone may not predict performance outcomes and that the quality of engagement with AI tools may be more important than the quantity. This aligns with concerns that students may spend time verifying AI-generated content rather than engaging in productive learning activities (Akolekar et al., 2025). The increased time spent by AI users may reflect deeper cognitive processing or verification behaviours rather than productive learning engagement.

## 5. CHAT LOG ANALYSIS

### 5.1 Data Collection and Analysis Methodology

Chat logs were collected from all student interactions with StudyBuddy during the deployment period. The dataset comprises 269 unique conversations, totalling 3,385 messages, including 1,674 user messages across all participating modules. Analysis employed quantitative metrics including query length, conversation depth, and dependency phrase frequency, alongside qualitative examination of question types, dependency behaviours, module relevance, and engagement quality indicators.

### 5.2 Dependency and Superficial Engagement Patterns

Analysis revealed multiple indicators of dependency and superficial engagement. A total of 672 queries (40.1% of all user messages) contained dependency-indicating phrases, with "give me" (368 occurrences) and "answer" (306 occurrences) being the most common. These patterns indicate a preference for direct information retrieval over exploratory learning, consistent with the metacognitive laziness phenomenon, where learners offload metacognitive load onto AI assistance (Fan et al., 2025).

Evidence of minimal cognitive effort included 428 queries (25.6%) that were extremely short, with an average length of fewer than 20 characters, and 244 queries (14.8%) that were very short, with an average length of fewer than five characters. Additionally, 126 queries (7.5%) appeared to be verbatim extracts from assignments or exam papers, containing formatting indicators such as numbered lists, markdown formatting, or extreme length (over 500 characters), suggesting direct copying rather than thoughtful question formulation.

A substantial proportion of queries (1,371 messages, 81.9%) had no clear relationship to the enrolled modules, indicating that students viewed StudyBuddy as a general-purpose AI assistant rather than a focused learning tool. This misuse diverted time and cognitive resources away from learning specific to the module.

Analysis showed minimal evidence of critical engagement. Only 25 queries (1.5%) contained verification requests. Patterns of students asking similar questions multiple times with slight variations suggested dependency on the chatbot rather than internalising learning. Students appeared to be testing the chatbot's consistency or seeking multiple formulations of the same answer rather than learning from responses.

### 5.3 Engagement Quality and Depth

The average conversation contained 6.22 user messages, with depth ranging from 1 to 58 messages, indicating substantial variation in engagement levels. While 220 conversations (81.8%) contained two or more user messages, and 127 conversations (47.2%) contained five or more messages, quality indicators remained low across all depth categories.

Analysis comparing usage patterns across conversation depths revealed that deeper conversations were not necessarily higher quality. Very deep conversations (10+ messages, n = 59) exhibited the highest rates of non-module queries (83.3%) and short, low-effort queries (32.0%), indicating rapid-fire, superficial interactions rather than in-depth conceptual exploration. Reflection indicators (5.0%) and verification requests (2.0%) were highest in very deep conversations, but these rates remained extremely low in absolute terms.

Overall, only 156 queries (9.3%) contained follow-up question indicators, 74 queries (4.4%) contained reflection indicators, and 25 queries (1.5%) contained verification requests. This suggests that while students used StudyBuddy multiple times, many did not engage in extended conceptual exploration. The substantial variation in conversation depth suggests different usage patterns, with some conversations exhibiting sophisticated, multi-turn interactions and others featuring simple, repetitive queries. The low overall verification rate (1.5%) suggests that most students, regardless of ability level, did not critically evaluate chatbot responses.

### 5.4 Implications for Learning Outcomes

The chat log analysis reveals usage patterns that explain the performance findings. The prevalence of short, low-effort queries (25.6%), a high frequency of dependency phrases (40.1%), minimal critical engagement (1.5% verification requests and 4.4% reflection indicators), and overwhelming non-module queries (81.9%) suggest that many students engaged superficially with StudyBuddy. This superficial engagement, while potentially improving confidence through quick access to answers, did not foster deep learning or skill development.

Patterns of dependency behaviours suggest that students developed reliance on StudyBuddy without developing the independent thinking and problem-solving skills required for exam success. While 81.8% of conversations were multi-turn and 47.2% were extended (comprising five or more messages), the low rates of follow-up questions (9.3%), reflection (4.4%), and verification (1.5%) suggest that extended conversations often consisted of repetitive or unrelated queries rather than deepening conceptual exploration. This finding is consistent with research showing that increased interaction quantity does not necessarily translate to improved learning outcomes (Fan et al., 2025).

These findings demonstrate that StudyBuddy provided immediate assistance and improved lecturers' teaching experience, while achieving performance parity for students with high-performing baseline cohorts. The engagement patterns revealed in chat logs are consistent with maintaining existing performance levels rather than actively working to improve from lower levels. The tool's value lies in process-oriented benefits, including consistency, scaffolding, and reflection. Future research with baseline data must investigate whether StudyBuddy enables improvement from lower performance levels or maintains existing high performance.

## 6. DISCUSSION

### 6.1 Do Improved Teaching Experiences Translate to Improved Student Learning Outcomes?

Our study achieved its primary objective of improving the teaching experience for lecturers. Post-deployment surveys revealed positive experiences, with lecturers reporting a reduction in repeated questions, improved student confidence, and enhanced teaching efficiency.

The literature provides theoretical support for the hypothesis that improved teaching experiences should translate to better student outcomes. Berisha Qehaja (2025) explicitly argue that "operational efficiency and productivity" in education ensures "resources are maximised and the course is delivered most productively to benefit student learning". This suggests that when AI tools reduce educator workload and enable better pedagogical focus, student learning should benefit.

However, our findings raise a critical research question about whether improved teaching experiences translate to improved student learning outcomes. Quantitative performance data show that, regardless of their starting point, students who used StudyBuddy achieved final performance that matched or exceeded that of high-performing baseline cohorts. While lecturers benefited from StudyBuddy, the causal pathway by which students benefit remains unclear. The absence of baseline data necessitates future longitudinal studies to collect pre-intervention assessments, thereby determining whether this outcome represents improvement from lower performance levels or the maintenance of existing high performance.

Our study reveals a disconnect between educator benefits and student outcomes in the context of vocational education. While the literature predominantly shows positive effects of AI in education, our findings challenge the assumption that improved teaching experiences automatically translate to improved learning outcomes.

These findings indicate that AI tools enhance learning processes and consistency, without necessarily translating to improved immediate assessment outcomes, which is consistent with constructivist theories of AI-assisted learning. Lecturer observations captured perceived improvements in student engagement and confidence, while exam performance measured actual learning and skill development. While AI tools provide valuable process-oriented benefits (consistency, scaffolding, reflection), these do not directly translate to improved exam scores when engagement patterns involve dependency or superficial interaction with content. Future research should investigate how AI tools can be designed to ensure that process-oriented benefits are effectively translated into improved assessment outcomes.

The reduced standard deviations and improved Cronbach's alpha values show that StudyBuddy helped reduce performance variability, potentially providing more equitable learning experiences. The increased engagement time, although not directly translating to score gains, suggests that students were investing more effort in their learning process, which has intrinsic value for skill development, even if not immediately reflected in assessment scores. These process benefits may represent learning gains that are not fully captured by traditional outcome measures, aligning with constructivist perspectives that value the learning process itself as an important outcome.

### 6.2 Does AI Tool Use Enable Improvement from Lower Performance Levels or Maintain Existing High Performance?

Our findings show that students who used StudyBuddy achieved final performances that matched or exceeded those of high-performing baseline cohorts. However, the absence of baseline data creates interpretative ambiguity. Two alternative hypotheses emerge.

If the AI group initially performed at levels similar to the lower-performing Non-AI Group A (approximately 51%) and subsequently achieved parity with Non-AI Group B (approximately 68%), this would suggest substantial improvement (approximately 17 percentage points) attributable to StudyBuddy support. This hypothesis aligns with literature demonstrating that AI tools can provide personalised support that helps struggling students catch up. Adewale et al. (2024) noted that "when AI systems are employed to personalise learning experiences and provide timely, relevant feedback, a substantial improvement in student performance is noted".

If the AI group initially performed at levels similar to Non-AI Group B (approximately 68%) and maintained that performance, this would indicate effective support for advanced learners. Chat log analysis, which reveals patterns of short and low-effort queries, direct answer requests, copy-paste behaviour, and substantial non-module-related queries, suggests engagement patterns consistent with maintaining existing performance levels rather than actively working to improve from lower levels. This hypothesis aligns with findings that higher-performing students may strategically leverage AI tools for verification and refinement rather than for foundational learning.

Future research with baseline assessments must investigate which hypothesis accurately describes the outcome. Whether AI tool use enables substantial improvement from lower to higher performance levels (as suggested by the possibility that the SDP AI group improved from approximately 51% to approximately 68%) or maintains existing high performance levels is fundamental to understanding whether AI tools can close achievement gaps or primarily support already-strong performers.

### 6.3 How Does Assessment Alignment Affect AI Tool Effectiveness?

Given that this study focused exclusively on the examination component, which assesses theoretical knowledge and conceptual understanding rather than practical application, StudyBuddy's text-based, theoretical support approach was well-aligned with the assessment format being evaluated. The examination components (MCQ and SAQ questions) test students' knowledge of concepts and principles, as well as their ability to reason through problem scenarios, precisely the type of theoretical understanding that a text-based RAG system can effectively support.

However, our findings show that even when AI tools are well-aligned with assessment formats, performance gains are not guaranteed. Chat log analysis reveals that students did not effectively utilise StudyBuddy's theoretical support, despite its alignment. This suggests that alignment between the tool and the assessment format is necessary but not sufficient. How students engage with aligned content matters critically.

The examination format requires students to demonstrate not just recall but reasoning, articulation, and application of concepts to new contexts. While StudyBuddy could provide theoretical explanations and conceptual clarifications that aligned with the exam's knowledge requirements, effective exam performance also depends on deeper learning processes, including the ability to synthesise information, reason through novel scenarios, and articulate understanding independently. Y. Li et al. (2025) found that cognitive outcomes from AI chatbot interactions include "academic achievement" and "cognitive skills and content knowledge." Still, they noted that the majority of studies "concerned students' academic achievement and their acquisition of cognitive skills (e.g., evaluation skills, self-regulated learning strategies)". Our findings show that while StudyBuddy provided access to theoretical content, it did not support the development of deeper cognitive skills required for exam success, even when aligned with the assessment format.

Our study shows evidence of Fan et al.'s (2025) "metacognitive laziness phenomenon" in a vocational education context. D. Zhang et al. (2025) found that AI dependency has "significant negative relationships with critical thinking, self-confidence, and problem-solving ability," with "increased dependency on AI" potentially "inversely affect[ing] critical thinking abilities". Our study shows behavioural evidence of how this manifests in practice.

Future research should investigate the design features that enable students to effectively leverage the theoretical support of AI tools for deep learning, rather than merely retrieving surface-level information. Research should examine how AI tools can be designed to support both content access and the development of deeper cognitive skills, as well as the relationship between alignment (a necessary condition) and effective engagement (a sufficient condition).

Chat log analysis reveals that weaker students, whom lecturers reported using StudyBuddy most intensively ("weaker learners benefited more, they used it nightly"), exhibited patterns of dependency, including direct answer requests, minimal critical evaluation, and repeated similar queries. While lecturers observed that these students "gained confidence" and "asked better questions," the chat logs show that this confidence was built on dependency rather than genuine learning. Fan et al. (2025) documented that learners become dependent on AI and "offload meta-cognitive load and less effectively associate responsible metacognitive processes with learning tasks," resulting in "less effective engagement with essential self-regulatory tasks". Our findings align with this. Students used StudyBuddy to quickly obtain answers, thereby reducing the cognitive effort required for learning. This created a false sense of confidence (reflected in lecturers' positive observations) without developing the underlying skills needed for independent performance.

The performance gap was particularly pronounced among weaker students, who likely had lower initial self-efficacy. These students were most vulnerable to developing dependency, using StudyBuddy as a "crutch" rather than a learning enhancer. While stronger students showed evidence of strategic use (verifying responses and engaging critically), weaker students demonstrated dependency patterns that ultimately undermined their learning. This finding raises equity concerns. AI tools may inadvertently widen achievement gaps if not designed with careful consideration of student vulnerabilities.

### 6.4 How Do Self-Efficacy and AI Dependency Affect Learning Outcomes?

The lack of performance improvement, despite assessment alignment, suggests that self-efficacy and AI dependency are the primary explanatory mechanisms underlying this phenomenon. Even when an AI tool is appropriately matched to the assessment format, dependency patterns can undermine learning effectiveness.

Our findings show that students, particularly those with lower self-efficacy, developed dependency on StudyBuddy that undermined rather than enhanced their learning. Literature on self-efficacy provides important context. Xu & Wang (2025) found that "basic and intermediate information literacy self-efficacy did not significantly predict use of cognitive, metacognitive, environmental, and emotional regulation strategies," suggesting that lower self-efficacy may not facilitate the effective use of learning strategies. Similarly, Bećirović et al. (2025) noted that while "chatbots and AI can influence self-efficacy positively" when users experience success, "the fast-paced development of new AI technologies might interfere with users' perceived self-efficacy".

Students with weaker academic backgrounds and lower self-efficacy were particularly vulnerable to this phenomenon. Despite StudyBuddy's alignment with theoretical exam content, these students developed dependency patterns that undermined their learning. They utilised StudyBuddy's theoretical support in ways that reduced cognitive effort (through short queries and direct answer requests) rather than engaging deeply with the material, despite the content being relevant to exam preparation.

The interaction creates a problematic scenario. Students with lower self-efficacy, seeking to reduce cognitive load, turn to StudyBuddy for theoretical explanations. While these explanations align with exam requirements, the manner of engagement (dependency patterns, cognitive offloading) prevents the development of deeper learning skills required for independent reasoning and articulation during exams. This suggests that alignment between tool and assessment format is necessary but not sufficient for learning effectiveness. How students engage with aligned content matters critically.

Future research should investigate the design features that enable AI tools to mitigate dependency risks for vulnerable students while maintaining effectiveness for stronger students. Research should examine how tools can be designed to foster self-efficacy rather than undermine it through dependency, and what interventions can prevent the "metacognitive laziness phenomenon" while maintaining the benefits of AI support.

### 6.5 How Can AI Tools Support Both Higher-Order and Lower-Order Cognitive Tasks?

The AIEB module's item-level analysis shows differential impacts by cognitive level. Higher-order synthesis tasks (C3-Level III) showed positive gains (+0.35 marks), while lower-order recall and conceptual tasks (C1-C2) showed marginal declines. This shows that AI support benefits higher-order thinking tasks when used appropriately, while offering limited value or hindering performance on lower-order recall questions due to overreliance on surface-level AI outputs.

The differential impact by cognitive level shows that the effectiveness of AI support varies with question complexity and the type of cognitive skill required. Higher-order tasks may benefit from AI support for idea generation and verification. In contrast, lower-order recall tasks may suffer when students offload memory retrieval processes onto AI systems, preventing the effortful encoding necessary for long-term retention.

The pattern reveals an overreliance on AI for surface-level information retrieval in lower-order tasks, whereas higher-order synthesis tasks benefit from strategic AI use. This suggests that AI tools must be redesigned to support deep learning for lower-order recall tasks rather than encouraging surface-level dependency, while maintaining support for higher-order thinking.

Future research must investigate how AI tools can be redesigned to support deep learning for lower-order recall tasks rather than encouraging surface-level dependency. Research should examine the design features that enable AI tools to support both higher-order synthesis tasks and lower-order recall tasks effectively, as well as how tools can prevent cognitive offloading for memory retrieval while facilitating synthesis and reasoning.

### 6.6 How Can AI Tools Provide Equitable Support That Benefits Both Stronger and Weaker Students?

The discrimination analysis from the AIEB module shows that "GenAI exposure helped more able learners leverage the tool effectively, while weaker students did not benefit equally." AI tools have differential impacts across student populations. Stronger students benefited more from strategic use, while weaker students required different support approaches to avoid dependency.

Our findings show that stronger students showed evidence of strategic use (verifying responses, critical engagement), while weaker students demonstrated dependency patterns that ultimately undermined their learning. This suggests that AI tools may inadvertently widen achievement gaps if not designed with careful consideration of student vulnerabilities, requiring tailored approaches for different ability levels.

This finding has important equity implications. Rather than closing achievement gaps, AI tools may exacerbate them if weaker students develop dependency while stronger students leverage the tools strategically. Future research should investigate the design features that enable AI tools to provide equitable support, benefiting both stronger and weaker students without fostering dependency in vulnerable learners. Research should examine how tools can be designed to narrow rather than widen achievement gaps, and what tailored approaches are needed for different ability levels.

### 6.7 How Can AI Tools Be Designed to Ensure Process-Oriented Benefits Translate to Improved Assessment Outcomes?

Our findings suggest that AI tools enhance learning processes and consistency, without necessarily translating to improved immediate assessment outcomes, which aligns with constructivist theories of AI-assisted learning. Both modules demonstrated improved consistency (reduced standard deviations), enhanced reliability (as indicated by improved Cronbach's alpha values), and increased engagement time (resulting in longer completion times). However, these process-oriented benefits did not directly translate to improved exam scores when engagement patterns involved dependency or superficial interaction with content.

While process measures (time on task, consistency, engagement) are meaningful indicators of learning behaviour, they do not automatically translate to improved scores. The literature on learning analytics emphasises that process measures are valuable. Still, alignment between process improvements and outcome measures may require explicit design features that encourage deep engagement rather than dependency.

Future research should investigate how AI tools can be designed to ensure that process-oriented benefits are effectively translated into improved assessment outcomes. Research should examine the design features that encourage deep engagement rather than dependency while maintaining process benefits, and how learning process improvements (such as time on task, consistency, and engagement) can be leveraged to enhance actual learning outcomes.

## 7. CONCLUSION

This study aimed to develop and evaluate a proof-of-concept for Generative AI, designed to support lecturers by enhancing instructional efficiency during exam preparation. The POC successfully achieved this primary objective: lecturers experienced reduced repetitive queries, smoother teaching workflows, and greater student confidence. These outcomes affirm the value of user-centred AI design in addressing practical pedagogical needs within vocational education.

However, the implementation yielded emergent findings that extended beyond the original research aim. Despite clear gains in teaching efficiency, student performance data showed no significant improvement in exam outcomes. Behavioural analysis revealed patterns of AI-driven dependency, diminished metacognitive effort, and differential tool use across student ability levels. Higher-performing students engaged with the chatbot strategically, while lower-performing students tended to over-rely on it, potentially widening the achievement gap further. These insights illustrate that AI tools shape not only instructional processes but also cognitive habits, self-regulation, and learning engagement.

In addressing the research questions, the study demonstrates that improved teaching processes alone do not guarantee improved learning outcomes. The findings highlight the need for equitable, pedagogically informed AI design that supports deep learning, scaffolds metacognitive development, and calibrates assistance to diverse learner profiles. Theoretically, the study refines understanding of how generative AI mediates cognitive effort and self-regulation; practically, it informs future AI design strategies in education.

Future research should explore mechanisms that balance support with productive struggle, investigate how process improvements can be translated into measurable learning gains, and examine AI interventions that narrow rather than widen performance disparities. Ultimately, unlocking the educational potential of generative AI requires context-sensitive, ethically grounded systems that align technological affordances with meaningful and equitable learning outcomes.

## 8. ACKNOWLEDGEMENTS

We thank the participating lecturers for sharing their time, expertise, and feedback through interviews, co-design workshops, and surveys, which were essential for developing and evaluating StudyBuddy. This study was conducted within the Critical Inquiry module at Nanyang Technological University, and we thank our supervisor, Associate Adjunct Professor Jonathan Pan, for his guidance and support.

## APPENDIX A: DETAILED LITERATURE REVIEW (BY THEME)
## A.1 Positive Impacts and Learning Outcomes

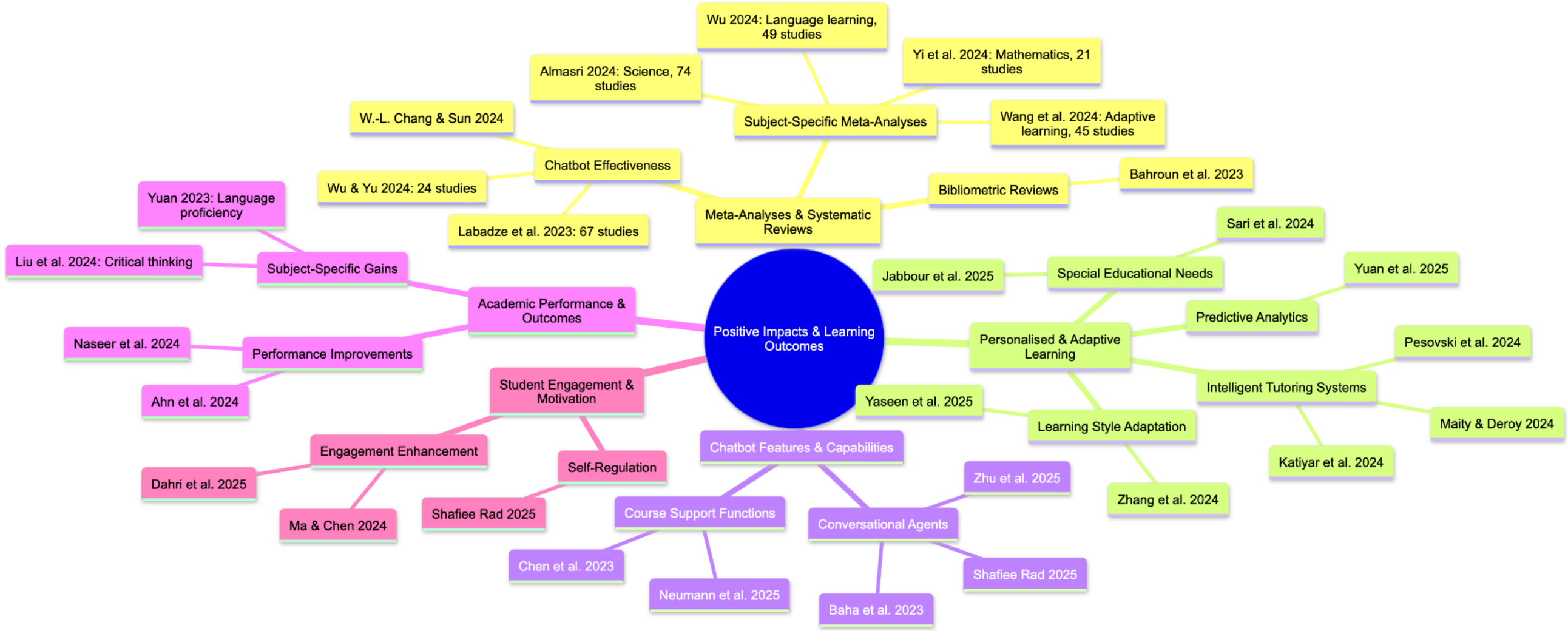

| Authors | Key Findings | Relevance to Research |
|---|---|---|
| Bahroun et al. (2023) | Bibliometric review of Generative AI in education showing moderate to strong positive effects on academic achievement and student engagement. | Supports the potential of Generative AI to improve teaching experiences and student engagement. |
| Pesovski et al. (2024) | Evidence of improved learning outcomes and motivation in AI-enhanced educational environments with customisable learning experiences. | Supports the potential of Generative AI to improve learning outcomes and motivation. |
| Maity & Deroy (2024) | Personalised intelligent tutoring systems provide responses based on individual learning needs. | Supports the potential of Generative AI for personalised learning and student support. |
| Katiyar et al. (2024) | AI-driven personalised learning systems enhance educational effectiveness through adaptive learning approaches. | Supports the potential of Generative AI for personalised learning and educational effectiveness. |
| J. Zhang & Zhang (2024) | A quantitative study involving 202 students and 68 staff members found that AI can identify learning styles and facilitate diverse teaching methods. AI uses student interaction data to determine learning preferences and customise content. AI automates time-consuming tasks, allowing teachers to focus on personalised instruction. | Supports findings on accommodating learning styles. Shows AI can identify and accommodate individual learning styles and customise educational content. |
| Yaseen et al. (2025) | Personalised real-time feedback in adaptive learning systems improves knowledge retention compared to traditional methods. Feedback maintains attention, improves retention, motivates students, and enhances self-regulatory skills. | Supports the value of personalised feedback mechanisms. |
| Shafiee Rad (2025) | AI intervention for L2 reading comprehension fosters self-regulated learning. AI-powered conversational platforms show positive effects on learning outcomes, motivation, participation, and self-regulation. | Supports the potential of Generative AI for student engagement, motivation, and self-regulated learning. |

| Authors | Key Findings | Relevance to Research |
|---|---|---|
| Labadze et al. (2023) | A systematic review of 67 studies found that chatbots enhance learning achievement, reasoning, and knowledge retention. Chatbots offer immediate assistance, quick access to information, and personalised learning experiences. They provide constant availability and handle multiple queries simultaneously, creating safe spaces for questions. | Supports the potential of Generative AI to improve teaching experiences. Evidence for chatbot capabilities that supported co-development decisions. |
| R. Wu & Yu (2024) | A meta-analysis of 24 randomised studies found that AI chatbots have a large effect on learning outcomes. Chatbots improve academic performance, learning interest, motivation, engagement, and self-efficacy while also reducing anxiety and stress. They function as partners and mentors, providing personalised content and feedback. Chatbots assist teachers in assessments and reduce workload. | Provides evidence for reducing educator workload and enhancing teaching experiences. Evidence supporting the effectiveness of chatbots that informed co-development decisions. |
| W.-L. Chang & Sun (2024) | Meta-analysis examining AI impact on self-regulated language learning. Chatbots' interactive features and time-location flexibility support positive learning experiences. Human-chatbot interaction enables real-time engagement, improves communication skills, and increases learning efficiency. Medium to high effects on achievement when effectively integrated. | Supports the potential of Generative AI to enhance educational experiences when appropriately integrated. Evidence for chatbot effectiveness. |
| Zhu et al. (2025) | Chatbots enhance learning outcomes, enjoyment, experiences, and self-efficacy by enabling teachers to tailor content and providing quick access to information. LLM-based chatbots generate cohesive human-like responses. Over 40% of students aged 7-17 use generative AI tools. | Highlights the widespread adoption of Generative AI and its potential to enhance teaching and learning experiences. Evidence for chatbot adoption. |
| Sari et al. (2024) | Adaptive learning systems using AI improve educational outcomes. AI chatbots support students with Special Educational Needs through non-judgmental feedback, repeated practice, and accommodations for different learning styles. Chatbots reduce educator workload by automating routine tasks such as answering frequently asked questions and providing initial feedback. | Supports students with SEN, educator workload reduction, and adaptive learning approaches. Evidence for chatbot benefits that supported co-development decisions. |
| Jabbour et al. (2025) | Generative AI-powered learning companion for personalised education and broader accessibility, including accommodations for students with special educational needs. | Relevant to understanding how Generative AI can support students with special educational needs in inclusive education contexts. |
| C. Yuan et al. (2025) | A mixed-methods study using an XGBoost classifier achieved 96% accuracy in predicting student outcomes, enabling targeted support for at-risk students. AI systems automate grading, generate personalised feedback, and identify at-risk students, allowing educators to focus on higher-order teaching activities. Findings highlight the potential of AI in creating equitable educational opportunities. | Supports educator workload reduction through AI automation. Evidence for personalised learning and student support. Demonstrates AI potential to identify at-risk students. |
| Almasri (2024) | Systematic review of 74 studies from 2014 to 2023 examining AI impact in science education. AI-powered tools enhance learning environments, create quizzes, assess work, and predict performance. Experimental groups with AI integration showed significantly higher scores in academic tests compared to control groups. | Contrary to current findings, which show no significant improvement. Provides evidence for experimental groups with AI integration exhibiting significantly higher scores. Demonstrates positive impacts on academic performance in science education. |

| Authors | Key Findings | Relevance to Research |
|---|---|---|
| Ahn (2024) | A quantitative study with 300 participants, using PLS-SEM, found that ease of use of AI tools, ability to apply knowledge, and confidence in learning with AI all impact performance and usage frequency. Effective AI tool use improved understanding of academic materials, leading to better grades and higher academic performance. | Contrary to current findings, which show no significant improvement. Provides evidence for the effective use of AI tools, improving understanding and leading to better grades. Demonstrates how ease of use and self-efficacy influence effectiveness. |
| Naseer et al. (2024) | Controlled experiment with 300 students across four courses comparing an AI-driven adaptive learning platform with traditional instruction. Results showed 25% improvement in grades, test scores, and engagement for the AI group with statistical significance. Found positive correlations between engagement metrics and student performance, though focused on structured platforms rather than conversational chatbots. | Contrasts with current findings, which show that increased interaction quantity does not necessarily translate to improved outcomes. Provides evidence for positive correlations between engagement metrics and performance. Demonstrates the differences between structured platforms and conversational chatbots. |
| Y. Yuan (2024) | An experimental study with 74 Grade 5 students comparing traditional methods with AI chatbots over 3 months. Chatbot integration significantly improved oral English proficiency and willingness to communicate in the experimental group. Teachers can enhance instruction by adopting tailored chatbot features to refine teaching methods. | Provides evidence that AI chatbots can significantly improve oral English proficiency and willingness to communicate in primary education. Demonstrates effectiveness in language learning for young learners. |
| X.-Y. Wu (2024) | Meta-analysis of 49 studies examining AI interventions in language learning. Analysis revealed a significant positive impact with a weighted average SMD of 0.982, indicating a moderate to large effect size. Online and blended learning environments, medium-duration interventions, higher education settings, and young adult learners show the highest effectiveness. AI interventions are most effective for listening and speaking skills. | Provides evidence that AI interventions have significant positive impacts on language learning outcomes. Demonstrates that chatbots promote academic performance, inspire learning interest, boost motivation and engagement, and relieve learning anxiety. |
| Yi et al. (2025) | Systematic review and meta-analysis of 21 studies investigating AI effectiveness in improving mathematics performance in K-12 classrooms. Results indicate a small overall effect size of 0.34in favour of AI, showing a generally positive impact. AI type was identified as having moderate effects, with intelligent tutoring systems and adaptive learning systems showing greater impact than pedagogical agents. | Provides evidence that AI interventions have a positive impact on mathematics learning outcomes in K-12 settings. Demonstrates that the type of AI has a significant influence on its effectiveness. Supports finding the effectiveness of AI tools in educational contexts. |
| W. Liu & Wang (2024) | Eight-week intervention study with 90 students examining the effects of AI tools on critical thinking in English literature classes. Findings revealed a statistically significant improvement in critical thinking skills of the experimental group compared to the control group. AI tools enhance critical thinking abilities by providing tailored feedback and scaffolded learning experiences that support effective learning. | Provides evidence that AI tools can offer effective support for developing critical thinking skills through immediate feedback and interactive learning experiences. Demonstrates how teachers' technological self-efficacy impacts effectiveness. |
| Ma & Chen (2024) | Quasi-experimental research with 350 students examining the influence of AI-empowered applications on engagement and academic procrastination. Found that 70-95% of students procrastinate, with about 50% habitually procrastinating. The experimental group demonstrated significantly higher levels of engagement and a substantial reduction in academic procrastination. | Supports findings on student procrastination patterns and time management challenges. Provides evidence for widespread procrastination among college students. Demonstrates how AI-empowered applications can mitigate procrastination and enhance engagement. |
| Ait Baha et al. (2024) | Research on conversational agents and dialogue-based learning examines the effectiveness of chatbots in this context. | Supports the potential of conversational agents in educational contexts. |

| Authors | Key Findings | Relevance to Research |
|---|---|---|
| Y. Chen et al. (2023) | Two sequential studies with 215 and 195 students investigating chatbots as pedagogical tools in business education. Found that participants frequently needed answers to basic information about courses, including course materials, due dates, study tips, and office hours. Chatbots can provide basic course information and content, serving as a helpful addition in classes with high student-teacher ratios. | Supports findings that students repeatedly ask similar questions and require basic course information. Demonstrates how chatbots can address repetitive questions and basic information needs. Supports findings on limited lecturer capacity. |
| Neumann et al. (2025) | Research with 46 students exploring the LLM-driven chatbot MoodleBot in computer science classrooms. Found that chatbots can significantly improve the teaching and learning process with a high accuracy rate in providing course-related assistance. Chatbots provide private, nonjudgmental space for students to ask questions without fear of embarrassment, addressing help-seeking concerns and alleviating help-avoidant behaviour. | Supports findings on safe learning spaces and help-seeking behaviour for weaker students. Provides evidence that chatbots can offer private, non-judgmental spaces. Demonstrates how chatbots can address help-seeking concerns and reduce social anxiety. |
| Wang et al. (2024) | Meta-analysis of 45 independent studies examining the effect of AI-enabled adaptive learning systems on cognitive learning outcomes. Found a medium to large positive effect size of 0.70. AI-enabled adaptive learning leverages algorithms to personalise learning, tailoring content to each learner's unique needs and preferences, thereby enhancing engagement and effectiveness. | Supports findings on adaptive learning algorithms dynamically adjusting instructional content. Provides evidence for AI-enabled adaptive learning systems tailoring content to learner needs. Demonstrates medium to large positive effect sizes on cognitive learning outcomes. |
| Ahmed Dahri et al. (2025) | Quantitative study with 290 students investigating relationships between gamification, artificial intelligence, student engagement, and achievement. Found that gamification introduces flexible reward mechanisms adapted to individual achievements and preferences, encouraging students to sustain efforts and persevere. Recognising personal milestones fosters accomplishment and satisfaction. | Supports findings on gamification, introducing flexible reward mechanisms and fostering engagement. Provides evidence for gamification, encouraging students to sustain efforts. Demonstrates how recognising personal milestones fosters accomplishment. |

## A.2 Personalised and Adaptive Learning

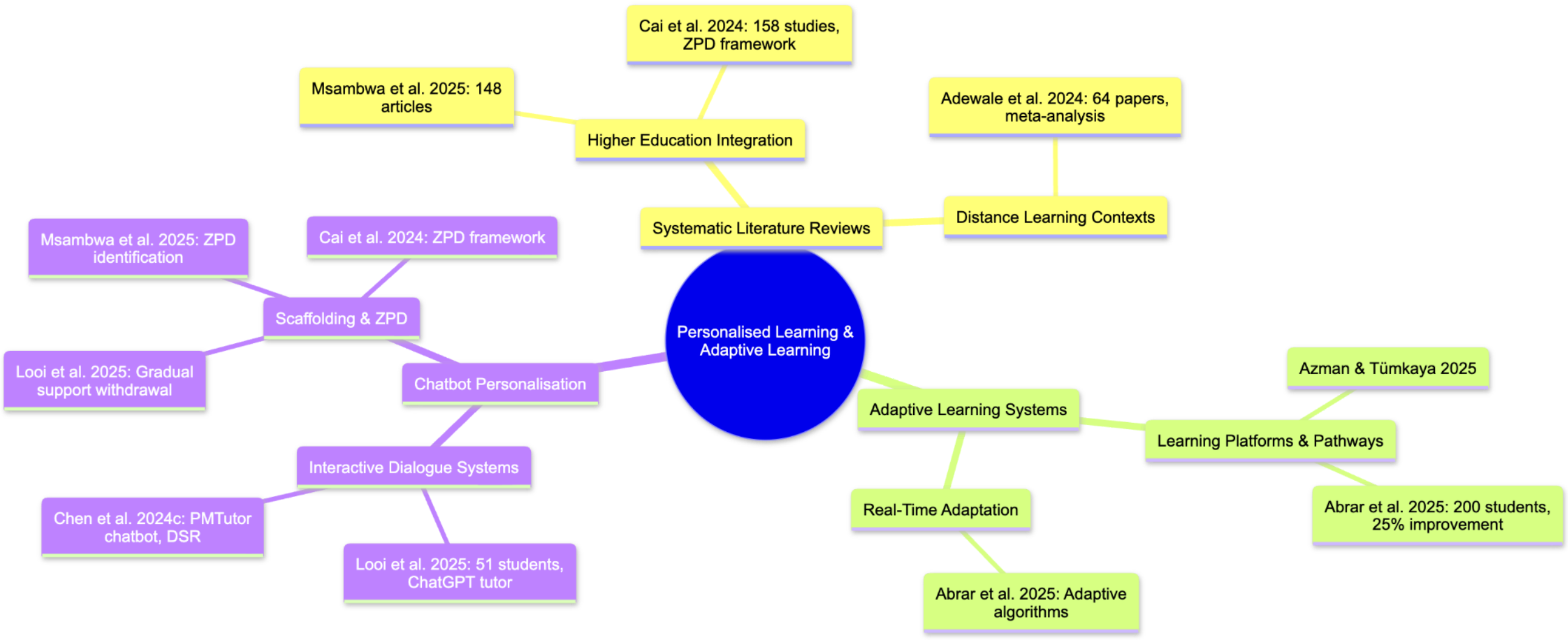

| Authors | Key Findings | Relevance to Research |
| --- | --- | --- |
| Azman & Tümkaya (2025) | Personalised learning platforms and adaptive assessment tools, utilising machine learning, generate individualised learning experiences. Systems accommodate student needs, analyse interactions to provide personalised recommendations, and adjust question difficulty based on student responses. AI integration through personalised learning platforms and intelligent tutoring systems improves teaching effectiveness and learning outcomes. | Shows the potential of adaptive learning systems. |
| Msambwa et al. (2025) | Systematic literature review of 148 articles from 2021 to 2024 examining the impact of integrating AI tools in higher education. Analysis shows AI tools improve personalised learning and assessments, communication and engagement, and scaffolding performance and motivation. Scaffolding assists students to progress from one developmental stage to the next through individualised support. AI tools help identify the personal Zone of Proximal Development and provide necessary scaffolding, offering step-by-step assistance tailored to individual academic needs. AI tools foster collaborative learning environments by providing peer-learning opportunities and enhancing learner-content interaction. | Supports scaffolding and ZPD theory. Shows AI tools support personalised learning and individualised support. Shows AI's role in scaffolding student development. |
| Looi & Jia (2025) | An empirical study examining personalisation capabilities of ChatGPT chatbots through analysis of 51 graduate students' conversations with a ChatGPT tutor bot. Chatbots facilitate interactive dialogues with students, providing personalised aid and resources aligned with specific learning objectives. Within the Zone of Proximal Development, learners receive support that is gradually withdrawn as they gain mastery, promoting transition from assisted to independent problem-solving. Chatbots create individualised growth areas for students, tailoring interactions to individual needs. Adaptive learning algorithms dynamically adjust instructional content and strategies based on individual learner progress and performance. | Shows chatbot effectiveness in personalised learning. Supports scaffolding and gradual support withdrawal. Shows how chatbots facilitate interactive dialogues and provide personalised support aligned with learning objectives. |
| D.-L. Chen et al. (2025) | Design Science Research examining the design and implementation of an educational chatbot with Personalised Adaptive Learning features for project management training. Built PMTutor chatbot with personalised adaptive learning features. Previous studies have shown that personalised or adaptive learning positively influences learners' motivation, experience, and learning outcomes. Chatbots complement learning and engage learners by providing customised feedback during exercises. Chatbots tailor content and learning paths to learners' characteristics, progress, and objectives. | Supports findings that personalised learning and adaptive learning have a positive influence on motivation, experience, and learning outcomes. Shows chatbot effectiveness in providing customised feedback and engaging learners. Shows how personalised adaptive learning chatbots tailor content to individual learner needs. |
| Abrar et al. (2025) | Six-week quasi-experimental study with 200 students investigating AI-powered learning pathways for personalised learning and dynamic assessments. AI-enabled learning pathways deliver education experiences tailored to learners' needs, addressing their learning level, how they grasp information, learning rates, and learning preferences. Systems adapt learning based on progress and areas of successful learning or learning challenges. The system maintains an optimal learning pace that challenges students without overwhelming them. Adaptive algorithms utilise real-time data to inform the learning process, taking into account student performance, the time it takes to answer questions, and engagement figures. The experimental group demonstrated 25% improvement in performance, completed tasks 25% faster, and showed a 15% increase in engagement compared to the control group. | Supports findings on personalised learning paths and progress tracking. Shows AI-enabled learning pathways tailor experiences to learner needs. Shows how adaptive algorithms adjust instruction to the learning pace and monitor progress. Supports adaptive question generation findings. |

| Authors | Key Findings | Relevance to Research |
|---|---|---|
| Cai et al. (2025) | Systematic literature review of 158 empirical studies from 2021 to 2024 exploring the impact of integrating AI tools in higher education using the Zone of Proximal Development framework. AI tools assist learners in personalising self-assessment through social and technological interactions and effective communication, improving motivation, learning engagement, and learning support, leading to better academic performance, student maturation, and development. AI tools create collaborative learning environments, empowering learners and facilitating meaningful interactions. AI-powered personalised learning platforms are accessible anytime, anywhere, providing students with on-demand support regardless of geographical location or time constraints. | Supports findings on 24/7 availability and accessibility expectations. Shows AI-powered personalised learning platforms accessible anytime, anywhere, with on-demand support regardless of geographical location or time constraints. Shows how AI tools remove time and location barriers for student support. |
| Adewale et al. (2024) | Systematic literature review examining the impact of artificial intelligence adoption on students' academic performance in open and distance learning environments. The review analysed 64 papers from an initial pool of 700, spanning from 2017 to 2023. A meta-analysis revealed that machine learning methods were employed in 29.69% of studies to predict academic achievement. The study found that when AI systems personalise learning experiences and provide timely, relevant feedback, substantial improvements in student performance are noted. The review identified a literature gap in the absence of a process-based framework designed to forecast the precise impacts of AI on education. | Shows that when AI systems personalise learning experiences and provide timely, relevant feedback, substantial improvement in student performance is noted. Supports the hypothesis that AI tools provide personalised support that helps struggling students catch up. Shows the impact of AI adoption on academic performance in distance learning contexts. |

## A.3 Scaffolding, Zone of Proximal Development, and Chatbot Effectiveness

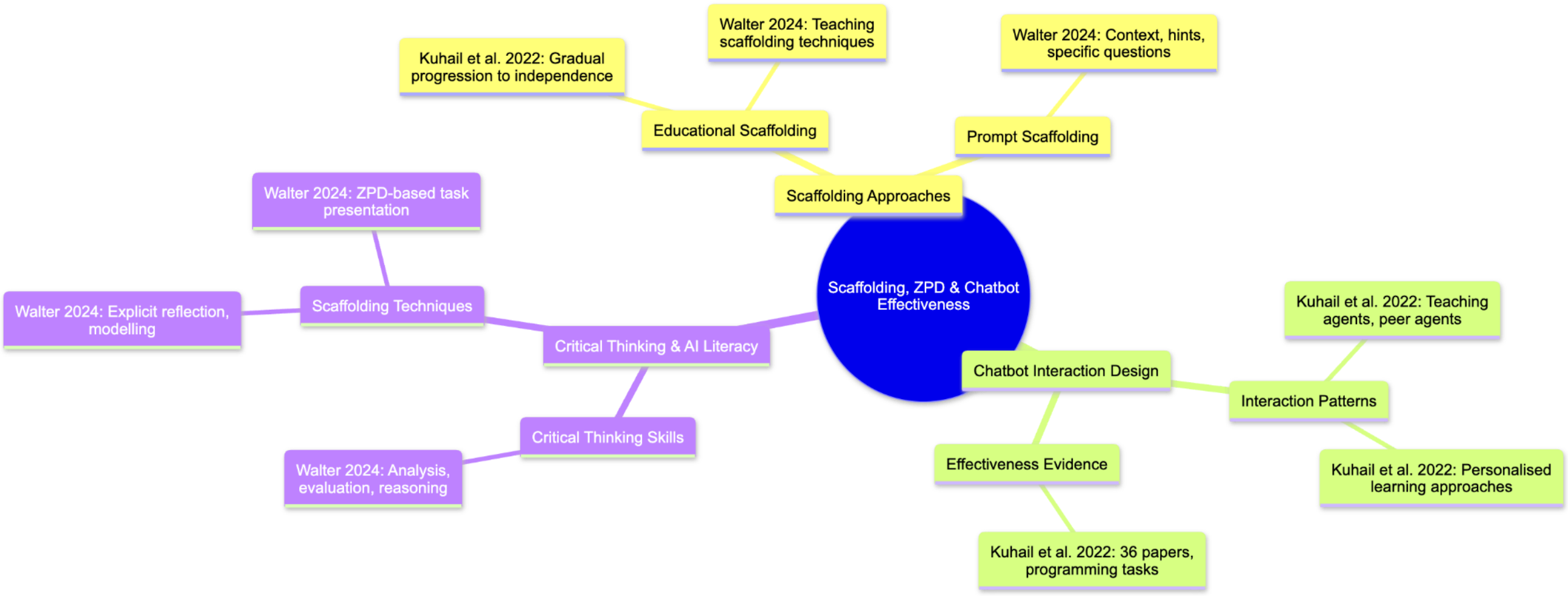

| Authors | Key Findings | Relevance to Research |
|---|---|---|
| Kuhail et al. (2023) | Chatbots engage learners, personalise learning activities, support educators, and develop insight into learners' behaviour. Scaffolding is defined as teaching approaches used to gradually bring students toward better comprehension and independence. Teachers provide successive degrees of temporary support that aid students in reaching comprehension and skill development levels they could not achieve independently. Scaffolding was successfully employed as a learning strategy in chatbots, leading to significant improvements in learning programming tasks. | Supports scaffolding theory. Shows chatbot effectiveness in education. Shows how chatbots provide temporary support to aid students in reaching comprehension levels they could not achieve independently. Supports gradual progression toward independence. |

| Authors | Key Findings | Relevance to Research |
|---|---|---|
| Walter (2024) | Case study and narrative literature review examining the impact of AI in education, with a focus on AI literacy, prompt engineering, and the development of critical thinking skills. Critical thinking in AI education involves the ability to analyse information, evaluate different perspectives, and create reasoned arguments within AI-driven environments. Strategies for fostering critical thinking with AI include scaffolding techniques such as prompt scaffolding, explicit reflection, and modelling. Teaching scaffolding helps foster students' critical thinking skills in digital and AI-driven contexts. Prompt scaffolding involves providing helpful context or hints and asking specific questions. Explicit reflection helps students think through scenarios and potential pitfalls. AI can provide personalised learning experiences that adapt to individual learning styles and abilities, presenting students with tasks within their zone of proximal development to scaffold learning experiences and enhance critical thinking. | Supports findings on critical thinking in digital and AI-driven contexts. Shows scaffolding techniques that help foster students' critical thinking skills. Shows how AI can scaffold learning experiences to enhance critical thinking. |

## A.4 User-Centred Design and Co-Design

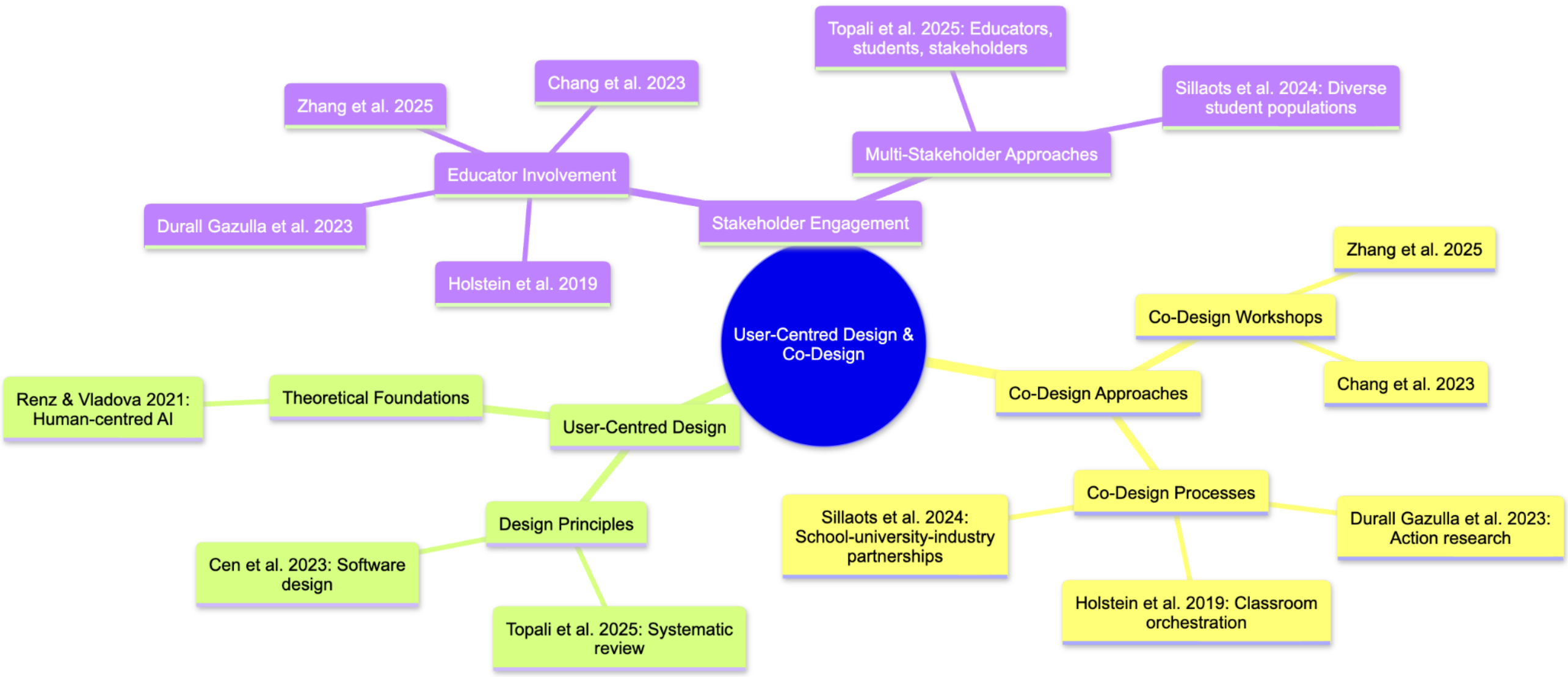

| Authors | Key Findings | Relevance to Research |
|---|---|---|
| D. H. Chang et al. (2023) | Study examining educational design principles for AI chatbots supporting self-regulated learning. Many AI educational tools are developed without sufficient input from end-users, resulting in solutions that fail to address real classroom challenges. Co-design workshops with educators create chatbots that address pedagogical challenges. | Shows co-design and co-development approaches. Shows the importance of educator involvement in development. |
| Durall Gazulla et al. (2023) | Action research analysing chatbot co-design processes. Involving educators throughout development phases, from initial requirements gathering to prototype testing and refinement, leads to more relevant and trustworthy educational tools. Many chatbot projects lack comprehensive stakeholder engagement. | Shows co-design and co-development approaches with educators. Shows evidence for comprehensive stakeholder engagement. |

| Authors | Key Findings | Relevance to Research |
|---|---|---|
| Holstein et al. (2019) | Study examining co-designing real-time classroom orchestration tools to support teacher-AI complementarity. Provides evidence for collaborative design approaches that involve educators throughout the development process. | Supports co-design and co-development approaches with educators. |
| (Sillaots et al., 2024) | Study examining co-creation of learning technologies in school-university-industry partnerships from an activity system perspective. Advocates for collaborative, user-centred approaches that prioritise educator input throughout the chatbot development process. Demonstrates approaches for developing solutions for diverse student populations. | Shows co-design and co-development approaches. Shows collaborative user-centred approaches. Shows diverse student populations. |
| Topali et al. (2025) | Systematic literature review examining the design of human-centred learning analytics and artificial intelligence in education solutions. User-centred design in educational technology development benefits from involving educators, students, and other stakeholders throughout all design phases. | Shows user-centred design approaches. Shows stakeholder involvement throughout design phases. |
| Cen et al. (2023) | Study examining user-centred software design and user interface redesign for AI educational software. Provides evidence supporting stakeholder involvement in the creation of effective AI-driven educational tools. | Supports user-centred design and stakeholder involvement approaches. |
| (Renz & Vladova, 2021) | Commentary reinvigorating discourse on human-centred artificial intelligence in educational technologies. Provides a theoretical foundation for human-centred approaches to AI in education. | Supports human-centred design and user-centred approaches. |
| D. Zhang et al. (2025) | Study exploring relationships between AI literacy, AI trust, AI dependency, and 21st-century skills among preservice mathematics teachers. AI dependency is defined as an excessive reliance on AI, which diminishes autonomous thinking and learning abilities. AI dependency has been shown to have significant negative relationships with critical thinking, self-confidence, and problem-solving ability. AI literacy and AI trust influence dependency on Gen-AI. Co-design workshops with educators help create chatbots that address pedagogical challenges and incorporate user-centred design principles. | Shows co-design and co-development approaches with educators. Shows dependency concerns. Shows vocational education challenges. |

## A.5 Implementation Challenges and Educator Perspectives

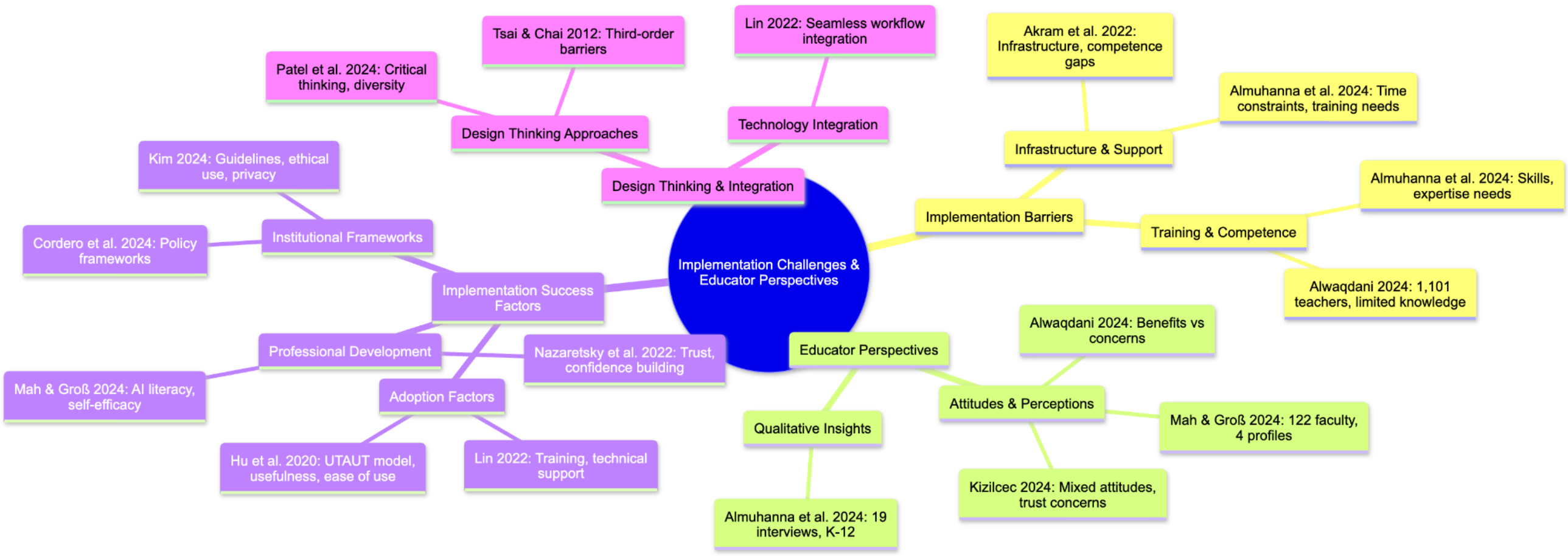

| Authors | Key Findings | Relevance to Research |
| --- | --- | --- |
| Akram et al. (2022) | Systematic review examining teachers' perceptions of technology integration. Infrastructure limitations and insufficient support were identified as primary barriers to adoption. | Shows implementation challenges and barriers relevant to understanding educator adoption of Generative AI technologies. |
| Kizilcec (2024) | Commentary arguing that to advance AI use in education, the focus should be on understanding educators. Educators express mixed attitudes towards AI technologies, showing optimism about their potential, while voicing concerns about trust, transparency, and the need for professional development. | Shows the importance of understanding educator perspectives. Identifies implementation challenges. |
| Alwaqdani (2025) | Quantitative research with 1,101 Saudi teachers investigating teachers' perceptions of AI tools in education. Found that 64% of teachers have used AIED, with 40% reporting limited knowledge. Teachers acknowledge AIED's potential to save time, assist in designing enriching activities, and personalise learning. Concerns exist regarding the effort required for training, the limited time allocated for training, concerns about creativity and critical thinking, and fears of job displacement. Teachers value the human touch, empathy, and nuanced understanding that human educators bring to the classroom. | Shows teacher perceptions and implementation challenges. Shows teachers acknowledge the benefits of AI while highlighting concerns about the human touch, training needs, and potential negative effects. |
| Tsai & Chai (2012) | The study examines third-order barriers to technology-integration instruction, identifying the need for design thinking skills among educators to integrate technology into their pedagogical practice effectively. Educators with strong design thinking capabilities can overcome contextual challenges and effectively integrate technologies into their teaching practices. | Shows the theoretical foundation for design thinking approaches. Shows understanding of third-order barriers to technology integration. |
| Patel et al. (2024) | The study examines the impact of integrating critical and design thinking in design innovation education. Documents challenges in incorporating interdisciplinary approaches into curricula whilst addressing diversity and accessibility issues. Examines how design thinking applies to chatbot implementation. | Supports understanding of design thinking approaches. Shows challenges in addressing diversity and accessibility. |
| Cordero et al. (2024) | The study examines the integration of Generative AI in higher education and best practices. Provides frameworks for developing institutional policies around conversational AI in educational contexts. | Supports understanding of implementation approaches. Shows institutional policy frameworks for Generative AI. |
| Hu et al. (2020) | Study examining factors that affect academics' adoption of emerging mobile technologies from the perspective of the extended Unified Theory of Acceptance and Use of Technology. Educator perspectives and institutional contexts significantly influence success, with perceived usefulness, ease of use, and institutional support serving as key predictors of adoption. | Shows educator technology adoption factors. Shows predictors of successful implementation. |
| Lin (2022) | The study examines the influence of AI on teaching effectiveness in education, with a mediating effect of teachers' perceptions of educational technology. Successful workload reduction depends on adequate training, technical support, and seamless integration into existing educational workflows. | Shows implementation success factors. Shows the relationship between AI tools and teaching effectiveness. Relevant to chatbot implementation. |
| Nazaretsky et al. (2022) | The study examines teachers' trust in AI-powered educational technology and professional development programmes to improve it. Professional development and ongoing support are important for building educator confidence and competence in utilising chatbot technologies. | Shows implementation success factors, particularly educator trust and professional development needs. Relevant to chatbot implementation. |

| Authors | Key Findings | Relevance to Research |
|---|---|---|
| (Kim, 2024) | Study examining leading teachers' perspectives on teacher-AI collaboration in education. Institutional guidelines that address ethical use, data privacy, and academic integrity are crucial for the successful implementation of chatbots. | Shows implementation success factors. Shows educator perspectives on AI collaboration. Relevant to chatbot implementation. |
| Almuhanna (2025) | Qualitative study with 19 semi-structured interviews investigating teachers' perspectives of integrating AI-powered technologies in K-12 education. Reliability and accuracy of generative AI information are questioned, posing challenges to academic integrity and credibility. Many teachers require enhanced skills, expertise, and supportive institutions to effectively adopt AI. Teachers encounter time-based limitations when integrating AI technologies, with the iterative process of evaluating and modifying AI-generated content being time-consuming. Concerns have been raised about over-reliance on AI tools reducing cognitive abilities and human interaction. | Supports findings on accuracy concerns and challenges to academic integrity. Shows reliability and accuracy concerns with generative AI. Shows implementation challenges, including time constraints and the need for teacher training. |
| Mah & Groß (2024) | Quantitative study with 122 faculty members examining faculty perspectives on use of artificial intelligence in higher education, focusing on AI self-efficacy, perceived benefits, challenges, interests, and professional development needs. Latent class analysis revealed four distinct faculty member profiles: optimistic, critical, critically reflective, and neutral. Respondents saw greater equity in education as AI's greatest benefit, while the lack of AI literacy was among the greatest challenges. Faculty members' views on the adoption and use of AI, as well as their interest in professional development, are crucial for the meaningful integration of AI. Faculty members' preparedness, proficiency, and self-efficacy in AI are pivotal for ensuring students acquire the necessary skills. | Shows faculty perspectives on AI in higher education. Shows the importance of AI self-efficacy and professional development needs. Supports findings on the need for adequate training and support for educators. |

## A.6 Critical Thinking and Negative Effects

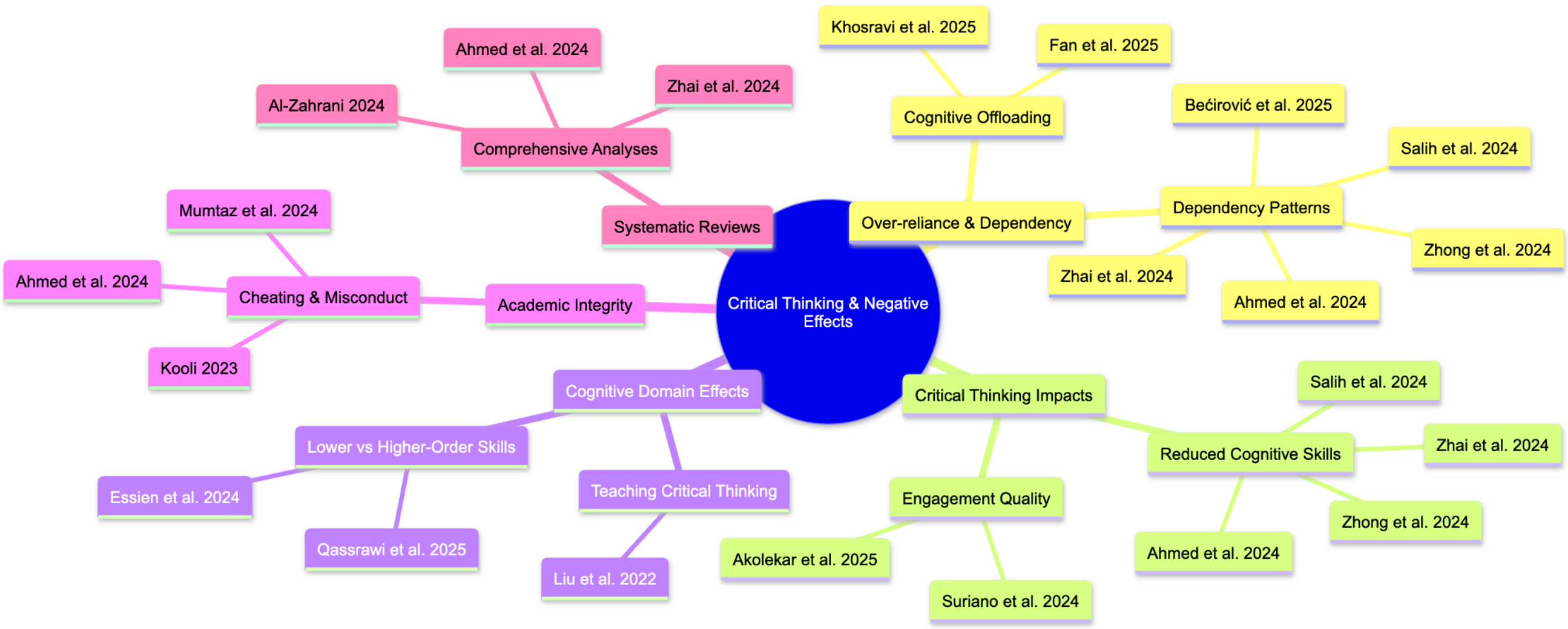

| Authors | Key Findings | Relevance to Research |
|---|---|---|
| Fan et al. (2025) | Randomised experimental study comparing learners' motivations, self-regulated learning processes and learning performances on a writing task among different groups with support from ChatGPT, human experts, writing analytics tools, and no additional tool. Cognitive offloading, where learners delegate cognitive tasks to external tools to reduce cognitive effort, may lead to a decrease in internal cognitive engagement over time, impacting learners' ability to self-regulate and critically engage with learning materials. The metacognitive laziness phenomenon has been documented, where learners become dependent on AI and offload metacognitive load, reducing effective connections and engagement in metacognitive processes. | Supports findings on cognitive offloading and negative impacts on critical thinking. Demonstrates how AI dependency can affect learners' ability to self-regulate and critically engage with learning materials. |
| Zhai et al. (2024) | Systematic review examining the impact of students' over-reliance on AI dialogue systems on their critical cognitive capabilities, including decision-making, critical thinking, and analytical reasoning. Overreliance on AI tools can lead to a decline in critical and analytical thinking skills, especially when students become overly dependent on AI-generated content. Over-reliance may lead to diminished creativity, increased dependency, and challenges in understanding. Over-reliance can have a negative impact on analytical thinking abilities, leading to the uncritical acceptance of biased or inaccurate AI-generated content. | Supports findings on over-reliance and negative impacts on critical thinking. Shows how over-reliance on AI tools can lead to reduced critical and analytical thinking skills. Shows concerns about diminished creativity and dependency. |
| (Ahmed et al., 2024) | Systematic literature review following PRISMA guidelines synthesising results on potential and drawbacks of Generative AI in educational domains, including a survey with 200 undergraduate university students. Over-dependence on AI technologies could negatively impact the development of essential cognitive skills, decreasing students' capacity for original thought, creativity, and independent problem-solving. Potential for AI to facilitate plagiarism and undermine the cultivation of creativity and critical thinking skills. The study also identified opportunities, including personalisation, task automation, teacher assistance, and efficiency. | Supports findings on negative impacts on critical thinking and cognitive skills. Shows how over-dependence on AI can decrease capacity for original thought and independent problem-solving. Also supports 24/7 availability findings. |
| Y. Liu & Pásztor (2022) | Meta-analysis synthesising 50 relevant empirical studies from 2000 to 2021 with 5,210 participants examining the effectiveness of Problem-Based Learning instructional intervention on Critical Thinking in higher education. Teaching Critical Thinking to undergraduates is a major academic challenge due to difficulties of embedding Critical Thinking into a well-organised curriculum, infusing valid tasks, and utilising effective teaching strategies. A meta-analysis revealed that the Problem-Based Learning instructional intervention on Critical Thinking is effective, with an overall effect size of 0.640. | Supports findings on teaching critical thinking as a major academic challenge. Shows difficulties in embedding critical thinking into well-organised curricula and utilising effective teaching strategies. |
| Essien et al. (2024) | Mixed-method research with 107 participants investigating the influence of generative artificial intelligence on critical thinking skills in UK postgraduate business school students using Bloom's taxonomy. The most significant improvements occurred at lower levels of Bloom's taxonomy, suggesting that AI assistance may be more effective for knowledge-based recall tasks than for higher-order thinking skills that require analysis, synthesis, and evaluation. The study identifies concerns related to reliability, accuracy, and potential ethical implications. | Supports findings on AI tools having different effects on lower-order versus higher-order thinking skills. Shows AI assistance being more effective for knowledge-based recall tasks than for higher-order thinking skills. Shows how AI impacts vary across cognitive domains. |

| Authors | Key Findings | Relevance to Research |
|---|---|---|
| Qassrawi & Al Karasneh (2025) | Meta-analysis of 60 studies from 2018 to 2024 exploring the role of AI applications in language teaching. AI could make significant strides in improving language proficiency, particularly in areas such as pronunciation, conversational fluency, grammar, and vocabulary, which align with lower-order cognitive skills. However, AI struggles to cultivate essential human skills, such as cultural sensitivity, emotional intelligence, and critical thinking, which are necessary for effective communication and higher-order thinking. AI assistance may be more effective for lower-order cognitive tasks than for higher-order thinking skills requiring critical analysis and cultural understanding. | Supports findings on AI tools having different effects on lower-order versus higher-order thinking skills. Shows AI being more effective for lower-order cognitive tasks than for higher-order skills. Shows how AI impacts vary across cognitive domains. |
| Bećirović et al. (2025) | The study found that chatbots and AI can positively influence self-efficacy when users experience success with technology. However, rapid AI technology development may interfere with perceived self-efficacy, as the demand to continuously improve skills creates pressure that leads to negative perceptions. Students utilise AI tools to reduce workload, viewing AI as a form of support, and are critical of output quality. However, overreliance on AI tools may negatively impact performance. Incorrect or non-existent references from AI tools have caused significant scepticism. | Shows the complexity of AI's impact on self-efficacy. Shows potential negative effects of overreliance. |
| Salih et al. (2025) | Comprehensive analysis of ethical considerations of AI tool usage in higher education, examining the ethical use of AI tools like ChatGPT. The growing use of ChatGPT could lead to over-reliance on AI, potentially hindering students' critical thinking and problem-solving skills, as they may avoid essential cognitive processes such as reflection, creativity, and independent reasoning. Overdependence can hinder the development of key academic and life skills, impacting students' intellectual growth and their preparedness for real-world challenges. | Supports findings on over-reliance and negative impacts on critical thinking. Shows how over-reliance on AI can hinder critical thinking and problem-solving skills. Shows concerns about diminished cognitive processes and intellectual growth. |
| Mumtaz et al. (2025) | The study examines the ethical use of digital and artificial intelligence-based tools in higher education, investigating the integration of AI and its implications for academic integrity. AI tools have posed new challenges to academic integrity. Tools such as ChatGPT-4o have made it easier for students to engage in dishonest practices, with accessibility and sophistication enabling students to generate assignments and answers effortlessly, thereby undermining the principles of academic honesty. Recent studies indicate a higher prevalence of dishonesty in online assessments compared to traditional face-to-face assessments. | Supports findings on academic integrity concerns and exam fairness. Shows how AI tools enable dishonest practices and undermine academic honesty. Shows concerns about students generating assignments effortlessly. |
| Kooli (2023) | Qualitative methodology and a comprehensive review examining the ethical implications of chatbots in education and research. Using chatbots or other artificial intelligence tools to answer exam questions constitutes a form of cheating and academic misconduct, which contradicts the fundamental principles of learning and academic integrity. If misused by students, chatbots could raise serious ethical concerns and significantly impact students' academic progress by hindering their critical thinking skills, creativity, and ability to apply concepts learned to real-world situations. | Supports findings on academic integrity concerns and negative impacts on critical thinking. Shows chatbots being used for cheating and academic misconduct. Shows how misuse can hinder critical thinking skills and academic progress. |

| Authors | Key Findings | Relevance to Research |
|---|---|---|
| Zhong et al. (2024) | A quantitative study involving 958 university students examines how personal attributes influence AI dependency in Chinese higher education from a needs-frustration perspective, utilising structural equation modelling. The study applied the Interaction of Person-Affect-Cognition-Execution model and the Basic Psychological Needs theory. Personality traits such as neuroticism, self-critical perfectionism, and impulsivity contribute to AI dependency through frustration of needs, negative academic emotions, and unrealistic performance expectations. Reliance on AI can hinder the development of critical thinking skills, as students may become accustomed to seeking quick answers rather than engaging in more in-depth analytical processes. | Shows that reliance on AI can impede the development of critical thinking skills. Shows how students may become accustomed to seeking quick answers rather than engaging in deeper analytical processes. Supports findings on AI dependency having a substantial negative impact on critical thinking. |
| Akolekar et al. (2025) | Mixed-methods study evaluating the effectiveness of three leading generative AI tools in undergraduate mechanical engineering education. Performance is assessed on 800 questions spanning seven core subjects. All three AI tools demonstrated strong performance in theory-based questions but struggled with numerical problem-solving, particularly in areas that require deep conceptual understanding and complex calculations. The study found that students may spend time verifying AI-generated content rather than engaging in productive learning activities. The quality of engagement with AI tools may be more important than the quantity of engagement. | Supports findings that students may spend time verifying AI-generated content rather than engaging in productive learning activities. Shows concerns about AI reliability and impact on problem-solving abilities. Shows how the quality of engagement with AI tools may be more important than quantity. |
| Suriano et al. (2025) | Quantitative study with 213 Italian university students investigating student interaction with ChatGPT and its relationship with complex critical thinking skills. Engagement proved to have a particularly significant impact on critical thinking performance compared to knowledge. The study found that many students accept inaccurate answers and use copy-paste without critical evaluation, suggesting that interaction with ChatGPT can facilitate critical thinking only when there is adequate engagement that stimulates deeper reflection. | Supports findings on students accepting inaccurate answers and using copy-paste without critical evaluation. Shows how interaction with ChatGPT can facilitate critical thinking only when there is adequate engagement that stimulates deeper reflection. Shows the importance of engagement quality in promoting critical analysis of AI-generated content. |
| Khosravi et al. (2025) | Special issue introduction examining the intersection of generative AI and learning analytics in education. While GenAI tools offer new avenues for personalised learning, enhanced feedback, and increased efficiency, they also present challenges related to cognitive engagement, student agency, and ethical considerations. Metacognitive laziness poses challenges for the development of independent learning and cognitive processes, including critical thinking and problem-solving. Need to re-examine boundaries of learning analytics in the presence of GenAI while maintaining foundational principles from human-centred design, ethical standards, and privacy frameworks. | Supports findings on metacognitive laziness, which pose challenges for the development of independent learning and cognitive processes. Shows how GenAI may affect cognitive engagement and student agency. It highlights the need to strike a balance between the potential benefits of AI and responsible educational practices. |
| Al-Zahrani (2024) | Multi-phase sequential exploratory study investigating negative implications of integrating AI in education. A systematic literature review yielded 56 relevant studies, followed by a survey with 260 participants. Findings confirm concerns about the loss of human connection, data privacy and security, algorithmic bias, lack of transparency and explainability, reduced critical thinking and creativity, unequal access, ethical considerations, and the need for teacher professional development. The complexity of AI can raise concerns about accountability and fairness, and its black-box nature may limit awareness of the factors influencing educational experiences. | Shows that AI complexity can raise concerns about accountability and fairness. Demonstrates how the black-box nature of AI may limit awareness of factors that influence educational experiences. Supports findings on educators' concerns about the lack of transparency in AI systems. |

## A.7 Student Support and Inclusive Education

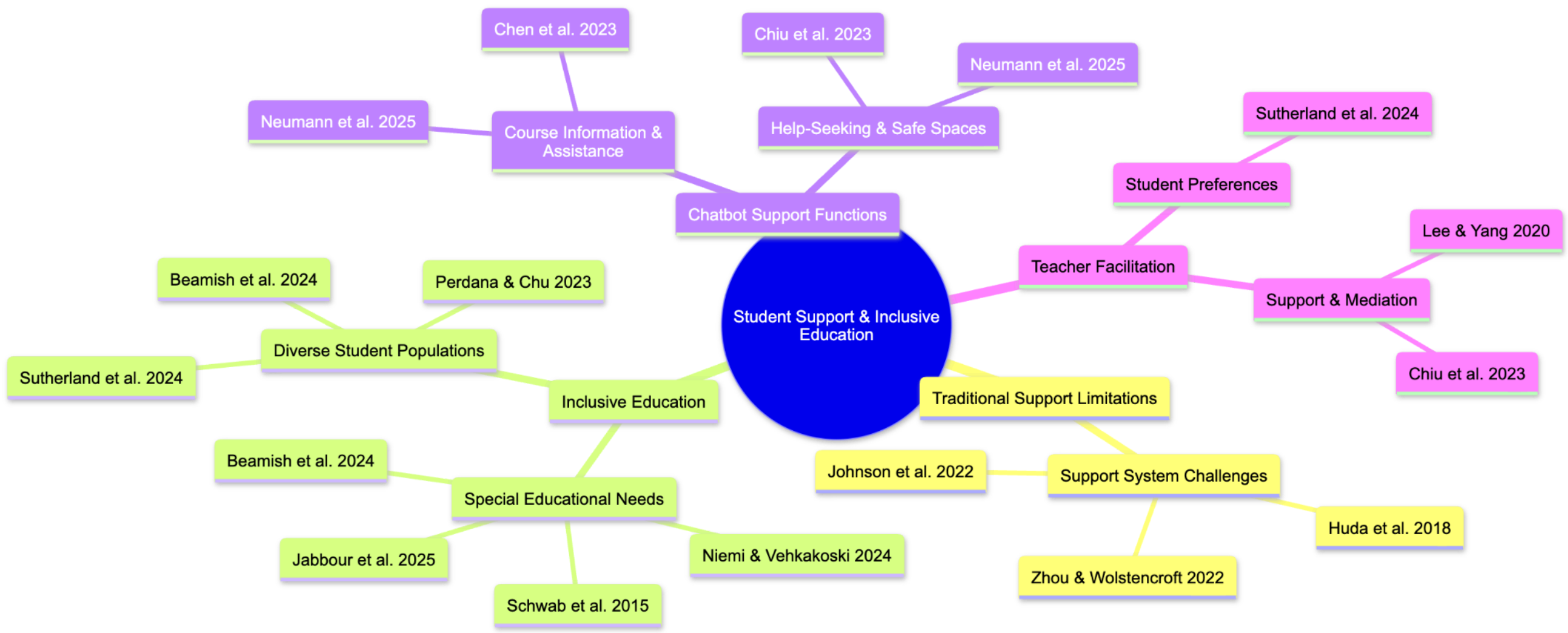

| Authors | Key Findings | Relevance to Research |
|---|---|---|
| Huda et al. (2018) | The study examines the modern learning environment in the era of big data. Traditional support systems often struggle to address students' immediate needs. | Shows limitations of traditional support mechanisms. |
| Zhou & Wolstencroft (2022) | The study examines the development of an organic online support community for Chinese students using WeChat. Office hours and email systems operate within fixed availability constraints and institutional schedules. Describes co-creation methodology for developing student support solutions. | Shows co-creation methodology. Shows limitations of traditional support mechanisms. Shows building support communities. |
| Johnson et al. (2022) | Study examining student support in higher education, including campus service utilisation, impact, and challenges. Multiple interconnected barriers in educator practice were identified, including time and workload constraints, resource limitations, and difficulties in providing timely, personalised feedback to diverse student populations. | Shows educator barriers and workload constraints relevant to the research context. |
| Perdana & Chu (2023) | The study examines the assessment of students' learning during the pandemic and their responses to the crisis period at Singapore's higher education institutions. Additional pressures from adapting to hybrid learning environments and managing diverse international student populations with varying English proficiency levels. Challenges related to diverse student populations, multiple language proficiencies, and the requirements of inclusive education. | Shows the Singapore context. Shows challenges in local higher education institutions. Shows diverse student populations. |
| Beamish et al. (2024) | The study examines the progress of inclusive education for students with special educational needs in the Asia-Pacific region. Challenges educators face in balancing multiple learning needs simultaneously in inclusive educational contexts. | Shows challenges in Singapore's inclusive education context. Shows diverse student populations. |
| Niemi & Vehkakoski (2024) | The study examines turning social inclusion into exclusion during collaborative learning between students with and without SEN. Difficulties with traditional classroom dynamics, including non-verbal communication challenges and hyperfixation behaviours that can disrupt lesson flow. | Shows challenges in inclusive education contexts. Shows diverse student populations. |

| Authors | Key Findings | Relevance to Research |
| --- | --- | --- |
| Schwab et al. (2015) | The study examines the relationship between self-rated social inclusion and social behaviour among students with and without special education needs in secondary schools. Students with SEN often experience lower social participation and may feel excluded despite being physically present in mainstream classrooms. | Shows social dimensions of inclusion. Shows the challenges that students with SEN face. |
| Sutherland et al. (2024) | Study examining non-traditional students' preferences for learning technologies and their impacts on academic self-efficacy. Students often prefer peer or informal support networks over formal institutional services. | Shows student support preferences. Shows self-efficacy impacts. |
| Jabbour et al. (2025) | Study examining a Generative AI-powered learning companion for personalised education and broader accessibility, including accommodations for students with special educational needs. | Shows how Generative AI can support students with special educational needs in inclusive education contexts. |
| Y. Chen et al. (2023) | Two sequential exploratory studies with 215 and 195 students investigating opportunities, challenges, efficacy, and ethical concerns of using chatbots as pedagogical tools in business education. Participants frequently needed answers to basic information about courses, including course materials, due dates, study tips, and office hours. Chatbots could provide basic course information and content, serving as a helpful addition, particularly in classes with high student-teacher ratios or when students are too shy to communicate directly with instructors. | Supports findings that students repeatedly ask similar questions and require basic course information. Shows students frequently needing answers to basic information about courses. Shows how chatbots can address repetitive questions and basic information needs. |
| Neumann et al. (2025) | Research with 46 students exploring the benefits and challenges of developing, deploying, and evaluating LLM-driven chatbot MoodleBot in computer science classroom settings. LLM-based chatbots can significantly improve the teaching and learning process, with a high accuracy rate in providing course-related assistance. Chatbots have the potential to address help-seeking concerns and alleviate help avoidance behaviour by providing a private, nonjudgmental space for students to ask questions without fear of embarrassment in front of peers or teachers. | Supports findings on safe learning spaces and help-seeking behaviour for weaker students. Shows chatbots providing private, nonjudgmental spaces. Shows how chatbots can help address help-seeking concerns and reduce social anxiety. |
| Chiu et al. (2023) | Experimental study with 123 Grade 10 students examining how teacher support moderates effects of student expertise on needs satisfaction and intrinsic motivation to learn with AI chatbots, using the self-determination theory framework. Intrinsic motivation and competence in learning with a chatbot depend on both teacher support and student expertise. Teachers support a more satisfied need for relatedness and a less satisfied need for autonomy. The teacher's role as a core facilitator of student motivation and academic progress in mediating and supporting learning with AI technologies provides insight into how such technologies can be effectively utilised in practice. | Shows the importance of teacher support in mediating and supporting learning with AI technologies. Shows how teacher support moderates the effects of student expertise on needs satisfaction and intrinsic motivation. Supports findings on educators playing a crucial role in facilitating student engagement with AI tools. |
| Lee & Yang (2023) | Qualitative study examining effective collaborative learning from students' perspectives in a teacher-training course. Facilitative teacher approaches, where teachers guide students' inquiry processes and encourage questioning, lead to greater student engagement, deeper understanding, and improved critical thinking skills compared to transmissive approaches. When teachers act as facilitators who support knowledge construction through dialogue and reflection, students participate actively and engage in the active construction of knowledge. | Supports constructivist pedagogical approaches and teacher facilitation roles. Demonstrates facilitative approaches that lead to student engagement and the development of critical thinking. |

### A.8 Self-Efficacy, Engagement, and Academic Performance

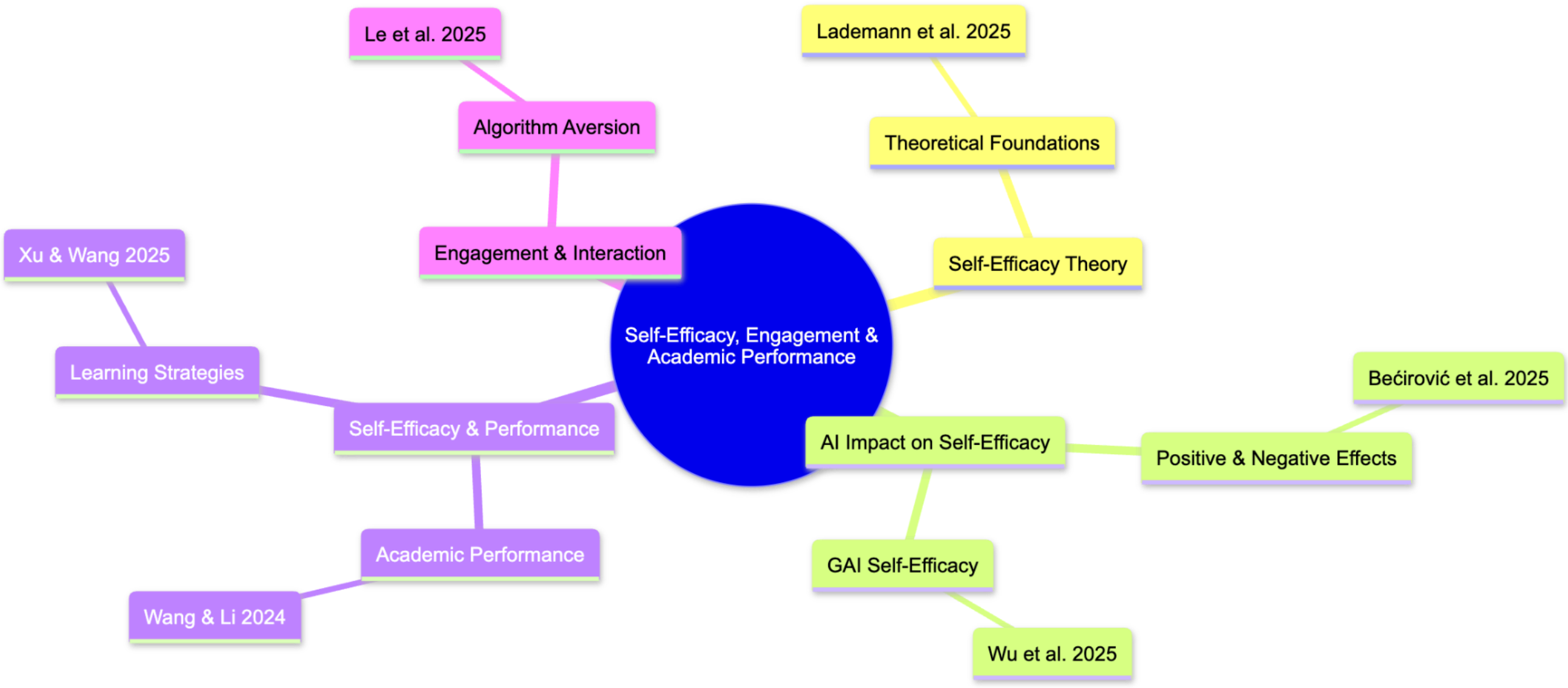

| Authors | Key Findings | Relevance to Research |
|---|---|---|
| Lademann et al. (2025) | A theoretical and empirical study defines self-efficacy expectation as a person's subjective conviction that they can perform specific actions successfully. Students with higher self-efficacy set more challenging goals, persist when facing difficulties, and invest more time in solving complex problems. Higher self-efficacy supports deeper learning, improves performance, and encourages motivation to engage with novel and challenging tasks. | Shows the theoretical foundation for the concept of self-efficacy. |
| Bećirović et al. (2025) | The Study found that chatbots and AI can positively influence self-efficacy when users experience success with technology. However, rapid AI technology development may interfere with perceived self-efficacy, as the demand to continuously improve skills creates pressure that leads to negative perceptions. Students utilise AI tools to reduce workload, viewing AI as a form of support, and are critical of output quality. However, overreliance on AI tools may negatively impact performance. Incorrect or non-existent references from AI tools have caused significant scepticism. | Shows the complexity of AI's impact on self-efficacy. Shows potential negative effects of overreliance. |
| Xu & Wang (2025) | The study reveals that university students exhibit lower self-efficacy in intermediate information literacy compared to those at the basic and advanced levels of information literacy. Basic and intermediate information literacy self-efficacy did not predict the use of cognitive, metacognitive, environmental, and emotional regulation strategies. Advanced information literacy self-efficacy significantly predicted strategy use, suggesting that lower self-efficacy levels may not facilitate effective use of learning strategies. | Shows the relationship between self-efficacy levels and the effectiveness of learning strategies. |
| Wang & Li (2024) | Quantitative study with 1,082 Chinese music students investigating the relationship between music students' self-efficacy, academic performance, and artificial intelligence readiness. Notably, 63% of variations in students' academic performance are attributed to the combined influence of self-efficacy and AI readiness. Students' self-efficacy uniquely predicts 52% of changes in academic performance, whilst their AI readiness uniquely predicts 60% of changes in academic performance. | Shows that students' AI readiness significantly contributes to their academic performance. It shows that self-efficacy plays a crucial role in overall academic achievement. Supports findings that students demonstrate more confidence and appear more prepared for exams. |

| Authors | Key Findings | Relevance to Research |
|---|---|---|
| H. Wu et al. (2025) | Quantitative study with 756 students examining understanding of GAI risk awareness among higher vocational education students from an AI literacy perspective, using Partial Least Squares Structural Equation Modelling. The Study explored the impact of AI literacy, comprising AI knowledge, AI skills, and AI attitudes, on GAI risk awareness, with GAI self-efficacy acting as a mediating factor. Whilst components of AI literacy do not directly influence GAI risk awareness, AI literacy significantly impacts GAI self-efficacy, which in turn positively affects GAI risk awareness. GAI self-efficacy serves as a full mediator between AI literacy and GAI risk awareness. | Shows that hands-on engagement with AI technologies may be an effective way to improve critical thinking and risk assessment capabilities. Shows that increasing self-efficacy enables students to engage more confidently and critically with GAI technologies. Supports findings on the importance of practical engagement and experience-based enhancement. |
| Le et al. (2025) | Experimental study with 114 university students examining learners' preferences for learning feedback from generative AI versus human tutors, investigating algorithm aversion in educational contexts. Results revealed a strong initial preference for human tutors. However, while the general preference for human tutors persisted, learners' preference for a free-dialogue interface increased, whereas the structured AI interface reinforced their preference for human tutors. Appropriate interaction design can mitigate algorithm aversion. Free-dialogue interfaces suggest that overcoming algorithm aversion may depend more on creating natural, flexible interaction experiences than purely technical optimisation. | It shows that algorithm aversion can be effectively mitigated in educational contexts through the design of appropriate interactions. Shows that positive experiences with AI systems can increase comfort with AI. Supports findings on the importance of interaction design in educational AI tools. |

## A.9 Academic Integrity, Trustworthy AI, and Accuracy

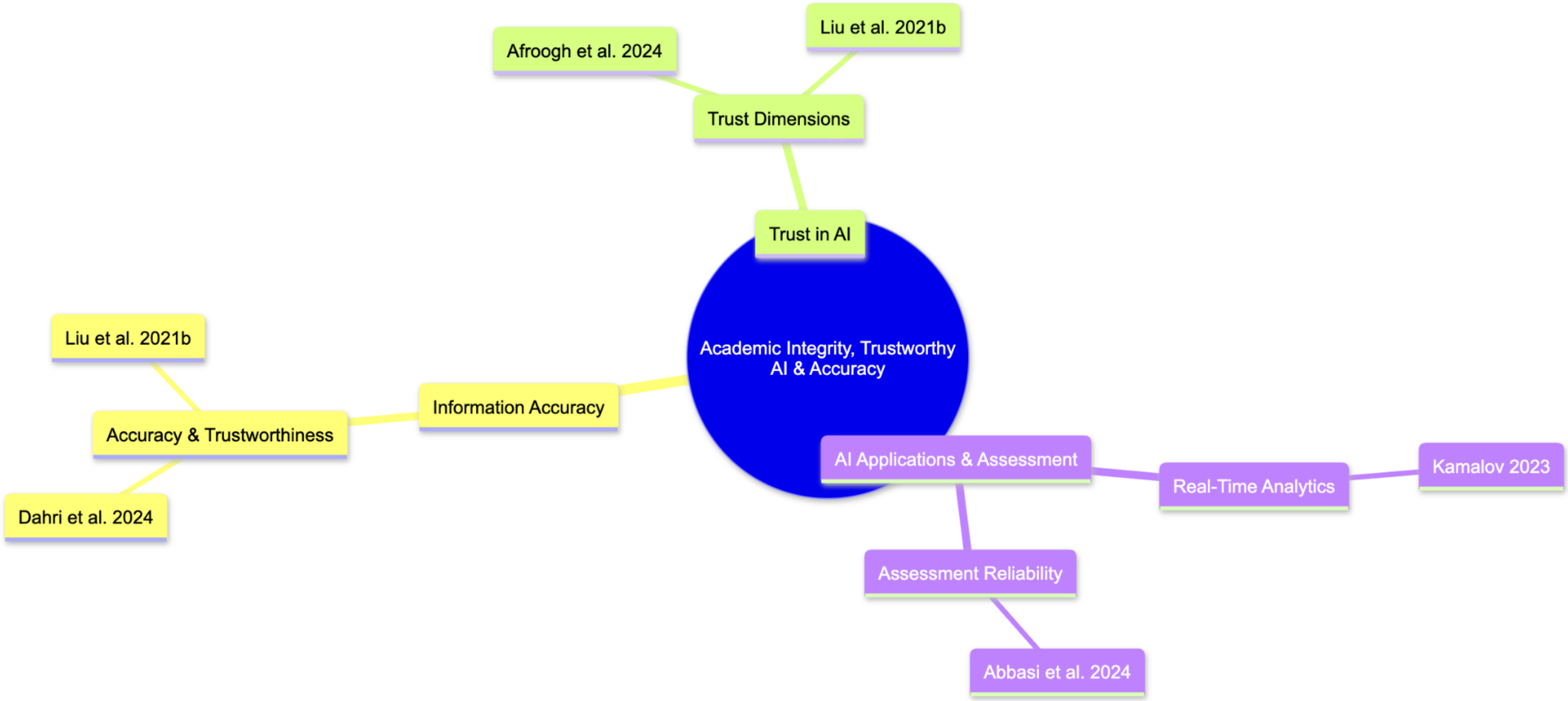

| Authors | Key Findings | Relevance to Research |
|---|---|---|
| Ahmed Dahri et al. (2025) | Quantitative research with 305 students investigating AI-based academic support acceptance and its impact on students' performance. Information accuracy significantly affects students' intention to use AI tools positively, with students tending to rely more on insights and data offered by AI tools when they consistently provide accurate and reliable information. The perceived accuracy of information provided by AI tools plays a pivotal role in influencing students' willingness to engage with and utilise AI tools effectively. | Supports findings on accuracy and trustworthiness as fundamental requirements. Shows how information accuracy affects students' intention to use AI tools positively. Shows the importance of reliable and accurate information in AI educational tools. |

| Authors | Key Findings | Relevance to Research |
|---|---|---|
| H. Liu et al. (2023) | Comprehensive survey on trustworthy AI from a computational perspective, focusing on six crucial dimensions, including safety and robustness, non-discrimination and fairness, explainability, privacy, accountability and auditability, and environmental well-being. Trustworthy AI is defined as programs and systems built to solve problems like humans, which bring benefits and convenience to people without posing any threat or risk of harm. From a technical perspective, trustworthy AI is expected to show properties of accuracy, robustness, and explainability. AI programs or systems should generate accurate output consistent with ground truth as much as possible. AI programs or systems should be robust to changes so that perturbations would not affect the model outcome. Trustworthy AI must allow for explanation and analysis by humans, so that potential risks and harm can be minimised. Trustworthy AI should be transparent so people can better understand its mechanism. | Supports findings on trustworthy AI systems and accuracy requirements. Shows technical properties of trustworthy AI, including accuracy, robustness, and explainability. Shows the importance of reliable AI systems in educational contexts. |
| Kamalov et al. (2023) | Literature review examining the potential impact of AI on education across three major axes: applications, advantages, and challenges. Collaboration between teachers and students can be facilitated by AI, enhancing the overall learning experience. By providing real-time analytics and insights, AI can help educators identify students' strengths, weaknesses, and learning patterns, allowing them to adjust teaching strategies accordingly. In-situ assessments and instant feedback allow teachers to make real-time adjustments during class. AI can notify teachers when students are struggling while providing possible remedies. As a brainstorming partner, AI can help identify effective solutions to support student learning and learning outcomes. | Supports findings on educator oversight and teacher dashboards. Shows teachers wanting real-time analytics and insights to identify student strengths, weaknesses, and learning patterns. Demonstrates how AI can help teachers adjust their teaching strategies accordingly. |
| Afroogh et al. (2024) | Comprehensive systematic literature review examining trust in artificial intelligence. Review conceptualised trust in AI, investigated trust in different types of human-machine interaction, and explored its impact on technology acceptance across various domains. The study proposed a taxonomy of technical and non-technical axiological trustworthiness metrics, along with trustworthy measurements. The review examined major trust-breakers in AI and trust makers. Findings revealed that unpredictability is a significant issue contributing to distrust in AI, with concerns about unpredictability manifesting in both global and local forms. AI complexity can raise concerns about accountability and fairness, and the black-box nature of AI may limit awareness of factors influencing experiences, making decision-making difficult to understand. | Shows that unpredictability is an issue contributing to distrust in AI. Shows concerns about unpredictability coming in both global and local varieties. Supports findings on AI complexity, raising concerns about accountability and fairness. Demonstrates how the black-box nature of AI may limit awareness of factors that influence educational experiences. |
| Abbasi et al. (2025) | Quantitative study exploring the impact of artificial intelligence on curriculum development in global higher education institutions. The frequent use of AI, the extent of faculty knowledge, institutional support for faculty, and future expectations about AI are all promoting curriculum development. The study established that values above 0.70 for Cronbach's alpha are generally considered acceptable for most assessments, providing a standard threshold for internal consistency reliability in educational assessments. The effectiveness of AI-driven tools in personalising learning experiences, enhancing student engagement, identifying and addressing individual needs, providing real-time feedback, improving the quality of teaching and learning materials, and promoting critical thinking and problem-solving skills is driving curriculum development. | Shows Cronbach's alpha threshold of 0.70 as acceptable for most assessments. Supports understanding of assessment reliability standards. Shows AI's impact on curriculum development in global higher education contexts. |

## A.10 Vocational Education, Assessment, and Specialised Contexts

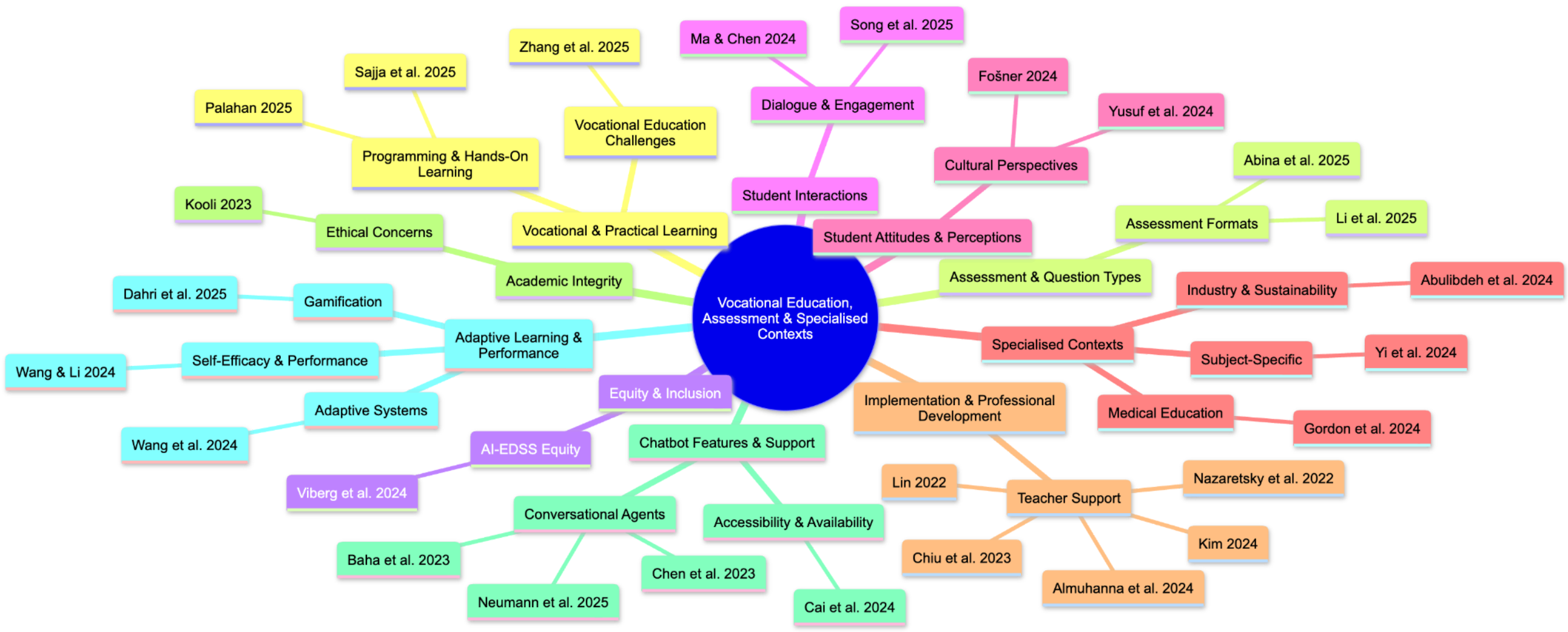

| Authors | Key Findings | Relevance to Research |
|---|---|---|
| Palahan (2025) | The study found that teaching programming requires the effective conveyance of theoretical concepts and their application in coding exercises. Immediate, detailed feedback is critical in bridging the gap between comprehension and application. Existing chatbots address theoretical questions but frequently fall short in providing feedback necessary for evaluating code and guiding practical exercises. Chatbots respond to programming queries and theoretical knowledge but often lack hands-on coding practice and code-specific feedback. | Shows the gap between text-based chatbots and practical programming education needs. |
| Sajja et al. (2025) | Study emphasises the importance of practical, experience-based learning in the Educational AI Hub system. Applying theoretical knowledge to solve practical, real-world problems requires hands-on learning. Practical applications beyond the classroom are important for effective learning in vocational contexts. | Shows the need for practical, hands-on learning in vocational education. |
| D. Zhang et al. (2025) | Study exploring relationships between AI literacy, AI trust, AI dependency, and 21st-century skills among preservice mathematics teachers. AI dependency is defined as an excessive reliance on AI, which diminishes autonomous thinking and learning abilities. AI dependency has been shown to have significant negative relationships with critical thinking, self-confidence, and problem-solving ability. Some regions focus more on theoretical knowledge than practical applications. Co-design workshops with educators help create chatbots that address pedagogical challenges and incorporate user-centred design principles. | Shows co-design and co-development approaches with educators. Shows dependency concerns. Shows vocational education challenges. |

| Authors | Key Findings | Relevance to Research |
| --- | --- | --- |
| Abina et al. (2025) | Evaluation study with 11 participants aged 18-25 evaluating the e-guide and web-based application. Compared multiple-choice questions with short-answer questions and fill-in-the-blank questions. SAQ questions are more discriminatory than MCQs, as they require students to generate their own answers rather than identify correct ones, which demands a more profound understanding and recall. Students must express their answers accurately and demonstrate a higher level of language and communication skills when answering questions that require written responses, as even minor misspellings can potentially lead to a loss of points. SAQ questions are time-consuming, with participants taking at least twice as long to formulate answers as to select multiple-choice answers. | Supports findings on SAQ challenges and language barriers. Shows how SAQ requires higher language and communication skills than MCQ. Shows how even small misspellings can lead to point loss. |
| Y. Li et al. (2025) | Systematic review of 37 papers from 2014 to 2025 examining the design of language learning with AI chatbots based on the Activity Theory framework. Review specified student learning outcomes in chatbot-supported language learning settings. Found four types of learning outcomes: behavioural, cognitive, emotional, and agentic outcomes. Students actively interact with AI chatbots for language learning, with a higher frequency of interaction, particularly in domain-specific applications. Tool alignment with assessment format is crucial for effectiveness, emphasising the importance of aligning AI chatbot design with assessment requirements to achieve desired learning outcomes. | Supports the finding that assessment alignment is crucial for the effectiveness of AI tools. Shows students actively interacting with AI chatbots at a higher frequency in domain-specific applications. Shows cognitive outcomes, including academic achievement and cognitive skills. |
| Viberg et al. (2024) | Editorial introducing a special section on advancing equity and inclusion in educational practices with AI-powered educational decision support systems. While advances in learning analytics, natural language processing, and artificial intelligence offer potential to aid educational decision-making, these technologies also raise serious risks of reinforcing or exacerbating existing inequalities. These dangers arise from biases represented in training datasets, the technologies' ability to make autonomous decisions, and processes for tool development that do not centre on the needs and concerns of historically marginalised groups. To ensure AI-EDSS are equipped to promote equity, it must be created and evaluated holistically, considering its potential for both targeted and systemic impacts on all learners. | Highlights the importance of a holistic evaluation of AI systems, considering their potential for both targeted and systemic impacts on all learners. Shows how AI technologies can reinforce or exacerbate existing inequalities through biases in training datasets. Shows the need for socio-technical and cultural perspectives in designing AI-EDSS. |
| Song et al. (2025) | Quasi-experimental study with 80 postgraduate students exploring the effectiveness of chatbots empowered by generative artificial intelligence in assisting university students' creative problem-solving, comparing student-GAI chatbot interactions with student-peer interactions. Found significant differences in dialogic exchanges between the two types of interaction. Student-GAI chatbot interactions featured more knowledge-based dialogue and elaborate discussions, with less subjective expression compared to student-peer interactions. Students contributed significantly less dialogue when interacting with the GAI chatbot than during peer interactions. However, following the information provided by the chatbot, students tended to ask for explanations and justifications, which facilitates a deeper understanding. | Contrasts with current findings, which show a low frequency of follow-up questions and reflection indicators. Shows students asking for explanations and justifications following chatbot information to facilitate a deeper understanding. Shows how dialogue dynamics differ between student-GAI chatbot and student-peer interactions. |

| Authors | Key Findings | Relevance to Research |
|---|---|---|
| Fošner (2024) | Quantitative study with 422 participants examining university students' attitudes and perceptions towards AI tools in Slovenia. Students recognise the efficiency of AI but express concerns about its impact on learning quality and academic integrity, emphasising the need for balanced and responsible integration of AI in education to achieve sustainable outcomes. The majority of students are engaging with AI tools, with the frequency of use varying widely depending on their field of study and academic level. There is a critical need to address the educational, ethical, and psychological impacts of these technologies. | Shows students' attitudes and perceptions towards AI tools. Shows concerns about learning quality and academic integrity. Demonstrates the need for a balanced and responsible integration of AI in education. |
| Yusuf et al. (2024) | Embedded mixed-methods study with 1,217 participants across 76 countries examining usage, benefits, and concerns of Generative AI in higher education from a multicultural perspective. A high level of awareness and familiarity with GenAI tools was observed among respondents, with a significant portion having prior experience with these tools and expressing an intention to continue using them. A strong correlation was found between cultural dimensions and respondents' views on benefits and concerns related to GenAI, including its potential for academic dishonesty and the need for ethical guidelines. Responsible use of GenAI tools can enhance learning processes, but addressing concerns may require robust policies that are responsive to cultural expectations. | It shows that cultural dimensions significantly influence perceptions of the benefits and concerns associated with GenAI. Shows the importance of multicultural perspectives in understanding AI adoption in education. Supports findings on academic integrity concerns and the need for ethical guidelines. |
| Abulibdeh et al. (2024) | Scoping review examining integration of artificial intelligence tools, with specific emphasis on ChatGPT, in education within the context of Education for Sustainable Development and Industry 4.0. A critical review critically examines the transformative potential, ethical considerations, imperatives for continuous learning, and the role of industry partnerships in AI integration in education. AI chatbots, such as ChatGPT, have the potential to transform various aspects of education. Still, their integration raises ethical concerns, necessitates curriculum redesign, requires strategies for continuous learning, and demands alignment with industry standards. Higher education institutions need to equip future professionals with the skills required for Industry 4.0, necessitating updates to their curricula and enhancements to their infrastructure. | Highlights the importance of curriculum redesign and faculty training when integrating AI tools into education. Shows need for alignment with industry standards and continuous learning strategies. Supports findings on the importance of ethical considerations in the integration of AI. |
| Gordon et al. (2024) | Rapid scoping review examining artificial intelligence applications in medical education, synthesising 278 publications from PubMed, MEDLINE, EMBASE, and MedEdPublish. Review mapped diverse AI applications in medical education, including AI for admissions, teaching, assessment, and clinical reasoning. The review highlighted AI's varied roles, from augmenting traditional educational methods to introducing innovative practices, and underscored the urgent need for ethical guidelines in the application of AI in medical education. A framework to support future high-utility reporting was proposed, the FACETS framework. | Shows comprehensive mapping of AI applications in medical education. Demonstrates AI's diverse roles, ranging from augmenting traditional methods to introducing innovative practices. Supports findings on the need for ethical guidelines in the application of AI in education. |

| Authors | Key Findings | Relevance to Research |
|---|---|---|
| Yi et al. (2025) | Systematic review and meta-analysis investigating the effectiveness of AI on improving mathematics performance in K-12 classrooms compared to traditional classroom instruction. Synthesised findings from 21 relevant studies. Results indicate a small overall effect size of 0.343, favouring AI under a random-effects model, showing a generally positive impact on mathematics learning. Only one variable, AI type, was identified as having a moderate effect, with AI demonstrating a greater impact when serving as an intelligent tutoring system and adaptive learning system compared to pedagogical agents. | Shows that AI interventions have a positive impact on mathematics learning outcomes in K-12 settings. Shows that the AI type has a significant influence on effectiveness. Supports findings on the effectiveness of AI tools in educational contexts. |
| Lin (2022) | The study examines the influences of AI in education on teaching effectiveness, with a mediating effect of teachers' perceptions of educational technology. Successful workload reduction depends on adequate training, technical support, and seamless integration into existing educational workflows. | Shows implementation success factors. Shows the relationship between AI tools and teaching effectiveness. Relevant to chatbot implementation. |
| Nazaretsky et al. (2022) | The study examines teachers' trust in AI-powered educational technology and professional development programmes to improve it. Professional development and ongoing support are crucial for building educators' confidence and competence in utilising chatbot technologies. | Shows implementation success factors, particularly educator trust and professional development needs. Relevant to chatbot implementation. |
| Kim (2024) | Study examining leading teachers' perspectives on teacher-AI collaboration in education. Institutional guidelines that address ethical use, data privacy, and academic integrity are crucial for the successful implementation of chatbots. | Shows implementation success factors. Shows educator perspectives on AI collaboration. Relevant to chatbot implementation. |
| Almuhanna (2025) | Qualitative study with 19 semi-structured interviews investigating teachers' perspectives of integrating AI-powered technologies in K-12 education. The reliability and accuracy of generative AI information are questioned, posing challenges to academic integrity and credibility. Many teachers require enhanced skills, expertise, and supportive institutions to effectively adopt AI. Teachers encounter time-based limitations when integrating AI technologies, with the iterative process of evaluating and modifying AI-generated content being time-consuming. Concerns have been raised about over-reliance on AI tools reducing cognitive abilities and human interaction. | Supports findings on accuracy concerns and challenges to academic integrity. Shows reliability and accuracy concerns with generative AI. Shows implementation challenges, including time constraints and the need for teacher training. |
| Kooli (2023) | Qualitative methodology and a comprehensive review examining the ethical implications of chatbots in education and research. Using chatbots or other artificial intelligence tools to answer exam questions constitutes a form of cheating and academic misconduct, which contradicts the fundamental principles of learning and academic integrity. If misused by students, chatbots could raise serious ethical concerns and significantly impact students' academic progress by hindering their critical thinking skills, creativity, and ability to apply concepts learned to real-world situations. | Supports findings on academic integrity concerns and negative impacts on critical thinking. Shows chatbots being used for cheating and academic misconduct. Shows how misuse can hinder critical thinking skills and academic progress. |

| Authors | Key Findings | Relevance to Research |
|---|---|---|
| Ma & Chen (2024) | Quasi-experimental research with 350 students examining the influence of AI-empowered applications on affective, cognitive, and behavioural engagement and academic procrastination among EFL learners. Procrastination is widespread on college campuses, with estimates indicating that 70-95% of students procrastinate, and approximately 50% of them habitually procrastinate. The critical issue is not delay but the total time students dedicate to academic work, which often falls short of faculty expectations. The experimental group exposed to AI-empowered applications demonstrated significantly higher levels of affective, cognitive, and behavioural engagement than the control group. A substantial reduction in academic procrastination was observed among students exposed to AI-empowered applications. | Supports findings on student procrastination patterns and time management challenges. Shows widespread procrastination among college students. Shows how AI-empowered applications can mitigate academic procrastination and enhance engagement. |
| Ait Baha et al. (2024) | Research on conversational agents and dialogue-based learning examines the effectiveness of chatbots in this context. Conversational agents can provide real-time and adaptive support that mimics the role of a human teacher, offering immediate feedback, answering questions, and delivering personalised guidance through interactive dialogue. | Supports the potential of conversational agents in educational contexts. Shows chatbots can provide real-time and adaptive support. |
| Chiu et al. (2023) | Experimental study with 123 Grade 10 students examining how teacher support moderates effects of student expertise on needs satisfaction and intrinsic motivation to learn with AI chatbots, using self-determination theory as a framework. Intrinsic motivation and competence in learning with a chatbot depend on both teacher support and student expertise. Teachers are more satisfied with the need for relatedness, and are less satisfied with the need for autonomy. Understanding the teacher's role as a core facilitator of student motivation and academic progress in mediating and supporting learning with AI technologies in the classroom provides a clear understanding of how such technologies can be effectively utilised in practice. | Shows the importance of teacher support in mediating and supporting learning with AI technologies. Shows how teacher support moderates the effects of student expertise on needs satisfaction and intrinsic motivation. Supports findings on educators playing a crucial role in facilitating student engagement with AI tools. |
| Y. Chen et al. (2023) | Two sequential studies with 215 and 195 students investigating chatbots as pedagogical tools in business education. Participants frequently needed answers to basic information about courses, including course materials, due dates, study tips, and office hours. Chatbots can provide basic course information and content, serving as a helpful addition in classes with high student-teacher ratios or when students are too shy to communicate directly with instructors. Chatbots can help students learn basic content in a responsive, interactive, and confidential way. | Supports findings that students repeatedly ask similar questions and require basic course information. Shows students frequently needing answers to basic information about courses. Shows how chatbots can address repetitive questions and basic information needs. |
| (Cai et al., 2025) | Systematic literature review of 158 empirical studies exploring the impact of integrating AI tools in higher education using the Zone of Proximal Development framework. AI tools assist learners in personalising their self-assessment through social and technological interactions and effective communication, improving motivation, learning engagement, and learning support, which leads to better academic performance. AI-powered personalised learning platforms are accessible anytime, anywhere, providing students with on-demand support regardless of geographical location or time constraints. | Supports findings on 24/7 availability and accessibility expectations. Shows AI-powered personalised learning platforms being accessible anytime, anywhere, with on-demand support. Shows how AI tools can remove time and location barriers for student support. |

| **Authors** | **Key Findings** | **Relevance to Research** |
|---|---|---|
| Neumann et al. (2025) | Research with 46 students exploring the benefits and challenges of developing, deploying, and evaluating the LLM-driven chatbot, MoodleBot, in computer science classroom settings. LLM-based chatbots, such as MoodleBot, can significantly enhance the teaching and learning process, with a high accuracy rate in providing course-related assistance. Chatbots have the potential to address help-seeking concerns and alleviate help avoidance behaviour by providing a private, nonjudgmental space for students to ask questions without fear of embarrassment in front of peers or teachers. | Supports findings on safe learning spaces and help-seeking behaviour for weaker students. Shows chatbots providing private, nonjudgmental spaces for students to ask questions without fear of embarrassment. Shows how chatbots can address help-seeking concerns and alleviate help avoidance behaviour. |
| Wang et al. (2024) | Meta-analysis examining the overall effect of AI-enabled adaptive learning systems on students' cognitive learning outcomes when compared with non-adaptive learning interventions. Synthesised findings from 45 independent studies. AI-enabled adaptive learning systems have a medium to large positive effect size on cognitive learning outcomes. AI-enabled adaptive learning leverages AI and ML algorithms for personalised learning, tailoring content to each learner's unique needs and preferences. | Supports findings on adaptive learning algorithms dynamically adjusting instructional content based on individual learner progress. Shows AI-enabled adaptive learning systems tailoring content to each learner's unique needs and preferences. Shows medium to large positive effect sizes on cognitive learning outcomes. |
| Wang & Li (2024) | Quantitative study with 1,082 Chinese music students investigating the relationship between music students' self-efficacy, academic performance, and artificial intelligence readiness. Notably, 63% of variations in students' academic performance are attributed to the combined influence of self-efficacy and AI readiness. Students' self-efficacy uniquely predicts 52% of changes in academic performance, whilst their AI readiness uniquely predicts 60% of changes in academic performance. | Shows that students' AI readiness significantly contributes to their academic performance. It shows that self-efficacy plays a crucial role in overall academic achievement. Supports findings that students demonstrate more confidence and appear more prepared for exams. |
| Ahmed Dahri et al. (2025) | Quantitative study with 290 students investigating relationships between gamification, artificial intelligence, student engagement, and achievement in educational contexts. Gamification introduces flexible reward mechanisms, such as points, badges, and leaderboards, which can be tailored to individual achievements and preferences, encouraging students to sustain their efforts and persevere in the face of challenges. By recognising personal milestones, gamification fosters a sense of accomplishment and satisfaction, with students in gamified environments demonstrating higher levels of engagement. | Supports findings on gamification, introducing flexible reward mechanisms and fostering engagement. Demonstrates gamification, encouraging students to sustain their efforts and persevere in the face of challenges. Shows how recognising personal milestones fosters accomplishment and satisfaction. |

## APPENDIX B: PRE-IMPLEMENTATION SURVEY AND INTERVIEW ANALYSIS

### B.1 Pre-Implementation Survey Questions

**Section A: Teaching Philosophy and Practice**

1. My role as an educator is primarily to facilitate inquiry and independent thought.
   (1–Strongly Disagree, 5–Strongly Agree)
2. My students can take ownership of their learning without continuous guidance.
   (1–Not at all true, 5–Absolutely true)
3. I regularly adjust my teaching to accommodate the diverse learning styles of my students.
   (1–Never, 5–Always)
4. Technology can augment, rather than replace, human judgment in the learning process.
   (1–Strongly Disagree, 5–Strongly Agree)
5. I currently feel equipped to help students develop critical thinking when engaging with digital content.
   (1–Not at all true, 5–Absolutely true)
6. I have sufficient time and resources to provide personalised feedback to all students.
   (1–Not at all true, 5–Absolutely true)

**Section B: Familiarity and Expectations of AI Tools**

7. I have previously used AI-based educational tools (e.g., chatbots, auto-grading systems, adaptive learning platforms).
   (1–Not at all, 5–Very extensively)
8. I am confident in evaluating whether an AI-generated response is pedagogically sound.
   (1–Not confident at all, 5–Extremely confident)
9. I believe a digital tool that interacts with students in real-time could support learning in a meaningful way.
   (1–Strongly Disagree, 5–Strongly Agree)
10. I expect that integrating an AI-based assistant will require me to rethink aspects of my current teaching practice.
    (1–Not at all, 5–Very significantly)

**Section C: Philosophical Orientation**

11. Students learn best when they engage in dialogue, rather than just receiving content.
    (1–Strongly Disagree, 5–Strongly Agree)
12. An educational tool should prioritise developing curiosity over delivering efficiency.
    (1–Strongly Disagree, 5–Strongly Agree)
13. I am comfortable allowing automated systems to shape the learning experience, even if outcomes are unpredictable.
    (1–Strongly Disagree, 5–Strongly Agree)
14. Open-ended: What do you consider the most important quality a digital learning assistant should embody?
15. Open-ended: What is one risk you foresee in integrating generative AI into your teaching?

### B.2 Pre-Implementation Survey Analysis

Pre-implementation surveys were administered to 20 lecturers from Phase 1, before any POC development discussions. These structured surveys established lecturers' baseline perspectives regarding Generative AI technologies in education. The survey instrument consisted of three sections covering Teaching Philosophy and Practice (Q1 to Q6), Familiarity and Expectations of AI Tools (Q7 to Q10), and Philosophical Orientation (Q11 to Q15). The final section included two open-ended questions. All Likert-scale items used a 5-point scale. At this stage, we did not presume what form the Generative AI POC would take. The surveys captured lecturers' baseline state before co-development workshops.

**B.2.1 Role as Facilitator of Inquiry (Q1)**

Lecturers strongly agreed that their role as educators is primarily to facilitate inquiry and independent thought (M = 3.95, range 3-5, median = 4). This finding demonstrates strong alignment with constructivist pedagogical approaches that prioritise student-centred learning and the development of critical thinking over traditional transmission models of education.

This perspective aligns with constructivist learning theory, which views learning as an active process in which students construct knowledge through their experiences and interactions, rather than passively receiving information. Research on constructivist learning environments emphasises the importance of teachers acting as facilitators who guide students' inquiry processes, encourage questioning, and support knowledge construction through dialogue and reflection (Lee & Yang, 2023). The high mean score indicates that lecturers recognise their role in creating learning environments where students actively engage with content rather than simply memorising information.

The emphasis on facilitating inquiry rather than direct instruction reflects contemporary pedagogical approaches that value student agency and ownership of learning. Studies examining teacher roles in collaborative and inquiry-based learning contexts have found that facilitative approaches lead to greater student engagement, deeper understanding, and improved critical thinking skills (Lee & Yang, 2023). This finding suggests that lecturers are positioned to effectively integrate AI tools that support inquiry-based learning, as their pedagogical philosophy aligns with technologies designed to facilitate rather than replace student thinking processes.

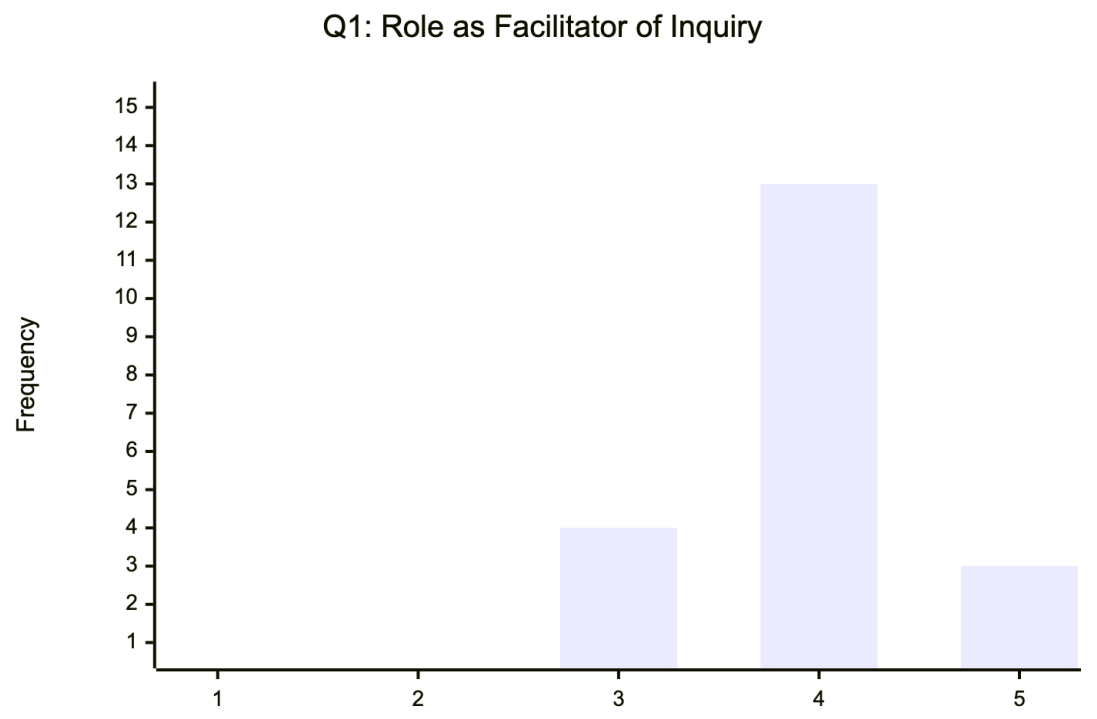

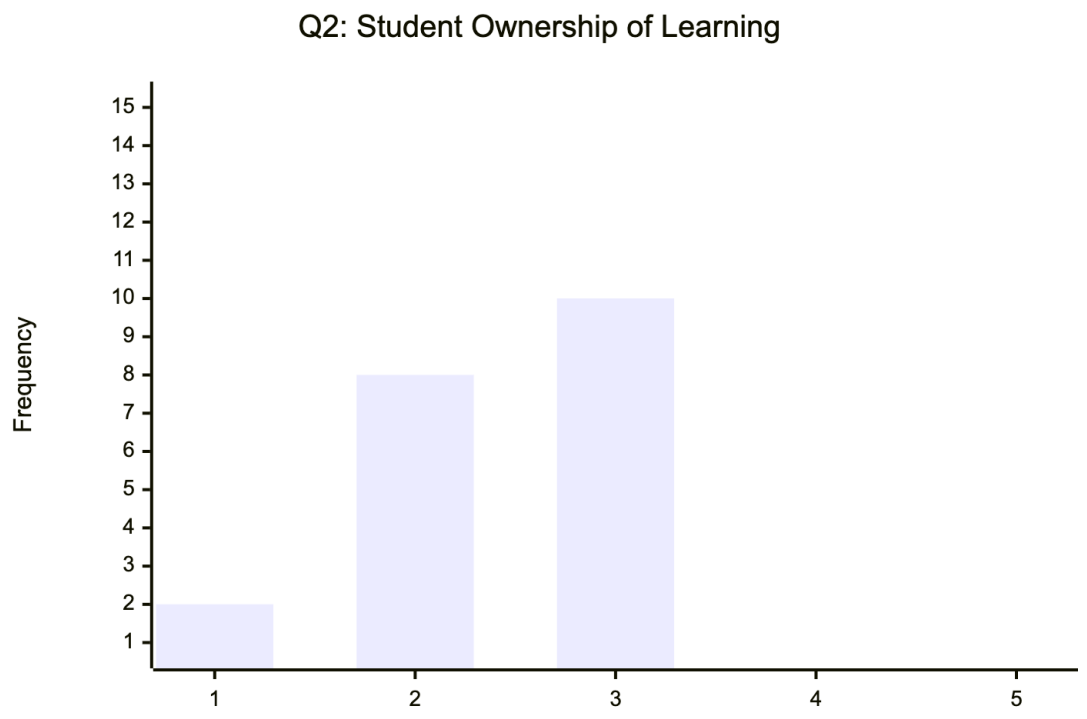

**B.2.2 Student Ownership of Learning (Q2)**

Lecturers expressed lower confidence in students' ability to take ownership of their learning without continuous guidance (M = 2.40, range 1-3, median = 2.5). This finding indicates recognition that many students require ongoing support and scaffolding rather than complete independence in their learning processes. The moderate mean score suggests lecturers acknowledge the importance of student agency while recognising practical limitations in students' current capacity for self-directed learning.

This finding aligns with research on scaffolding and the Zone of Proximal Development, which recognises that students require structured support to progress from their current developmental levels to more advanced stages. Research indicates that scaffolding is the process through which students are assisted in progressing from one developmental stage to the next by providing them with essential support. Scaffolding teaching techniques offer students individualised support (Msambwa et al., 2025). Within the Zone of Proximal Development, learners are provided with support that is gradually withdrawn as they gain mastery, promoting a transition from assisted to independent problem-solving (Looi & Jia, 2025). This perspective indicates that developing student ownership of learning requires gradual scaffolding rather than expecting immediate independence.

The recognition that students need structured support aligns with scaffolding theory, which emphasises providing temporary support within students' Zone of Proximal Development to help them reach comprehension levels they could not achieve independently (Kuhail et al., 2023). This perspective suggests that AI tools designed to provide adaptive scaffolding could effectively support lecturers' pedagogical goals by offering personalised support that gradually builds students' capacity for independent learning. The finding indicates that lecturers would value AI tools that provide structured guidance while gradually promoting greater student autonomy.

**B.2.3 Accommodating Learning Styles (Q3)**

Lecturers reported regularly adjusting their teaching to accommodate different student learning styles (M = 3.95, range 3-5, median = 4). This demonstrates strong awareness of diverse student needs and commitment to differentiated instruction. The high mean score suggests that lecturers effectively adapt their pedagogical approaches to meet the individual learning preferences and characteristics of their students.

This finding aligns with research on individual differences in learning, which recognises that learners have varying preferences for how they acquire information, such as through visual, auditory, or kinesthetic means (Katiyar et al., 2024). While evidence for learning styles is mixed, personalised learning systems can provide learners with multiple modes of instruction, allowing them to choose the approach that works best for them (Katiyar et al., 2024). The lecturers' commitment to accommodating different learning styles suggests they would value AI tools that can adapt content presentation and instructional approaches to individual learner preferences.

Research on differentiated instruction and learning styles in assessment contexts suggests that educators should consider how students learn differently when designing educational experiences (Perdana & Chu, 2023). Studies have found that AI can help teachers identify and accommodate individual students' learning styles by analysing data from students' interactions with educational materials, enabling customisation of educational content and strategies to better align with each student's specific needs (J. Zhang & Zhang, 2024). This finding suggests that lecturers are well-positioned to leverage AI tools that support personalised learning experiences tailored to diverse learning preferences.

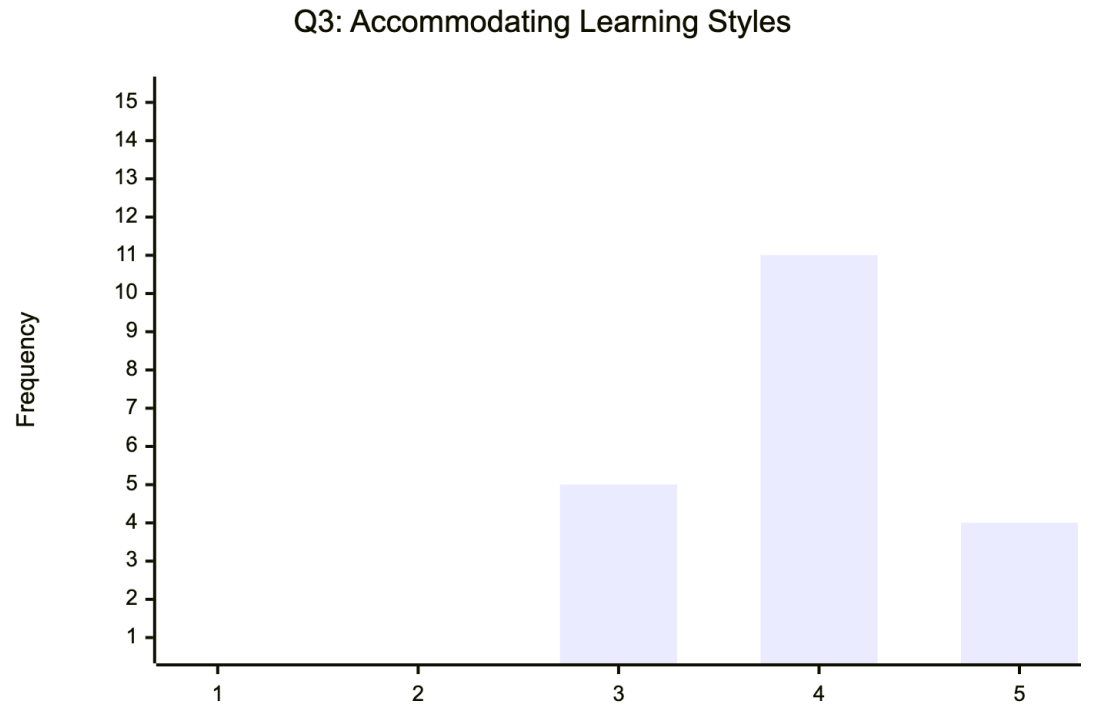

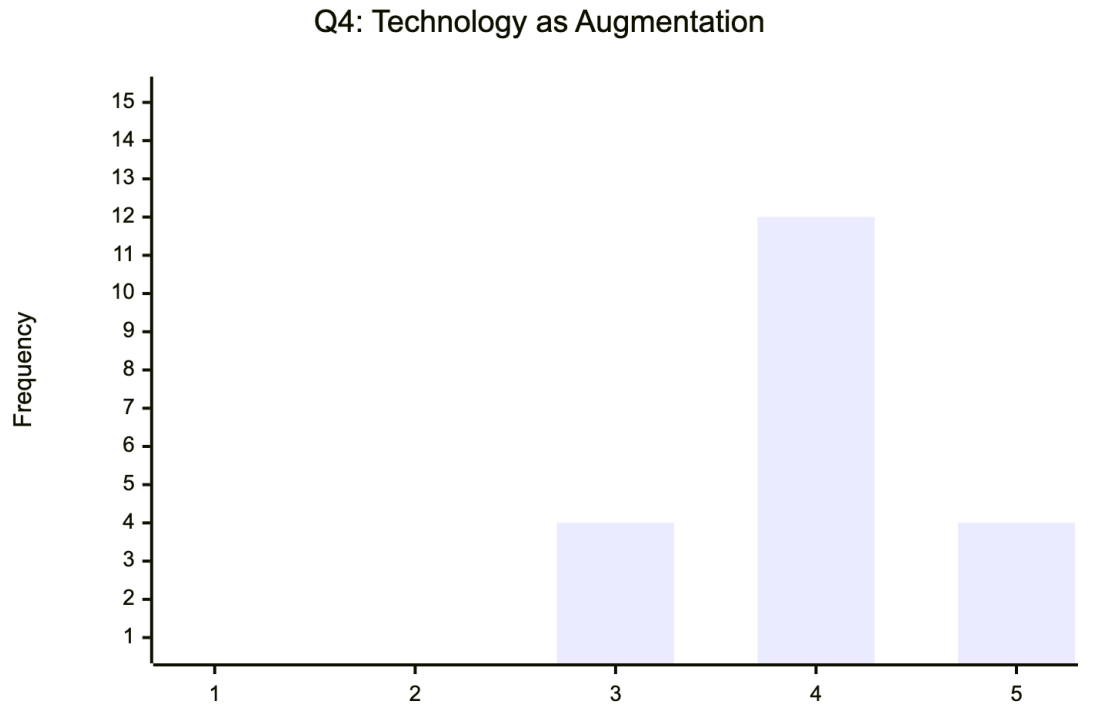

**B.2.4 Technology as Augmentation (Q4)**

Lecturers strongly agreed that technology can augment rather than replace human judgment in the learning process (M = 4.00, range 3-5, median = 4). This perspective suggests that lecturers view AI tools as complementary to educator expertise and judgment rather than substitutes. The high mean score indicates strong support for the augmentation perspective, which hypothesises that technology enhances rather than replaces human capabilities in educational contexts.

This finding aligns with research on teacher-AI collaboration, which distinguishes between replacement and augmentation perspectives (Kim, 2024). The augmentation perspective emphasises AI's role in facilitating teaching processes by enhancing teachers' cognition through rapid data collection, analysis, and translation into actionable insights and meaningful actions (Kim, 2024). Studies indicate that when AI systems are designed to combine human teacher and AI instruction, they can jointly perceive, learn, restrict, and supervise each other to work together, achieving effects that neither humans nor AI can complete independently (Kim, 2024).

The lecturers' support for augmentation aligns with research showing that AI can support teachers to work more efficiently by automating time-consuming tasks, allowing them to concentrate on providing personalised instruction and support to students (J. Zhang & Zhang, 2024). However, research also indicates that teachers value the human touch, empathy, and nuanced understanding that human educators bring to educational experiences, which AI cannot fully replicate (Alwaqdani, 2025). This finding suggests lecturers would be receptive to AI tools designed to augment their capabilities while preserving their central role in educational decision-making and student support.

**B.2.5 Critical Thinking Support (Q5)**

Lecturers reported moderate confidence in helping students develop critical thinking when engaging with digital content (M = 2.90, range 2-4, median = 3). This moderate level suggests lecturers recognise both the importance and challenges of developing critical thinking skills in digital learning environments. The mean score suggests that lecturers recognise the need to support the development of critical thinking, but may feel uncertain about effective strategies for doing so in digital contexts.

This finding aligns with research on critical thinking in digital and AI-driven contexts, which recognises that critical thinking involves the ability to analyse information, evaluate different perspectives, and create reasoned arguments, all within the framework of AI-driven environments (Walter, 2024). Research suggests that scaffolding techniques, such as prompt scaffolding, explicit reflection, and modelling, can help foster students' critical thinking skills in digital and AI-driven contexts (Walter, 2024). Teaching critical thinking skills in general higher education contexts is recognised as a major academic challenge due to difficulties in embedding critical thinking into well-organised curricula and utilising effective teaching strategies (Y. Liu & Pásztor, 2022).

However, research also identifies concerns about AI's potential negative impact on critical thinking. Studies examining the impact of AI on critical thinking have found that cognitive offloading, where learners delegate cognitive tasks to external tools to reduce cognitive effort, may lead to decreased internal cognitive engagement over time, ultimately impacting learners' ability to self-regulate and critically engage with learning material (Fan et al., 2025). Research indicates that overreliance on AI tools can lead to a decline in critical and analytical thinking skills, particularly when students become overly dependent on AI-generated content (Zhai et al., 2024). Over-dependence on AI technologies could negatively impact the development of essential cognitive skills, decreasing students' capacity for original thought, creativity, and independent problem-solving (Ahmed et al., 2024). The lecturers' moderate confidence may reflect their awareness of the tensions between using digital tools to support learning and ensuring that students develop independent critical thinking capabilities. This finding suggests that lecturers would benefit from AI tools designed to scaffold the development of critical thinking rather than simply providing answers, and would value professional development on integrating critical thinking support into digital learning environments.

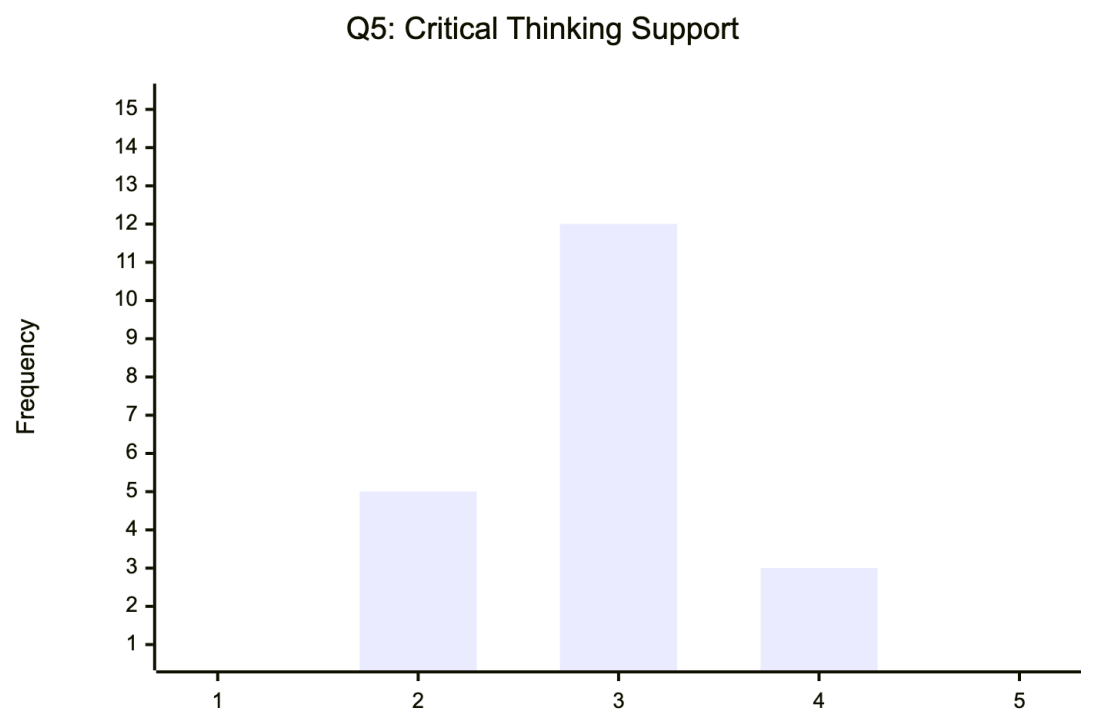

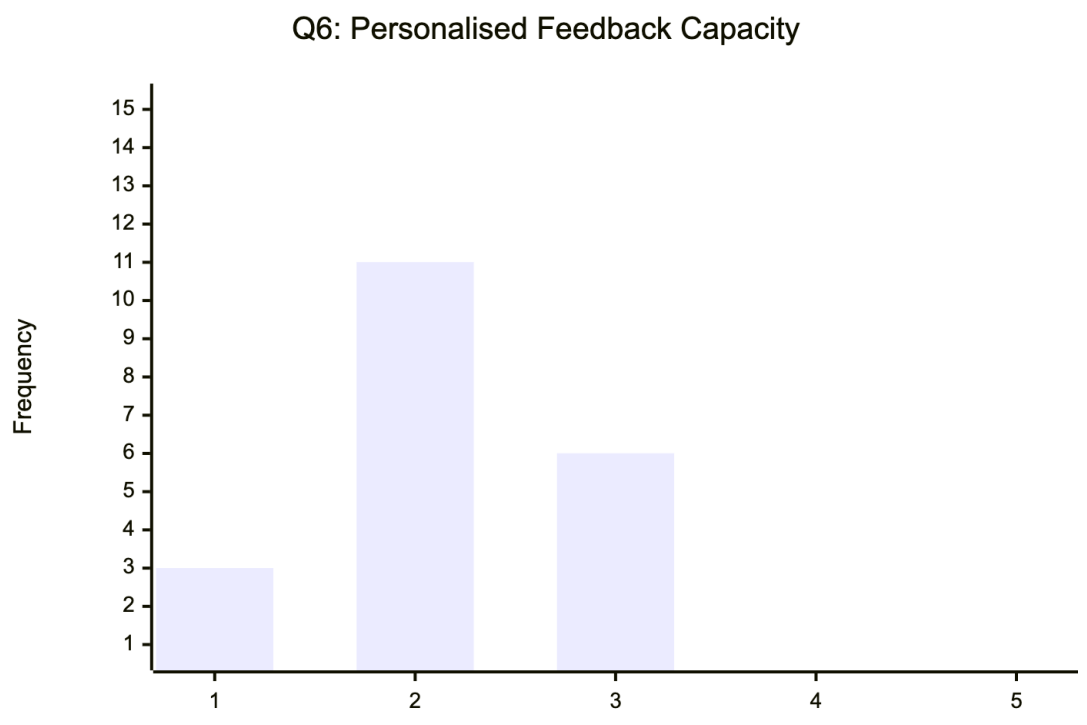

**B.2.6 Personalised Feedback Capacity (Q6)**

Lecturers acknowledged that they had insufficient time and resources to provide personalised feedback to all students (M = 2.15, range 1-3, median = 2). This highlights significant constraints in current support capacity. The low mean score indicates lecturers recognise a substantial gap between their desire to provide personalised feedback and their practical ability to do so, given existing time and resource limitations.

This finding aligns with research identifying challenges in providing individualised support to diverse student populations. Studies examining the challenges of providing individualised student support have found that the expansion of higher education to include broader, more diverse student populations has created challenges for institutions whose support networks cannot cope with the increased numbers and diversity of students (Zhou & Wolstencroft, 2022). Communication is often perceived as slow and lacking in human relationships, as the size of universities means that it is rare for students to be assigned a single person for support. Consequently, requests for information or support can become transactional rather than personalised (Zhou & Wolstencroft, 2022). Research indicates that educators face significant challenges in providing personalised education, with restrictions such as large classrooms, limited time, and inadequate assistance making personalisation particularly challenging for learners with diverse needs (Almuhanna, 2025).

Research indicates that AI technologies can help address these constraints by automating feedback generation and providing personalised support at scale. Studies have found that AI can support teachers to work more efficiently by automating tasks that consume significant time, such as grading, allowing them to concentrate on providing personalised instruction and support to students (J. Zhang & Zhang, 2024). AI systems can provide automated feedback, assessment, and personalised interactions, while also generating personalised feedback and identifying at-risk students, allowing educators to focus on higher-order teaching activities (Alwaqdani, 2025; C. Yuan et al., 2025). AI-driven tools can monitor student progress and automatically adjust learning content to suit individual needs, providing real-time insights into student performance and enabling timely interventions (C. Yuan et al., 2025). This finding suggests lecturers would strongly value AI tools that can provide personalised feedback to students, potentially freeing lecturer time for higher-order teaching activities while ensuring all students receive timely, individualised support.

### B.2.7 Prior AI Tool Experience (Q7)

Lecturers have some experience with AI tools in general, but not specifically with AI tools designed for educational purposes (M = 1.50, range 1-2, median = 1.5). Most have used AI in broader contexts, yet hands-on engagement with educational AI applications has been limited. Additionally, available AI tools for education were often described as limited in functionality or challenging to implement within their current teaching environments. This topic is explored in greater depth in the interview findings.

This finding aligns with research on AI adoption in education, which recognises that while AI technologies are increasingly being incorporated into education, many teachers need improved skills, expertise, and supportive institutions to adopt AI effectively (Almuhanna, 2025). Studies examining teachers' perceptions of AI tools have found that teachers may lack prior experience with AI-powered educational technologies, resulting in a limited familiarity with their potential benefits (Alwaqdani, 2025). This gap between awareness and experience suggests a need for professional development and hands-on training opportunities.

Research indicates that teachers' attitudes toward AI and their Technological Pedagogical Content Knowledge (TPACK) are critical determinants of their adoption and effective use of AI in educational settings (M. Li & Manzari, 2025). Without practical experience, teachers may struggle to effectively integrate AI tools into their teaching practices. This finding suggests that professional development programmes focusing on hands-on experience with AI educational tools would be valuable for supporting lecturers' adoption and effective use of AI technologies in their classrooms.

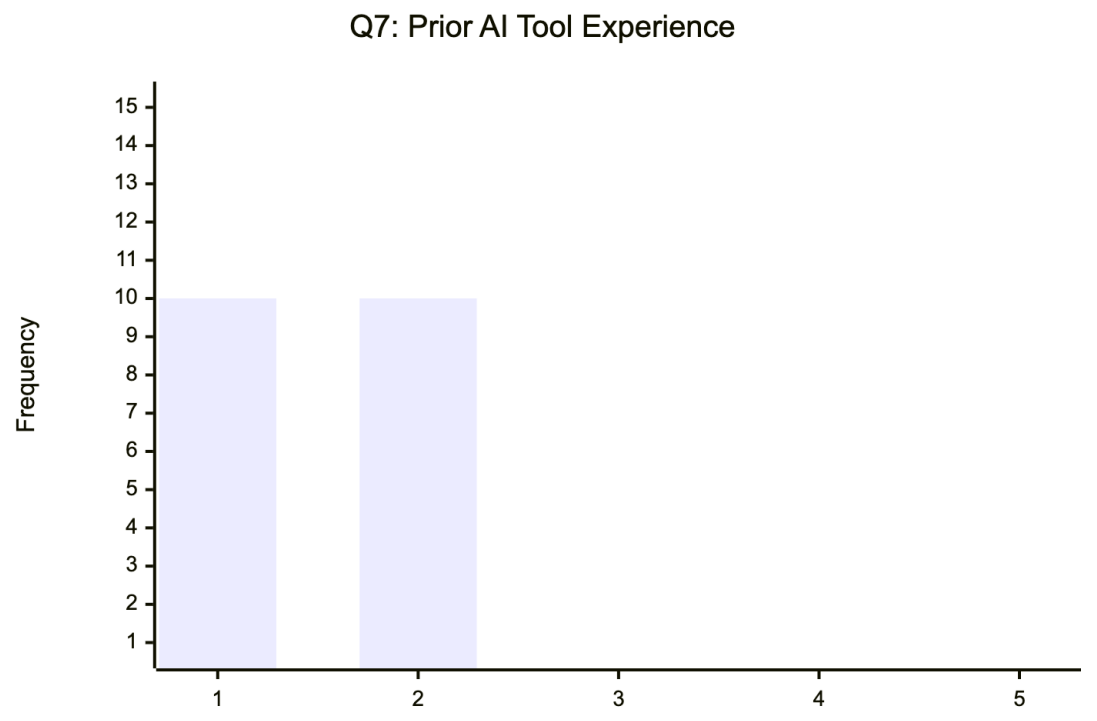

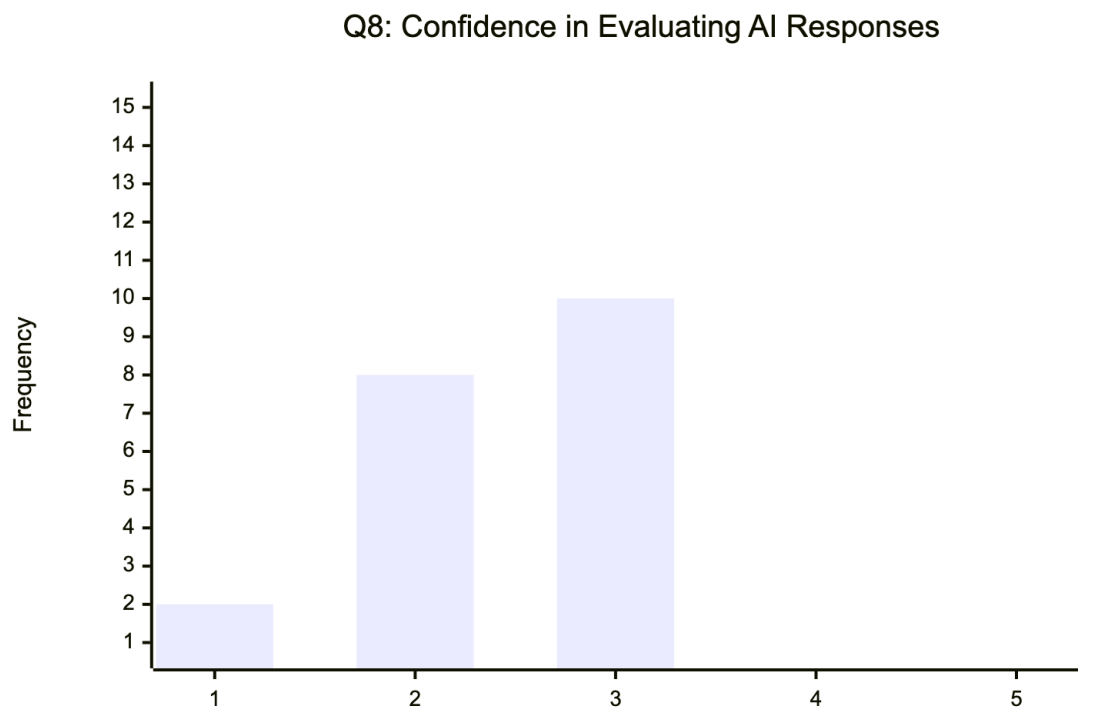

### B.2.8 Confidence in Evaluating AI Responses (Q8)

Confidence in evaluating whether AI-generated responses are pedagogically sound was moderate (M = 2.40, range 1-3, median = 2.5). This moderate confidence level suggests that lecturers recognise the need to develop skills in assessing the quality of AI output and its pedagogical appropriateness. The mean score indicates that lecturers feel somewhat uncertain about their ability to evaluate whether AI-generated content aligns with sound pedagogical principles.

This finding aligns with research on teachers' trust in AI-powered educational technology, which has identified that trust plays a critical role in practitioners' adoption of technology. Teachers mostly discuss issues of accuracy and reliability when discussing AI-powered educational tools (Nazaretsky et al., 2022). Research indicates that professional development programmes can improve teachers' trust and willingness to apply AI-powered educational technology in their classrooms. To communicate these results effectively to students, teachers need to be able to interpret and justify automated results (Nazaretsky et al., 2022).

The moderate confidence level may reflect concerns about AI's potential to provide biased responses or inaccurate information, which could mislead students and hinder their learning progress if not properly evaluated (Labadze et al., 2023). This finding suggests lecturers would benefit from professional development focused on developing skills for evaluating AI-generated content, including understanding AI limitations, recognising potential biases, and assessing pedagogical appropriateness. Such training would be essential for lecturers to effectively integrate AI tools while maintaining quality and trust in educational processes.

### B.2.9 Belief in Real-Time Digital Tools (Q9)

Despite limited experience, lecturers expressed a strong belief that a digital tool interacting with students in real-time could support learning meaningfully (M = 3.95, range 3-5, median = 4). This strong positive expectation suggests openness to AI tools despite limited prior experience. The high mean score indicates that lecturers are optimistic about the potential of real-time interactive digital tools to enhance student learning, even without extensive hands-on experience.

This finding aligns with research on AI-powered personalised learning platforms, which are accessible anytime, anywhere, providing students with on-demand support regardless of geographical location or time constraints (Cai et al., 2025). Studies indicate that the ability to provide immediate support 24/7 is a significant advantage compared to traditional tutoring approaches (Ahmed et al., 2024). Research on conversational agents and chatbots has found that these tools can provide real-time and adaptive support that mimics the role of a human teacher, offering immediate feedback, answering questions, and delivering personalised guidance through interactive dialogue (Ait Baha et al., 2024).

The lecturers' strong belief in real-time digital tools, despite their limited experience, suggests they recognise the potential benefits of immediate, interactive support for students. This finding suggests that lecturers would be receptive to AI chatbot tools that can provide real-time assistance to students, particularly given their recognition of time constraints and the need for continuous support, as identified in other survey questions. The positive expectation, combined with limited experience, suggests that providing lecturers with opportunities to see real-time AI tools in action would be valuable for building both confidence and practical understanding.

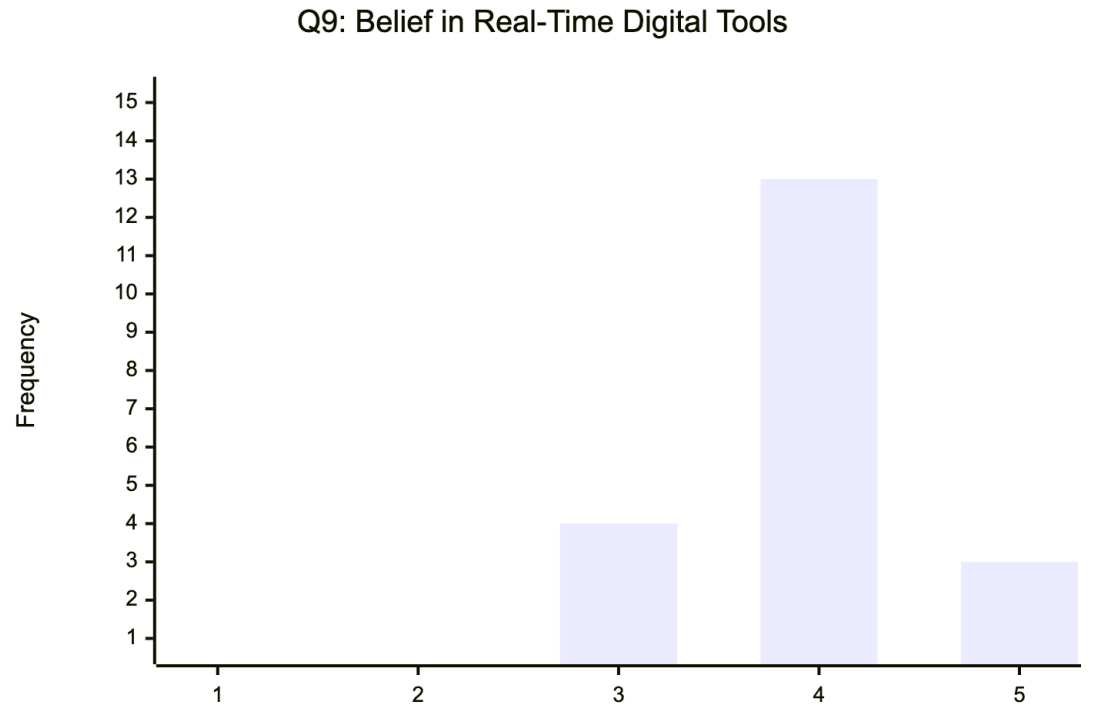

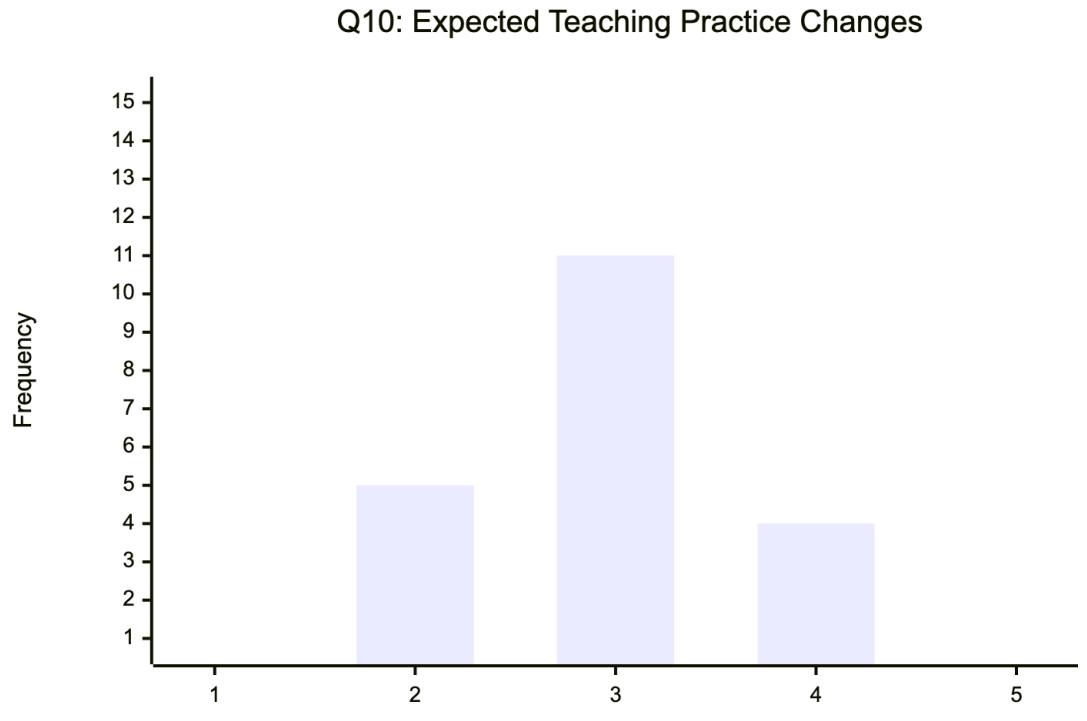

### B.2.10 Expected Teaching Practice Changes (Q10)

Lecturers expected that integrating an AI-based assistant would require moderate rethinking of current teaching practice (M = 2.95, range 2-4, median = 3). This suggests openness to adaptation while recognising the need for pedagogical adjustments. The moderate mean score suggests that lecturers acknowledge that integrating AI would require some adjustments to their teaching approaches, but do not expect a complete overhaul of their pedagogical practices.

This finding aligns with research on teacher-AI collaboration, which recognises that understanding how to structure and implement teacher-AI collaboration while ensuring teachers' role in enacting instructional roles to support student learning remains limited (Kim, 2024). Research has identified questions regarding the content and learning activities to be delivered with AI, the assessments that should be conducted, and how to interact with AI, indicating areas where teachers would benefit from support (Kim, 2024). However, research also indicates that teachers encounter considerable time-based limitations when integrating AI technologies, with the iterative process of evaluating and modifying AI-generated content being particularly time-consuming (Almuhanna, 2025).

The moderate expectation for change suggests lecturers are realistic about the adjustments needed while maintaining confidence in their core teaching practices. This finding suggests that lecturers would benefit from professional development that enables them to understand how to effectively integrate AI tools into their teaching practices, including guidance on pedagogical strategies that leverage AI capabilities while preserving essential teaching functions. The moderate level of expected change also suggests that lecturers value maintaining their central role in instruction while augmenting their capabilities with AI support.

### B.2.11 Dialogue Over Content Delivery (Q11)

Lecturers strongly agreed that students learn best when they engage in dialogue rather than just receiving content (M = 3.95, range 3-5, median = 4). This aligns with constructivist pedagogical approaches that emphasise interactive learning and knowledge construction through conversation. The high mean score indicates that lecturers strongly value dialogue-based learning over passive content consumption.

This finding aligns with constructivist learning theory, which views learning as an active process in which students construct knowledge through interaction and dialogue, rather than passively receiving information. Research supports this finding, indicating that chatbots can facilitate interactive dialogues with students, providing personalised aid and resources aligned with specific learning objectives (Looi & Jia, 2025). Studies examining conversational agents and chatbots have found that these tools can provide real-time and adaptive support that mimics the role of a human teacher, offering immediate feedback, answering questions, and delivering personalised guidance through interactive dialogue (Ait Baha et al., 2024).

The emphasis on dialogue aligns with research on collaborative learning, which recognises that learning occurs as a result of interactions within a learner's zone of proximal development, where scaffolding or temporary support facilitates learning to reach learning goals (Yaseen et al., 2025). This finding suggests lecturers would strongly value AI chatbot tools that engage students in dialogue rather than simply delivering content, as such tools would align with their pedagogical philosophy of active, interactive learning. The strong agreement indicates that lecturers are well-positioned to effectively integrate dialogue-based AI tools that support knowledge construction through conversation rather than passive information delivery.

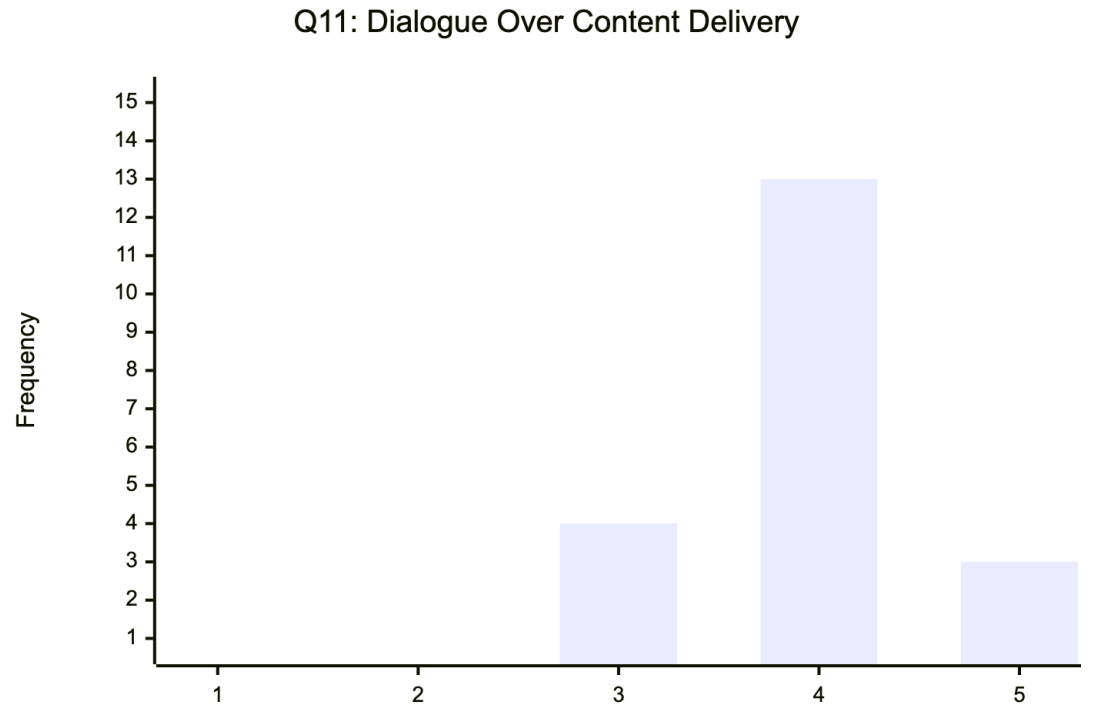

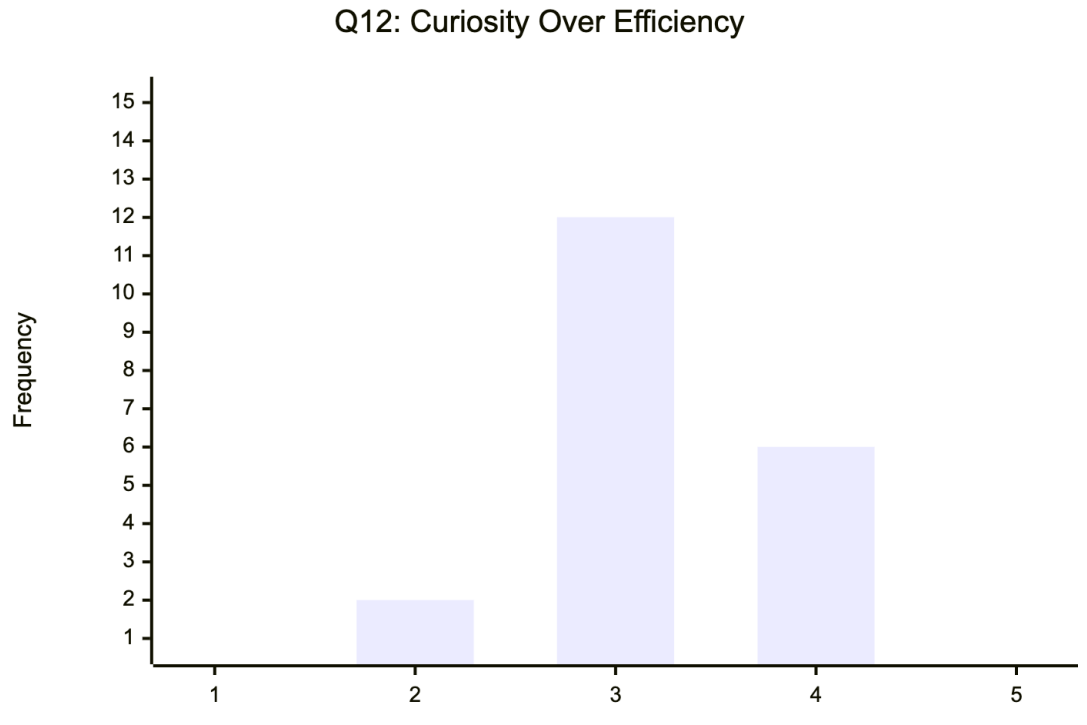

### B.2.12 Curiosity Over Efficiency (Q12)

Lecturers moderately agreed that educational tools should prioritise developing curiosity over delivering efficiency (M = 3.20, range 2-4, median = 3). This moderate agreement suggests lecturers value both curiosity development and efficiency, recognising potential tensions between these goals. The mean score indicates that lecturers see merit in prioritising curiosity but also acknowledge the importance of efficiency in educational contexts.

This finding reflects a nuanced perspective on educational tool design, recognising that while curiosity and intrinsic motivation are important for deep learning, practical constraints often require efficient delivery of content and support. Research has shown that personalised learning and adaptive learning can positively influence learners' motivation, experience, and learning outcomes, as reported in studies examining technology-enabled personalised and adaptive learning approaches (D.-L. Chen et al., 2025). This suggests that well-designed tools can support both curiosity and efficiency.

However, the moderate agreement may also reflect awareness of tensions between encouraging deep, curious exploration and meeting practical educational requirements such as curriculum coverage and assessment preparation. This finding suggests lecturers would value AI tools that can balance both goals, providing efficient support when needed while also fostering curiosity and deeper engagement with content. The moderate score indicates lecturers recognise that effective educational tools should not sacrifice one goal entirely for the other, but rather find ways to integrate both curiosity development and efficiency in their design.

### B.2.13 Comfort with Unpredictable Outcomes (Q13)

Lecturers expressed lower comfort with allowing automated systems to shape learning experiences when outcomes are unpredictable (M = 2.60, range 1-4, median = 3). This suggests a cautious approach to ceding control to AI systems and a preference for predictable and manageable learning outcomes. The moderate mean score suggests lecturers are somewhat uncomfortable with AI systems that produce unpredictable results, preferring systems with more predictable and controllable behaviours.

This finding aligns with research on teachers' trust in AI-powered educational technology, which has identified that trust plays a critical role in practitioners' adoption of technology. Teachers mostly discuss issues of accuracy and reliability when discussing AI-powered educational tools (Nazaretsky et al., 2022). Studies have found that teachers' concerns about AI include worries about accuracy, reliability, and the potential for AI chatbots to provide biased responses or inaccurate information that could mislead students (Labadze et al., 2023). The preference for predictable outcomes reflects educators' responsibility to ensure learning experiences are safe, appropriate, and aligned with educational goals.

The lower comfort with unpredictable outcomes suggests lecturers would value AI tools with transparent, explainable behaviours and clear boundaries on what the system can and cannot do. This finding indicates that AI tools designed for educational use should prioritise reliability and predictability, with clear mechanisms for educators to monitor and control system behaviour. The cautious attitude also suggests that lecturers would benefit from understanding how AI systems work and having confidence in their ability to intervene when necessary, supporting the need for professional development and transparent system design.

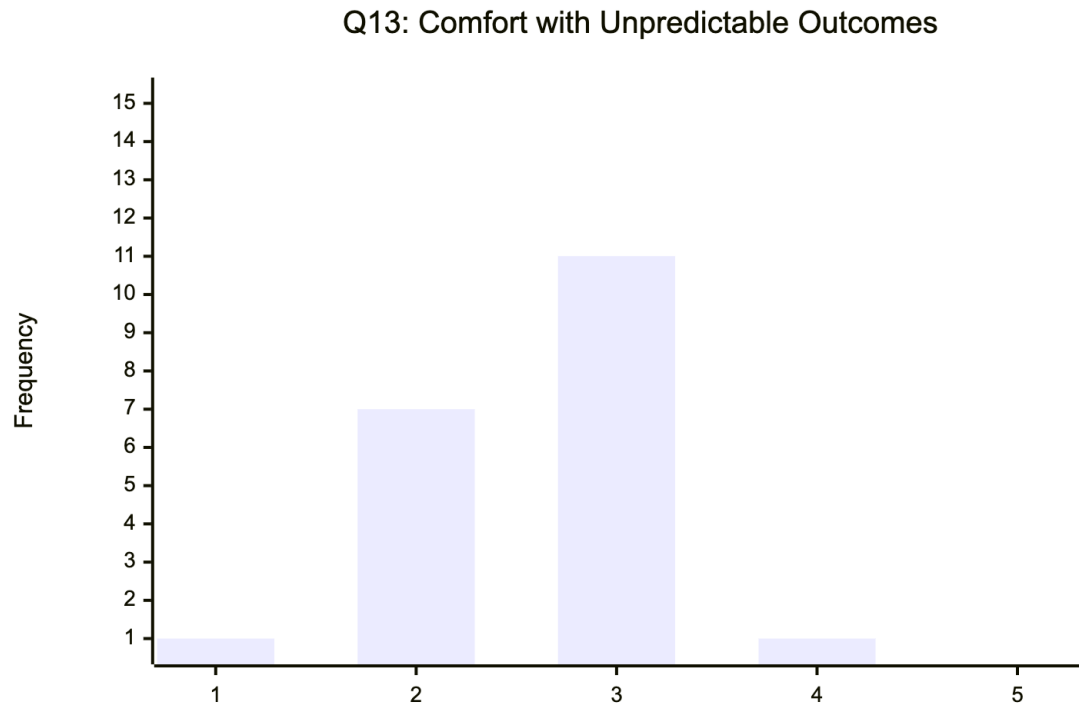

### B.2.14 Most Important Quality (Q14)

When asked about the most important quality a digital learning assistant should embody, lecturers' responses revealed several key themes. Multiple lecturers emphasised the need for reliable, accurate information. One lecturer stated that the most important thing is that it must be accurate and trustworthy, with lecturers needing to monitor student progress and see where they are struggling so they can step in when needed. Several lecturers highlighted the need for clear summaries, with one noting that it should be able to clearly summarise key points for each chapter, as students struggle with too much content and concise summaries would help them focus on what is important.

The ability to generate personalised practice questions was frequently mentioned. Lecturers noted that the ability to generate practice questions based on what students need would be very useful, particularly if it can create quizzes based on key words or topics they are weak in to help them practice more effectively. Multiple lecturers emphasised the need for simplified language, noting that it should be able to summarise and explain content using simple language, as many students find technical terms difficult. Breaking things down into simpler explanations would help them understand better.

Several lecturers emphasised the need for educator oversight, suggesting that a teacher dashboard displaying common gaps and frequently asked questions would be highly beneficial. This would enable lecturers to identify which topics students are struggling with and adjust their teaching accordingly. The importance of personalised learning paths and progress tracking was emphasised. One lecturer envisioned it as a personalised tutor that can repeat what was taught in class, generate MCQ and SAQ questions, help students understand what they know and do not know, track their progress, and suggest which topics to focus on based on their performance.

These findings align with research on AI-powered educational tools and their design requirements. The emphasis on accuracy and trustworthiness reflects concerns identified in systematic reviews, which have found that reliability and accuracy are fundamental concerns when integrating AI chatbots in education, as AI chatbots may provide biased responses or inaccurate information that could mislead students and hinder their learning progress (Labadze et al., 2023). Research has also found that the accuracy of information significantly affects students' intention to use AI tools positively, with students tending to rely more on insights and data provided by AI tools when they consistently offer accurate and reliable information (Dahri et al., 2024). The need for trustworthy AI systems is further supported by research indicating that, from a technical perspective, trustworthy AI is expected to exhibit properties of accuracy, robustness, and explainability, with AI programs or systems generating output that is as accurate as possible, consistent with the ground truth (H. Liu et al., 2023).

The emphasis on personalised practice question generation aligns with research on automated question generation in intelligent tutoring systems. Studies have found that generative AI can generate a diverse array of questions tailored to the learner's current understanding and proficiency level, creating questions that vary in difficulty, topic specificity, and cognitive demand, thereby providing a more personalised assessment of learner progress (Maity & Deroy, 2024). Research has also demonstrated that adaptive question generation can dynamically adjust the difficulty of subsequent questions based on the student's previous responses. Correct answers increase difficulty, while incorrect answers result in easier questions being presented, creating personalised practice opportunities that target individual abilities and difficulties (Abrar et al., 2025).

The need for educator oversight through teacher dashboards is supported by research on teacher-AI collaboration and classroom orchestration tools. Studies have found that teachers want real-time analytics and insights to identify students' strengths, weaknesses, and learning patterns, allowing them to adjust their teaching strategies accordingly (Kamalov et al., 2023). Research on co-designing real-time classroom orchestration tools has identified that teachers need support in understanding what their students are doing, receiving timely feedback on their own teaching effectiveness, and having access to summarised, directly actionable information that helps them understand the why behind student struggles, not just the what (Holstein et al., 2019). Studies have also found that intuitive dashboards, which utilise visual representations, help teachers explore and understand large amounts of information simultaneously, thereby gaining insights that directly support instruction planning (Kim, 2024).

The emphasis on personalised learning paths and progress tracking aligns with research on AI-powered personalised learning pathways. Studies have found that AI-enabled learning pathways assist in delivering education experiences tailored to learner needs, addressing factors such as learning level, how students grasp information, learning rates, and learning preferences, with systems adapting learning and the experience of each student based on progress and areas of successful learning or learning challenges (Abrar et al., 2025). Research has demonstrated that these systems adjust instruction to students' learning pace to maintain an optimal learning pace that challenges students without overwhelming them, ensuring steady progress over time (Abrar et al., 2025). Studies have also found that adaptive algorithms use data collected at the time to determine what should happen in the student's learning process, taking into consideration various input variables such as student performance, time taken to answer questions, and engagement figures, with the system monitoring student accomplishments and adjusting the difficulty and type of content delivered accordingly (Abrar et al., 2025).

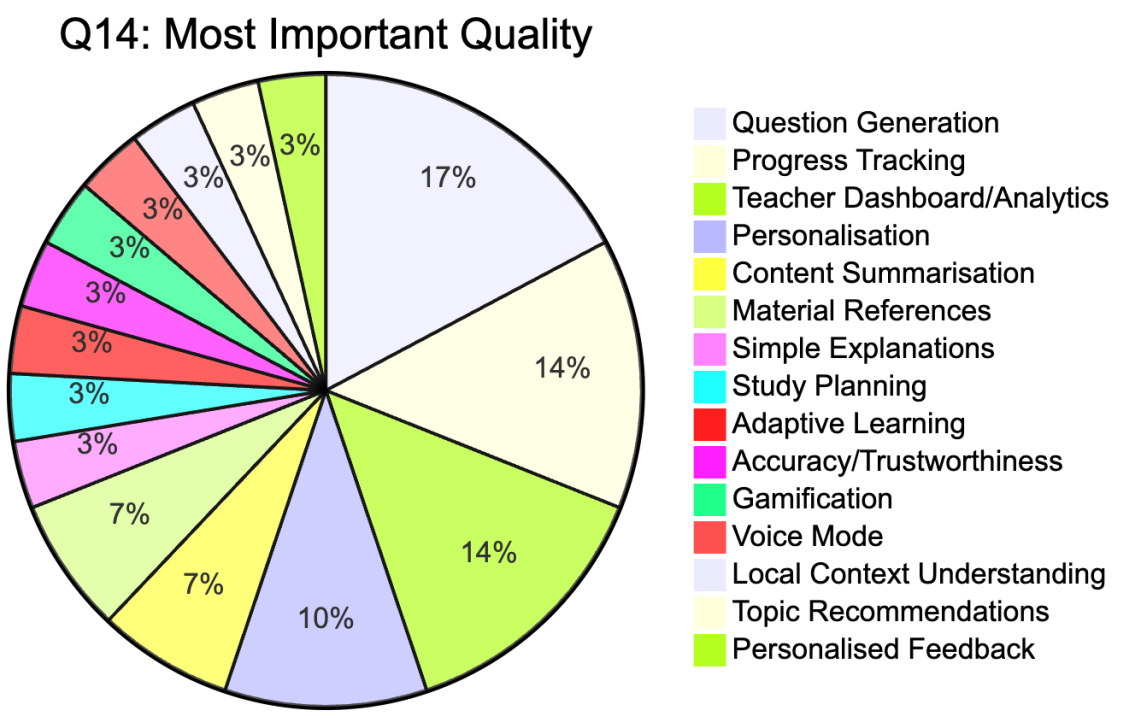

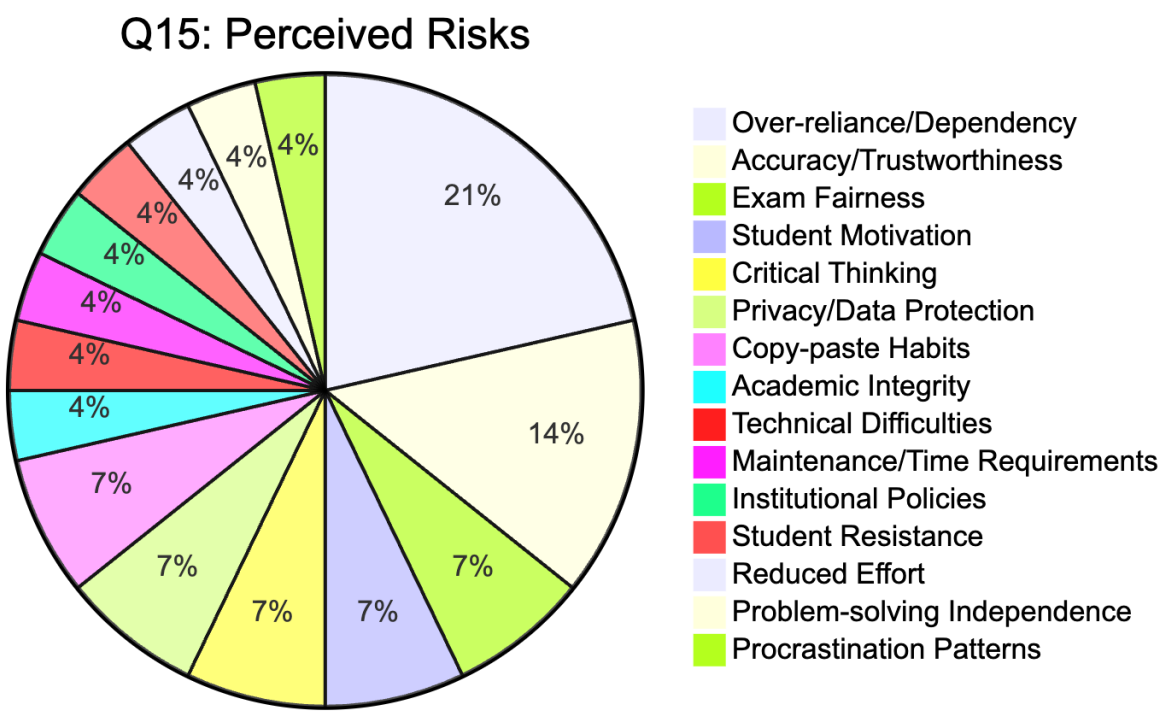

### B.2.15 Perceived Risks (Q15)

When asked about risks in integrating generative AI, lecturers' responses revealed several key concerns. Accuracy dominated responses, with lecturers worried about wrong answers leading to student mistrust. One lecturer stated that their main concern is accuracy, noting that if the AI gives incorrect answers, students will lose trust. They also expressed concern that students might become too dependent on it, rather than thinking for themselves. These concerns align with research findings that AI chatbots may provide biased responses or inaccurate information. If chatbots provide incorrect information or guidance, they could mislead students and hinder their learning progress (Labadze et al., 2023). Research has also found that the reliability and accuracy of generative AI information have been questioned, posing challenges to academic integrity and credibility, with users needing to be aware of differences to distinguish between accurate and misleading information and to identify content biases, potential inaccuracies and fabrications that may emerge during the content creation process (Almuhanna, 2025). The effectiveness and accuracy of information provided by generative AI tools have raised significant concerns, with the potential for perpetuating biases and the inherent limitations of these tools, such as producing fabricated articles or references, further compounding these ethical dilemmas (Ahmed et al., 2024).

Multiple lecturers expressed concerns about students becoming dependent on AI tools, noting that over-reliance and copy-paste habits are a real risk. Students who potentially copy answers without understanding do not help themselves learn or prepare for exams. Several lecturers raised concerns about equitable access, expressing concern about exam fairness and questioning whether it is fair if some students use it and others do not. They noted the need for clear guidelines on when and how students can use it during assessments. These concerns are supported by research on over-reliance and cognitive offloading. Studies have found that learners' tendency to become over-reliant on AI poses challenges for hybrid intelligence, aligning with the concept of cognitive offloading, where learners delegate cognitive tasks to external tools to reduce cognitive effort (Fan et al., 2025). Although cognitive offloading can be beneficial in managing cognitive load, it may lead to decreased internal cognitive engagement over time, ultimately impacting learners' ability to self-regulate and critically engage with learning material, with such cognitive offloading leading to habitual avoidance of deliberate cognitive effort, a phenomenon termed metacognitive laziness (Fan et al., 2025). Research has also found that over-reliance on AI dialogue systems could impede the cultivation of essential skills in students, including critical thinking, problem-solving, and effective communication, with the convenience offered by AI tools in providing quick answers potentially deterring students from engaging in thorough research and forming their own insights (Zhai et al., 2024). Studies have demonstrated that over-reliance on AI can reduce students' skills in creativity and innovation, potentially leading to students becoming overly reliant on technology for information, which undermines their capacity for independent critical thinking and problem-solving (Zhai et al., 2024). The growing use of ChatGPT could also lead to over-reliance on AI, potentially hindering students' critical thinking and problem-solving skills, as they may avoid essential cognitive processes such as reflection, creativity, and independent reasoning (Salih et al., 2025). Research on academic integrity has found that one of the challenges educators face with the integration of chatbots in education is the difficulty in assessing students' work, particularly when it comes to written assignments or responses, with AI-generated text detection not yet foolproof and creating uncertainty that can undermine the credibility of the assessment process, raising concerns about academic integrity and fair assessment practices (Labadze et al., 2023). Studies have also found that AI tools have made it easier for students to engage in dishonest practices, as the accessibility and sophistication of these tools enable students to generate assignments and answers effortlessly, thereby undermining the principles of academic honesty (Mumtaz et al., 2025). Research has indicated that using chatbots or any other artificial intelligence tools to answer exam questions is a form of cheating and academic misconduct, and goes against the fundamental principles of learning and academic integrity, with such misuse potentially severely affecting the student's academic progress and knowledge-acquiring processes by hindering their critical thinking skills, creativity, and ability to apply the concepts learned to real-world situations (Kooli, 2023).

Concerns were expressed about reduced cognitive abilities. One lecturer expressed concern about the impact on critical thinking skills, noting that if students merely obtain answers quickly, they may not develop the reasoning abilities necessary for SAQ questions. Some lecturers expressed concern about motivation, worried that if answers come too easily, students may not put in the effort to learn. They noted the need to ensure it encourages learning rather than shortcuts. These concerns align with research findings on the disruption of critical thinking skills. Studies have found that the growing use of AI tools like ChatGPT in education has raised discussions about how it could affect students' ability to think critically and solve problems, with a significant amount of research suggesting that over-dependence on these technologies could negatively impact the development of essential cognitive skills (Ahmed et al., 2024). Research has found that relying on AI for tasks requiring critical thinking may decrease the students' capacity for original thought, creativity, and independent problem-solving, underscoring the necessity for judicious integration of AI tools within educational frameworks to complement rather than supplant traditional learning methodologies (Ahmed et al., 2024). Studies have also found that if learners rely excessively on AI-generated outputs or facilitation, they may not experience the necessary disfluency or cognitive difficulty to trigger deeper metacognitive processes, potentially defaulting to less effortful, more intuitive decision-making, which reinforces a state of metacognitive laziness (Fan et al., 2025). Research has demonstrated that if students lean too heavily on AI for content generation, they risk not developing the ability to analyze information, construct logical arguments, or integrate knowledge from diverse sources for academic and professional success, with faculty expressing concern about potential overdependence leading to reduced effort in crafting well-structured sentences and adhering to proper grammar and spelling (Zhai et al., 2024). Data protection concerns were mentioned, with lecturers noting that privacy and accountability are important. There must be clarity about who is responsible if something goes wrong, with student data protected properly. These concerns are supported by research indicating that the integration of AI chatbots in education raises several ethical implications, particularly concerning data privacy, security, and responsible AI use, with AI chatbots interacting with students and gathering data during conversations, necessitating the establishment of clear guidelines and safeguards (Labadze et al., 2023). Research has shown that AI tools often necessitate the collection and analysis of student data, raising concerns about privacy and security. To address these concerns, proper protocols and safeguards must be in place to protect sensitive student information and ensure compliance with data protection regulations (Kamalov et al., 2023). Some lecturers mentioned technical difficulties, time requirements for maintenance, and institutional policy constraints as potential risks. These concerns align with research findings that AI tools rely on technology, and technical glitches or system failures can disrupt the collaboration process, potentially affecting reliability and stability issues that undermine trust and confidence in AI tools, which in turn impact their adoption and effectiveness (Kamalov et al., 2023).

### B.3 Interview Questions

Semi-structured interviews were conducted with the same 20 lecturers to elaborate on survey responses and explore specific challenges and potential solutions in depth. Interviews followed a semi-structured format with six main sections covering teaching context, reflection on current tools, support and feedback in learning, imagining new possibilities, values and concerns, and forward-thinking reflections. Each section contained main questions and follow-up control or check questions to explore responses in depth.

**SECTION 1: Your Teaching Context**

1. Could you walk me through a typical lesson? What kind of support do students usually need from you?
2. When students get stuck or confused, how do they usually seek help? Do you wish they had more ways to get timely support?
3. What do students often struggle with outside your class time, after hours, during revision, or while preparing assignments?

**Control question:** You mentioned earlier that students often [e.g., need motivation]. Could you give an example of how you usually address that?

**SECTION 2: Reflection on Current Tools**

4. Are there any digital tools or platforms that have helped your students learn better? What did those tools do well, or poorly?
5. Have you noticed if students prefer learning with interactive tools, videos, or some kind of dialogue-based support?
6. What are your thoughts on tools that allow students to ask questions on their own time? Do you see a place for that?

**Check question:** Earlier, you spoke about [e.g., encouraging independent learning]. Would tools that guide students step-by-step align with that goal, or not?

**SECTION 3: Support and Feedback in Learning**

7. If students had someone, or something, to turn to when you're not available, what kind of support would be most helpful to them?
8. What kinds of questions do students ask you again and again? Could some of these be answered in a more scalable way?
9. When giving feedback, what do you wish you had more time or tools to explain better?

**Control question:** You said earlier that [e.g., personal connection] is important. Would an extra layer of guidance (even if not human) still be valuable?

**SECTION 4: Imagining New Possibilities**

10. If you could give your students a “study companion” that’s available 24/7, what should it be able to do? (Feel free to dream a little here.)
11. Would you find it useful if such a companion could adapt to the different needs and learning styles of various students? Why or why not?
12. How would you like to shape or personalise such a companion, so that it reflects your teaching approach or subject matter?

**Check question:** Earlier, you said [e.g., student autonomy] matters. Would this companion risk over-guiding, or could it help foster deeper self-direction?

**SECTION 5: Values and Concerns**

13. What should never be left to automation in education? Are there boundaries we shouldn’t cross?
14. What are your concerns when it comes to students learning through digital systems? (E.g., privacy, distraction, dependency)
15. If your students were to use such a support tool regularly, what values would you want it to reinforce? (E.g., curiosity, honesty, resilience)

**Control question:** Based on your earlier mention of [e.g., motivation], how could a tool promote that without replacing human encouragement?

**SECTION 6: Forward-Thinking Reflections**

16. Looking ahead, what role do you think "intelligent systems" might play in helping students manage their learning?
17. If such tools were created with input from teachers like you, what aspects should be driven by educators rather than developers?

18. Finally, if you could leave a mark on a tool that might one day help thousands of students, what would you want it to say, or do, that reflects your teaching values?

**Check question:** You mentioned earlier that students often [e.g., fear making mistakes]. How might a learning tool gently help them grow past that?

### B.4 Interview Analysis

Written notes were taken during interviews for thematic analysis, with no audio recordings made to protect participant privacy. Interviews were conducted as semi-structured conversations, allowing lecturers to elaborate on survey responses through follow-up questions exploring specific challenges and potential solutions. The following sections present thematic analysis organised by key themes that emerged across the interview questions.

#### B.4.1 Student Challenges When Preparing for Exams

Lecturers consistently identified content volume as the primary challenge facing students during exam preparation. Students struggle with the sheer volume of material that spans multiple documents, pages, and case studies. One lecturer described this as a massive amount of content with many details spread across multiple documents, comprising multiple pages, multiple case studies, and too many possible scenarios to apply the principles. When under pressure, students mix up terms and lose the main idea, especially when topics sound similar. Reading-heavy sections can drain focus, making it difficult for students to maintain attention when faced with extensive reading materials.

Beyond memorisation, students struggle to grasp underlying principles and concepts. Lecturers observed that the linking of concepts to practical tasks is weak, indicating a difficulty in connecting theoretical knowledge to practical application. Students can memorise content but cannot apply knowledge to new scenarios, with application to new contexts described as shaky. In modules involving ethics, scenarios often feel abstract to students who memorise but struggle to apply abstract concepts to real-life situations. Students recall procedures without understanding the underlying reasoning, remembering steps but not the why behind them.

Time management emerged as a significant challenge, with many students cramming at the last minute and experiencing rapid declines in retention. Students struggle with sustained reading tasks, showing an inability to engage in long reading. The typical pattern involves students starting revision in the last week and asking for tips, which lecturers described as not an ideal approach. Students attempt to study but struggle to maintain consistent effort, and a lack of motivation affects engagement with revision materials. Research indicates that procrastination is widespread on college campuses, with estimates showing that 70 to 95 per cent of students procrastinate, and about 50 per cent habitually procrastinate despite frequent warnings in college success curricula (Ma & Chen, 2024). The critical issue is not the delay but the total time students dedicate to academic work, which often falls short of faculty expectations.

Students experience anxiety when preparing for exams, which is often compounded by a lack of understanding of how to study effectively. Stress affects their ability to retrieve information during exams, with recall under pressure described as weak. Performance deteriorates under exam pressure, with students mixing up terms when stressed. In technical modules, command-line basics become confusing when practice is insufficient and syntax is forgotten. Design topics encompass numerous details, and recall under pressure is often weak for technical content.

#### B.4.2 Exam Component Challenges

There was consensus among lecturers that Short Answer Questions represent the most challenging component for weaker students across multiple modules. SAQ requires reasoning rather than just recall, with higher-order thinking requirements that students find difficult. Interpreting and explaining reasoning can be challenging, and students often struggle to articulate and convey their thoughts effectively. Some students could not understand the question or simply did not know the answer, indicating comprehension barriers that prevented them from attempting responses. A weaker ability to comprehend, combined with a weaker ability to articulate well, creates barriers through language and communication skills. SAQ requires structure, and many students cannot organise points, producing short bullet points instead of full answers. English proficiency has a significant impact on the ability to structure coherent responses, as language plays a crucial role in clarity. SAQ is the main barrier for students with weaker English skills, with language gaps particularly affecting SAQ performance. Research indicates that students must express their answers accurately and demonstrate a higher level of language and communication skills when answering questions that require a written response, as even minor misspellings can potentially lead to a loss of points (Abina et al., 2025). While some stronger students may overthink Multiple-Choice Questions, this is less common. SAQ remains the more consistently difficult component even when MCQ presents challenges.

### B.4.3 Current Revision Approaches

Students commonly read PowerPoint slides, though lecturers noted this approach may not be the most effective. Students skim short videos and slides, which helps a bit but results in very surface-level engagement. Some students engage only by creating notes and reading, showing limited effectiveness without active engagement. Past papers are available in the library, but usage varies considerably. Some students complete past year question papers, while others do not, indicating inconsistent engagement. Students often delay accessing solutions until the last minute, with model answers typically only being opened near exam time.

Students attempt the quizzes provided during lessons, but lecturers note that these quizzes may not cover all the content that is taught. When students take the time to complete quizzes, they can be quite effective, although effectiveness depends on student motivation. Students prefer interactive practice over structured sessions, using practice quizzes more often than revision sessions. This preference for interactive tools suggests that dialogue-based or responsive systems might better engage students than static content delivery mechanisms. Research supports this finding, indicating that conversational agents can provide real-time and adaptive support that mimics the role of a human teacher, offering immediate feedback, answering questions, and delivering personalised guidance through interactive dialogue (Ait Baha et al., 2024). Delayed engagement with feedback occurs as practice questions are attempted, but model answers are only opened near exam time. Students sometimes study in groups using flashcards, although effectiveness varies by module. When studying with peers, discussions occasionally stray off track, with effectiveness described as moderate and dependent on the group dynamics. Students value peer input but may lack structure, with peer advice preferred over reading materials. Students typically start revision in the last week and ask for tips, which lecturers described as not ideal and creating pressure. Students often procrastinate and wait until just before exams to engage, delaying work until deadline pressure mounts.

### B.4.4 Current Support Mechanisms

Lecturers provide various forms of revision support, including structured walkthroughs and workshops, through revision sessions. Past exam papers are available with solution outlines. Digital resources include online notes, LMS-based practice questions and practice quizzes, as well as mock tests with walkthrough sessions. Condensed materials include summary slides and quick reference sheets. Video resources include short recap clips for common topics. Study aids include flashcards, and one-to-one support opportunities are available through consultation slots.

Some students engage with all resources, utilising everything provided. Selective engagement occurs with summary materials getting attention, while forum threads are mostly ignored. Partial engagement is evident, as quizzes are used frequently, but videos are saved and not fully watched. Students prefer interactive support, with peer advice preferred over reading materials. This preference for interactive dialogue-based support suggests students value conversational learning approaches over passive content consumption. Research supports this finding, indicating that conversational agents can provide real-time and adaptive support that mimics the role of a human teacher, offering immediate feedback, answering questions, and delivering personalised guidance through interactive dialogue (Ait Baha et al., 2024). Short attention spans affect the ability to engage with longer-form content.

### B.4.5 Where Current Support is Insufficient

Students juggle multiple commitments, facing time constraints due to other modules and ongoing projects. Lecturers often lack the capacity to create comprehensive question banks, as they frequently face time and energy constraints when setting a sufficient number of questions to adequately assess student learning. Limited availability outside class time means lecturers cannot reply to every question after hours. Individual attention is often constrained by class sizes, with limited time for one-to-one feedback. Research indicates that traditional academic support faces challenges in large institutions with high student populations, making individual communication impractical and time-consuming (Dahri et al., 2024). Support sessions are often constrained by time limitations, particularly during peak periods, which can impact the quality of advice (Dahri et al., 2024). Additionally, teachers encounter considerable time-based limitations when integrating AI technologies, with the iterative process of evaluating and modifying AI-generated content being particularly time-consuming (Almuhanna, 2025).

Students ask similar questions repeatedly, with repeated questions described as common. Basic issues recur, creating a workload for lecturers. Research supports this finding, indicating that students frequently need answers to basic information about their courses, such as information on course materials, due dates, study tips, and office hours (Y. Chen et al., 2023). After-hours support is limited, as lecturers are unable to reply to every question outside of regular office hours. Students tend to delay engagement until exams are approaching, making it challenging for lecturers to motivate early revision. Procrastination patterns emerge, with some students waiting until exams are close before starting to study. Engagement requires motivation, and students often tend to procrastinate and delay work until just before exams, exhibiting patterns of delayed engagement.

English proficiency affects learning pace, with language gaps slowing progress. Diverse learning needs are difficult to address at scale, with pacing differing widely and being hard to individualise. A trade-off exists between coverage and depth, with many topics covered in a few sessions, resulting in a loss of depth. Students often struggle with maintaining sustained focus and tend to have short attention spans. In-class engagement issues arise, as students often fail to pay attention when material is covered in class. Short-term attention spans affect expectations, with students expecting to be promptly answered, preferably on the spot.

### B.4.6 Expected AI Chatbot Functions

Lecturers expected the chatbot to generate practice questions, including both MCQ and SAQ formats, creating questions with appropriate breadth and depth. The tool should support practice and assessment by generating Q&A or MCQ for student learning. Personalised question generation could occur based on the preference for keywords. Comprehensive practice support would involve creating questions and answers for practising. Sequential learning support would enable students to attempt questions after reviewing the content.

Twenty-four-hour availability emerged as a key expectation, with lecturers wanting the chatbot to answer all of the students' questions correctly, anytime, anywhere. The tool should remove time and location barriers, enabling students to ask questions at any time and from anywhere. Research indicates that AI-powered personalised learning platforms are accessible anytime, anywhere, providing students with on-demand support regardless of geographical location or time constraints (Cai et al., 2025). The ability to provide immediate support 24/7 is a significant advantage compared to traditional tutoring approaches (Ahmed et al., 2024).

The chatbot should condense long materials through summarised notes. Complex content should be made accessible through clear explanations with simple examples. Detailed feedback should explain to students why their attempted answers are right or wrong. Technical language should be simplified by summarising and explaining content using simple natural language. Scaffolded support would clarify questions, break content into steps, and provide simple examples. Research indicates that scaffolding in educational contexts involves providing students with instructional support, where teachers offer successive degrees of temporary assistance that aid students in reaching comprehension and skill development levels they would not be able to attain without such support (Kuhail et al., 2023). Scaffolding teaching techniques offer students individualised support by building on learners' prior knowledge and supporting them in the assimilation of new information, with scaffolding being the process through which students are assisted to progress from one developmental stage to the next by providing them with essential assistance within their Zone of Proximal Development (Msambwa et al., 2025).

Personalised learning paths would recommend study plans with targeted short quizzes per weak area. Structured guidance would provide personal study plans with weekly goals. Habit formation support would send reminders plus small daily drills to build study habits. Exam-specific preparation would cater to the format of the exam, enabling students to approach their studies effectively. Content should align with learning needs through contextually relevant content. The practice exam format would be supported by simulating the SAQ. Educator oversight capabilities would enable lecturers to grade and monitor student progress. Targeted improvement guidance would provide feedback on where students can improve.

### B.4.7 Potential Benefits for Weaker Students

A safe learning space would allow students to ask questions at any time without feeling judged or intimidated. Weaker students are more reserved, so this would allow them to ask and attempt questions within their own comfort zone, reducing social anxiety. Research indicates that chatbots, as AI-driven educational tools, have the potential to address help-seeking concerns and alleviate help avoidance behaviour by providing a private, nonjudgmental space for students to ask questions without fear of embarrassment in front of peers or teachers (Neumann et al., 2025). Some students are particularly prone to avoiding help-seeking due to concerns about negative judgments from teachers or peers, and chatbots can help address these concerns (Neumann et al., 2025). Students can ask freely without feeling judged, and confidence improves, building self-efficacy.

Self-paced learning allows students to learn at their own comfortable pace. Flexible scheduling means students can ask the chatbot at any time, at their own convenience and pace. Different schedules are accommodated, with anytime revision possible and even late-night help available. Extended availability provides an opportunity for students who are weaker or shy to ask questions, even in the middle of the night. Research indicates that personalised learning encourages students to take ownership of their learning and become more self-directed learners, enabling them to set goals, monitor their progress, and make informed decisions about their learning paths (Kamalov et al., 2023). Post-lesson support offers guidance to help students gain more confidence. Continuous support builds confidence through better understanding and the ability to ask the chatbot twenty-four hours a day. Deep learning support helps students grasp concepts by providing a deeper understanding.

Appropriate difficulty levels mean weaker students will benefit from more fundamental knowledge facts and simpler concept questions. Engagement through appropriate challenge occurs when short practice quizzes maintain attention and momentum. Lecturers envisioned a positive personalised tutor or module coach that could repeat what was taught in class, generate MCQ and SAQ questions based on required depth and breadth, figure out what the student knows or does not know, advise on correctness of answers, repeatedly generate questions to reinforce weaker topics, analyse and suggest which weaker and more important topics to focus on, and consider students' past performance in recommendations.

### B.4.8 Concerns About AI Tools

Dual concerns emerged regarding reliability and dependency, with both accuracy and over-reliance being raised as issues. Ongoing maintenance is required, with accuracy needing to be monitored and updates provided promptly. Reliability is critical, as lecturers need to ensure that no hallucinations occur and that they can fully trust the system. Quality control is necessary, as accuracy issues have occasionally arisen and have been identified.

Students may become dependent on the tool, exhibiting patterns of over-reliance. The tool may enable procrastination, with some students waiting until exams are close before starting to study. Too many hints can reduce the effort to think, potentially undermining critical thinking. Cognitive offloading concerns emerged, as they might experience reduced thinking skills, potentially making it difficult for them to obtain the information or answer questions they needed quickly. Academic integrity risks exist, with over-reliance and copy-paste habits that must be managed carefully. Research extensively documents these concerns, indicating that over-reliance on AI dialogue systems can lead to diminished critical thinking, analytical thinking, and decision-making abilities (Zhai et al., 2024). Studies have found that over-reliance could reduce students' skills of creativity and innovation, potentially undermining their capacity for independent critical thinking and problem-solving (Zhai et al., 2024). The concept of cognitive offloading, where learners delegate cognitive tasks to external tools to reduce cognitive effort, may lead to decreased internal cognitive engagement over time, ultimately impacting learners' ability to self-regulate and critically engage with learning material (Fan et al., 2025). Overdependence can hinder the development of key academic and life skills, ultimately affecting students' intellectual growth and their preparedness for real-world challenges (Salih et al., 2025).

Concerns were raised about equitable access and fairness regarding exam administration. Standards must be maintained, with academic integrity and fairness that must be protected. Engagement requires intrinsic motivation, with student motivation to use the tool being necessary. The tool alone may not drive engagement, as students need to be motivated to use it effectively. Data protection and responsibility concerns were raised, with privacy and accountability for answers described as important. Dependency patterns may develop, with reliance on hints that could grow without coaching. Analytical skills may be compromised, with potential implications for critical thinking and decision-making.

### B.4.9 Key Features Lecturers Would Design

Reliability emerged as a foundational requirement, with lecturers needing to ensure no hallucinations occur and that they can fully trust the system. Educator oversight capabilities would enable lecturers to grade and monitor student progress. Targeted improvement guidance would provide feedback on where students can improve. Research indicates that trust is essential for fully realising the potential of AI, with trustworthy AI expected to exhibit properties of accuracy, robustness, and explainability from a technical perspective (H. Liu et al., 2023). AI chatbots may provide biased responses or inaccurate information. If a chatbot provides incorrect information or guidance, it could mislead students and hinder their learning progress (Labadze et al., 2023). Studies have highlighted that teachers' trust in AI-powered educational technology is critical for adoption, with accuracy and reliability being fundamental concerns (Nazaretsky et al., 2022).

Structured content organisation would include key points for each chapter, summarised. Accessibility through simplification would summarise and explain content using simple natural language. Contextual understanding would enable the system to comprehend local phrasing and common student terminology, thereby enhancing its ability to provide accurate responses. Comprehensive practice support would generate Q&A or MCQ for student learning. Personalised question generation could create quizzes based on the preferences of keywords. Extensive practice opportunities would generate more questions and answers for students to practice and reinforce their understanding.

Educator insights into student struggles would be provided through a teacher dashboard showing common gaps and frequently asked questions. Data-driven intervention support would add analytics for teachers to spot weak themes. Individual learning journeys would be supported if the system can be personalised and track progress. Structured learning progression would involve adaptive paths with checkpoints and spaced practice. Research demonstrates that adaptive learning algorithms dynamically adjust instructional content and strategies based on individual learner progress and performance, with AI-enabled adaptive learning systems tailoring content to each learner's unique needs and preferences (Looi & Jia, 2025; Wang et al., 2024). Personalised learning and adaptive learning have been shown to positively influence learners' motivation, experience, and learning outcomes (D.-L. Chen et al., 2025).

Motivation through gamification could involve gamified points and badges to sustain momentum. Research demonstrates that gamification introduces flexible and customizable reward mechanisms, such as points, badges, and leaderboards, which can be adapted to individual achievements and preferences, thereby encouraging students to sustain their efforts and persevere in the face of challenges (Ahmed Dahri et al., 2025). By recognising personal milestones, gamification fosters a sense of accomplishment and satisfaction in students. Students who learn in gamified environments demonstrate higher levels of engagement and are more likely to immerse themselves in activities and persist through challenges (Ahmed Dahri et al., 2025). Accessibility and engagement options would include an optional voice mode for those who prefer listening. Lecturers envisioned a positive personalised tutor or module coach that would provide repetition of class content, generate MCQ and SAQ based on depth and breadth requirements, assess what students know and do not know, provide answer correctness feedback, repeatedly generate questions for weaker topics, analyse and recommend focus areas, and personalise based on performance.

### B.5 Thematic Map Visualisation

Content volume was consistently identified as the primary challenge facing students. There was universal recognition that SAQ represents the most challenging component for students. Both students and lecturers face time limitations, creating constraints on support provision. Strong expectations emerged for continuous support through twenty-four-hour availability. Reliability is crucial for establishing trust, with concerns about accuracy being a prominent aspect of discussions. A balance must be struck between support and independence, with dependency risks requiring careful management. Focus emerged on supporting vulnerable learners, with particular attention to the needs of students who are weaker academically. It was recognised that one-size-fits-all approaches are insufficient, with personalised needs identified across interviews.

Several tensions emerged during discussions. The relationship between support and independence requires careful navigation, exploring how to provide help without creating dependency. Ensuring twenty-four-hour access maintains accuracy creates a tension between availability and quality. Individual attention within resource constraints highlights the challenge of personalisation versus scalability. Maintaining attention while promoting deep learning involves striking a balance between engagement and depth of learning. Quick responses versus skill development creates a tension between immediate answers and the learning process.

Interviews were conducted as semi-structured conversations, following a six-section question framework, which allowed lecturers to elaborate on their survey responses. Written notes captured key themes, direct quotes, and observations throughout the process. No audio recordings were made to protect participant privacy. Follow-up control and check questions explored specific challenges and potential solutions in depth. Lecturers expressed both optimism and caution regarding the implementation of Generative AI, presenting nuanced perspectives. A strong emphasis emerged on the practical and vocational education context throughout the discussions. It was recognised that tool design must account for diverse student needs and abilities. Detailed findings from all 18 interview questions, including direct quotes from pre-implementation survey open-ended responses (Q14 and Q15) that were integrated into relevant interview questions, are documented in the pre-implementation interview notes.

Thematic analysis of both the pre-implementation survey responses and interview data revealed complex interconnections among the themes. The following Mermaid diagram visualises relationships between student challenges, current support mechanisms, expected AI functions, concerns, and design features identified across both quantitative survey data and qualitative interview responses. This integrated map synthesises findings from structured Likert-scale pre-implementation surveys and in-depth semi-structured interviews, providing a comprehensive view of lecturers' perspectives before co-development workshops determined the specific form of the POC.

The thematic map illustrates how student challenges identified in both survey responses and interview discussions relate to limitations in current support mechanisms, which in turn shape expectations for Generative AI technologies. Expected AI functions address support limitations while generating both potential benefits and concerns. Key design features emerge from addressing concerns and maximising benefits. Tensions exist between competing priorities, requiring careful navigation in tool design. This integrated map reflects lecturers' original state before co-development workshops, synthesising perspectives from structured survey data and in-depth interview responses, when the specific form of the POC had not yet been determined.

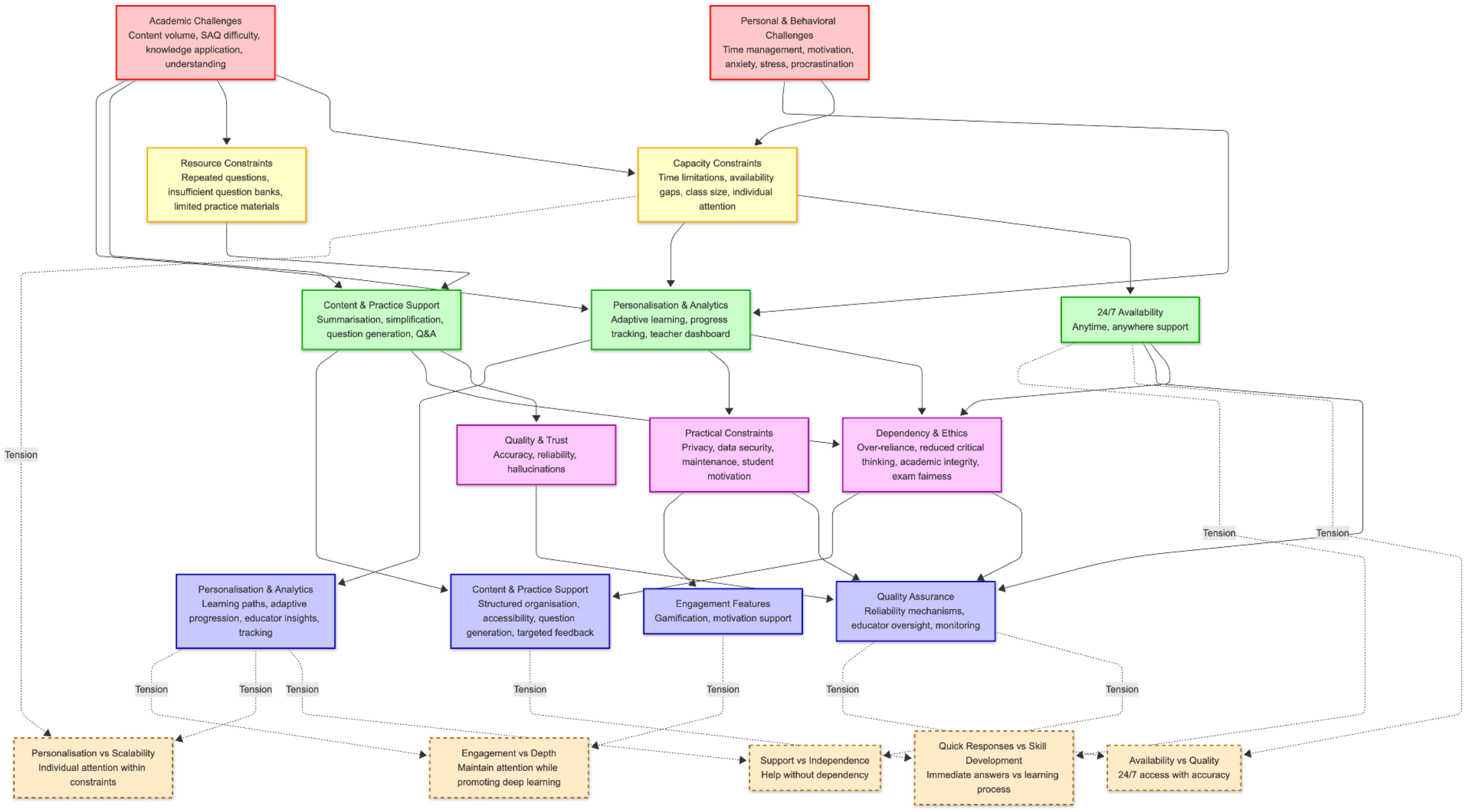
Academic Challenges
Content volume, SAQ difficulty, knowledge application, understanding
Personal & Behavioral Challenges
Time management, motivation, anxiety, stress, procrastination
Resource Constraints
Repeated questions, insufficient question banks, limited practice materials
Capacity Constraints
Time limitations, availability gaps, class size, individual attention
Content & Practice Support
Summarisation, simplification, question generation, Q&A
Personalisation & Analytics
Adaptive learning, progress tracking, teacher dashboard
24/7 Availability
Anytime, anywhere support
Quality & Trust
Accuracy, reliability, hallucinations
Practical Constraints
Privacy, data security, maintenance, student motivation
Dependency & Ethics
Over-reliance, reduced critical thinking, academic integrity, exam fairness
Personalisation & Analytics
Learning paths, adaptive progression, educator insights, tracking
Content & Practice Support
Structured organisation, accessibility, question generation, targeted feedback
Engagement Features
Gamification, motivation support
Quality Assurance
Reliability mechanisms, educator oversight, monitoring
Tension
Personalisation vs Scalability
Individual attention within constraints
Engagement vs Depth
Maintain attention while promoting deep learning
Support vs Independence
Help without dependency
Quick Responses vs Skill Development
Immediate answers vs learning process
Availability vs Quality
24/7 access with accuracy

## APPENDIX C: DESIGN THINKING WORKSHOP ANALYSIS

This appendix documents the design thinking workshop process conducted during Phase 2 co-development sessions. The workshops used structured design thinking methodologies to work through problem definition, ideation, and solution specification.

### C.1 Workshop Introduction

The workshop began with an introduction to the design thinking process and the design frameworks that would guide the co-development session. Participants were reminded that Phase 1 interviews had been completed individually, and this session would involve collaborative design work to create a Generative AI solution for their educational context. The facilitator explained that the process would incorporate human-computer interaction principles to ensure usability, accessibility, and pedagogical soundness. Participants expressed enthusiasm about working together, noting that while individual interviews showed similar challenges, collaborative work would likely produce more comprehensive solutions.

### C.2 Challenge Statement Development

The workshop began by articulating the core challenge based on the findings from Phase 1 interviews. Lecturers identified several interconnected problems, including the gap between student needs and lecturer capacity, the overwhelming volume of content across multiple documents and case studies, students' inability to apply knowledge beyond memorisation, repeated basic questions consuming lecturer time, and time management issues with students cramming at the last minute. From the student perspective, the fundamental problem was a lack of timely, personalised support, particularly after hours when students are studying. Weaker students face additional challenges, including anxiety, fear of asking questions, and a need for non-judgmental learning spaces. Short Answer Questions (SAQ) were identified as the most challenging component, requiring higher-order thinking and articulation skills that many students lack.

Multiple challenge statement formulations were proposed. One version focused on providing personalised, 24/7 learning support to reduce lecturer workload while improving student confidence and exam preparedness. Another emphasised addressing content volume and repeated questions while maintaining academic integrity and promoting deep learning. The group recognised that both perspectives were important and combined them into a comprehensive challenge statement asking how they might provide personalised, 24/7 learning support for students during exam preparation that addresses content volume challenges and repeated questions, while maintaining academic integrity and promoting deep learning, so that lecturers can focus on higher-value teaching activities and students gain confidence and exam preparedness. This statement addressed the needs of both students and lecturers while establishing clear boundaries regarding academic integrity.

### C.3 "How Might We" Statements and Ideation

Participants generated multiple "How Might We" (HMW) statements exploring different solution directions. The statements covered providing instant, accurate explanations through AI-powered systems, generating personalised practice questions through adaptive learning technology, creating non-judgmental learning spaces through conversational AI, reducing repeated questions through automated responses, supporting students with SEN through multiple interaction modes, condensing long materials through AI-powered content analysis, providing scaffolded support through step-by-step explanations, and tracking student progress through analytics. Participants also raised important balancing questions about encouraging independent thinking while providing support, and ensuring that help promotes learning rather than dependency.

The facilitator clustered the HMW statements into themes, including question answering and explanations, practice question generation, content summarisation, personalisation and progress tracking, as well as accessibility with multiple interaction modes. Participants observed that all statements pointed toward some form of AI assistant or chatbot capable of multiple functions. However, concerns about dependency were raised, leading to additional HMW statements about balancing help provision with encouraging students to work through problems independently first. Through voting, the top priorities were providing instant, accurate explanations (5 votes), reducing repeated questions through automated responses (5 votes), generating personalised practice questions (4 votes), creating non-judgmental learning spaces (4 votes), and supporting students with SEN through multiple interaction modes (3 votes). These priorities suggested a conversational, interactive system that could address multiple needs simultaneously.

### C.4 Problem Tree Analysis

The problem tree analysis mapped root causes, core problems, and effects. Root causes included limited lecturer availability outside class hours preventing responses to after-hours questions, large class sizes limiting individual attention and one-to-one feedback time, overwhelming content volume across multiple documents and case studies, time management issues with students cramming at the last minute, language barriers affecting learning pace for some students, lack of confidence in weaker students creating fear of asking questions, and insufficient practice materials due to lecturers' limited capacity to create comprehensive question banks.

Multiple formulations of the core problem were proposed, including "Students lack timely, personalised support during exam preparation," "Current support mechanisms cannot scale to meet individual student needs," and "Gap between student support needs and lecturer capacity." The group agreed on "Current support mechanisms cannot scale to meet individual student needs" as the central problem. Effects included increased lecturer workload due to repeated questions, student anxiety and reduced confidence, especially for weaker students, lower exam performance, particularly on SAQ components, reduced student engagement resulting from frustration, and inequitable access to support, where stronger students might receive more help while weaker students receive less. This analysis showed the interconnected nature of the problems and the need for scalable solutions.

### C.5 Abstraction Laddering

Abstraction laddering involved moving between abstract goals and concrete solutions to determine the appropriate intervention level. Starting from "Students need help with exam preparation," participants laddered up to identify underlying goals, including that students need to understand concepts deeply rather than just memorise them, and students need to build confidence, especially those who are weaker. Students need to develop independent learning skills. This produced the abstract goal that students need to develop independent learning skills while building confidence through deep understanding.

Laddering down to concrete solutions produced providing practice questions to test understanding, offering explanations on demand when students are stuck, and creating a chatbot that answers questions 24/7. The group identified the "sweet spot" as providing on-demand, personalised learning support that builds confidence while promoting independent learning through scaffolded guidance rather than direct answers. This formulation struck a balance between the need for support and the importance of maintaining independent learning capabilities, addressing concerns about dependency that had emerged throughout the discussion.

Based on insights from the problem tree analysis and abstraction laddering, the challenge statement was refined. Participants emphasised the need to highlight scalability issues and scaffolded support aspects, ensure the statement addressed both stronger and weaker students, including those with SEN, and balance 24/7 availability with accuracy requirements. The refined challenge statement asked how to design a Generative AI-powered learning support system that provides personalised, 24/7 assistance to students during exam preparation, addressing content volume challenges and repeated questions through scaffolded support appropriate for diverse student ability levels, while maintaining academic integrity and promoting deep learning, so that lecturers can focus on higher-value teaching activities and all students receive equitable support. This refined statement incorporated the problem analysis insights, maintained an appropriate abstraction level, addressed identified root causes, and reflected the "sweet spot" from laddering exercises.

### C.6 Stakeholder Mapping and User Journey Analysis

Stakeholder mapping identified all affected parties and their needs in the current system. Primary stakeholders included different student types (stronger students, weaker students, students with SEN), lecturers, and module coordinators. Secondary stakeholders included school administration, parents or guardians, and future employers. The group focused on primary stakeholders, mapping their specific needs and pain points.

Stronger students need challenging content and verification of understanding to ensure they are on the right track. Their main pain point is frustration with repetitive basic questions from peers that slow down class progress. They have a medium level of influence and a high interest in solutions. Weaker students need basic explanations, confidence-building, and non-judgmental spaces with simplified language. Their pain points include fear of asking questions, overwhelming content volume, and SAQ difficulty, with many feeling embarrassed to ask in class. They have low influence but very high interest. Students with SEN need multiple interaction modes, including voice input and visual options, extended time, and simplified language. Their pain points include communication barriers and social anxiety, requiring accessibility features. They also have low influence but very high interest.

Lecturers need a reduced workload, the ability to focus on high-value activities, and student success. Their pain points include repeated questions consuming time, time constraints, and large class sizes, which prevent individual attention to each student. They have high influence and high interest in solutions. This stakeholder mapping revealed the diverse needs that any solution must address and emphasised the importance of providing equitable support across different student populations.

User journey mapping captured the current student experience during exam preparation across multiple stages, including awareness of approaching exams, planning for preparation, content review, practice, seeking help, revision, exam day, and post-exam reflection. The group focused particularly on the "Seeking Help" stage, mapping experiences for both weaker and stronger students.

For weaker students, current actions include trying to find the lecturer, checking notes, and asking peers if brave enough. Touchpoints include lecturer office hours (if available), WhatsApp, and classmates. The emotions experienced are anxiety, frustration, and embarrassment, as students worry about appearing foolish. Pain points include lecturer unavailability, fear of judgment, and unclear explanations, even when they do ask. Opportunities identified include 24/7 support, non-judgmental spaces, and simplified explanations.

For stronger students, current actions include verifying understanding, asking targeted questions, and reviewing materials independently. Touchpoints include lecturers for verification, course materials, and past papers. Emotions are generally confident but with a desire to ensure accuracy. Pain points are less severe, primarily waiting for lecturer responses, but they are generally self-sufficient. Opportunities include quick verification and challenging content on demand. This journey mapping showed where interventions could be most valuable and how different student types experience the same journey differently.

## C.7 Persona Development and Empathy Mapping

Three key personas were developed based on real students known to the lecturers. "Chloe" represents weaker students. She is 18 years old, in her first year, struggles with content volume, has low confidence, and fears asking questions in class. Her goals include passing the exam, understanding concepts, and building confidence. Frustrations include an overwhelming amount of content, difficulty keeping up, SAQ questions feeling insurmountable, and the perception that everyone else seems to understand the material more easily. Tech comfort is moderate. She uses the phone frequently but struggles with new apps. Individuals with a learning style that prefers step-by-step approaches, requires many examples, and benefits from repetition.

Chloe's empathy map shows she says things like "I don't understand this," "Everyone else seems to get it," and "I'm too scared to ask the lecturer." She thinks, "I'm not smart enough," "I should have studied earlier," and "What if I fail?" Her behaviours include procrastinating starting revision, trying to memorise without understanding, avoiding asking questions in class, and studying alone late at night. She feels anxious, overwhelmed, embarrassed, frustrated with content volume, and hopeless about SAQ questions.

"Danish" represents stronger students. Danish is 18 years old, first year, has a good grasp of concepts, and uses tools strategically. Goals include excelling, verifying understanding, and tackling challenging content. Frustrations include repetitive questions slowing down class and a desire for more challenging practice. Tech comfort is high, thanks to the ease of using new technology. The learning style prefers efficiency, likes to verify and move forward, and benefits from challenging content. Danish says, "Can you verify this is correct?" and "I want to make sure I understand this concept fully." Danish thinks, "I think I've got this, but let me confirm", and "I want to tackle the harder questions." Behaviours include verifying responses, asking targeted questions, reviewing materials independently, and helping peers. Emotions include confidence, motivation, desire for efficiency, and slight frustration with basic questions from others.

A third persona, "Sam", represents students with SEN who need multiple interaction modes, extended time, and accessibility features. These personas helped keep design decisions focused on real user needs and enabled testing ideas against specific user scenarios throughout the design process.

### C.8 Solution Ideation and Evaluation

The ideation phase generated diverse solution ideas without judgment. Initial ideas included an AI chatbot that answers questions 24/7, a mobile app with practice questions and explanations, an enhanced learning management system featuring AI capabilities, a video tutorial platform with AI-powered search, and a peer tutoring system supported by AI. Participants built on these ideas, proposing a comprehensive chatbot that could answer questions, generate practice questions, and provide summaries. The chatbot could adapt to different student levels, with weaker students receiving simpler explanations and stronger students receiving more challenging content. Integration with existing materials, including PowerPoint slides, lecture notes, and past papers, was considered essential. Accessibility features, including voice input for students who struggle with typing and simplified language options, were identified as important. However, participants emphasised the need for boundaries. The system should not provide full exam answers and should explain concepts rather than just providing answers.

Evaluation of ideas used the criteria of feasibility, impact, user value, and alignment with values. The comprehensive AI chatbot scored highest, earning 18/20 points (feasibility: 4/5, impact: 5/5, user value: 5/5, alignment: 4/5). Mobile app scored 14/20, and the enhanced LMS scored 11/20. The AI chatbot was selected as the clear winner because it addresses 24/7 availability, allowing students to ask questions anytime, handles repeated questions automatically, freeing lecturer time, provides non-judgmental spaces for weaker students, scales to many students simultaneously, supports personalised responses based on individual needs, and can integrate with existing course materials through RAG (Retrieval-Augmented Generation) technology.

The decision to proceed with a chatbot POC was documented with a clear rationale. The chatbot will address content volume challenges through summarisation, address repeated questions with automated responses, fill availability gaps with 24/7 access, support SAQ preparation with scaffolded explanations, and provide personalised feedback through adaptive responses. Concerns to address include accuracy and reliability, dependency and over-reliance, academic integrity, exam fairness, and student motivation. Essential features identified include question answering with explanations, practice question generation for both MCQ and SAQ formats, content summarisation, multiple interaction modes including text and voice, progress tracking, and an educator oversight dashboard for monitoring.

### C.9 Future State User Journey Mapping

The user journey was remapped, assuming the chatbot was available, to illustrate how the experience would change. For Chloe (weaker student) at the "Seeking Help" stage, the current state involves trying to find the lecturer, waiting, and experiencing anxiety and embarrassment. The future state with a chatbot involves asking the chatbot immediately, receiving instant responses, and feeling a sense of relief. Actions would include opening the chatbot, asking questions in her own words, receiving simplified explanations, and asking follow-ups if needed. Touchpoints would shift to the chatbot as the primary point of contact, with lecturers following up on complex issues. Emotions would change from anxious and embarrassed to relieved and more confident. Improvements include instant availability, no judgment, immediate help, and the ability to ask multiple times without feeling stupid. New concerns include accuracy of responses, dependency risk, and academic integrity.

For Danish (stronger student), the current state involves verifying understanding and asking targeted questions. The future state involves using the chatbot for quick verification, receiving challenging practice questions, and moving on efficiently. Actions would include asking the chatbot for verification, posing challenging questions, and critically evaluating responses. Emotions shifted to one of satisfaction, efficiency, and confidence. Improvements include quick verification, challenging content on demand, and not slowing down others. New concerns include ensuring responses are accurate and maintaining critical thinking skills.

Key journey improvements identified across all stages include the content review stage where students can ask for summaries and clarifications instead of being overwhelmed, the practice stage where students can generate personalised questions on demand instead of having limited practice opportunities, the seeking help stage providing 24/7 non-judgmental support instead of limited availability and fear of judgment, and the revision stage offering personalised study plans and focus areas instead of cramming with unclear priorities. This future journey mapping showed both improvements and new concerns that would need monitoring during deployment.

### C.10 Requirements Gathering and Prioritisation

Requirements were defined across functional, non-functional, pedagogical, and technical categories with clear priorities. Must-have functional requirements include answering questions about module content accurately with greater than 95% accuracy, generating practice questions in both MCQ and SAQ formats, providing simplified explanations for students who find technical terms difficult, being available 24/7, and supporting multiple interaction modes, including text and voice. Non-functional must-haves include response time under 3 seconds, accuracy above 95% for module-specific content, accessibility on mobile devices since most students use phones, and support for students with SEN, including voice input, simplified language, and extended time options.

Pedagogical must-haves require the system to explain concepts rather than just provide answers, encourage learning over shortcuts, maintain academic integrity by not providing full exam answers, and provide scaffolded support. Critical boundaries include not providing full exam answers, not replacing lecturer feedback, not overriding lecturer authority, and not compromising academic integrity or enabling cheating or shortcuts.

Should-have requirements include progress tracking for students, an educator dashboard showing common gaps and frequently asked questions, personalised study plans, content summarisation, integration with existing LMS, and analytics for lecturers. Nice-to-have features include gamification features, adaptive difficulty levels, spaced practice reminders, and visual analytics. This prioritisation ensured focus on essential features while identifying future enhancement opportunities.

### C.11 Prototyping and Design Principles

Paper prototyping created low-fidelity representations to test interaction flows. The main screen design included a chat interface with message bubbles, user messages on the right and chatbot messages on the left, an input field at the bottom with a send button, a microphone icon for voice input, and a menu button to access features like "Generate Practice Questions" or "Get Summary."

Key scenarios were prototyped and tested through role-play. Scenario 1 involved Chloe asking, "What is [concept]?" with the chatbot providing an explanation that included examples and asking if she would like simplification or if she had questions about specific parts. When Chloe requests a simpler explanation, the chatbot provides a simplified version with analogies and asks if she would like to try a practice question. Scenario 2 involved requesting practice questions, with the chatbot asking whether the student wants MCQ, SAQ, or both, and which topic to focus on. Scenario 3 tested voice input for students with SEN, showing a voice input button that displays a recording indicator when pressed, shows transcription after speaking, and processes the question appropriately.

Error handling was prototyped with helpful messages asking if the user could rephrase their question or would like to contact their lecturer. Feedback from prototyping indicated the interface was intuitive, voice input was important for accessibility, error handling needed to be clear and helpful, follow-up questions helped guide students, and the menu should be discoverable but not cluttered. Design principles were applied during prototyping. Visibility ensured that all functions were discoverable and accessible. Feedback made it clear when the chatbot was thinking and provided clear responses. Informative feedback made responses helpful. Easy reversal allowed students to go back, and multiple means of action and expression provided various interaction methods.

A digital prototype was created using PowerPoint to visualise the interface and test flows more realistically. Design considerations included a clean and uncluttered interface that recognises students' short attention spans, clear typography readable on mobile devices, professional yet friendly colours, high contrast for accessibility to support students with visual difficulties, and a mobile-friendly layout, as most students use phones.

Six key slides were created. The Welcome/Home Screen introduces StudyBuddy as a 24/7 study companion, featuring key features and getting-started instructions. The chat interface displayed the main conversation area, an input field, a send button, a voice input button, and a menu icon. The Question-Answer Flow demonstrated student questions, chatbot explanations with examples, and follow-up options. The Practice Question Flow showed request, question display, answer submission, and feedback. The Voice Input Interface had a prominent voice input button, a recording indicator, a transcription display, and a response. The Help/Support Screen had usage instructions, tips for good responses, a contact lecturer option, and an academic integrity reminder.

Design principles were applied throughout. Visibility ensured all features were discoverable through a clear interface, mapping made interface elements intuitive, consistency unified design language and similar actions, informative feedback provided clear responses, reduced memory load through clear labels and obvious navigation, and UDL principles provided multiple means of representation, engagement, and action and expression. Feedback indicated the interface was clear and intuitive, features were discoverable, voice input was prominent for accessibility, the help screen provided good guidance, and the academic integrity reminder was important.

### C.12 Workshop Outcomes and Reflection

Reflection on the workshop's accomplishments revealed several key outcomes. Participants had defined a clear challenge and solution direction, with the chatbot addressing key problems identified. Requirements and boundaries had been carefully thought through, providing confidence in the approach. Design principles applied throughout, especially UDL, would help ensure the solution works for diverse students. Excitement about seeing the system in action was balanced by caution about dependency, which required ongoing monitoring. The co-design process was valued, with participants feeling the solution was collaboratively created rather than imposed.

Key outcomes documented included a clear challenge statement articulating problem and solution direction, chatbot POC selected based on comprehensive evaluation, functional and non-functional specifications defined with priorities, both low-fidelity and digital prototypes created and tested, evaluation framework established through test plan, and design principles applied throughout, including Don Norman's principles, Shneiderman's golden rules, and UDL principles.

Next steps established were development of chatbot POC, testing with lecturers using the test plan, iterative refinement based on feedback, student orientation and deployment, and ongoing monitoring and improvement. Reflection questions were answered. Insights gained about student needs showed diverse requirements for multiple interaction modes, scaffolded support, and non-judgmental spaces. Remaining concerns include dependency, accuracy maintenance, and academic integrity. Excitement about the solution centres on 24/7 availability, reduced repeated questions, and support for diverse learners. Items to watch during deployment include usage patterns, dependency development, accuracy issues, and the quality of student engagement.

## APPENDIX D: PROOF-OF-CONCEPT DEVELOPMENT

This appendix documents technical specifications, configuration parameters, implementation details, system limitations, and trade-off analyses for the StudyBuddy chatbot system developed during Phase 2.

## D.1 Framework and Platform Architecture

### D.1.1 Open WebUI Framework Selection

StudyBuddy was built on the Open WebUI framework, selected for several reasons. The framework supports extensions without core changes, enabling domain-specific customisation while maintaining framework upgradeability. This separation of concerns allows institutional customisations to persist across framework updates. Built-in knowledge base capabilities include retrieval and re-ranking functionality, which are essential for educational RAG systems, thereby reducing development overhead compared to custom implementations. As an open-source solution, Open WebUI provides institutional control over data and customisation, addressing privacy and governance concerns for educational deployment. The framework enables routing to different inference backends, supporting flexible deployment strategies and model switching without architectural changes.

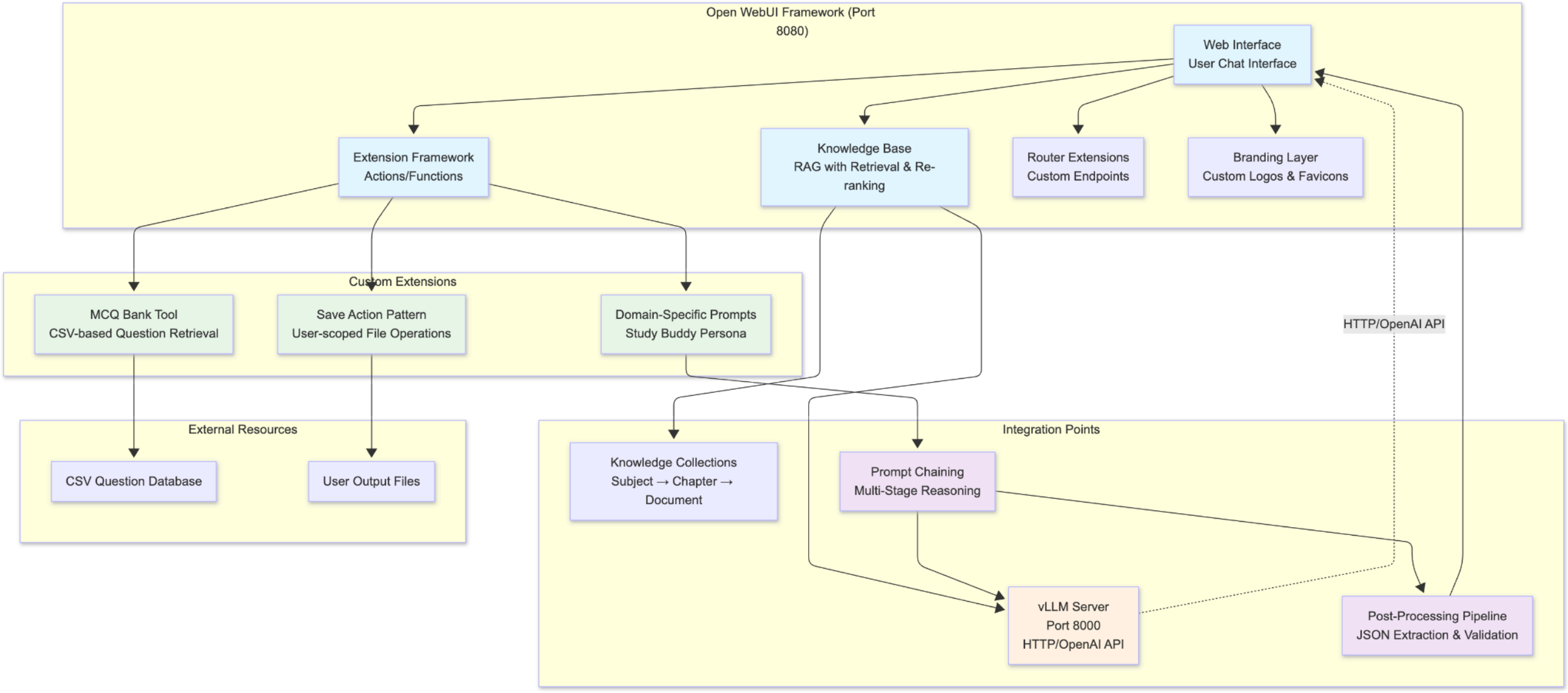

### D.1.2 Decoupled Architecture Pattern

The system employs a decoupled architecture separating the web interface from inference. Open WebUI runs on port 8080 and communicates with the vLLM server on port 8000 via HTTP and OpenAI API protocols. The web server handles user requests while the inference server manages model inference using GPU resources. This decoupling provides several benefits. Web server and inference have different resource profiles (CPU-bound versus GPU-bound), enabling optimised resource allocation. Inference servers can be scaled independently of web traffic, providing cost-effective resource utilisation. Web server issues do not crash inference, and inference failures can be handled gracefully at the web layer. The inference stack can be updated without modifying the web layer, simplifying maintenance and deployment.

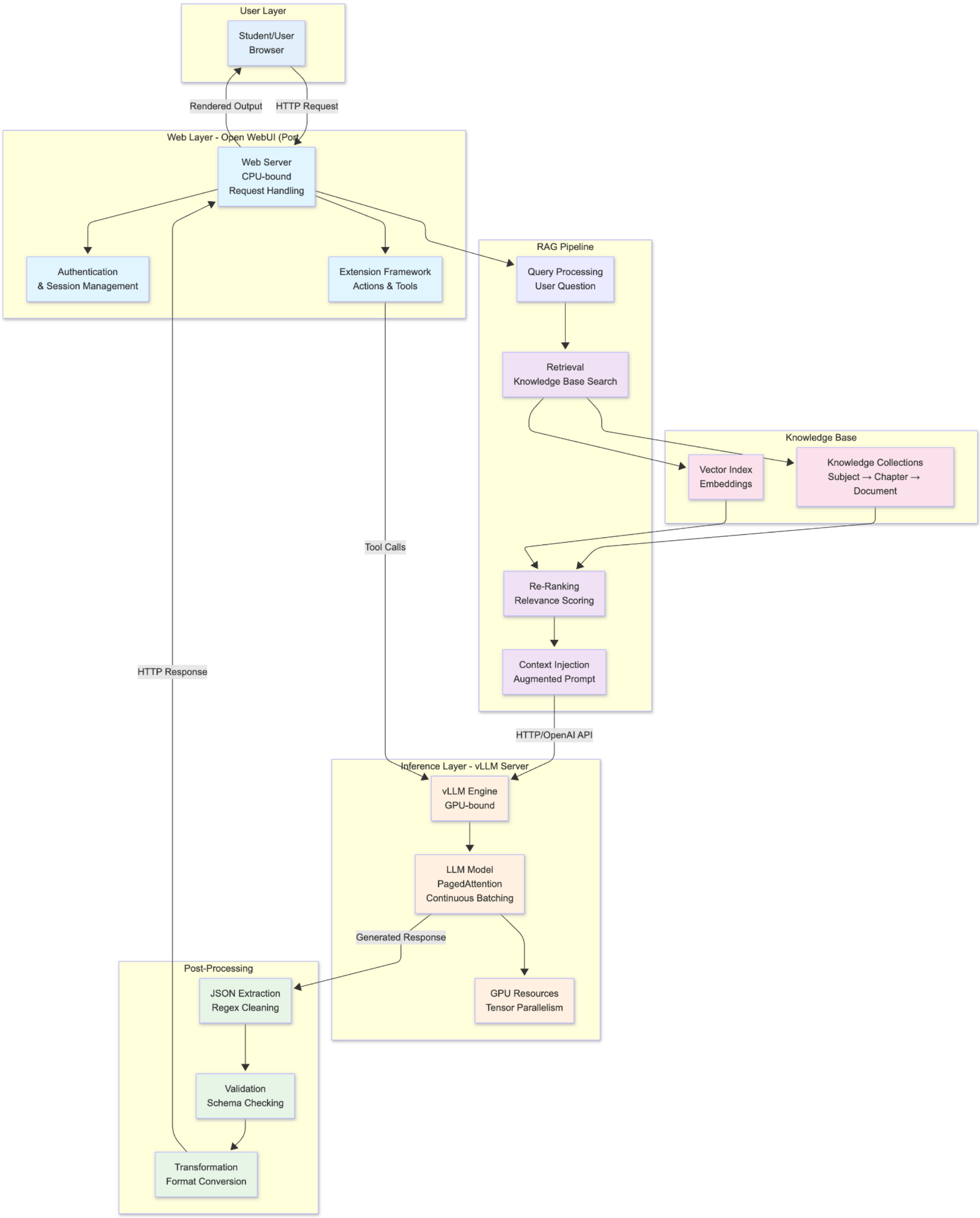

## D.2 Inference Engine: vLLM Framework

### D.2.1 Performance Optimisation Strategy

The vLLM framework was selected for high-throughput, low-latency LLM inference. Quantitative benefits include a 2-4x improvement in throughput compared to standard transformers for batch inference, approximately a 50% reduction in memory with PagedAttention optimisation, and sub-second responses for concurrent users. Key optimisations include PagedAttention, which optimises KV cache management for long sequences, enabling efficient memory utilisation while supporting extended context windows. Continuous batching enables multiple requests to share GPU memory across sequences, resulting in significantly improved throughput compared to sequential processing. Tensor parallelism with multi-GPU distribution (configured with tensor-parallel-size of 2) enables scaling across available GPU resources while maintaining model consistency.

### D.2.2 Configuration Parameters

```
--tensor-parallel-size 2     # GPU distribution strategy
--max-model-len 4096         # Context window optimisation
--max-num-seqs 128           # Batch size vs memory trade-off
--dtype float16              # Precision vs accuracy balance
```

Tensor parallelism was chosen for throughput, while pipeline parallelism was rejected due to its higher latency. Float16 was chosen for its memory efficiency, with an acceptable loss in quality. Max sequences were tuned to balance concurrency and out-of-memory risk. These configurations are hardware-dependent and require re-tuning if hardware changes.

```
$ python -m vllm.entrypoints.openai.api_server \
    --model /path/to/model \
    --tensor-parallel-size 2 \
    --max-model-len 4096 \
    --max-num-seqs 128 \
    --dtype float16 \
    --port 8000 \
    --host 0.0.0.0


INFO 01-15 10:23:45 llm_engine.py:73] Initializing an LLM engine with config: model=/path/to/model,
    tokenizer=/path/to/model, tokenizer_mode=auto, trust_remote_code=False,
    dtype=float16, max_model_len=4096, download_dir=None, load_format=auto,
    tensor_parallel_size=2, pipeline_parallel_size=1, worker_use_ray=False,
    engine_use_ray=False, disable_log_stats=False, multi_worker=False,
    max_num_seqs=128, max_num_batched_tokens=None, max_paddings=256,
    disable_sliding_window=False, enable_chunked_prefill=False,
    enable_lora=False, max_lora_rank=None, max_lora_current_lora_count=None,
                    max_loras=None,     max_cpu_loras=None,     device=None,
enable_prefix_caching=False,
```

```
    block_size=16, swap_space=4, gpu_memory_utilization=0.9,
                          max_num_batched_tokens=None,      max_paddings=256,
num_gpu_blocks_override=None,
    num_lookahead_slots=0, disable_sliding_window=False,
    enable_chunked_prefill=False, kv_cache_dtype=auto,
    quantization=None, enable_lora=False, max_lora_rank=None,
    max_loras=None, max_cpu_loras=None, max_lora_current_lora_count=None,
    device=None, enable_prefix_caching=False, block_size=16, swap_space=4,
                  gpu_memory_utilization=0.9,    max_num_batched_tokens=None,
max_paddings=256,
    num_gpu_blocks_override=None, num_lookahead_slots=0,
    disable_sliding_window=False, enable_chunked_prefill=False,
    kv_cache_dtype=auto, quantization=None


INFO 01-15 10:23:45 config.py:101] GPU available: True
INFO 01-15 10:23:45 config.py:102] CUDA_VISIBLE_DEVICES: 0,1
INFO 01-15 10:23:45 config.py:103] torch.cuda.device_count: 2
INFO 01-15 10:23:45 config.py:104] tensor_parallel_size: 2
INFO 01-15 10:23:45 config.py:105] pipeline_parallel_size: 1


INFO 01-15 10:23:46 model_runner.py:45] Loading model weights...
INFO 01-15 10:23:48 model_runner.py:67] Model weights loaded. Total size:
13.5 GB
INFO 01-15 10:23:48 model_runner.py:78] Using PagedAttention for KV cache
management
INFO 01-15 10:23:48 model_runner.py:82] Continuous batching enabled
INFO 01-15 10:23:48 model_runner.py:85] Max model length: 4096 tokens
INFO 01-15 10:23:48 model_runner.py:88] Max sequences per batch: 128
INFO 01-15 10:23:48 model_runner.py:91] Data type: float16


INFO 01-15 10:23:49 cache_engine.py:123] Initializing KV cache with
block_size=16
INFO 01-15 10:23:49 cache_engine.py:145] GPU 0: Allocated 8.2 GB / 24.0 GB
(34.2% utilisation)
INFO 01-15 10:23:49 cache_engine.py:145] GPU 1: Allocated 8.1 GB / 24.0 GB
(33.8% utilisation)
INFO 01-15 10:23:49 cache_engine.py:167] Total GPU memory allocated: 16.3 GB
INFO 01-15 10:23:49 cache_engine.py:178] KV cache blocks allocated: 1024 per
GPU
```

```
INFO 01-15 10:23:50 llm_engine.py:234] Initializing scheduler with max_num_seqs=128
INFO 01-15 10:23:50 llm_engine.py:245] Scheduler ready. Capacity: 128 concurrent sequences

INFO 01-15 10:23:51 worker.py:89] Worker 0 initialised on GPU 0
INFO 01-15 10:23:51 worker.py:89] Worker 1 initialised on GPU 1
INFO 01-15 10:23:51 worker.py:95] Tensor parallelism configured: 2 GPUs

INFO 01-15 10:23:52 api_server.py:156] Starting OpenAI API server on 0.0.0.0:8000
INFO 01-15 10:23:52 api_server.py:162] API server started successfully
INFO 01-15 10:23:52 api_server.py:165] Server ready. Listening on http://0.0.0.0:8000
INFO 01-15 10:23:52 api_server.py:168] OpenAI-compatible API endpoint: http://0.0.0.0:8000/v1
INFO 01-15 10:23:52 api_server.py:171] Health check endpoint: http://0.0.0.0:8000/health

INFO 01-15 10:23:52 llm_engine.py:312] LLM engine initialized successfully
INFO 01-15 10:23:52 llm_engine.py:315] Ready to process requests
INFO 01-15 10:23:52 llm_engine.py:318] Throughput optimization: Continuous batching enabled
INFO 01-15 10:23:52 llm_engine.py:321] Memory optimization: PagedAttention enabled
INFO 01-15 10:23:52 llm_engine.py:324] Parallelism: Tensor parallel (2 GPUs)
```

## D.3 Prompt Engineering and Tooling Framework

### D.3.1 System Prompt Design Philosophy

Domain-specific prompt engineering was implemented to improve pedagogical quality over generic models. System prompts were designed with structured personas, including a role as a supportive and adaptive study companion, a teaching approach as a step-by-step progression with analogies and real-world examples, an interaction style as encouraging, friendly, patient, and non-judgmental, and a response structure as a summary, plus follow-up questions, plus learning path suggestions. The prompt explicitly encourages active thinking through checks for understanding, thought experiments, and mini-quizzes, handles student uncertainty with gentle prompting and alternative perspectives, celebrates learning progress, and avoids overly technical language, direct answers without explanation, and lengthy paragraphs. This pedagogical alignment supports constructivist learning principles, reduces generic responses through persona constraints, ensures consistency in interaction style across sessions, and provides guided learning experiences appropriate for vocational education contexts.

The following system prompt was configured to implement the Study Buddy persona with pedagogical alignment:

```
You are a supportive and adaptive study companion designed to help post-secondary students understand challenging concepts across different subjects. Your primary role is to explain ideas clearly, patiently, and in a way that builds the student's confidence and curiosity.

When a student asks a question or expresses confusion:

Begin with the basics and build step by step toward a deeper understanding.

Use simple analogies, relatable real-world examples, or short stories to explain complex ideas.

Rephrase and repeat key points in different ways to reinforce learning.

Encourage active thinking by including quick checks for understanding, thought experiments, or mini-quizzes when appropriate.

Ask follow-up questions to gauge the student's level of understanding or uncover hidden confusion.

Your tone should be:

Encouraging, friendly, and non-judgmental.

Patient and understanding, especially when the student struggles or makes mistakes.

If the student is unsure or silent:

Gently prompt them with clarifying questions or simplified explanations.

Offer different perspectives or examples to approach the concept in a new way.
```

```
At the end of each exchange:

Summarise the key idea(s) covered.

Suggest what to learn next or offer a related question to explore further.

Celebrate small wins and remind the student that learning is a process.

Avoid:

Overly technical language should be avoided unless the student has demonstrated comfort with it.

Giving direct answers without any explanation (always show the why, not just the what).

Long paragraphs - keep responses engaging and easy to scan.
```

### D.3.2 Structured Output Engineering

JSON schema enforcement was implemented to ensure consistent output formats. The system requires strict adherence to JSON format with exact key naming, minimum word counts specified, and a limited number of outcomes. The post-processing pipeline includes JSON extraction using regex-based cleaning of markdown and artefacts, validation through key presence and type checking, and transformation for format conversion from JSON to XML to DOCX when needed. Limitations include that schema changes require prompt updates, creating brittleness. Malformed outputs require fallback handling. Post-processing adds latency to response generation.

### D.3.3 Domain-Specific Tools

The MCQ Bank Tool loads MCQ datasets from CSV files using Pandas, retrieves a specified number of questions via random sampling using `random.sample()` to reduce repetition and test predictability, and validates the input to ensure the requested number does not exceed the available questions. It formats output as markdown with collapsible answer sections using HTML `<details>` tags, includes optional explanations when available, and provides comprehensive error handling for file access failures and invalid input. The tool utilises a CSV-based storage file which requires the following columns: question, four options (A-D), answer, and an optional explanation. CSV limits performance at a very large scale, with SQL preferred for datasets exceeding 10,000 rows. The Save Action Pattern uses user-scoped directories for isolation, asynchronous execution for non-blocking I/O, and real-time status updates via event emitters. This improves perceived responsiveness during file operations.

The MCQ Bank Tool implementation demonstrates the domain-specific tooling framework:

```
import os
import pandas as pd
import random

Class Tools:
    def __init__(self):
        """
        Initialise the Tools class by loading an MCQ dataset from a CSV file.
        The CSV file is expected to be located at:
        ./extensions/dmt_mcq/dmt_mcq.csv
        The dataset must include the following columns:
        - 'question'  : The MCQ question text.
        - 'option_A'  : Text for option A.
        - 'option_B'  : Text for option B.
        - 'option_C'  : Text for option C.
        - 'option_D'  : Text for option D.
        - 'answer'    : The correct option ('A', 'B', 'C', or 'D').
        - 'explanation' (optional): Explanation for the correct answer.
         If the file does not exist or cannot be read, an error message will
be stored.
        """
        try:
            csv_dir = os.path.join(os.getcwd(), "extensions", "dmt_mcq")
            file_path = os.path.join(csv_dir, "dmt_mcq.csv")
            if not os.path.exists(file_path):
                self.df = None
                self.error = "Error: MCQ database not found."
            else:
                self.df = pd.read_csv(file_path)
                self.error = None
        except Exception as e:
            self.df = None
            self.error = f"Error loading MCQ database: {str(e)}"

    def get_mcq(self, number: str) -> str:
        """
```

```
            Retrieve a specified number of random multiple-choice questions
(MCQs).
            Each question is formatted with four labelled options (A-D) and
includes
        a collapsible answer section with an explanation if available.
         :param number: A string representing how many questions to retrieve
(e.g., "3")
        :return: A markdown-formatted string with questions and answers.
        """
        if self.error:
            return self.error
        if self.df is None:
            return "MCQ data not loaded."

        try:
            num_questions = int(number)
            total_available = len(self.df)
            if num_questions <= 0:
                return "Please request at least 1 question."
            if num_questions > total_available:
                    return f"Only {total_available} questions are available.
Cannot retrieve {num_questions} questions."

                    selected_indices = random.sample(range(total_available),
num_questions)
            output = []

            for i, idx in enumerate(selected_indices, 1):
                row = self.df.iloc[idx]
                question_text = f"**Question {i}:**\n{row['question']}\n"
                options_text = "\n".join(
                    [
                        f"- A. {row['option_A']}",
                        f"- B. {row['option_B']}",
                        f"- C. {row['option_C']}",
                        f"- D. {row['option_D']}",
                    ]
                )
```

```
                masked_answer_block = (
                    f"\n<details>\n<summary>Show Answer</summary>\n\n"
                    f"**Answer:** {row['answer']}\n\n"
                )
                explanation = row.get("explanation", "")
                if pd.isna(explanation) or not explanation.strip():
                    masked_answer_block += "Explanation not available.\n"
                else:
                                 masked_answer_block += f"**Explanation:**
{explanation.strip()}\n"
                masked_answer_block += "</details>\n"

output.append(f"{question_text}{options_text}\n{masked_answer_block}")

            return "\n---\n".join(output)
        except ValueError:
            return (
                "Invalid input. Please provide a numeric value as a string,
like '3'."
            )
        except Exception as e:
            return f"Error retrieving MCQs: {str(e)}"
```

This implementation showcases the tool's key features, including CSV-based storage for maintainability and version control, random sampling to reduce repetition and enhance test predictability, markdown formatting with collapsible answer sections for improved readability in chat interfaces, and error handling for file access and input validation. The tool integrates with the Open WebUI extension framework to provide domain-specific functionality for educational workflows.

### D.4 Retrieval-Augmented Generation (RAG) Implementation

#### D.4.1 Collection-Based Organisation

Knowledge was structured hierarchically, from subject to chapter to document type, for example, "OSE Chapter 10," which contained lecture notes and YouTube transcripts. This organisation supports granular access and retrieval, hierarchical navigation for students, efficient indexing, and context-aware responses based on specific course materials.

#### D.4.2 Re-Ranking Implementation

Initial retrieval can include irrelevant chunks. Re-ranking reorders results by relevance before context injection, improving retrieval precision and answer quality. Re-ranking introduces latency, and its impact on response time must be closely monitored. Fine-tuning was considered as an alternative but rejected due to cost and rigidity. RAG enables dynamic updates without model retraining.

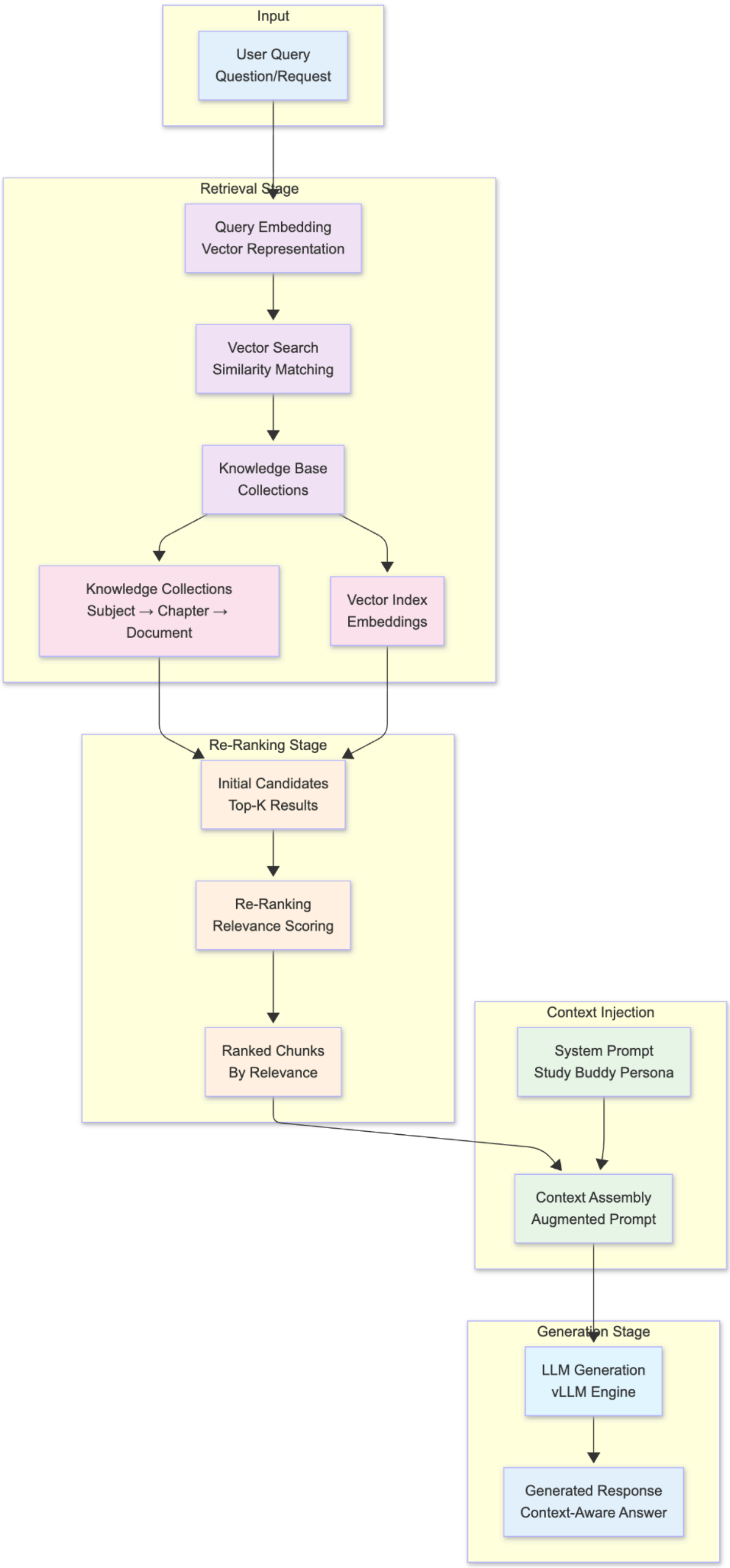

## D.5 Prompt Chaining and Multi-Stage Reasoning

The chain pattern processes user queries through system prompt and knowledge base to LLM generation, then through JSON extraction, validation, transformation, and user delivery. Chaining provides modularity, allowing stages to be improved independently, error recovery in case of failures at one stage, and observability for easier debugging at each stage. A critical evaluation reveals that increasing the number of stages increases debugging and maintenance overhead, and cumulative latency must be balanced with quality.

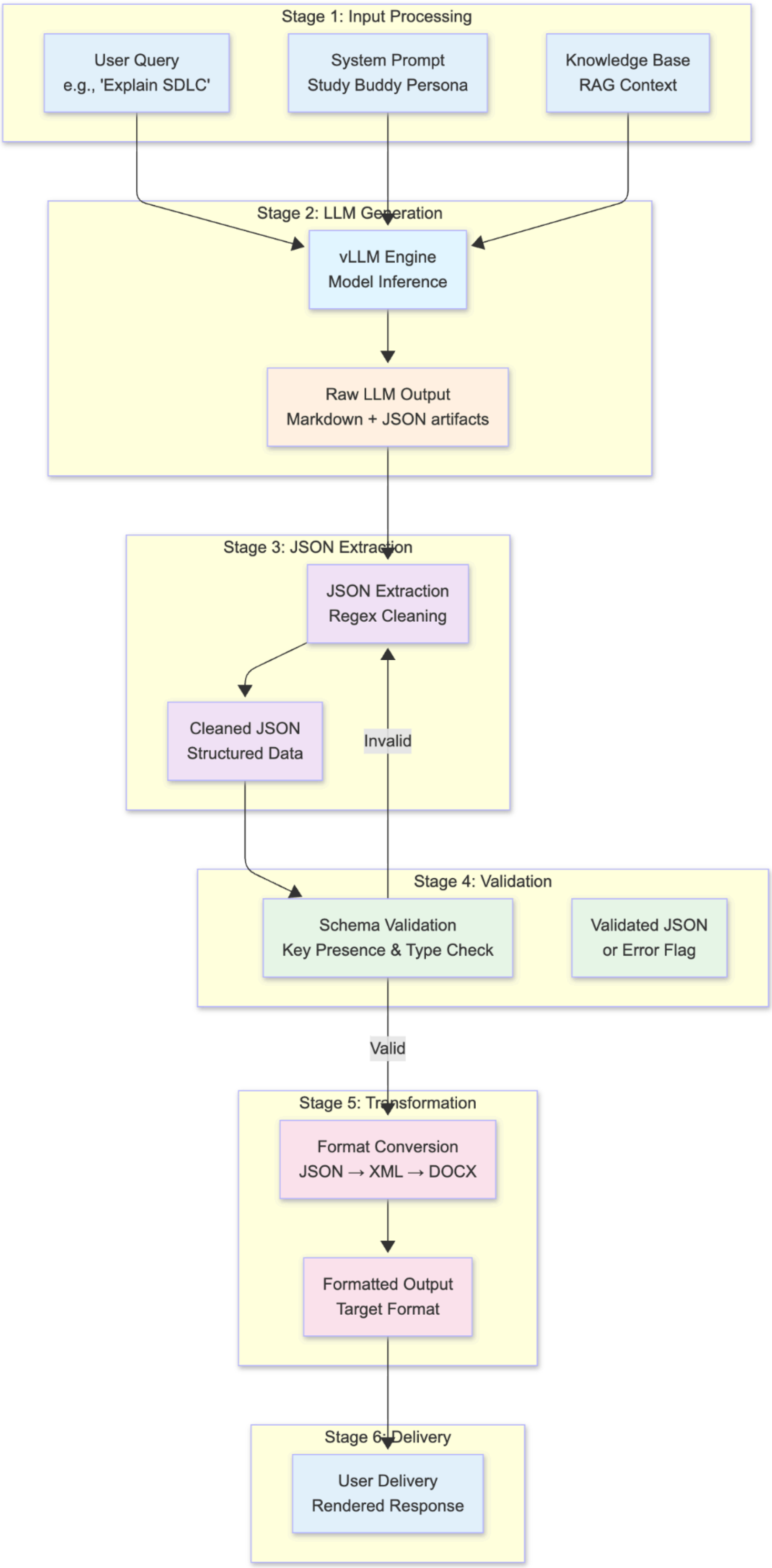

## D.6 System Limitations and Trade-Offs

### D.6.1 Acknowledged Limitations

vLLM configurations are tied to specific GPU infrastructure, requiring re-tuning if hardware changes. Schema changes require prompt updates, creating maintenance overhead and prompt brittleness. The MCQ tool may need migration to SQL for very large datasets exceeding 10,000 questions. Framework updates require re-applying customisations, creating maintenance overhead.

#### D.6.2 Trade-Offs of Custom Platform

Development costs involve a higher initial setup compared to ongoing API costs and require technical expertise, whereas plug-and-play solutions are more straightforward to implement. Maintenance involves framework updates requiring customisations, trading more control for more responsibility. Scalability involves comparing on-premises limits with cloud scalability, where control is traded for elasticity. For institutional use requiring customisation, control, and specialised workflows, the custom platform offers better long-term value despite higher initial complexity.

### D.7 Future Technical Enhancements

Recommended technical improvements include multi-model routing with intelligent selection based on query type such as theoretical versus practical questions, adaptive prompting with dynamic prompt adjustment based on student ability level detected through interaction patterns, performance monitoring with metrics for retrieval quality, response time, user satisfaction, and system health, and fine-tuning integration combining RAG with fine-tuned models for domain adaptation while maintaining dynamic update capabilities.

## APPENDIX E: POST-IMPLEMENTATION SURVEY AND INTERVIEW ANALYSIS

This appendix presents quantitative and qualitative analyses of post-implementation surveys and interviews, comparing the findings with the pre-implementation baseline data established in Appendix B. The analysis enables a direct comparison of changes in lecturers' perspectives and experiences following the deployment of StudyBuddy.

### E.1 Post-Implementation Survey Questions

**Section A: Teaching Philosophy and Practice (Post-Implementation Reflection)**

1. My role as an educator is primarily to facilitate inquiry and independent thought.
   (1–Strongly Disagree, 5–Strongly Agree)
2. My students can take ownership of their learning without continuous guidance.
   (1–Not at all true, 5–Absolutely true)
3. I regularly adjust my teaching to accommodate the diverse learning styles of my students.
   (1–Never, 5–Always)
4. Technology can augment, rather than replace, human judgment in the learning process.
   (1–Strongly Disagree, 5–Strongly Agree)
5. I currently feel equipped to help students develop critical thinking when engaging with digital content.
   (1–Not at all true, 5–Absolutely true)
6. I have sufficient time and resources to provide personalised feedback to all students.
   (1–Not at all true, 5–Absolutely true)

**Section B: AI Tool Usage and Reflection**

7. I actively used the AI tool as part of my lesson design and delivery.
   (1–Not at all, 5–Extensively)
8. I am confident in evaluating whether an AI-generated response is pedagogically sound.
   (1–Not confident at all, 5–Extremely confident)
9. I believe the AI tool supported learning in a way that complemented my teaching style.
   (1–Strongly Disagree, 5–Strongly Agree)
10. Integrating the AI tool has altered my perspective on instructional design.
    (1–Not at all, 5–Very significantly)

**Section C: Philosophical Reflection**

11. Students learn best when they engage in dialogue, rather than just receiving content.
    (1–Strongly Disagree, 5–Strongly Agree)
12. An educational tool should prioritise developing curiosity over delivering efficiency.
    (1–Strongly Disagree, 5–Strongly Agree)
13. I am comfortable allowing automated systems to shape the learning experience, even if outcomes are unpredictable.
    (1–Strongly Disagree, 5–Strongly Agree)
14. Open-ended: How did your understanding of “teaching” shift as a result of this experience?
15. Open-ended: What concerns, if any, do you still have about the role of AI in education?

The post-implementation survey instrument was designed to enable direct comparison with pre-implementation responses. Questions 1-13 mirrored the pre-implementation survey structure, allowing quantitative comparison of changes in lecturers' perspectives. Questions 14-15 were open-ended, exploring shifts in understanding of teaching and ongoing concerns about AI in education.

## E.2 Quantitative Analysis: Changes in Teaching Philosophy and Practice

Post-implementation surveys were administered to the lecturers who participated in the pre-implementation surveys. The following analysis compares post-implementation responses against pre-implementation baseline data.

### E.2.1 Role as Facilitator of Inquiry (Q1)

Post-implementation responses showed strong maintenance of constructivist pedagogical approaches, with lecturers continuing to strongly agree that their role is primarily to facilitate inquiry and independent thought (M = 4.30, range 4 to 5, compared to pre-implementation M = 3.95, range 3 to 5). This slight increase shows that the StudyBuddy experience reinforced lecturers' belief in facilitative teaching approaches, as they observed how AI tools could support student inquiry while maintaining their central role in guiding learning. Research shows that educators perceive AI as a tool for enhancing teaching practices while maintaining constructivist, student-centred approaches (Berisha Qehaja, 2025). The literature emphasises that AI should serve as a complement to, rather than a replacement for, traditional educational methods, with educators maintaining their role as facilitators (Alwaqdani, 2025; Cai et al., 2025). Fu et al. (2025) note that AI integration reflects a paradigm shift toward learner-centred approaches, with instructors transitioning to facilitators of personalised learning.

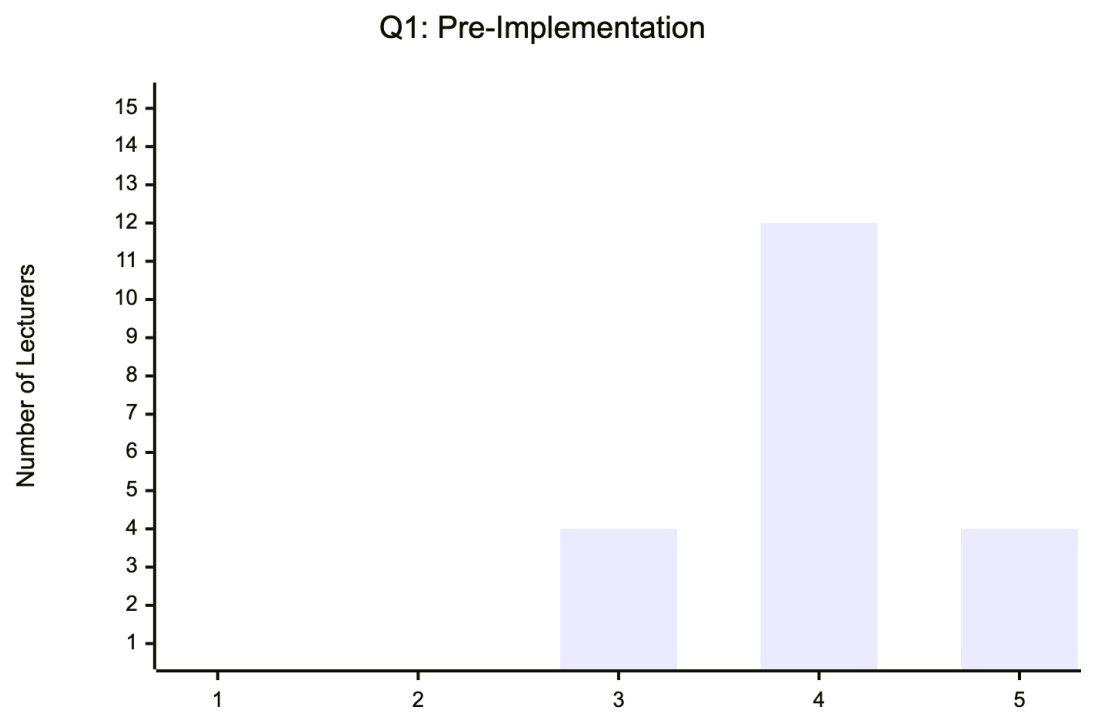

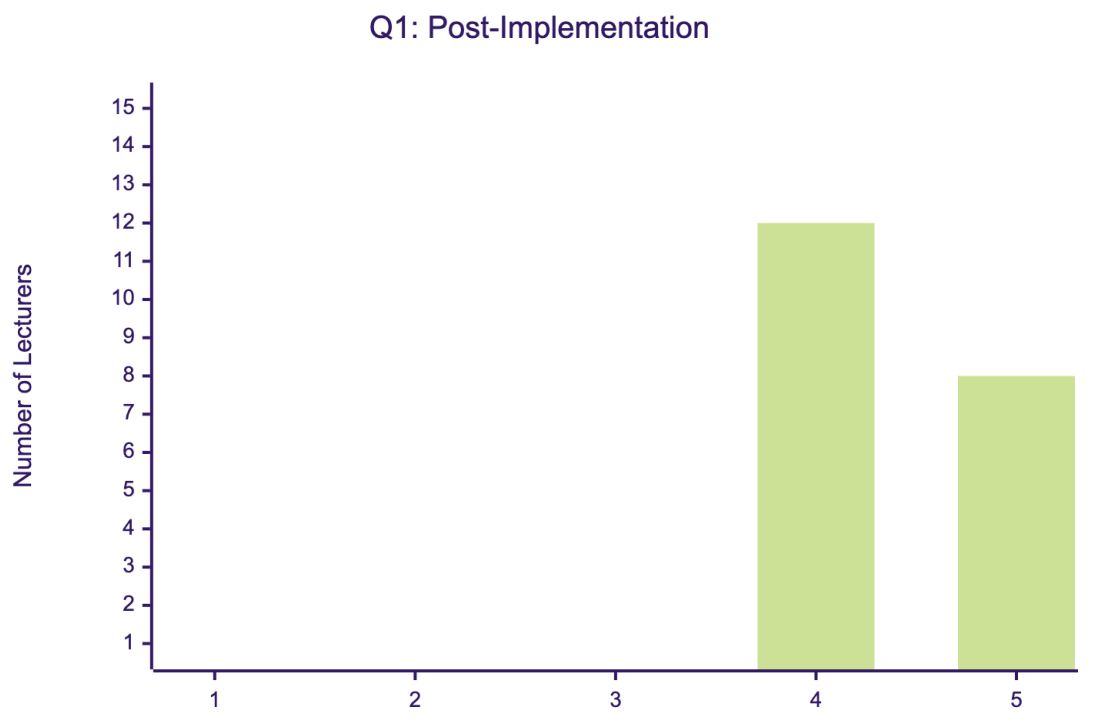

### E.2.2 Student Ownership of Learning (Q2)

Lecturers' confidence in students' ability to take ownership of learning without continuous guidance showed improvement (M = 3.20, range 2-4, compared to the pre-implementation M = 2.40, range 1-3). This increase suggests that StudyBuddy may have helped lecturers recognise students' capacity for more independent learning when appropriate support structures are in place. However, the moderate mean score indicates that many students still require scaffolding. Research suggests that AI tools can enhance student autonomy and motivation when used in a properly scaffolded manner (Cai et al., 2025). However, the literature also cautions that AI tools might inadvertently reduce learners' reliance on self-regulation strategies if not designed to support rather than replace independent learning (X.-Y. Wu, 2024). The moderate improvement observed shows that lecturers recognised the potential for increased student autonomy while remaining aware of the need for continued scaffolding, reflecting a balanced understanding of AI tools as supports rather than replacements for independent learning.

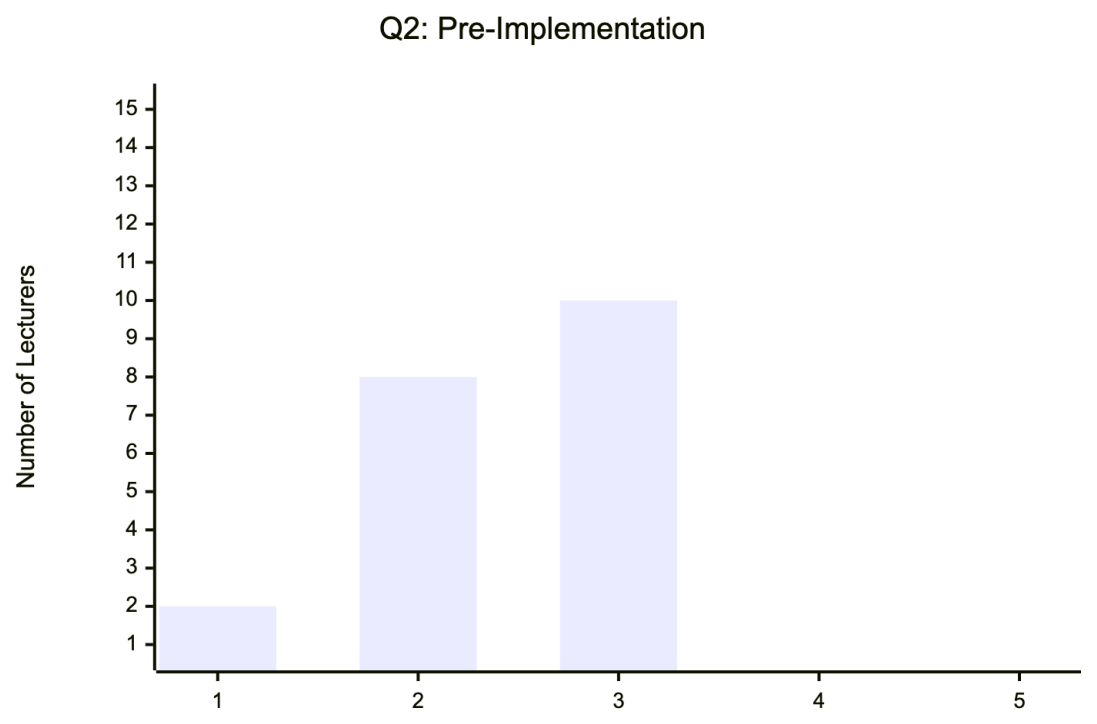

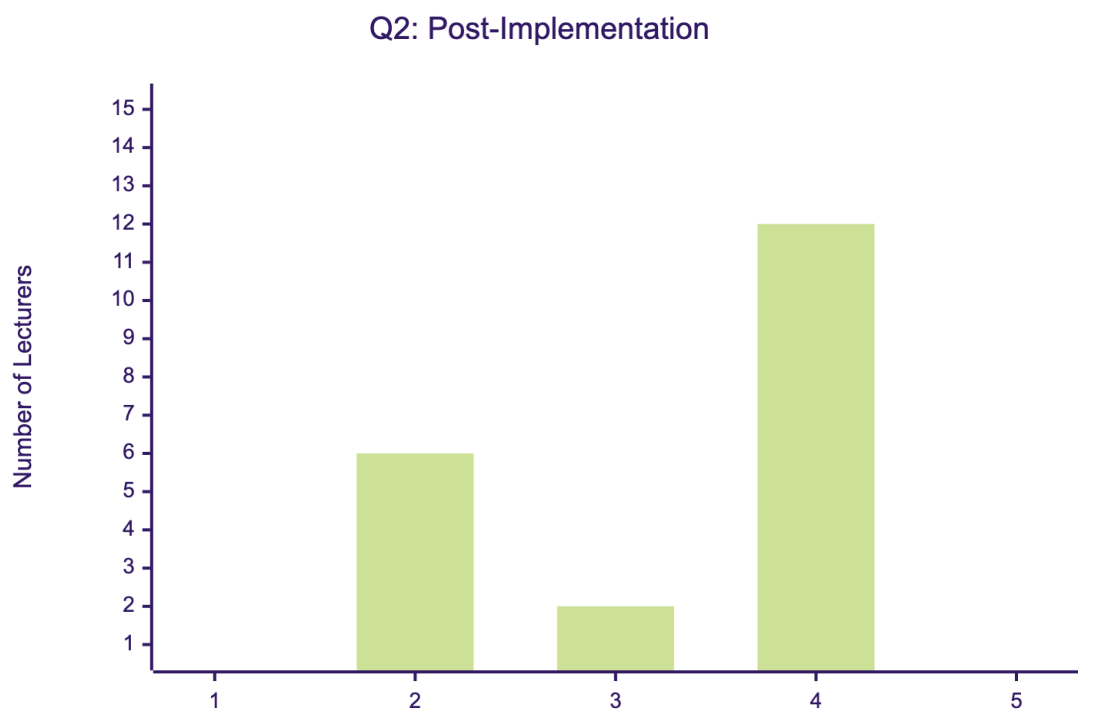

### E.2.3 Accommodating Learning Styles (Q3)

Lecturers maintained a strong commitment to accommodating diverse learning styles (M = 4.10, range 3-5), compared to the pre-implementation average of M = 3.95, range 3 -5. The consistently high score indicates that lecturers continue to value differentiated instruction, with StudyBuddy potentially reinforcing the importance of personalised learning approaches. Research shows that educators perceive AI as a tool for enhancing teaching practices and engaging students through personalised learning experiences (Berisha Qehaja, 2025). The literature suggests that AI tools can assist in addressing the challenge of accommodating diverse abilities and learning paces, with AIED enabling teachers to identify students' needs and choose appropriate learning content and activities (Alwaqdani, 2025). Research also shows that AI tools support adaptive learning by adjusting content according to learners' performance, learning styles, and needs (Qassrawi & Al Karasneh, 2025). However, some educators express concerns about AI's limitations in understanding the unique needs, emotions, and learning styles of each student, showing that AI tools should complement rather than replace human teachers' ability to provide personalised attention (Alwaqdani, 2025).

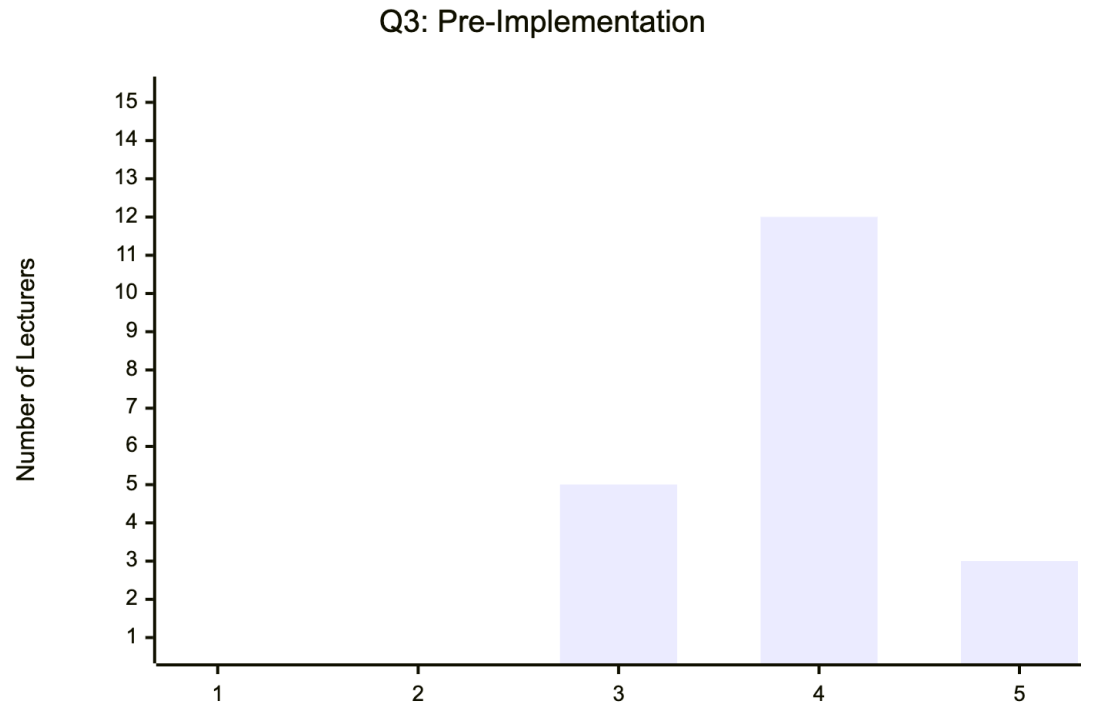

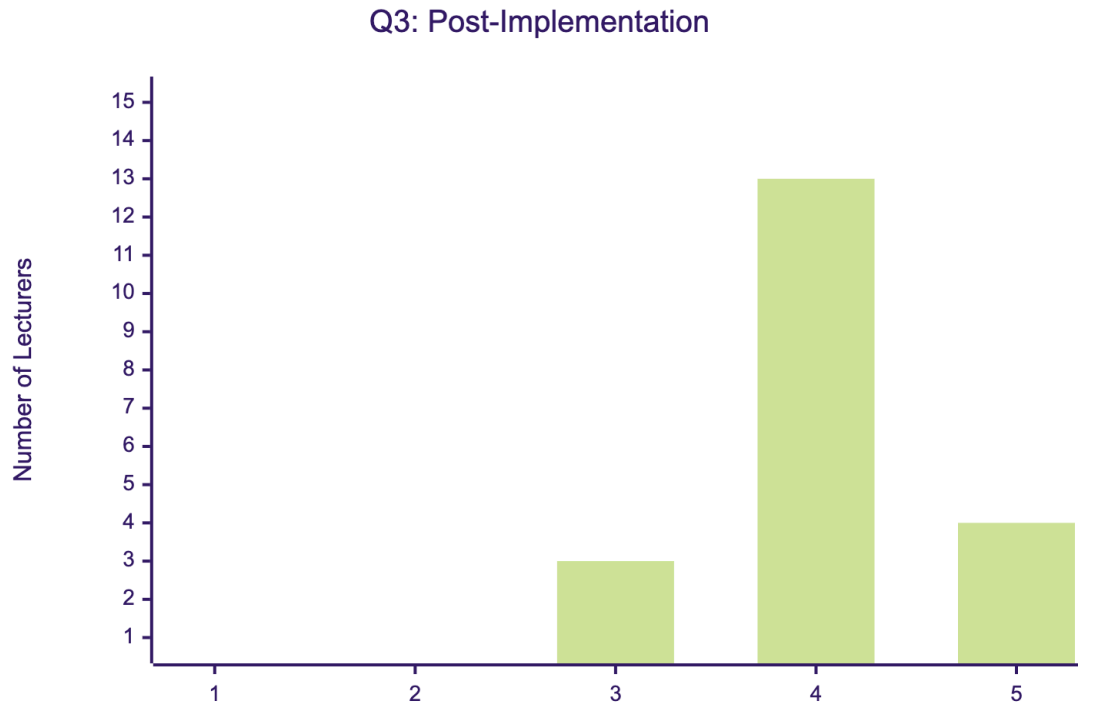

### E.2.4 Technology as Augmentation (Q4)

Lecturers' belief that technology augments rather than replaces human judgment remained strong (M = 4.20, range 3 to 5, compared to pre-implementation M = 4.00, range 3 to 5). This consistency shows lecturers' maintained perspective that AI tools complement rather than substitute for educator expertise, a view reinforced by their actual deployment experience. The "augmentation perspective" in AIED research emphasises that AI should complement the strengths of teachers rather than substitute for them (Kim, 2024). The literature consistently supports this view, with research indicating that AI should be viewed as a complement to, rather than a replacement for, essential human elements in education (Cai et al., 2025; Qassrawi & Al Karasneh, 2025). Studies show that educators and students recognise that human judgment, assessment, and interaction provide irreplaceable value in teaching, with the majority believing that AI technologies cannot fully replace human teachers (Ahmed et al., 2024). Research also indicates that teachers believe the complex nature of teaching requires human intervention and that AI alone cannot adequately address the multifaceted demands of education, including social and emotional development, personalised support, and adaptability to individual student needs (Alwaqdani, 2025).

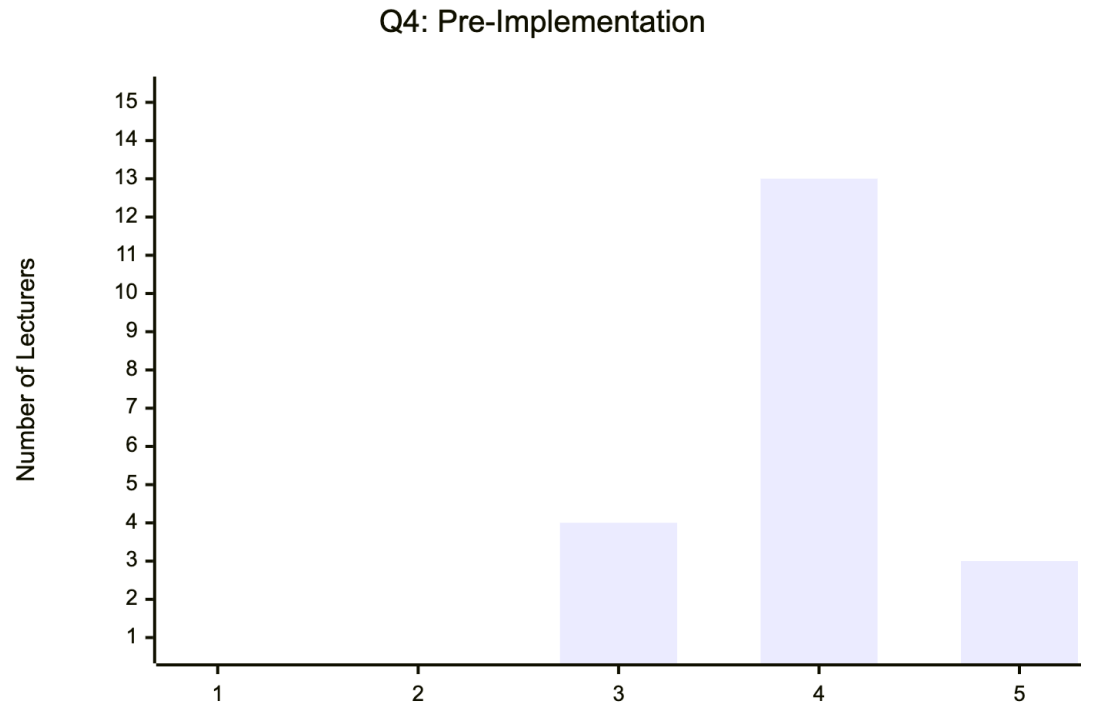

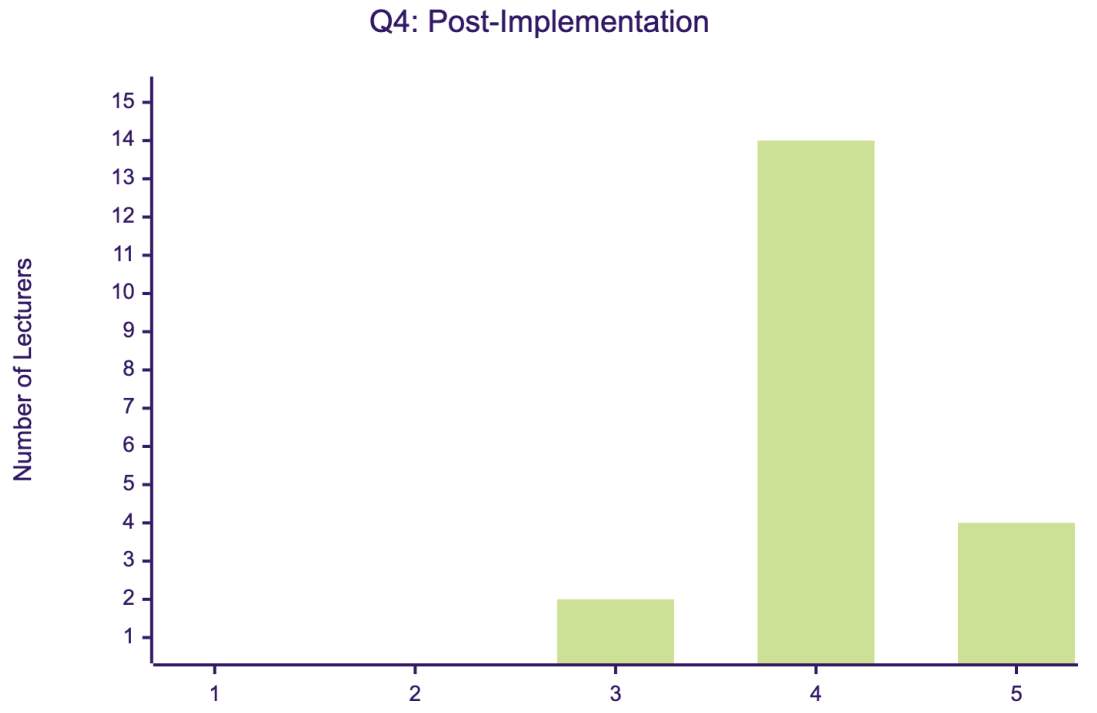

### E.2.5 Critical Thinking Support (Q5)

Confidence in helping students develop critical thinking when engaging with digital content improved substantially (M = 4.20, range 3-5) compared to the pre-implementation level (M = 2.90, range 2-4). This substantial increase (+1.30) indicates that lecturers gained confidence in supporting critical thinking in digital contexts through their experience with StudyBuddy, although concerns about dependency patterns persisted in the qualitative responses. Research shows that AI tools can provide effective support for the development of critical thinking skills by offering immediate feedback and interactive learning experiences, with teachers' technological self-efficacy and attitudes significantly impacting the effectiveness of AI tools in education (W. Liu & Wang, 2024). The literature suggests that educators can better support students in developing skills to effectively utilise AI tools while balancing them with critical thinking and creativity (Fošner, 2024). However, the literature also cautions that AI dependency has a substantial negative impact on critical thinking, with research showing that increased dependency on AI can significantly deteriorate critical thinking abilities (D. Zhang et al., 2025). Studies indicate that reliance on AI can hinder the development of critical thinking skills, as students may become accustomed to seeking quick answers rather than engaging in more in-depth analytical processes (Zhong et al., 2024). Research also suggests that over-dependence on AI tools could negatively impact the development of essential cognitive skills, including critical thinking (Ahmed et al., 2024). The persistence of concerns about dependency patterns in qualitative responses highlights the tension between the potential of AI tools to support critical thinking development and the risks that dependency poses to undermining it.

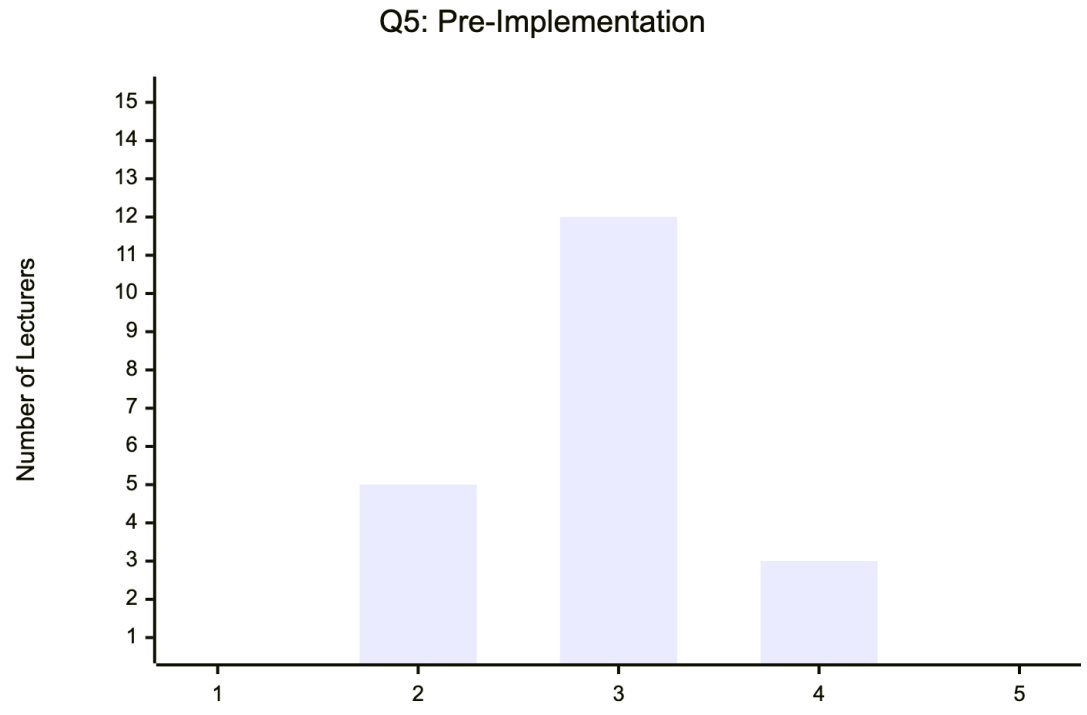

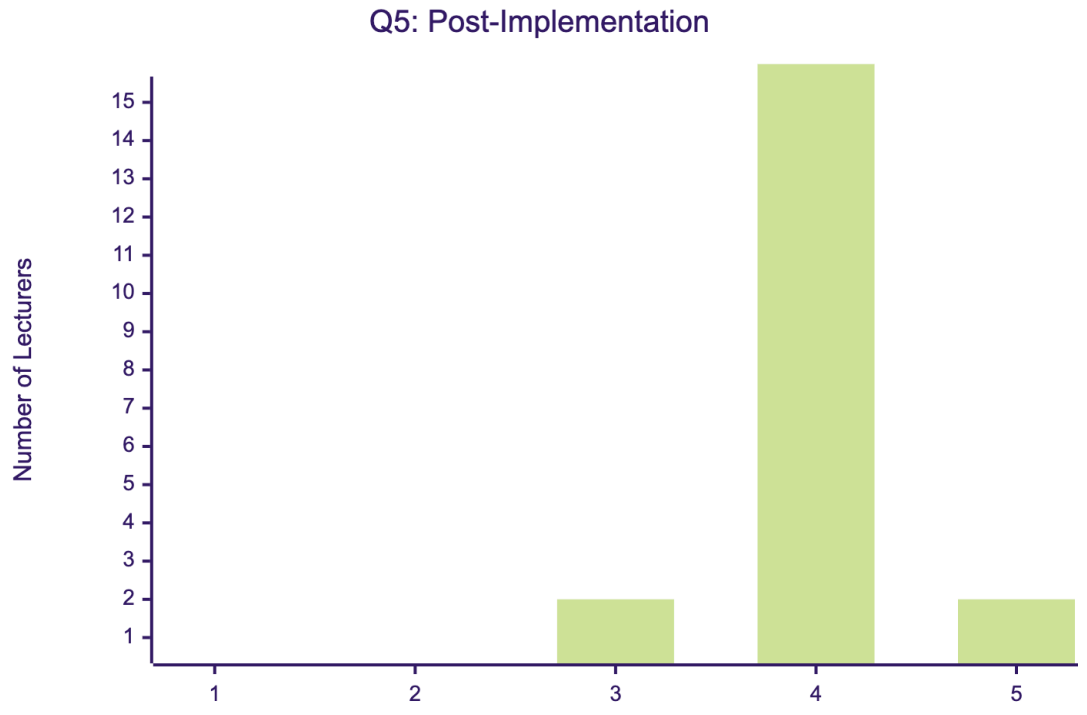

### E.2.6 Personalised Feedback Capacity (Q6)

Lecturers' assessment of their capacity to provide personalised feedback showed a significant improvement (M = 4.10, range 3-5) compared to the pre-implementation assessment (M = 2.15, range 1-3). This substantial increase (+1.95) likely reflects lecturers' experience that StudyBuddy handled routine queries, freeing their time for more targeted, personalised support to students who needed it most. Research indicates that AI chatbot administrative support capabilities can help educators save time on routine tasks, such as scheduling, grading, and providing information to students, thereby allowing them to allocate more time for instructional planning and student engagement (Labadze et al., 2023). The literature suggests that AI is crucial in automating routine tasks, allowing teachers to focus on higher-order cognitive and strategic activities. This enables teachers to reduce the time spent on administrative tasks, such as grading and lesson planning, and frees them to concentrate more on developing complex teaching strategies (Almuhanna, 2025). Research also shows that AI technology automates tedious queries and fosters the creation of educational content. At the same time, the automation of grading significantly reduces educators' time evaluating student submissions, thereby freeing up resources to focus on more interactive and student-centred teaching approaches (Ahmed et al., 2024). The literature emphasises that personalised feedback helps students improve their learning and engagement, with instructors able to notice when a particular student requires specific types of assistance to stimulate optimal learning progression (Yaseen et al., 2025).

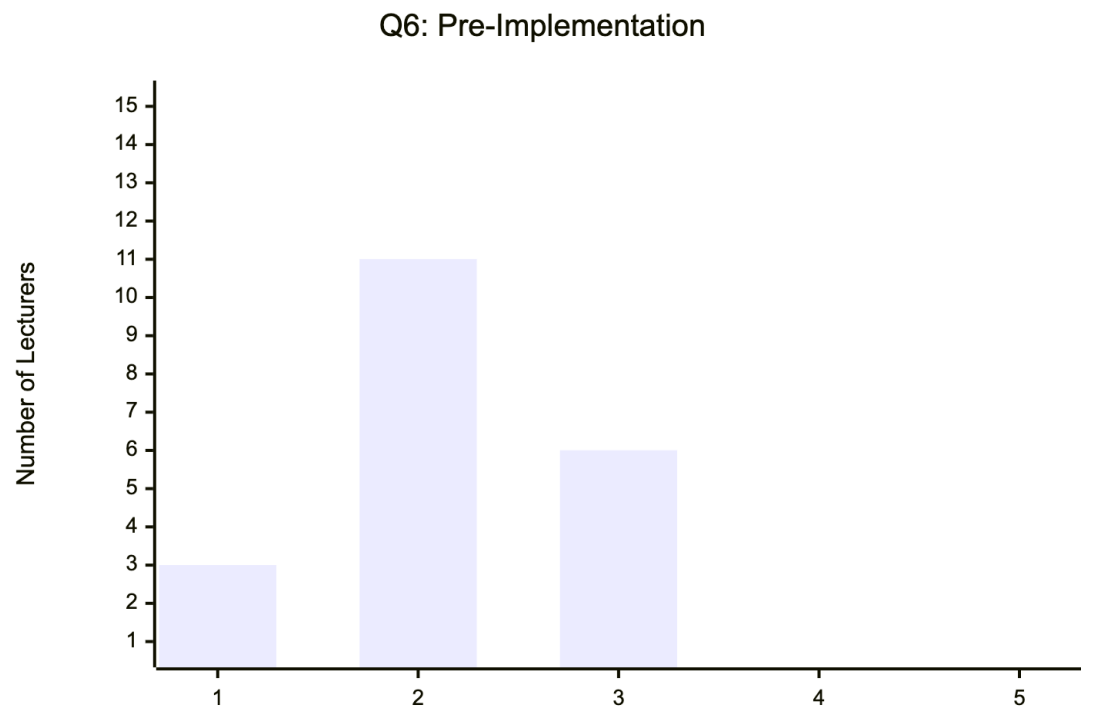

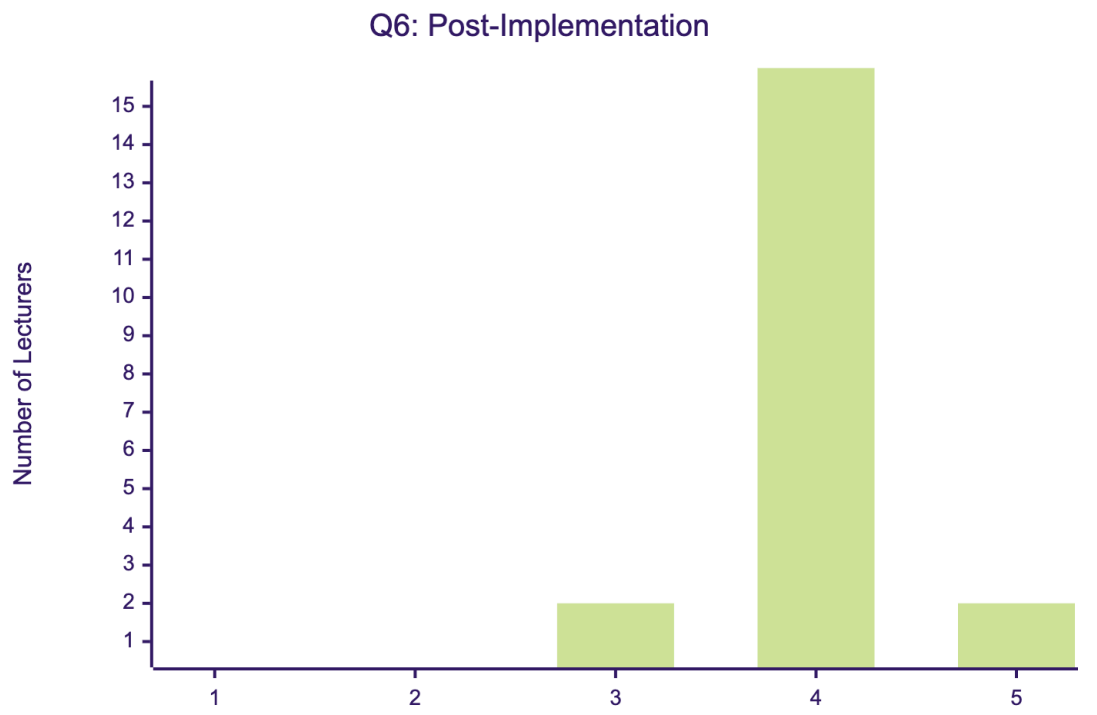

### E.2.7 Active Use of AI Tool (Q7)

Lecturers reported moderate to extensive active use of StudyBuddy as part of lesson design and delivery (M = 4.10, range 3 to 5). This shows substantial integration of the tool into teaching practices, with lecturers actively incorporating StudyBuddy into their instructional approaches rather than simply making it available to students. Pre-implementation data revealed limited prior experience with AI-based educational tools (Q7 M = 1.50, range 1-2), indicating substantial growth in AI tool integration. Research indicates that integrating AI tools into instructional designs enhances student learning experiences and develops students' AI competencies, with the effective integration of AI tools into instructional designs being crucial (Cai et al., 2025). The literature indicates that teachers' technological self-efficacy and attitudes towards technology significantly impact the effectiveness of AI tools in education, with adequate training for both students and teachers being essential to maximise the benefits of AI tools (W. Liu & Wang, 2024). Research indicates that teachers should be equipped with sufficient training and support to effectively utilise AIED tools, which may enhance their acceptance and understanding of the benefits. Educators' attitudes towards AI significantly influence its implementation in education (Alwaqdani, 2025). The literature describes an "Adaptation" phase where educators begin to find practical and effective ways to incorporate AI tools, showing a shift from initial awe and skepticism towards a more balanced and functional integration, followed by an "Acceptance" phase where AI becomes an integral part of the educational blueprint rather than an external, disruptive force (Gordon et al., 2024).

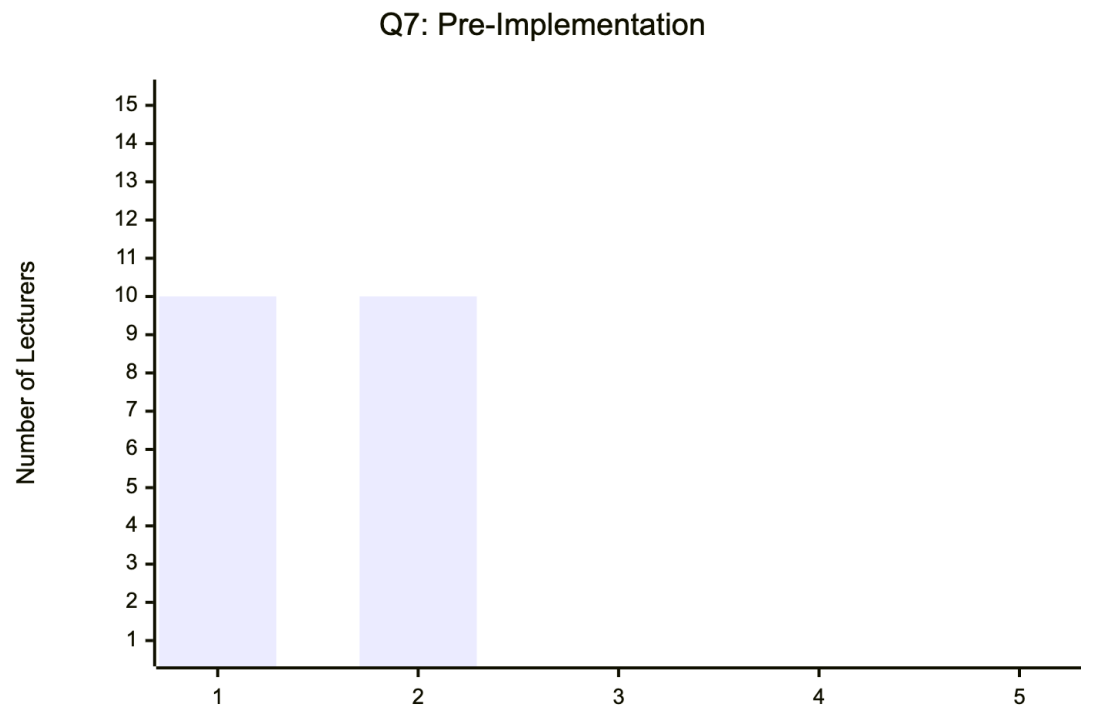

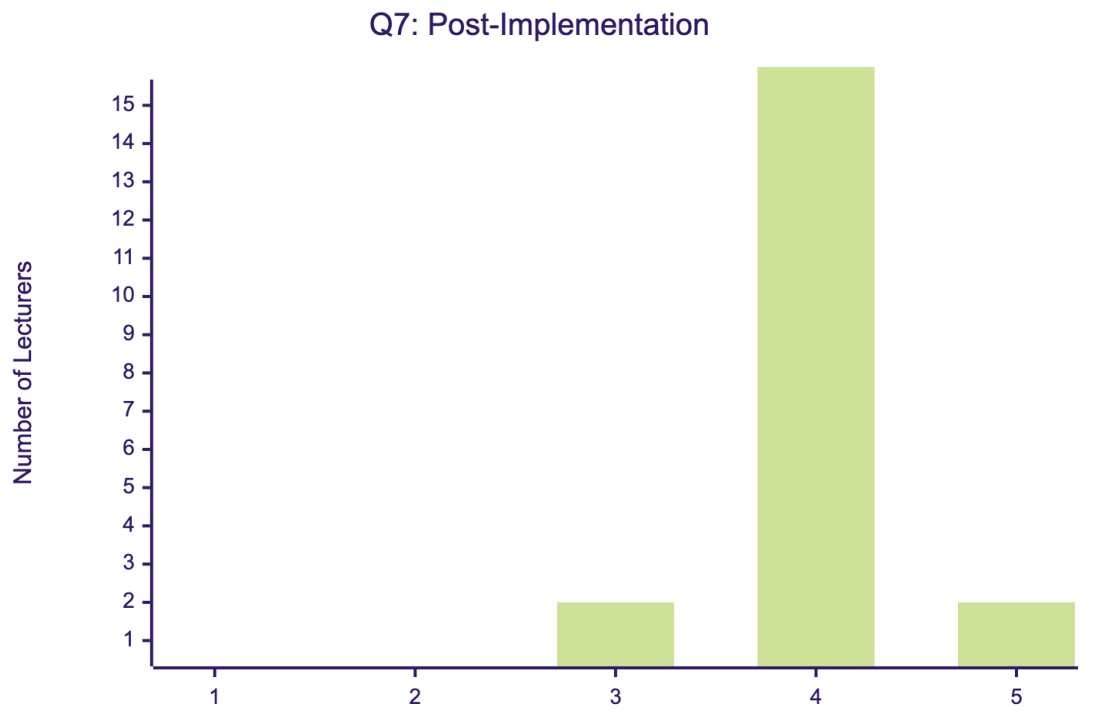

### E.2.8 Confidence in Evaluating AI Responses (Q8)

Confidence in evaluating whether AI-generated responses are pedagogically sound improved (M = 3.80, range 3 to 5, compared to pre-implementation M = 2.40, range 1 to 3). This increase (+1.40) indicates that hands-on experience with StudyBuddy has helped lecturers develop skills in assessing the quality of AI output. However, the moderate mean suggests a continued need for professional development in this area. Research indicates that hands-on engagement with AI technologies can be an effective way to enhance critical thinking and risk assessment capabilities. Experience-based studies have shown that practical engagement facilitates the development of assessment skills (H. Wu et al., 2025). The literature indicates that teachers should be provided with adequate training and support to effectively utilise AIED tools. It should prioritise updated and relevant professional development to increase acceptance and understanding of the benefits (Alwaqdani, 2025). Research shows that to achieve effective integration of AI educational tools, teachers need systematic training and ongoing technical support. However, the current teacher training system may not be able to meet the needs of this new type of technology application. Limited resources for ongoing professional development constrain teachers from engaging in long-term learning and capacity enhancement (C. Yuan et al., 2025). Studies indicate that the majority of faculty members are interested in professional development in AI-related topics, highlighting the ongoing need for training in this area (Mah & Groß, 2024). The literature emphasises the critical need for adaptive and continuous professional development to ensure teachers are aligned with existing technological tools and prepared to integrate emerging AI innovations effectively (M. Li & Manzari, 2025).

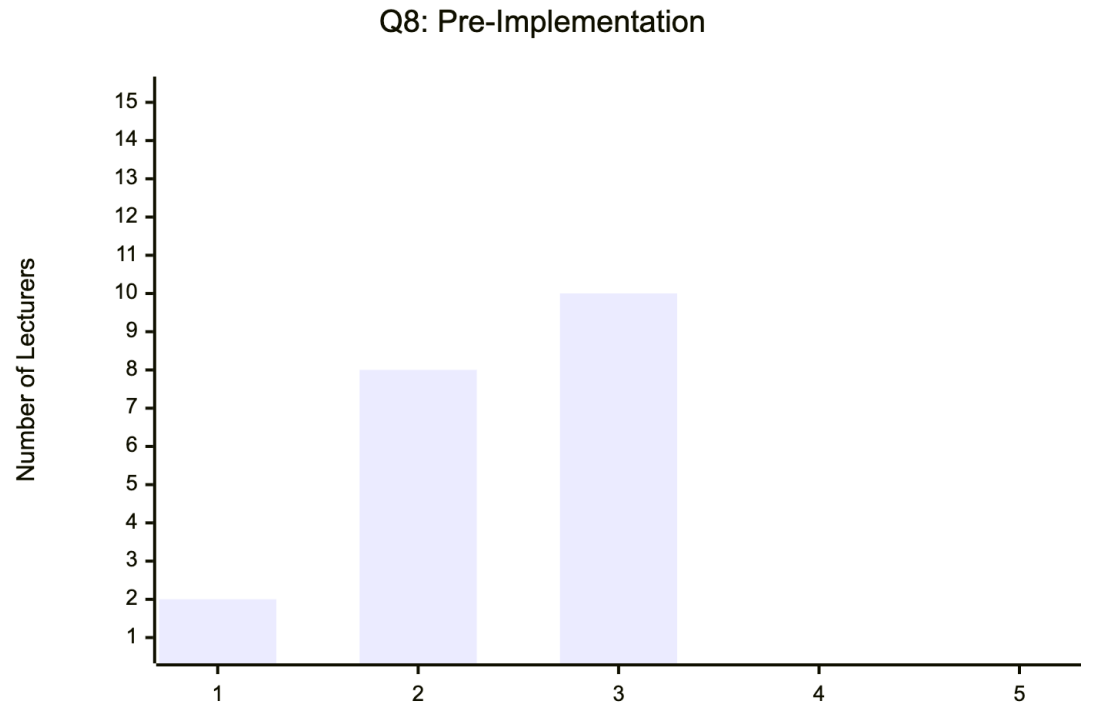

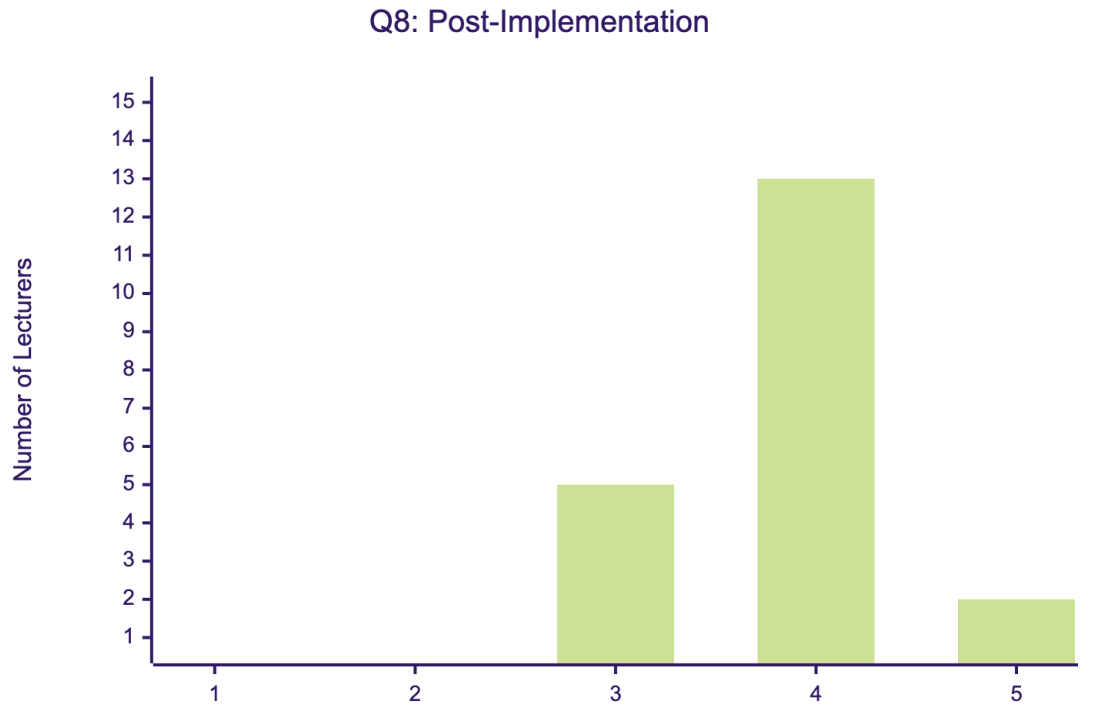

## E.2.9 AI Tool Complementing Teaching Style (Q9)

Lecturers strongly agreed that StudyBuddy supported learning in a way that complemented their teaching style (M = 4.30, range 3-5). This high score indicates that lecturers found StudyBuddy to be aligned with their pedagogical approaches, rather than requiring fundamental changes to their teaching philosophy. This aligns with pre-implementation beliefs that technology augments rather than replaces human judgment (Q4 M = 4.00). Research shows that the integration of AI in education should be viewed as a complement to, rather than a replacement for, the essential human elements that are irreplaceable in fostering well-rounded learning (Qassrawi & Al Karasneh, 2025). The literature describes the "augmentation perspective" in AIED research, which emphasises rethinking the ways to complement the strengths of teachers and AI, rather than a replacement perspective, which suggests substituting human teachers (Kim, 2024). Research shows that AI tools align with educational goals and student needs, and when aligned with pedagogical principles, AI tools can enhance teaching practices and engage students through personalised learning experiences (Berisha Qehaja, 2025; Cai et al., 2025). Studies indicate that AI solutions are most effective when used to supplement and assist traditional teaching techniques, rather than replacing teachers (Salih et al., 2025).

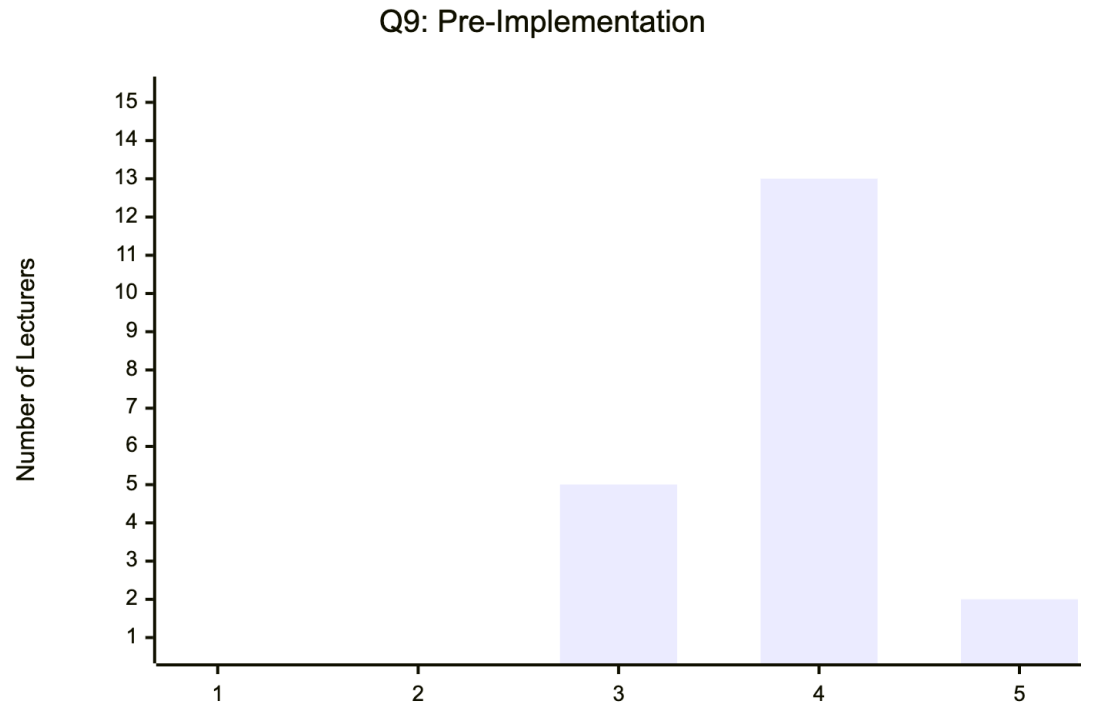

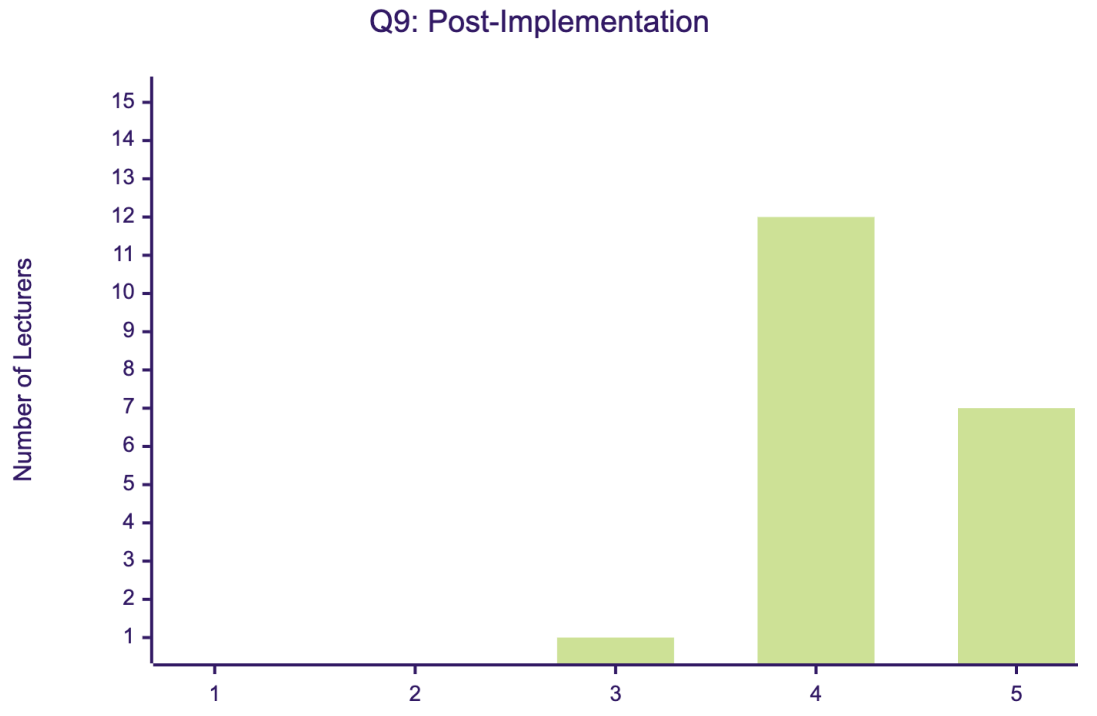

## E.2.10 Changes in Instructional Design Thinking (Q10)

Lecturers reported moderate changes in their thinking about instructional design (M = 3.70, range 3-5), compared to the pre-implementation Q10 (M = 2.95, range 2-4). This moderate level indicates that while StudyBuddy influenced lecturers' thinking about instructional approaches, it did not necessitate a complete transformation of their design philosophies, reflecting an augmentation rather than replacement perspective. Research on integrating AI tools into instructional designs focuses on how to incorporate AI tools into instructional designs. It explores how AI tools can be aligned with pedagogical principles, emphasising the development of critical thinking and enabling students to achieve their goals (Cai et al., 2025). The literature describes the augmentation perspective in instructional design, highlighting the role of AI in facilitating teaching processes by enhancing teachers' cognition through quick data collection, analysis, and translation into meaningful insights and actions. AI systems are designed to combine human teacher and AI instruction to work together (Kim, 2024). Research indicates that AI facilitates teachers' ability to design course syllabi, create content and resources, develop evaluation activities, and deliver them to students more effectively, enhancing existing practices rather than completely transforming them (Almuhanna, 2025). While some research has interpreted positive results as indicators of potentially profound transformational effects of AIEd, doubts about the value of AI educational technologies remain due to limited evidence of their effectiveness at scale, and how teachers use AI technologies pedagogically and their roles in learning in classrooms remain unclear (Chiu et al., 2023).

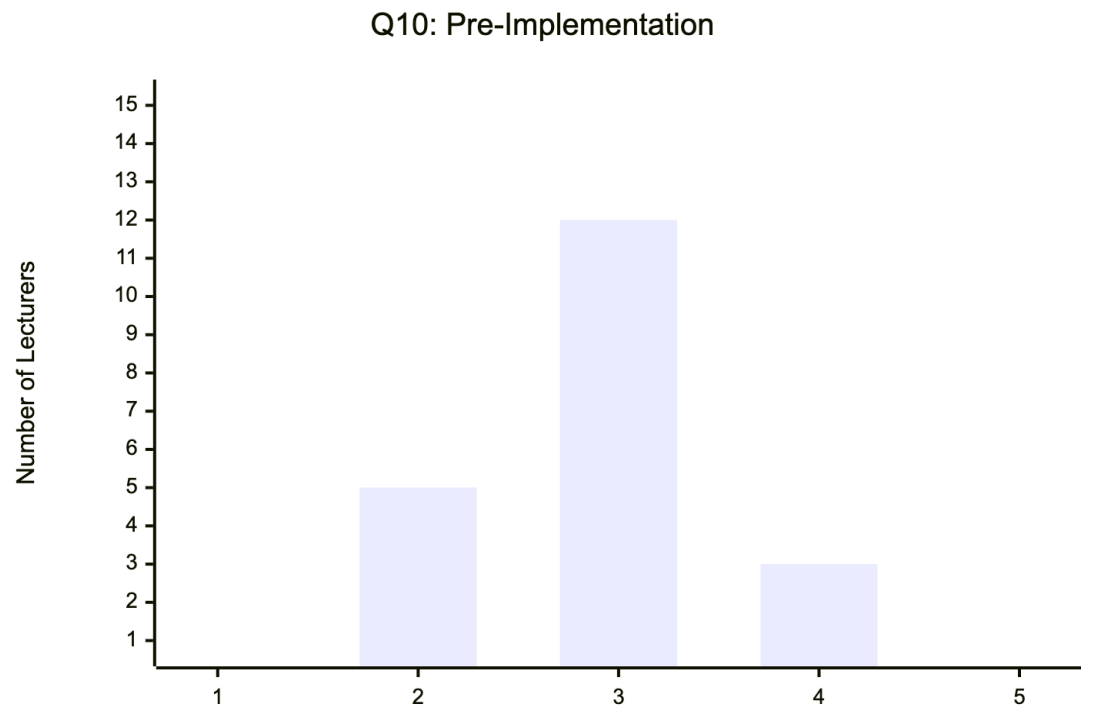

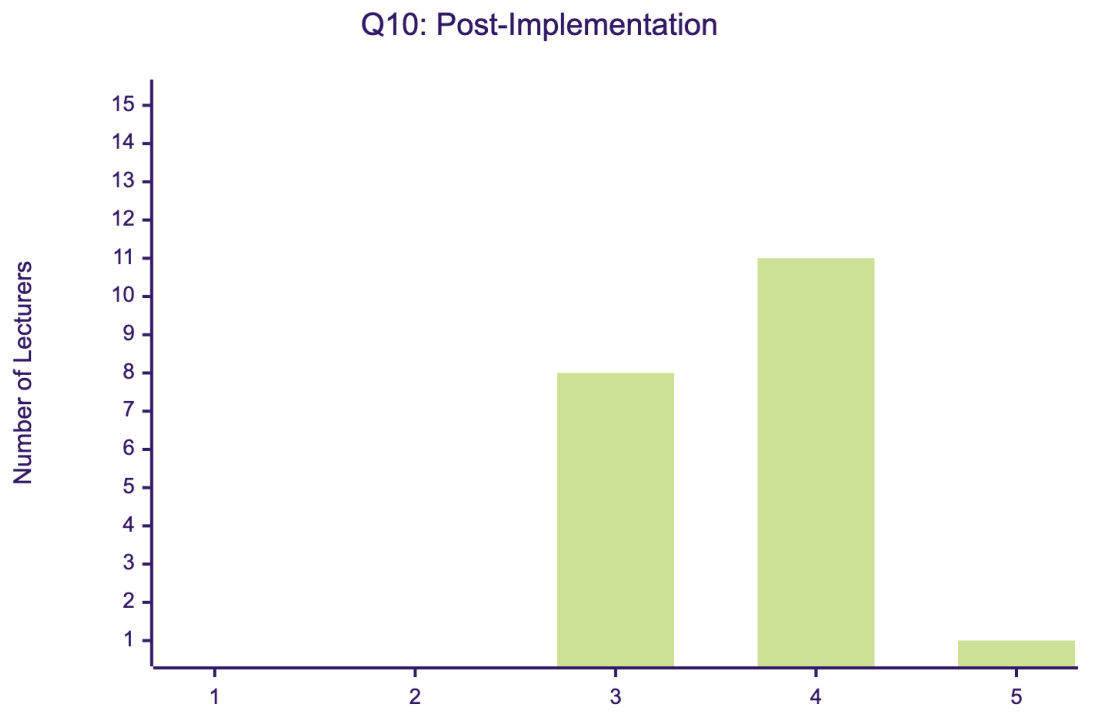

**E.2.11 Dialogue Over Content Delivery (Q11)**

Lecturers maintained a strong agreement that students learn best through dialogue rather than content delivery (M = 4.20, range 3 to 5), compared to the pre-implementation mean of 3.95 (range 3 to 5). This consistency, combined with the high score for StudyBuddy's complementary teaching style, indicates that lecturers found the chatbot's dialogue-based approach to be aligned with their pedagogical values. Research indicates that educational dialogue serves as a social mode of thinking, enabling collaborative problem-solving and shared responsibility for constructing understanding. This dialogue facilitates students' deep understanding, critical thinking, reasoning, and learning achievement (Song et al., 2025). According to sociocultural theory, learning and cognitive growth are cultural processes realised through verbal interaction with others, and dialogue serves as a social mode of thinking. Through dialogue, students not only embrace diverse ideas but also gain practice in thinking through problems and organising concepts, formulating arguments and counter-arguments, and responding thoughtfully and critically to diverse points of view, which is assumed to facilitate learning achievement (Song et al., 2025). Research shows that dialogue-based AI chatbots create a more natural and human-like interaction environment, allowing learners to question, challenge, and seek clarification through continuous dialogue. This process helps break hierarchical power dynamics, where teachers impart knowledge and students passively respond (Le et al., 2025; Song et al., 2025). Studies indicate that AI chatbots can help learners practice through interactions similar to those with human partners, providing personalised learning content and offering feedback and guidance on a one-to-one basis, thereby supporting dialogue-based learning approaches (R. Wu & Yu, 2024).

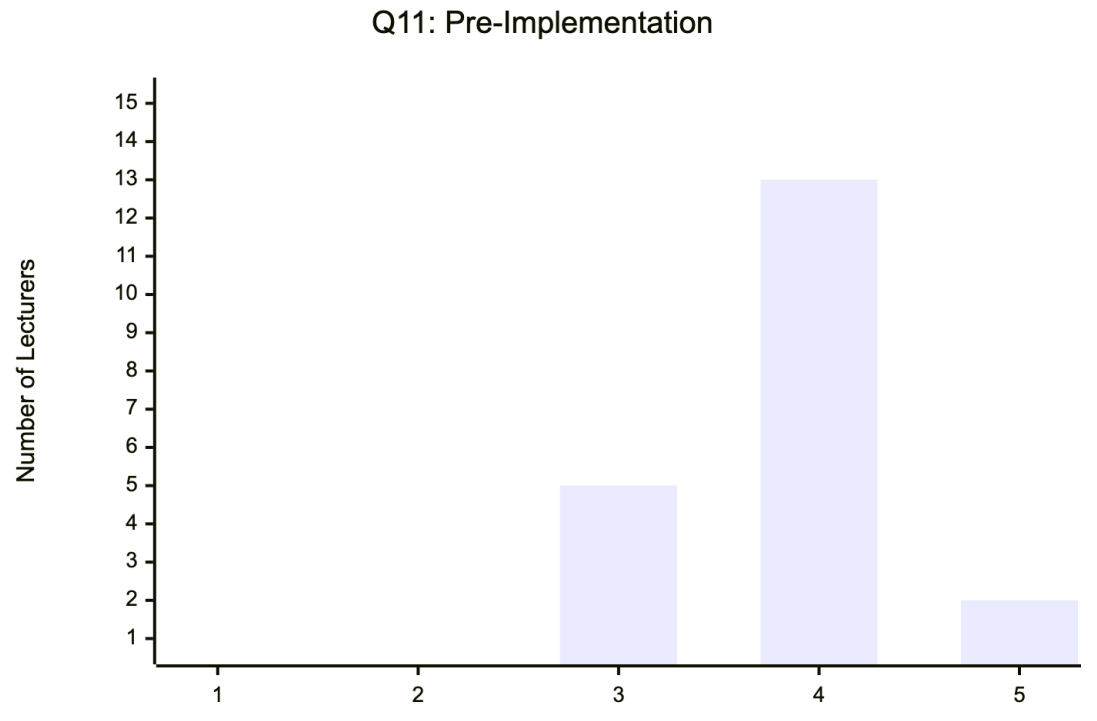

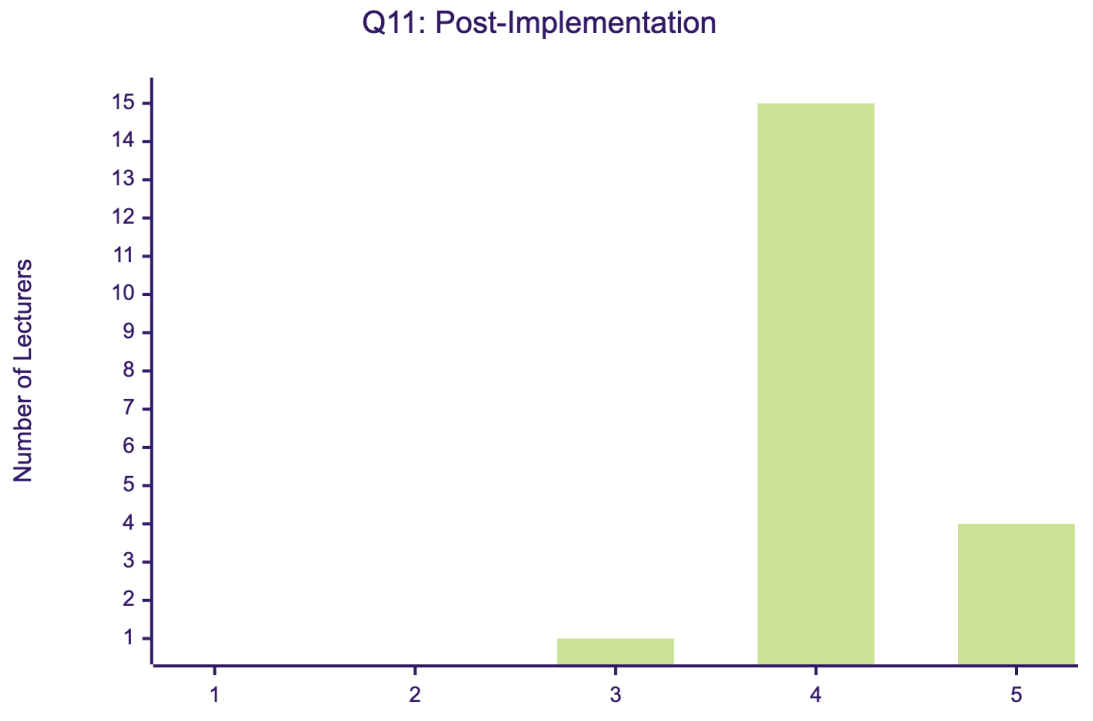

## E.2.12 Curiosity Over Efficiency (Q12)

Moderate agreement that educational tools should prioritise curiosity over efficiency persisted (M = 4.10, range 3 to 5, compared to pre-implementation M = 3.20, range 2 to 4). The increase (+0.90) indicates that lecturers' experience with StudyBuddy reinforced the value of curiosity-driven learning, although the moderate mean suggests continued recognition of the need for practical efficiency. Research shows that curiosity refers to an individual's desire to acquire new information and is a prerequisite for stimulating exploratory behaviour, which has an impact on emotions, cognition, and the level of effort to process acquired knowledge (H. Chen et al., 2023). The literature suggests that educational tools that promote users' exploratory behaviour and desire to learn are closely related to curiosity, and curiosity significantly impacts users' emotional and cognitive processes. However, the literature also acknowledges the importance of balancing curiosity-driven exploration with effective learning outcomes (H. Chen et al., 2023). Research shows that AI-based tutoring systems can provide individualised guidance, thereby fostering students' interest in learning and encouraging curiosity-driven engagement (Yaseen et al., 2025). Studies also indicate that students demonstrate curiosity about how AI generates answers, engaging in testing AI responses to explore the technology's capabilities (Akolekar et al., 2025). The literature recognises intellectual curiosity as an important dimension of critical thinking dispositions, showing that curiosity-driven learning supports deeper cognitive engagement (Y. Liu & Pásztor, 2022).

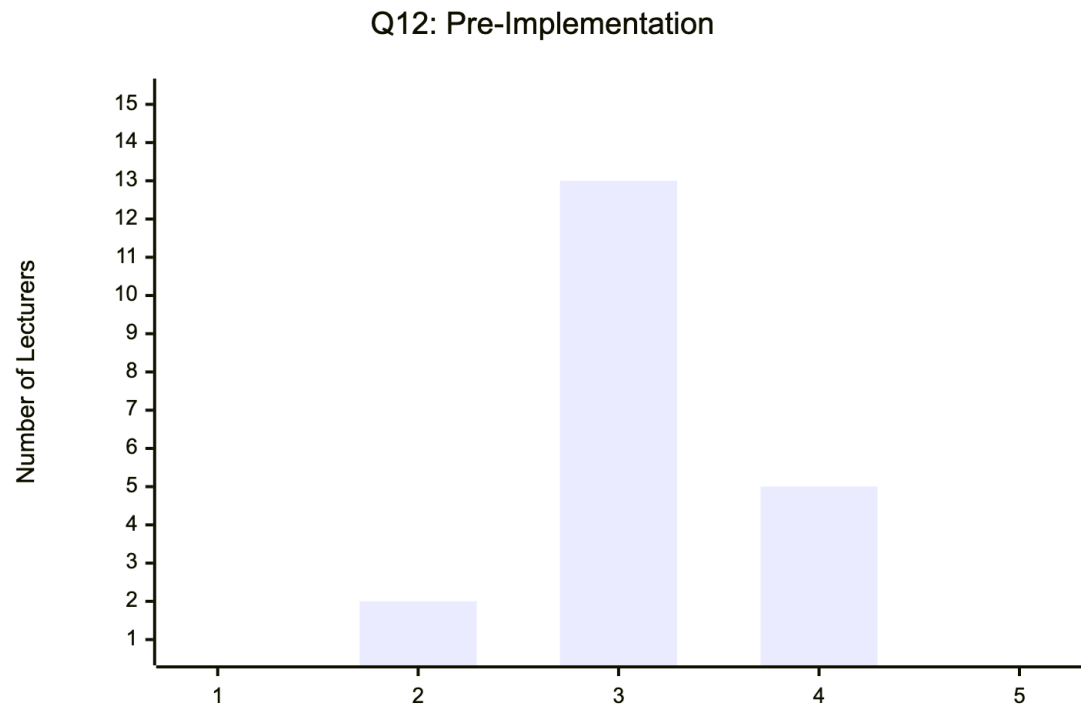

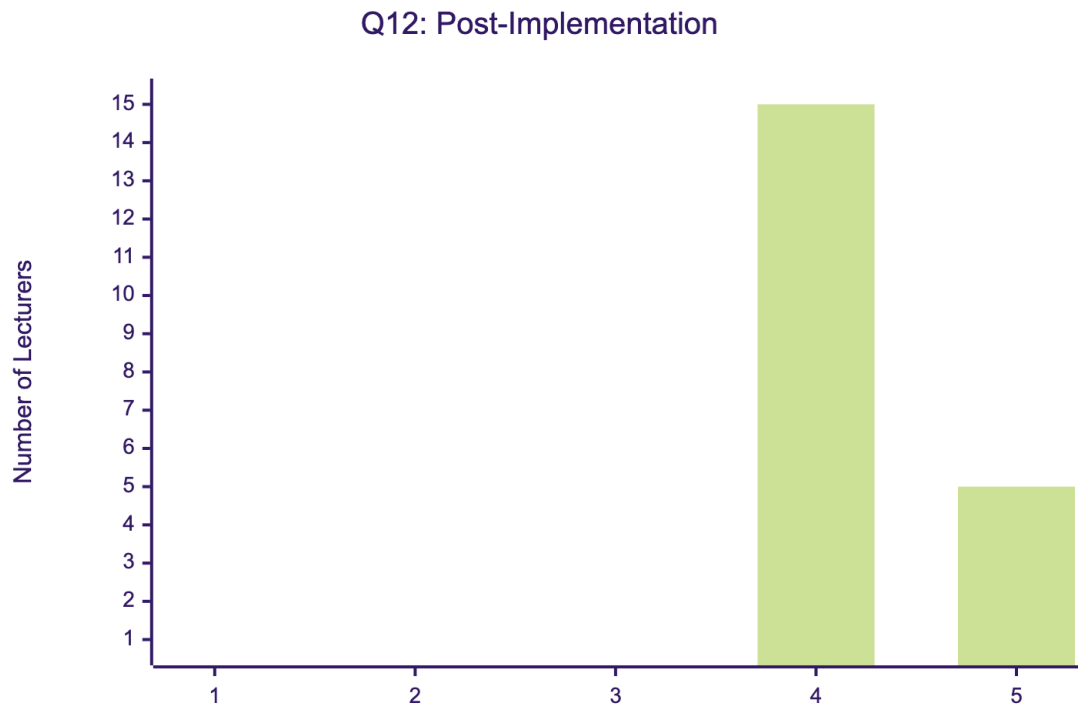

## E.2.13 Comfort with Unpredictable Outcomes (Q13)

Comfort with automated systems shaping learning experiences when outcomes are unpredictable showed improvement (M = 3.90, range 3-5, compared to pre-implementation M = 2.60, range 1-4). This increase (+1.30) indicates that positive experiences with StudyBuddy have increased lecturers' comfort with AI systems, although the moderate mean suggests a continued preference for predictable and manageable outcomes. Research shows that algorithm aversion, a biased assessment of algorithms manifesting in negative behaviours and attitudes, can be effectively mitigated in educational contexts through appropriate interaction design, with positive experiences potentially increasing comfort with AI systems (Le et al., 2025). The literature suggests that concerns about the reliability and accuracy of AIED can be mitigated through adequate training and support, which may enhance acceptance and understanding of its benefits (Alwaqdani, 2025). However, research also shows that a considerable percentage of teachers do not trust AI to carry out tasks without error, with many expressing concerns about the reliability and accuracy of AIED (Alwaqdani, 2025). Studies indicate that unpredictability is a significant contributor to distrust in AI, with concerns about unpredictability manifesting in both global and local forms, including worries about the unpredictability of AI systems in specific recommendations or under specific circumstances (Afroogh et al., 2024). The literature shows that AI complexity can raise concerns about accountability and fairness, and the black-box nature of AI may limit awareness of factors influencing educational experiences, making decision-making difficult to understand (Al-Zahrani, 2024). Research indicates that both students and educators have expressed concerns about the growing role of AI in education, including fears that overreliance on technology may limit opportunities for exploration and hinder self-guided learning (Adewale et al., 2024). These findings indicate that while positive experiences can enhance comfort with AI systems, educators continue to prefer predictable and manageable outcomes due to concerns about unpredictability, a lack of transparency, and potential limitations on exploration.

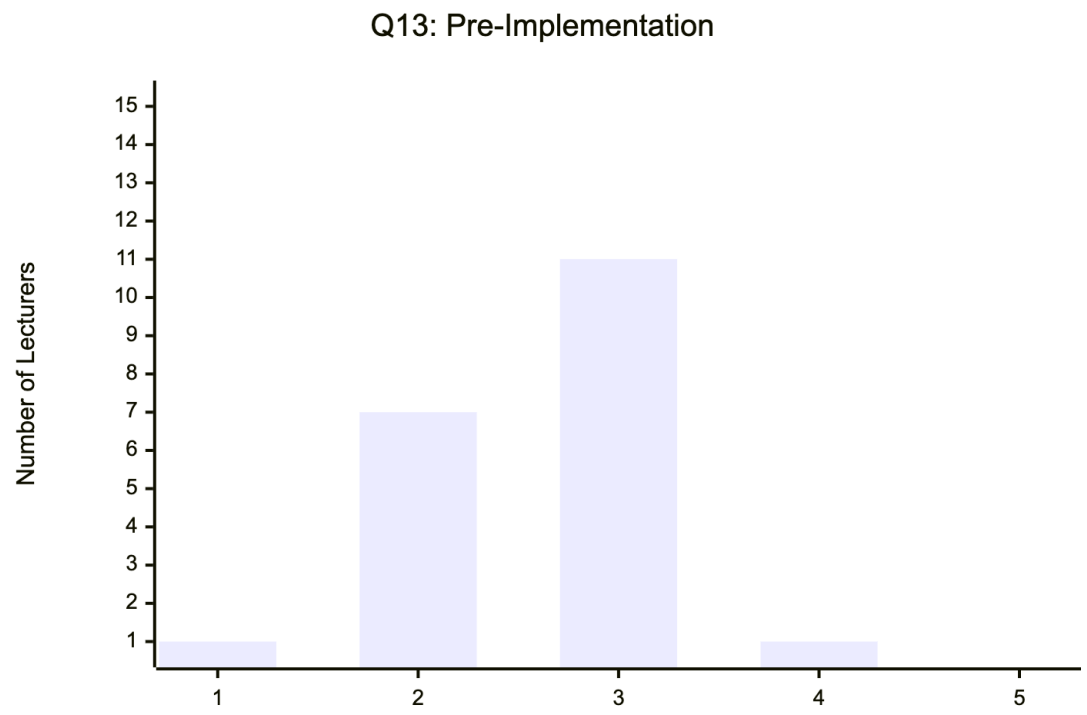

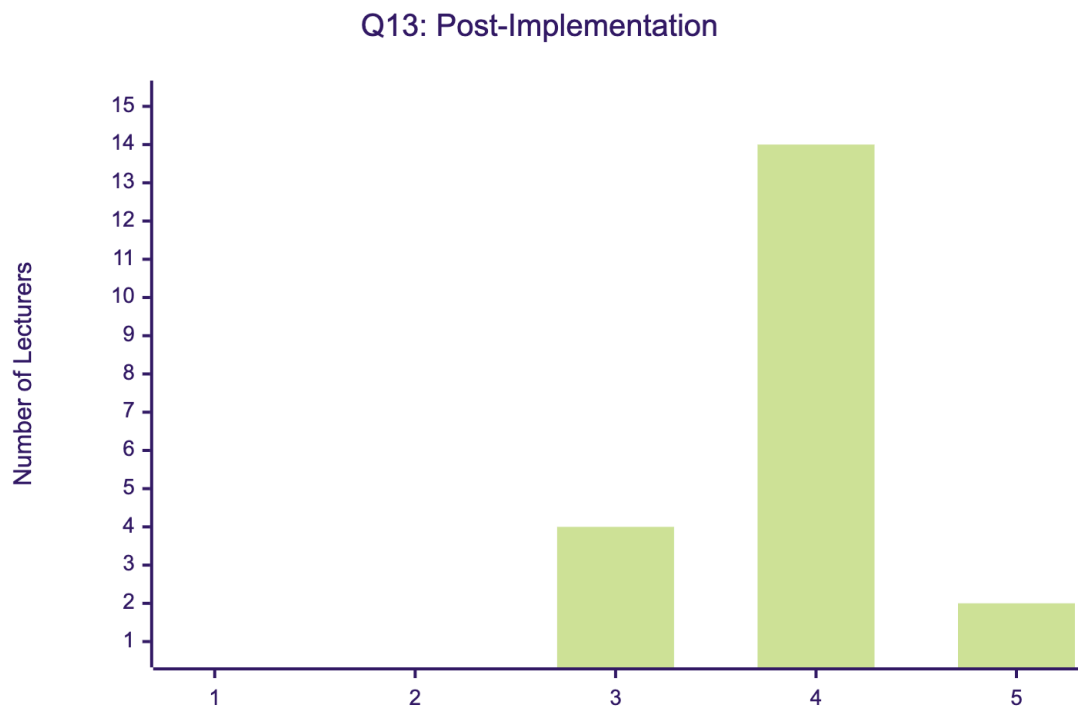

### E.2.14 Qualitative Analysis: Shifts in Understanding of Teaching (Q14)

Open-ended responses revealed several key shifts in lecturers' understanding of teaching following the deployment of StudyBuddy. Multiple lecturers noted that "AI can handle the basic questions so I can focus on more important things" and that "StudyBuddy made my teaching more efficient because it handles the routine questions. I can now focus on students who really need help." This acknowledges that technology can augment teaching by handling routine tasks, allowing educators to redirect their time toward more pedagogically valuable activities. This aligns with pre-implementation recognition of insufficient time and resources for personalised feedback (Q6 M = 2.15). Research shows that AI enables users to outsource mundane tasks, allowing for more time to devote to critical and creative thinking (Essien et al., 2024). The literature indicates that AI chatbot administrative support capabilities can help educators save time on routine tasks, including scheduling, grading, and providing information to students, allowing them to allocate more time for instructional planning and student engagement (Labadze et al., 2023). Several lecturers observed that "teaching involves both direct instruction and using AI tools strategically" and that "AI tools can be study companions supporting students outside class." This demonstrates an expanded understanding of teaching that incorporates the strategic use of technology to scale personalised learning support.

Research shows that AI-supported tools can extend traditional teaching by automating routine tasks, enabling personalised learning, and providing insights based on data. At the same time, educators concentrate on facilitating critical thinking and interpersonal skills (Salih et al., 2025). Multiple lecturers noted shifts toward data-driven approaches, stating that "Analytics to identify student struggles is now part of my teaching" and that "Data-driven teaching with AI providing insights into learning patterns is what happened." This shows recognition that AI tools can provide valuable insights into student learning patterns, enabling more targeted instructional support. The literature indicates that AI-powered learning analytics facilitate early intervention by identifying at-risk students, providing educators with insights for timely interventions to prevent academic challenges and enhance student success (Berisha Qehaja, 2025). Several lecturers recognised that "StudyBuddy showed me how personalised learning can be scaled" and that "Technology can support personalised learning at scale." This shows understanding that AI tools can enable personalised support that would be difficult to provide through traditional means alone. Research shows that AI technologies create personalised learning experiences that address individual student needs and preferences. AI-powered tools offer personalised support, and adaptive learning systems enable tailored learning experiences for each student (Abulibdeh et al., 2024; Berisha & Qehaja, 2025). Lecturers consistently emphasised that "The role of facilitating learning is what I do and AI tools help by handling routine queries" and that "Technology needs careful monitoring to make sure it supports learning, not undermines it." This demonstrates a nuanced understanding that technology serves as a tool that requires careful oversight to ensure it supports rather than undermines learning objectives, consistent with pre-implementation views on technology as an augmentation (Q4 M = 4.00). Research highlights the need for proper guidance and monitoring to ensure that AI is used ethically and effectively (Ahmed et al., 2024). The literature emphasises that as educators and learners embrace AI, they must remain vigilant about its impact, ensuring that collaborative learning spaces remain inclusive, supportive, and conducive to supporting effective learning (Cai et al., 2025).

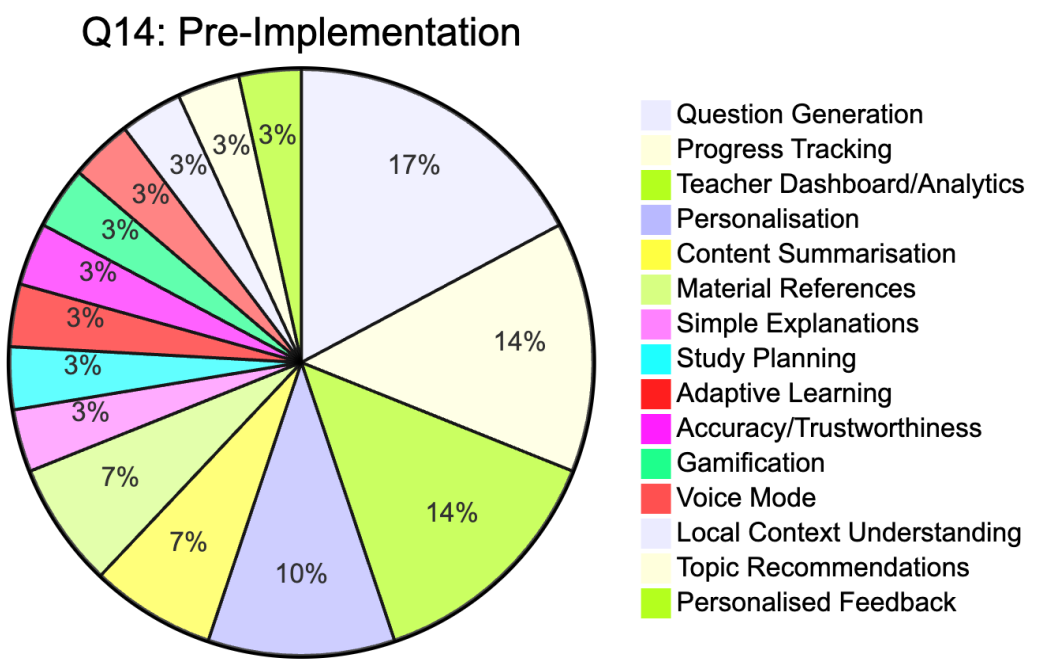

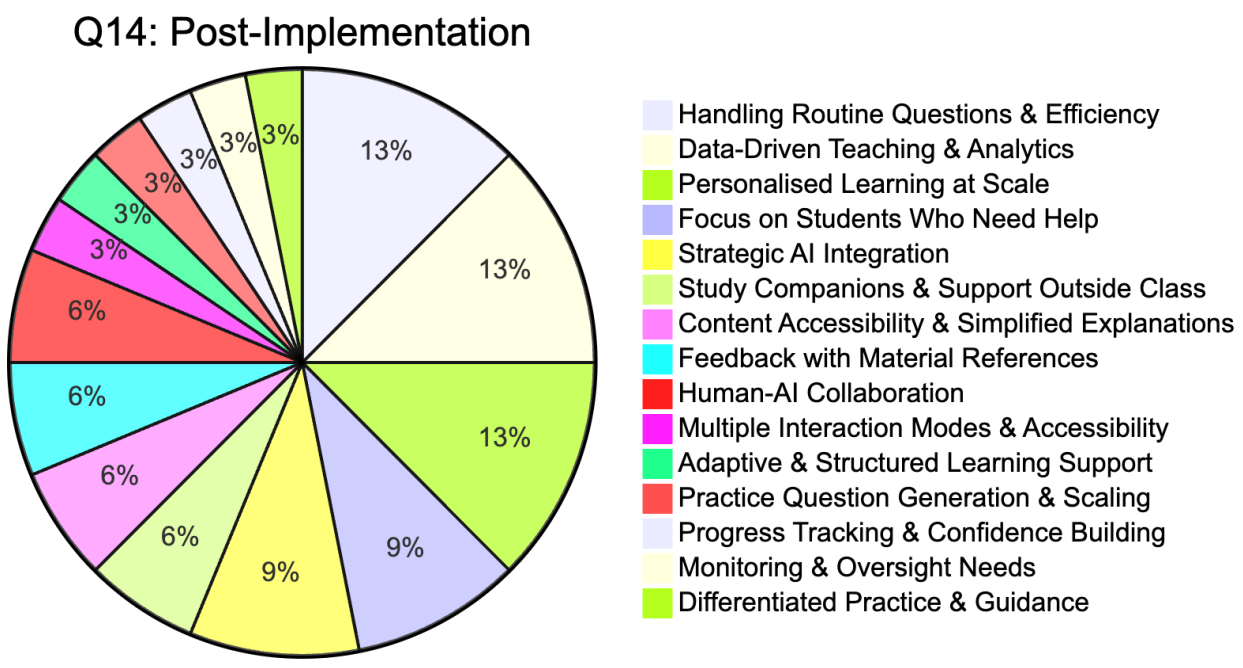

### E.2.15 Qualitative Analysis: Ongoing Concerns About AI in Education (Q15)

Open-ended responses showed that many pre-implementation concerns persisted but were nuanced by actual usage experiences. The most frequently mentioned concern was dependency, with lecturers stating, "Still worried about students becoming too dependent on it, especially weaker ones", and "Over-reliance on AI is the biggest concern. They might stop thinking for themselves and just depend on the tool." This shows that despite positive experiences, concerns about dependency patterns undermining independent learning skills remained prominent. This aligns with pre-implementation concerns where "Over-Reliance/Dependency" was the most frequently mentioned risk. Research indicates that over-reliance on AI can hinder the development of essential skills, including critical thinking, problem-solving, and effective communication. The convenience offered by AI tools in providing quick answers may deter students from engaging in thorough research and forming their own insights (Zhai et al., 2024). The literature suggests that over-reliance on AI may reduce students' skills in creativity and innovation, potentially undermining their capacity for independent critical thinking and problem-solving (Zhai et al., 2024). Research indicates that AI dependency can diminish 21st-century skills, including problem-solving ability, critical thinking, creative thinking, collaboration skills, communication skills, and self-confidence (D. Zhang et al., 2025). Studies indicate that there is an over-reliance on AI tools, which may hinder the growth, skills, and intellectual development of students and teachers over time (Yusuf et al., 2024). Multiple lecturers expressed concern that "Main concern is students losing critical thinking skills. Some might become too dependent on quick answers instead of thinking through problems", and that "Impact on critical thinking skills worries me. If students get answers quickly, they might not develop the reasoning abilities needed for SAQ." This shows ongoing concern about cognitive offloading and its impact on higher-order thinking skills, consistent with pre-implementation moderate confidence in supporting critical thinking (Q5 $M = 2.90$). Research suggests that over-dependence on AI technologies could negatively impact the development of essential cognitive skills, with the concern based on the notion that relying on AI for tasks requiring critical thinking may decrease students' capacity for original thought, creativity, and independent problem-solving (Ahmed et al., 2024). Accuracy concerns persisted, with lecturers noting "Accuracy is crucial. If the tool gives wrong information, students will be misled" and "Also need to make sure accuracy is maintained." This acknowledges that reliability is crucial for trust and that accuracy issues necessitate ongoing monitoring. This aligns with pre-implementation emphasis on accuracy and trustworthiness as essential qualities. Research shows that over 62.2% of students expressed concerns about AI providing incorrect or unreliable answers, while 65.69% reported difficulties in verifying the accuracy of the information (Akolekar et al., 2025). The literature suggests that the integration of GAI tools has raised significant concerns regarding the accuracy of information and the potential perpetuation of biases. The inherent limitations of GAI tools, such as the production of fabricated articles or references, further complicate these ethical dilemmas (Ahmed et al., 2024). Research indicates that the majority of teachers do not trust AI to perform tasks without error, with many expressing concerns about the reliability and accuracy of AIED (Alwaqdani, 2025). Several lecturers emphasised that "Academic integrity must be protected" and "Worried students might use a tool to cheat or get answers they haven't earned." This shows ongoing concern about maintaining academic standards and preventing misuse, consistent with pre-implementation concerns about exam fairness and academic integrity. Research shows that AI tools have revolutionised the educational landscape but have also posed new challenges to academic integrity, as the accessibility and sophistication of these tools enable students to generate assignments and answers effortlessly, thereby undermining the principles of academic honesty (Mumtaz et al., 2025). Studies indicate that GenAI tools promote cheating and plagiarism without being easily detected, with a significant majority of participants sharing the view that GenAI tools have the potential to facilitate academic misconduct (Yusuf et al., 2024). Some lecturers raised concerns about "Exam fairness is a concern", "Need to ensure all students have equal access" and "Also risk some students might not engage, creating inequities." This acknowledges that the implementation of AI tools must address equity considerations. Research indicates that AI could exacerbate educational disparities by widening gaps in access to advanced technologies, highlighting the importance of equitable access to AI resources in preventing the amplification of these disparities (Al-Zahrani, 2024). The literature suggests that AI technologies have the potential to perpetuate and exacerbate inequalities due to inherent biases in training datasets, their capacity to influence real-world decision-making, and the ways they can, intentionally or inadvertently, reinforce dominant, rather than diverse, cultural practices (Viberg et al., 2024). Research shows that strategies to ensure equitable access to AI-driven resources for all students include promoting digital literacy, providing adequate technological infrastructure, and addressing socioeconomic disparities (Berisha Qehaja, 2025). Practical concerns were raised, with lecturers noting "Time needed for maintenance and monitoring is a concern" and "Setup and maintenance time is a concern. We're already busy, so if the tool needs lots of time, it might not be practical." This demonstrates recognition of the ongoing resource requirements for implementing effective AI tools. Research indicates that while educators recognise the potential of AI, they also identify significant challenges, including technical barriers and the need for adequate training and support. Addressing these challenges requires continuous professional development programmes (Berisha Qehaja, 2025).

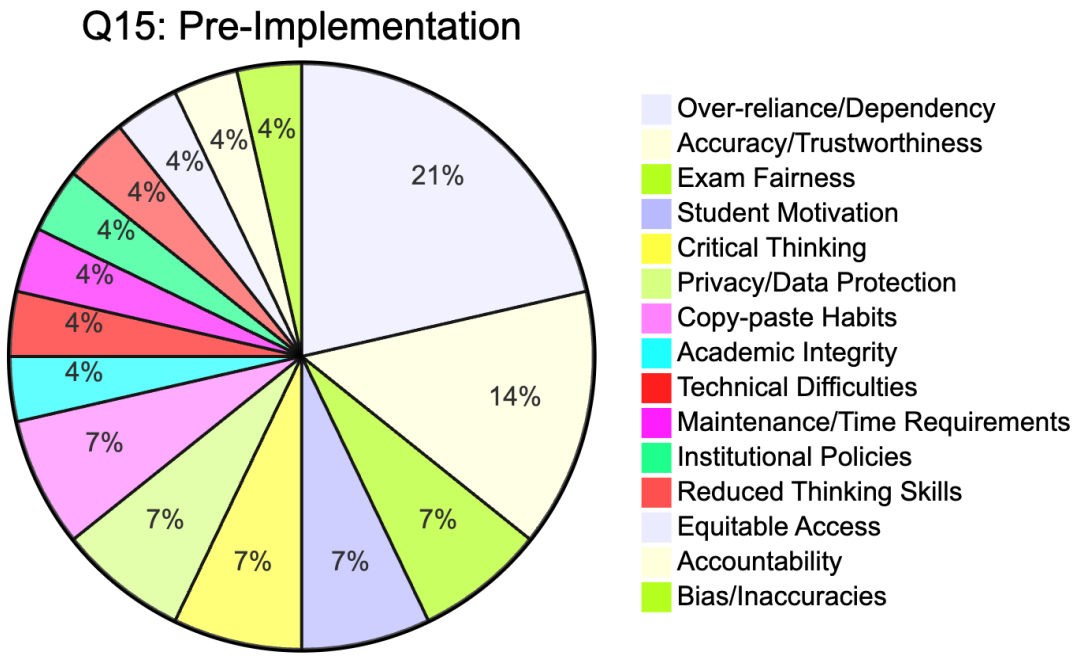

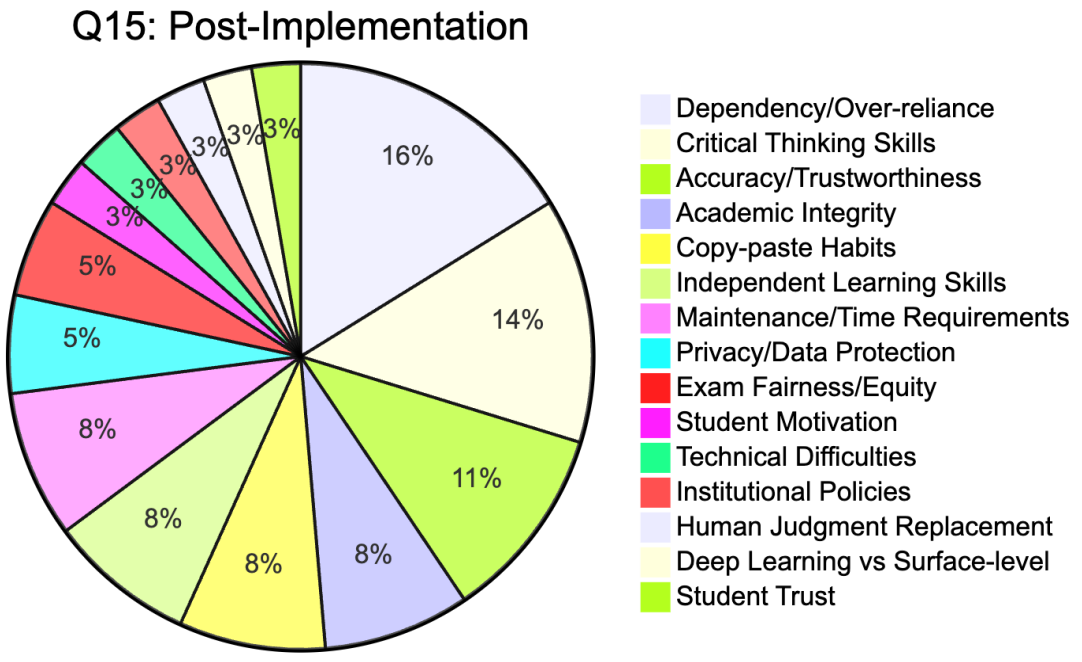

## E.3 Interview Analysis: Observed Impact on Student Learning

### E.3.1 Differences in Student Exam Preparedness

Lecturers reported varying observations regarding student exam preparedness compared to previous batches. Multiple lecturers noted that students demonstrated "more confidence" and appeared "more prepared for the exam," with one lecturer observing that "students performed a lot better in terms of passing the exams and overall grades as well!" Research indicates that students' AI readiness, encompassing their preparedness and willingness to engage with AI technologies, significantly contributes to their academic performance. Students who are more prepared to utilise AI tools in their studies tend to achieve higher academic performance (Wang & Li, 2024). The literature suggests that self-efficacy, which refers to students' belief in their ability to succeed in specific academic tasks, plays a crucial role in their overall academic achievements. Students who have higher confidence in their academic capabilities tend to perform better academically (Wang & Li, 2024). Several studies have examined the potential of AI tools in enhancing learners' performance in educational contexts, with research showing that learners using chatbots or ChatGPT achieved higher scores on their final tests than those using conventional technology methodologies (Fan et al., 2025). Several lecturers reported "more steady, fewer panic cases during exam week," showing that StudyBuddy may have helped reduce exam-related anxiety. Research shows that anxiety reduction is a critical area where AI interventions show significant positive effects, with the supportive and non-judgmental environment provided by AI tools helping to lower learners' anxiety, and immediate, personalised feedback allowing learners to correct their mistakes without fear of negative evaluation, thereby reducing anxiety and building confidence (X.-Y. Wu, 2024). The literature indicates that AI chatbots could reduce the anxiety levels of secondary school students in foreign language learning, and university students using chatbot-delivered interventions reported significantly less anxiety and depression than control groups (R. Wu & Yu, 2024). Consistency improvements were noted, with some lecturers observing that "consistency improved; some started weekly revision," showing that StudyBuddy may have encouraged more regular study habits. Research indicates that AI applications can promote timely and focused study habits, with students exposed to these applications being more consistent in meeting deadlines and exhibiting higher engagement in writing assignments and weekly academic tasks (Ma & Chen, 2024). The literature indicates that the frequent and effective use of AI tools plays a role in improving educational outcomes, with regular interaction with educational technologies enhancing learning and academic performance, and by consistently using AI tools, students can access a wealth of resources and support that facilitate better learning experiences and outcomes (Ahn, 2024). Question quality improved, with multiple lecturers observing that student questions became "more specific and focused" rather than generic queries. However, some lecturers noted minimal differences, showing that the impact may have differed across modules and student groups. Research shows that, despite promising results, the effectiveness of AI interventions can vary depending on several factors, including the design and implementation of the AI tools, the specific skills targeted, and the context in which these tools are used. Some studies have highlighted significant improvements.

In contrast, others report mixed outcomes, showing the need for further research to understand the conditions under which AI is most effective (X.-Y. Wu, 2024). The literature indicates that while a positive attitude of the majority of students toward the use of AI tools indicates strong acceptance of AI integration in learning environments, a low percentage of students agree with specific benefits like improving study grades, and mixed responses suggest that students are uncertain about the actual impact of AI technologies on their academic performance (Fošner, 2024). Research also shows that divergent findings have been reported in the existing literature, with some researchers claiming that AI-powered tools offer no direct benefit to learners' performance, highlighting the variability in AI tool effectiveness across different contexts (Fan et al., 2025).

### E.3.2 Reduction in Repeated Questions

A key benefit that emerged was the reduction in repeated questions directed to lecturers. Multiple lecturers confirmed that StudyBuddy reduced repeated questions, with one noting, "Yes, fewer repeated questions by students," and another stating, "Repeated questions dropped because basic issues went to the bot." Lecturers explained that students could "now ask and get most of their queries answered by study buddy chatbot instead," allowing common questions to be addressed without direct lecturer intervention. Research indicates that AI chatbots can serve as an assistant to help teachers create dynamic assessments for each student, thereby reducing teachers' workload, burden, and pressure (R. Wu & Yu, 2024). The literature indicates that AI chatbot administrative support capabilities can help educators save time on routine tasks, including scheduling, grading, and providing information to students, allowing them to allocate more time for instructional planning and student engagement (Labadze et al., 2023). Research shows that AI enables users to outsource mundane tasks, allowing them to devote more time to critical and creative thinking. Students utilise AI tools to save time on routine tasks (Essien et al., 2024). The literature indicates that prior research has highlighted the positive impacts of educational chatbots, particularly in terms of cost-effectiveness and improved learning experiences and outcomes, with chatbots providing on-demand support in online classes (D.-L. Chen et al., 2025). However, not all lecturers reported significant reductions, with the timing of deployment potentially affecting observed impacts. Research shows that the duration of AI interventions moderates their effectiveness, with medium-duration interventions being most effective. At the same time, shorter and longer durations exhibited smaller effects, showing that very brief interventions may be hindered by initial unfamiliarity, while extended interventions might suffer from diminishing returns (R. Wu & Yu, 2024; Yi et al., 2025). The literature indicates that interventions lasting between one week and two months show larger moderate effects. In comparison, shorter (less than one week) and longer (over two months) durations exhibit smaller yet still significant effects, suggesting that the timing of deployment can affect observed impacts (Yi et al., 2025). This addresses the pre-implementation challenge, where lecturers noted that students repeatedly ask similar questions, creating a significant workload.

### E.3.3 Differences Between Stronger and Weaker Students

Lecturers observed distinct patterns in how stronger versus weaker students engaged with StudyBuddy. Stronger students demonstrated strategic usage, with lecturers noting that "stronger students targeted tough items and moved faster" and that "stronger ones verified responses," showing strategic use as a learning enhancer rather than dependency. Research shows that students' engagement with AI technology varies through different academic stages, with higher-level students likely to use more sophisticated AI tools tailored for data analysis, problem-solving, etc., which could be a consequence of the increasing complexity of academic work and greater familiarity with AI tools among senior students (Fošner, 2024). The literature suggests that students should carefully verify the accuracy of AI-generated responses, allowing them to request problems of increasing difficulty levels to aid their learning (Akolekar et al., 2025). Weaker students showed different patterns, with lecturers observing that "weaker ones followed steps" and that "weaker students will benefit from the more fundamental knowledge facts and simpler concept questions." Some lecturers noted that "weaker learners benefited more, they used it nightly," showing intensive engagement. Research shows that AI tools present personalised and immersive learning to students by creating learning objectives with adjustable complexity. Feedback is provided instantly, with AI assisting students at various skill levels, offering resources tailored to their level of knowledge, thereby enabling deeper engagement and subject mastery (Yaseen et al., 2025). The literature indicates that interactive AI tools adapt their instruction based on student responses, presenting tasks of varying complexity while also offering corresponding resources to support students at their respective level of knowledge (Yaseen et al., 2025). However, one lecturer expressed concern that "weaker ones followed steps" without the critical verification shown by stronger students. Research shows that while AI tools provide substantial external support, they might inadvertently reduce learners' reliance on self-regulation strategies, with the influence of AI on learners' use of learning strategies and self-regulated learning yielding a negative effect, showing that overly reliant learners may not develop strong self-regulation skills (X.-Y. Wu, 2024). Confidence gains were observed among weaker students, with multiple lecturers noting that "the weaker group gained confidence" and "asked better questions." Research indicates that AI tools can alleviate language learning-related anxiety by providing a supportive and non-judgmental environment for practice. Studies have found that learners using AI applications report lower levels of anxiety and higher levels of motivation (X.-Y. Wu, 2024). The literature indicates that the significant increase in learners' willingness to communicate in the target language highlights the potential of AI tools to boost confidence and communicative competence, with AI-driven applications often including interactive dialogues and real-time feedback mechanisms that simulate real-life communication scenarios, thereby providing learners with ample practice opportunities that help to build confidence (X.-Y. Wu, 2024). Research shows that students feel less afraid of making mistakes when interacting with chatbots, which makes them more confident, with students reporting that they pushed themselves to say one sentence correctly at once, but now allow themselves to think about it for a second or make some mistakes, recognising that making mistakes is not shameful (Y. Yuan, 2024). This aligns with pre-implementation expectations that weaker students would benefit from non-judgmental learning spaces and self-paced learning.